\documentclass[12pt]{article}
\newcommand{\topstar}[1]{\setlength{\unitlength}{1mm}
\begin{picture}(2,0)(-1,-1.4)
   \put(0,0){\makebox(0,0){$#1$}}
   \put(0,2.4){\makebox(0,0){\mbox{\tiny$\star$}}}
\end{picture}}

\newcommand{\eqa}{\begin{eqnarray}}
\newcommand{\ena}{\end{eqnarray}}
\renewcommand{\thefootnote}{\fnsymbol{footnote}}
\usepackage{epsfig}
\usepackage{tocloft}
\begin{document}
\begin{center}
{\Huge\bf A Non-Metric Approach to Space,\\\ \\ Time and Gravitation}
\end{center} 
{\vspace*{4cm}}
\begin{center}
\bf by
\end{center}
{\vspace*{0.5cm}}
\begin{center}
\bf Dag {\O}stvang
\end{center}
{\vspace*{2cm}}
\begin{center}
{\Large\bf Thesis submitted for the degree of \\\ \\
DOCTOR SCIENTIARUM}
\end{center}
{\vspace*{2.5cm}}
\begin{center}\bf
NORWEGIAN UNIVERSITY OF SCIENCE \\ AND TECHNOLOGY, TRONDHEIM
\end{center}
{\vspace*{0.5cm}}
\begin{center}\bf
Faculty of Physics, Informatics and Mathematics \\
Department of Physics
\end{center}
{\vspace*{5mm}}
\begin{center}
\bf 2001
\end{center}

\newpage                                        
\thispagestyle{empty}
\vspace*{5cm}
\newpage
\thispagestyle{empty}
\begin{center}
{\Large {\bf Forord}}
\end{center}
$\\ [2mm]$

Dette arbeidet har blitt utf{\o}rt ved Institutt for fysikk, NTNU med professor
K{\aa}re Olaussen som hovedveileder og professor Kjell Mork som formell 
veileder. Avhandlingen er resultatet av flere {\aa}rs utvikling fra noe 
som startet som temmelig vage, kvalitative ideer. Veien fram til en 
konkretisering av disse ideene har v{\ae}rt temmelig t{\o}ff og blindsporene 
har v{\ae}rt mange. Sikkert er det i alle fall at uten st{\o}tte fra min 
faglige veileder professor K{\aa}re Olaussen, som kom til unnsetning da 
utsiktene var som m{\o}rkest, ville dette arbeidet aldri ha blitt fullf{\o}rt. 
Jeg synes det var modig gjort av ham {\aa} satse tid og prestisje p{\aa} noe 
som kunne fortone seg som temmelig usikkert, og jeg er ham stor takk skyldig. 

Ellers har en rekke personer vist interesse for prosjektet. Av disse vil 
jeg spesielt nevne Amir Ghaderi, Terje R. Meisler og Tommy {\O}verg{\aa}rd.

Det skal dessuten nevnes at professor Steve Carlip har v{\ae}rt s{\aa} vennlig 
{\aa} komme med detaljkritikk via e-post. Jeg vil ogs{\aa} takke mine foreldre 
for moralsk st{\o}tte.

{\vspace*{1cm}}
\begin{center}
Trondheim, april 2001
\end{center}
{\vspace*{5mm}}

\begin{center}
Dag {\O}stvang
\end{center}

\newpage
\thispagestyle{empty}
\vspace*{5cm}
\newpage
\thispagestyle{empty}
\vspace*{-2cm}
\addcontentsline{toc}{section}
\protect\numberline{}
{\protect\textbf{{\hspace*{-0.3cm}}Notations}}
{\protect{ \hspace*{12.99cm}{\bf 1}}}
\addtocontents{toc}{\protect\vspace*{5ex}} \\
\vspace*{0.5cm}
\addcontentsline{toc}{section}
\protect\numberline{}
{\protect\textbf{{\hspace*{-7.8cm}}Prologue}}
{\protect{\hspace*{13.4cm}{\bf 5}}} \\
\vspace*{0.1cm}
\\
\addcontentsline{toc}{section}
\protect\numberline{}
{\protect\textbf{{\hspace*{-7.65cm}}Preliminaries: Some Basic Mathematical 
Concepts}}{\protect{\hspace*{4.98cm}{\bf 9}}} \\
\addcontentsline{toc}{section}
\protect\numberline{}
{\protect\rm{{\hspace*{-7.8cm}}
1{\ }{\ }{\ }{\ }{\ }Introduction . . . . . . . . . . . . . . . . . . . . .
. . . . . . . . . . . . . . . . . .{\ }{\ }{\,}{\,}}}{\protect 11} \\
\addcontentsline{toc}{section}
\protect\numberline{}
{\protect\rm{{\hspace*{-7.8cm}}
2{\,}{\ }{\ }{\ }{\ }{\ }From 3-geometry to 4-geometry{\,} 
. . . . . . . . . . . . . . . . . . . . . . . . . . .{\ }{\ }{\,}}
{\protect 11} \\
\addcontentsline{toc}{section}
\protect\numberline{}
{\protect\rm{{\hspace*{-7.8cm}}
3{\ }{\ }{\ }{\ }{\ }Connection and curvature
. . . . . . . . . . . . . . . . . . . . . . . . . . . . . . .{\ }{\ }}
{\protect 14} \\
\addcontentsline{toc}{section}
\protect\numberline{}
{\protect\rm{{\hspace*{-7.8cm}}
4{\ }{\ }{\ }{\ }{\ }Two simple examples{\hspace*{-0.43mm}} . . . . . . .
. . . . . . . . . . . . . . . . . . . . . . . . . . .{\ }{\ }{\,}}
{\protect 15} \\
\addcontentsline{toc}{section}
\protect\numberline{}
{\protect\rm{{\hspace*{-7.8cm}}
{\ }{\ }{\ }{\ }{\ }{\ }References{\,} . . . . . . . . . . . 
. . . . . . . . . . . . . . . . . . . . . . . . . . . . .{\ }{\ }}
{\protect 16} \\
\addtocontents{toc}{\protect\vspace*{1cm}}
\vspace*{0.5cm}
\\
\addcontentsline{toc}{section}
\protect\numberline{}
{\protect\textbf{{\hspace*{-7.75cm}}Chapter One: A Non-Metric Approach to
Space, Time and Gravitation{\ }}}{\protect{\hspace*{0.55cm}{\bf 17}}} \\
\addcontentsline{toc}{section}
\protect\numberline{}
{\protect\rm{{\hspace*{-7.8cm}}
1{\ }{\ }{\ }{\ }{\ }Introduction . . . . . . . . . . . . . . . . . . . . .
. . . . . . . . . . . . . . . . . .{\ }{\ }{\,}{\,}}}{\protect 19} \\
\addcontentsline{toc}{section}
\protect\numberline{}
{\protect\rm{{\hspace*{-7.8cm}}
2{\ }{\ }{\ }{\ }{\ }An alternative framework of space and time{\,}{\,} 
. . . . . . . . . . . . . . . . . . .{\ }{\ }{\,}}
{\protect 23} \\
\addcontentsline{toc}{section}
\protect\numberline{}
{\protect\rm{{\hspace*{-7.8cm}}
3{\ }{\ }{\ }{\ }{\ }The role of gravity
. . . . . . . . . . . . . . . . . . . . . . . . . . . . . . . . . . .{\ }{\ }}
{\protect 28} \\
\addcontentsline{toc}{section}
\protect\numberline{}
{\protect\rm{{\hspace*{-7.77cm}}
4{\ }{\ }{\ }{\ }{\ }{\,}Kinematics and non-kinematics of the FHSs{\,}
. . . . . . . . . . . . . . . . . . . .{\ }{\ }{\,}}
{\protect 36} \\
\addcontentsline{toc}{section}
\protect\numberline{}
{\protect\rm{{\hspace*{-7.8cm}}
5{\ }{\ }{\ }{\ }{\ }Test particle motion revisited{\,}{\,}
. . . . . . . . . . . . . . . . . . . . . . . . . . . .{\ }{\ }}
{\protect 39} \\
\addcontentsline{toc}{section}
\protect\numberline{}
{\protect\rm{{\hspace*{-7.8cm}}
6{\ }{\ }{\ }{\ }{\ }Discussion{\,}{\,}{\,}{\,} . . . . . . . . . . . .
. . . . . . . . . . . . . . . . . . . . . . . . . . .{\ }{\ }{\,}}
{\protect 41} \\
\addcontentsline{toc}{section}
\protect\numberline{}
{\protect\rm{{\hspace*{-7.8cm}}
{\,}{\ }{\ }{\ }{\ }{\ }{\ }References . . . . . . . . . 
. . . . . . . . . . . . . . . . . . . . . . . . . . . . . . .{\ }{\ }}
{\protect 42} \\
\addtocontents{toc}{\protect\vspace*{1cm}}
\vspace*{0.5cm}
\\
\addcontentsline{toc}{section}
\protect\numberline{}
{\protect\textbf{{\hspace*{-7.75cm}}Chapter Two: On the Non-Kinematical
Evolution of the 3-geometry{\ }}}{\protect{\hspace*{0.95cm}{\bf 43}}} \\
\addcontentsline{toc}{section}
\protect\numberline{}
{\protect\rm{{\hspace*{-7.8cm}}
1{\ }{\ }{\ }{\ }{\ }Introduction . . . . . . . . . . . . . . . . . . . . .
. . . . . . . . . . . . . . . . . .{\ }{\ }{\,}{\,}}}{\protect 45} \\
\addcontentsline{toc}{section}
\protect\numberline{}
{\protect\rm{{\hspace*{-7.8cm}}
2{\ }{\ }{\ }{\ }{\ }The quasi-metric framework{\,}{\,} 
. . . . . . . . . . . . . . . . . . . . . . . . . . . . .{\ }{\ }{\,}}
{\protect 46} \\
\addcontentsline{toc}{section}
\protect\numberline{}
{\protect\rm{{\hspace*{-7.8cm}}
{\,}{\,}{\ }{\ }{\ }{\ }{\ }{\ }2.1{\ }{\ }{\ }{\ }{\ }Uniqueness and 
existence of the global time function
. . . . . . . . . . .{\ }{\ }}
{\protect 54} \\
\addcontentsline{toc}{section}
\protect\numberline{}
{\protect\rm{{\hspace*{-7.8cm}}
{\,}{\ }{\ }{\ }{\ }{\ }{\ }2.2{\ }{\ }{\ }{\ }{\ }The question of general 
covariance. . . . . . . . . . . . . . . . . . . . . .{\ }{\ }}
{\protect 57} \\
\addcontentsline{toc}{section}
\protect\numberline{}
{\protect\rm{{\hspace*{-7.8cm}}
{\,}{\ }{\ }{\ }{\ }{\ }{\ }2.3{\ }{\ }{\ }{\ }{\ }The metric approximation 
{\,}{\,}. . . . . . . . . . . . . . . . . . . . . . . . . .{\ }{\ }}
{\protect 58} \\
\addcontentsline{toc}{section}
\protect\numberline{}
{\protect\rm{{\hspace*{-7.8cm}}
3{\ }{\ }{\ }{\ }{\ }Equations of evolution{\,}{\,}{\,}
. . . . . . . . . . . . . . . . . . . . . . . . . . . . . . . .{\ }{\ }}
{\protect 63} \\
\addcontentsline{toc}{section}
\protect\numberline{}
{\protect\rm{{\hspace*{-7.8cm}}
{\,}{\ }{\ }{\ }{\ }{\ }{\ }3.1{\ }{\ }{\ }{\ }{\ }Projections{\,} 
{\,}. . . . . . . . . . . . . . . . . . . . . . . . . . . . . . . . . . .
{\ }{\ }}{\protect 66} \\
\addcontentsline{toc}{section}
\protect\numberline{}
{\protect\rm{{\hspace*{-7.8cm}}
{\,}{\ }{\ }{\ }{\ }{\ }{\ }3.2{\ }{\ }{\ }{\ }{\ }The field equations{\,} 
{\,}{\,}. . . . . . . . . . . . . . . . . . . . . . . . . . . . . .{\ }{\ }}
{\protect 70} \\
\addcontentsline{toc}{section}
\protect\numberline{}
{\protect\rm{{\hspace*{-7.8cm}}
{\,}{\ }{\ }{\ }{\ }{\ }{\ }3.3{\ }{\ }{\ }{\ }{\ }The quasi-metric 
initial-value problem
{\,}. . . . . . . . . . . . . . . . . . .{\ }{\ }}
{\protect 77} \\
\addcontentsline{toc}{section}
\protect\numberline{}
{\protect\rm{{\hspace*{-7.8cm}}
{\,}{\ }{\ }{\ }{\ }{\ }{\ }3.4{\ }{\ }{\ }{\ }{\ }Equations of motion
{\,}. . . . . . . . . . . . . . . . . . . . . . . . . . . . . .{\ }{\ }}
{\protect 80} \\
\addcontentsline{toc}{section}
\protect\numberline{}
{\protect\rm{{\hspace*{-7.76cm}}
4{\ }{\ }{\ }{\ }{\ }Spherically symmetric vacua
. . . . . . . . . . . . . . . . . . . . . . . . . . . . .{\ }{\ }{\,}}
{\protect 82} \\
\addcontentsline{toc}{section}
\protect\numberline{}
{\protect\rm{{\hspace*{-7.76cm}}
{\,}{\ }{\ }{\ }{\ }{\ }{\ }4.1{\ }{\ }{\ }{\ }{\ }The conformally Minkowski 
case {\,}. . . . . . . . . . . . . . . . . . . . . . .{\ }{\ }}
{\protect 83} 
\newpage
\thispagestyle{empty}
{\vspace*{-2cm}
\addcontentsline{toc}{section}
\protect\numberline{}
{\protect\rm{{\hspace*{-0.29cm}}
{\ }{\ }{\ }{\ }{\ }{\ }4.2{\ }{\ }{\ }{\ }{\ }Testing the NKE-paradigm 
{\,}. . . . . . . . . . . . . . . . . . . . . . . . . .{\ }{\ }}
{\protect 86} \\
\addcontentsline{toc}{section}
\protect\numberline{}
{\protect\rm{{\hspace*{-7.85cm}}
5{\ }{\ }{\ }{\ }{\ }Conclusion{\,}{\,}. . . . . . . . . . . .
. . . . . . . . . . . . . . . . . . . . . . . . . . . .{\ }{\ }}
{\protect 92} \\
\addcontentsline{toc}{section}
\protect\numberline{}
{\protect\rm{{\hspace*{-7.76cm}}
{\ }{\ }{\ }{\ }{\ }{\ }{\ }References . . . . . . . . .
. . . . . . . . . . . . . . . . . . . . . . . . . . . . . . . .{\ }{\ }}
{\protect 93} \\
\addcontentsline{toc}{section}
\protect\numberline{}
{\protect\rm{{\hspace*{-7.85cm}}
{\,}{\,}{\ }{\ }{\ }{\ }{\ }{\ }Appendix A: Derivation of a static line
element . . . . . . . . . . . . . . . . . .{\ }{\ }}
{\protect 93} \\
\addcontentsline{toc}{section}
\protect\numberline{}
{\protect\rm{{\hspace*{-7.85cm}}
{\,}{\ }{\ }{\ }{\ }{\ }{\ }Appendix B: The Newtonian limit . . . . . . . 
. . . . . . . . . . . . . . . . . . .{\ }{\ }}
{\protect 95} \\
\addcontentsline{toc}{section}
\protect\numberline{}
{\protect\rm{{\hspace*{-7.85cm}}
{\,}{\,}{\ }{\ }{\ }{\ }{\ }{\ }Appendix C: Isotropic cosmological models with 
matter . . . . . . . . . . . . .{\ }{\ }}
{\protect 96} \\
\addcontentsline{toc}{section}
\protect\numberline{}
{\protect\rm{{\hspace*{-7.8cm}}
{\,}{\,}{\ }{\ }{\ }{\ }{\ }{\ }Appendix D: Symbols and acronyms . . . . . . 
. . . . . . . . . . . . . . . . . . .{\ }{\ }}
{\protect 98} \\
\addtocontents{toc}{\protect\vspace*{1cm}}
\vspace*{0.5cm}
\\
\addcontentsline{toc}{section}
\protect\numberline{}
{\protect\textbf{{\hspace*{-7.75cm}}Chapter Three: The Gravitational Field
outside a Spherically \\ {\ }{\ }{\ }{\ }{\ }{\ }{\hspace*{1.75mm}}
Symmetric, Isolated Source in a Quasi-Metric Theory of Gravity{\ }}}
{\protect{\hspace*{1.2cm}{\bf 101}}} \\
\addcontentsline{toc}{section}
\protect\numberline{}
{\protect\rm{{\hspace*{-7.8cm}}
1{\ }{\ }{\ }{\ }{\ }Introduction . . . . . . . . . . . . . . . . . . . . .
. . . . . . . . . . . . . . . . . .{\ }{\ }{\,}{\,}}}{\protect 103} \\
\addcontentsline{toc}{section}
\protect\numberline{}
{\protect\rm{{\hspace*{-7.83cm}}
2{\ }{\ }{\ }{\ }{\,}General equations of motion{\,}{\,}{\,} 
. . . . . . . . . . . . . . . . . . . . . . . . . . . .{\ }{\ }{\,}}
{\protect 104} \\
\addcontentsline{toc}{section}
\protect\numberline{}
{\protect\rm{{\hspace*{-7.83cm}}
{\,}{\ }{\ }{\ }{\ }{\ }{\ }2.1{\ }{\ }{\ }{\ }{\ }Special equations of
motion{\,}{\,}{\,} . . . . . . . . . . . . . . . . . . . . . . . .{\ }{\ }{\,}}
{\protect 107} \\
\addcontentsline{toc}{section}
\protect\numberline{}
{\protect\rm{{\hspace*{-7.8cm}}
3{\ }{\ }{\ }{\ }{\ }The dynamical problem
. . . . . . . . . . . . . . . . . . . . . . . . . . . . . . . .{\ }{\ }}
{\protect 110} \\
\addcontentsline{toc}{section}
\protect\numberline{}
{\protect\rm{{\hspace*{-7.8cm}}
4{\ }{\ }{\ }{\ }{\ }Discussion{\,}{\,} . . . . . . . . .
. . . . . . . . . . . . . . . . . . . . . . . . . . . . . .{\ }{\ }{\,}}
{\protect 113} \\
\addcontentsline{toc}{section}
\protect\numberline{}
{\protect\rm{{\hspace*{-7.8cm}}
{\,}{\,}{\ }{\ }{\ }{\ }{\ }{\ }4.1{\ }{\ }{\ }{\ }{\ }Expanding space and 
the solar system . . . . . . . . . . . . . . . . . . .{\ }{\ }}
{\protect 114} \\
\addcontentsline{toc}{section}
\protect\numberline{}
{\protect\rm{{\hspace*{-7.8cm}}
{\,}{\,}{\ }{\ }{\ }{\ }{\ }{\ }References{\hspace*{-0.1mm}} . . . . . . . . . 
. . . . . . . . . . . . . . . . . . . . . . . . . . . . . . .{\ }{\ }}
{\protect 118} \\
\addtocontents{toc}{\protect\vspace*{1cm}}
\vspace*{0.5cm}
\\
\addcontentsline{toc}{section}
\protect\numberline{}
{\protect\textbf{{\hspace*{-7.75cm}}Chapter Four: An Assumption of a Static
Gravitational Field \\ {\ }{\ }{\ }{\ }{\ }{\ }{\hspace*{1.75mm}}
Resulting in an Apparently Anomalous Force{\ }}}
{\protect{\hspace*{5.5cm}{\bf 119}}} \\
\addcontentsline{toc}{section}
\protect\numberline{}
{\protect\rm{{\hspace*{-7.8cm}}
1{\ }{\ }{\ }{\ }{\ }Introduction . . . . . . . . . . . . . . . . . . . . .
. . . . . . . . . . . . . . . . . .{\ }{\ }{\,}{\,}}}{\protect 121} \\
\addcontentsline{toc}{section}
\protect\numberline{}
{\protect\rm{{\hspace*{-7.83cm}}
2{\ }{\ }{\ }{\ }{\,}{\,}A quasi-metric model{\,}{\,}{\,} . . . .
. . . . . . . . . . . . . . . . . . . . . . . . . . . .{\ }{\ }{\,}}
{\protect 123} \\
\addcontentsline{toc}{section}
\protect\numberline{}
{\protect\rm{{\hspace*{-7.83cm}}
{\,}{\ }{\ }{\ }{\ }{\ }{\ }2.1{\ }{\ }{\ }{\ }{\ }Cosmic expansion and the
PPN-formalism{\,}{\,}{\,} . . . . . . . . . . . . . . . .{\ }{\ }{\,}}
{\protect 128} \\
\addcontentsline{toc}{section}
\protect\numberline{}
{\protect\rm{{\hspace*{-7.8cm}}
3{\ }{\ }{\ }{\ }Conclusion{\,}{\,} . . . . . . .
. . . . . . . . . . . . . . . . . . . . . . . . . . . . . . . .{\ }{\ }}
{\protect 130} \\
\addcontentsline{toc}{section}
\protect\numberline{}
{\protect\rm{{\hspace*{-7.8cm}}
{\ }{\ }{\ }{\ }{\ }{\ }References{\,} . . . . . . . . .
. . . . . . . . . . . . . . . . . . . . . . . . . . . . . . .{\ }{\ }}
{\protect 130} \\
\addtocontents{toc}{\protect\vspace*{1cm}}
\vspace*{0.5cm}
\\
\addcontentsline{toc}{section}
\protect\numberline{}
{\protect\textbf{{\hspace*{-7.75cm}}Chapter Five: On the Non-Kinematical
Expansion of \\ {\ }{\ }{\ }{\ }{\ }{\ }{\hspace*{1.25mm}}
Gravitationally Bound Bodies{\ }}}
{\protect{\hspace*{8.5cm}{\bf 131}}} \\
\addcontentsline{toc}{section}
\protect\numberline{}
{\protect\rm{{\hspace*{-7.8cm}}
1{\ }{\ }{\ }{\ }{\ }Introduction . . . . . . . . . . . . . . . . . . . . .
. . . . . . . . . . . . . . . . . .{\ }{\ }{\,}{\,}}}{\protect 133} \\
\addcontentsline{toc}{section}
\protect\numberline{}
{\protect\rm{{\hspace*{-7.8cm}}
2{\ }{\ }{\ }{\ }{\,}{\,}A brief survey of the quasi-metric theory . . . . 
. . . . . . . . . . . . . . . . . .{\ }{\ }}
{\protect 134} \\
\addcontentsline{toc}{section}
\protect\numberline{}
{\protect\rm{{\hspace*{-7.8cm}}
3{\ }{\ }{\ }{\ }{\ }Metrically static, spherically symmetric interiors
. . . . . . . . . . . . . . . . .{\ }{\ }}
{\protect 136} \\
\addcontentsline{toc}{section}
\protect\numberline{}
{\protect\rm{{\hspace*{-7.8cm}}
{\,}{\,}{\,}{\ }{\ }{\ }{\ }{\ }3.1{\ }{\ }{\ }{\ }{\,}Analytical calculations 
and a numerical recipe . . . . . . . . . . . . . . .{\ }{\ }}
{\protect 137} \\
\addcontentsline{toc}{section}
\protect\numberline{}
{\protect\rm{{\hspace*{-7.8cm}}
4{\ }{\ }{\ }{\ }{\,}Discussion{\,}{\,} . . . . . . . . .
. . . . . . . . . . . . . . . . . . . . . . . . . . . . . .{\ }{\ }{\,}}
{\protect 140} \\
\addcontentsline{toc}{section}
\protect\numberline{}
{\protect\rm{{\hspace*{-7.8cm}}
{\ }{\ }{\ }{\ }{\ }{\ }References{\,}{\,} . . . . . . . .
. . . . . . . . . . . . . . . . . . . . . . . . . . . . . . .{\ }{\ }}
{\protect 142} 

\newpage

\setcounter{page}{1}

\section*{Notations}
This is a table of symbols and acronyms used in the thesis. Sign conventions 
and notations are as in the classic book {\em Gravitation} by C.W. Misner, 
K.S. Thorne and J.A. Wheeler if otherwise not stated. In particular,
Greek indices may take integer values $0-3$ and Latin indices may take 
integer values $1-3$.

\begin{table}[h]
\begin{tabular}{lp{11.1cm}}
{\bf SYMBOL} & {\bf NAME/EXPLANATION} \\ 
KE/NKE & kinematical/non-kinematical evolution (of the FHSs) \\
FOs/FHSs & fundamental observers/fundamental hypersurfaces \\
NKR & non-kinematical redshift \\
GTCS & global time coordinate system \\
HOCS & hypersurface-orthogonal coordinate system \\
WEP/EEP & weak/Einstein equivalence principle \\
GWEP/SEP & gravitational weak/strong equivalence principle \\
LLI/LPI & local Lorentz/position invariance \\
SR/GR & special/general relativity \\
SC & Schiff's conjecture \\
$t=t(x^{\mu})$ & global time function, conventionally we define $t=x^{0}/c$
when using a GTCS (where $x^0$ is a global time coordinate) \\
${\cal M}$/${\overline{\cal M}}$ & metric space-time manifold obtained by 
holding $t$ constant \\
${\cal N}$ & quasi-metric space-time manifold $({\cal N},{\bf g}_t)$ \\
${\overline{\cal N}}$ & shorthand notation for $({\cal N},{\bf {\bar g}}_t)$ \\
${\bf g}_t$ or $g_{(t){\mu}{\nu}}$ & one-parameter family of
space-time metric tensors \\
${\bf {\bar g}}_t$ or ${\bar g}_{(t){\mu}{\nu}}$ & auxiliary one-parameter family 
of space-time metric tensors (${\bf {\bar g}}_t$ is found as a solution of the 
field equations) \\
${\bf n}_t$ or $n_{(t)}^{\mu}$ & unit vector field family normal to the FHSs 
in ${\cal N}$ \\
${\bf {\bar n}}_t$ or ${\bar n}_{(t)}^{\mu}$ & unit vector field family normal to 
the FHSs in ${\overline{\cal N}}$ \\
${\bf h}_t$ or $h_{(t)ij}$ & metric tensor family intrinsic to the FHSs 
in ${\cal N}$ \\
${\bf {\hat h}}_t$ or ${\hat h}_{(t)ij}$ & defined as 
${\frac{t_0^2}{t^2}}{\bf h}_t$ \\
${\bf {\bar h}}_t$ or ${\bar h}_{(t)ij}$ & metric tensor family intrinsic to 
the FHSs in ${\overline{\cal N}}$ \\
${\bf {\tilde h}}_t$ or ${\tilde h}_{(t)ij}$ & defined as 
${\frac{t_0^2}{t^2}}{\bar N}_t^{-2}{\bf {\bar h}}_t$ \\
$N(x^{\mu})$ & lapse function field in ${\cal N}$ \\
${\bar N}_t(x^{\mu},t)$ & lapse function field family in 
${\overline{\cal N}}$ \\
\end{tabular}
\end{table}
\clearpage
\begin{table}[t]
\begin{tabular}{lp{10.6cm}}
${\frac{t_0}{t}}N^j_{(t)}(x^{\mu},t)$ & components of the shift vector field 
family in ${\cal N}$ \\
${\frac{t_0}{t}}{\bar N}^j_{(t)}(x^{\mu},t)$ & components of the shift vector 
field family in ${\overline{\cal N}}$ \\
${\Gamma}^{\alpha}_{(t){\beta}{\gamma}}$ & components of the family of metric 
connections ${\nabla}_{t}$ compatible with ${\bf g}_{t}$ \\
${\bar {\Gamma}}^{\alpha}_{(t){\beta}{\gamma}}$ & components of the family of metric 
connections ${\bar {\nabla}}_{t}$ compatible with ${\bf {\bar g}}_{t}$ \\
${\topstar{\Gamma}}^{{\,}{\alpha}}_{t{\gamma}}$,
${\topstar{\Gamma}}^{{\,}{\alpha}}_{{\beta}{\gamma}}$ & components of the
5-dimensional connection ${\topstar{\nabla}}$ associated with the
family ${\bf g}_{t}$ \\
${\topstar{\bar {\Gamma}}}^{{\,}{\alpha}}_{t{\gamma}}$,
${\topstar{\bar {\Gamma}}}^{{\,}{\alpha}}_{{\beta}{\gamma}}$ & components of the 
5-dimensional connection ${\stackrel{\star}{\bf {\bar {\nabla}}}}$ 
associated with the family ${\bf {\bar g}}_{t}$ \\
${\bf a}_t$ or ${a}_{(t)}^{\mu}$ & family of metric 4-accelerations in $\cal M$
(any observer) \\
$()_{{\bar *}{\alpha}}$ & coordinate expression for a covariant derivative 
obtained from ${\stackrel{\star}{\bf {\bar {\nabla}}}}$ and
compatible with ${\bf {\bar g}}_t$ \\
$()_{{\mid}j}$ & coordinate expression for a spatial covariant derivative 
compatible with ${\bf h}_t$ or ${\bf {\bar h}}_t$ as appropriate 
(holding $t$ constant) \\
$()_{;{\alpha}}$ & coordinate expression for a metric covariant derivative 
compatible with ${\bf g}_t$ or ${\bf {\bar g}}_t$ as appropriate 
(holding $t$ constant) \\
${\perp}$/${\bar {\perp}}$ & projection symbols (projection 
onto the normal direction to the FHSs) \\
{\pounds}$_{{\bf y}_t}$/{\pounds}$_{{\bf {\bar y}}_t}$ & Lie derivative with respect 
to ${\bf y}_t/{\bf {\bar y}}_t$ in ${\cal M}/{\overline{\cal M}}$ \\
${\topstar{\pounds}}_{{\,}{\bf y}}/{\,}{\topstar{\pounds}}_{{\,}{\bf {\bar y}}}$ & Lie 
derivative with respect to ${\bf y}/{\bf {\bar y}}$ in 
${\cal N}/{\overline{\cal N}}$ \\
${\cal L}_{{\bf y}_t}/{\cal L}_{{\bf {\bar y}}_t}$ & projected Lie derivative with 
respect to ${\bf y}_t/{\bf {\bar y}}_t$ in ${\cal M}/{\overline{\cal M}}$ \\
${\topstar{\cal L}}_{{\,}{\bf y}}/{\,}{\topstar{\cal L}}_{{\,}{\bf {\bar y}}}$ 
& projected Lie derivative with respect to 
${\bf y}/{\bf {\bar y}}$ in ${\cal N}/{\overline{\cal N}}$ \\
${\bf a}_{\cal F}$/${\bf {\bar a}}_{\cal F}$ or 
$a^i_{\cal F}$/${\bar a}^i_{\cal F}$ & four-acceleration of the FOs
in $\cal M$/${\overline{\cal M}}$ \\
${\bf {\bar x}}_{\cal F}$ or ${\bar x}^i_{\cal F}$ & ``local distance vector from 
the centre of gravity'' generalized from the spherically symmetric case \\
${\bf {\bar e}}_{\cal F}$/${\bf {\bar {\omega}}}_{\cal F}$ or 
${\frac{t_0}{t}}{\bar e}_{\cal F}^i$/${\frac{t}{t_0}}{\bar {\omega}}_{{\cal F}i}$ &
unit 3-vector/covector field in the ${\bf {\bar x}}_{\cal F}$-direction \\
${\bf w}_t/{\bf {\bar w}}_t$ or $w^i_{(t)}/{\bar w}^i_{(t)}$ & 3-velocity family 
of test particles (or fluid sources) with respect to the local FOs in 
${\cal M}/{\overline {\cal M}}$ \\
${\bf u}_t$ or $u^{\mu}_{(t)}$ & 4-velocity family 
(of test particles) in $\cal M$ \\
${\bf {\bar u}}_t$ or ${\bar u}^{\mu}_{(t)}$ & 4-velocity family
(of test particles or fluid sources) in ${\overline{\cal M}}$ \\
${\bf v}_t$ or $v^{\mu}_{(t)}$ & 3-vector field family determining 
norm-preserving transformations ${\bf {\bar Y}}_t{\rightarrow}{\bf Y}_t$ of 
tensor field families (any rank) \\
${\stackrel{\star}{\bf a}}$ or ${\topstar{a}}^{\mu}$ & degenerate 
4-acceleration in $\cal N$ (any observer) \\
$m_t/M_t$ {\hspace*{2.6cm}} & active masses (scalar fields measured 
dynamically) \\
\end{tabular}
\end{table}
\newpage
\begin{table}[t]
\begin{tabular}{lp{9.5cm}}
${\bf {\bar R}}_t$/${\bf {\bar G}}_t$ or ${\bar R}_{(t){\mu}{\nu}}$/
${\bar G}_{(t){\mu}{\nu}}$ & Ricci/Einstein tensor family in 
${\overline{\cal M}}$ \\
${\bf {\bar C}}_t$ or ${\bar C}_{(t){\alpha}{\beta}{\mu}{\nu}}$ & Weyl tensor family 
in ${\overline{\cal M}}$ \\
${\bf {\bar H}}_t$/${\bf {\tilde H}}_t$ 
or ${\bar H}_{(t)ij}$/${\tilde H}_{(t)ij}$ & Einstein tensor families intrinsic 
to the FHSs in ${\overline{\cal M}}$ \\
${\bf {\bar P}}_t$ or ${\bar P}_{(t)ij}$ & Ricci tensor family intrinsic to the 
FHSs in ${\overline{\cal M}}$ \\
${\bar P}_t$ & Ricci scalar family intrinsic to the FHSs in
${\overline{\cal M}}$ \\
${\bf {\bar K}}_t$ or ${\bar K}_{(t)ij}$ & extrinsic curvature 
tensor family of the FHSs in ${\overline{\cal M}}$ \\
${\bf {\bar Q}}_t$ or ${\bar Q}_{(t){\mu}{\nu}}$ & family of foliation-defined 
gravitational tensors in ${\overline{\cal M}}$ \\
${\bf {\hat Z}}_t$ & the space-time tensor family ${\bf Z}_t$ fully projected 
onto the FHSs (several different projections are possible) \\
${\bar H}_t$/${\bar y}_t$ & global+local/local measure of the NKE \\
${\bar x}_t$ & measure of the KE \\
$d{\tau}_t$/$\overline{d{\tau}}_t$ & proper time interval (any observer) in 
$\cal N$/${\overline{\cal N}}$ \\
$d{\tau}_{\cal F}$/$\overline{d{\tau}}_{\cal F}$ {\hspace*{2.3cm}} & proper 
time interval for a FO in $\cal N$/${\overline{\cal N}}$ \\
${\bf T}_t$ or $T^{{\mu}{\nu}}_{(t)}$ {\hspace*{2.3cm}} & 
total stress-energy tensor family as an active source of gravitation (in 
${\overline{\cal M}}$) \\
${\cal T}_t$ or ${\cal T}^{{\mu}{\nu}}_{(t)}$ {\hspace*{2.3cm}} & 
total passive stress-energy tensor family (in ${\cal M}$) \\
${\bf T}^{\rm (EM)}_t$ or $T^{{\rm (EM)}{\mu}{\nu}}_{(t)}$ {\hspace*{2.3cm}} & 
electromagnetic stress-energy tensor family as an active source of gravitation 
(in ${\overline{\cal M}}$) \\
${\bf T}^{\rm (MA)}_t$ or $T^{{\rm (MA)}{\mu}{\nu}}_{(t)}$ {\hspace*{2.3cm}} & 
stress-energy tensor family for a fluid of material particles as an active 
source of gravitation (in ${\overline{\cal M}}$) \\
$G^{\rm S}/{\kappa}^{\rm S}{\equiv}{\frac{8{\pi}G^{\rm S}}{c^4}}
$ {\hspace*{1.3cm}} & gravitational coupling 
``constant'' for material particles \\
$G^{\rm B}/{\kappa}^{\rm B}{\equiv}{\frac{8{\pi}G^{\rm B}}{c^4}}
$ {\hspace*{1.3cm}} & gravitational coupling 
``constant'' for the electromagnetic field \\
$G^{\rm eff}$ {\hspace*{2.3cm}} & effective gravitational coupling ``constant''
measured in a given experiment \\
${\bar F}_t{\equiv}{\bar N}_tct$ {\hspace*{2.3cm}} & scale factor family 
of the FHSs in ${\overline{\cal N}}$ \\
${\Psi}_t{\equiv}{\bar F}_t^{-1}$ {\hspace*{2.3cm}} & a scalar field 
describing how atomic time units vary in space-time \\
${\varrho}_{\rm m}$ {\hspace*{2.3cm}} & passive mass density in local rest 
frame of the source \\
$p$ {\hspace*{2.3cm}} & passive pressure \\
${\tilde {\varrho}}_{\rm m}$ {\hspace*{2.3cm}} & active mass density in local 
rest frame of the source \\
${\tilde p}$ {\hspace*{2.3cm}} & active pressure \\
${\bar {\varrho}}_{\rm m}$ {\hspace*{2.3cm}} & properly scaled active mass
density defined as ${\frac{t^2}{t_0^2}}{\bar N}_t^2{\tilde {\varrho}}_{\rm m}$ \\
${\bar p}$ {\hspace*{2.3cm}} & properly scaled active pressure
defined as ${\frac{t^2}{t_0^2}}{\bar N}_t^2{\tilde p}$ \\
\end{tabular}
\end{table}
\clearpage
\hspace*{1cm}

\newpage

\setcounter{page}{5}

\section*{Prologue}

$\\ [2mm]$
Observationally, the Hubble law in its most famous form may be found by 
measuring spectral shifts of ``nearby" galaxies and thereby inferring their 
motion; it relates the average recessional velocity $v{\ll}c$ of 
such galaxies to their distance $d$ via the equation
\eqa
v=Hd, {\nonumber}
\ena
where $H$ is the Hubble parameter. We see that the Hubble law does not depend 
on direction. This is merely a consequence of the fact that it is an empirical 
law; had the observations suggested the existence of anisotropic recessional 
velocities, the Hubble law could still be formulated but in an anisotropic 
version. 

The question now is if this apparent lack of anisotropy may follow from some 
hitherto undiscovered physical principle rather than being a consequence of 
some rather special cosmic initial conditions. Once one suspects this, it 
is natural to assume that the potential new fundamental law is local in nature.
If so, the Hubble law should be important not only for cosmological scales
but it should also be relevant for local gravitational scales. For this to make
sense, a local version of the Hubble law should follow as a natural consequence
of some general geometrical property of a relativistic space-time framework.

Since the Hubble parameter is a scalar and not a tensor (as it would have to 
be in an anisotropic version of the Hubble law), one may suspect that the 
Hubble parameter in fact may be expressed as a piece of the affine connection 
obtained from some kind of spatial {\em scale factor} somewhat similar to that 
present in the Robertson-Walker (RW) models in standard cosmology. Howevever, 
the particular geometrical structure of the RW-models is merely due to the high
symmetry present in these isotropic and homogeneous universe models. This means
that a spatial scale factor is not and cannot be any fundamental general 
constituent of any space-time geometrical framework based on a 
pseudo-Riemannian manifold. Contrary to this, a spatial scale factor must be a 
fundamental constituent of any alternative space-time geometrical framework 
where a local version of the Hubble law is required to hold in general. That
is, such a space-time framework should not be based on a pseudo-Riemannian 
manifold. Moreover, within such a framework, the {\em physical interpretation}
of the Hubble law is expected to differ from its interpretation in standard 
cosmology.

In standard cosmology, the Hubble law applies only to a particular set of 
observers associated with the smeared-out motion of the galaxies (i.e., said
observers are at rest with respect to the ``cosmic rest frame''). This 
suggests that a hypothetical local version of the Hubble law should also be 
associated with a privileged class of observers. At least parts of the 
interrelationship between nearby privileged observers should consist of an 
``expansion". For reasons explained in the thesis, this kind of expansion is 
called ``non-kinematical". Furthermore, to define the local version of the 
Hubble law, there must exist a ``preferred'' foliation of space-time into space
and time. That is, the one-parameter family of 3-dimensional spatial 
hypersurfaces defined by the privileged observers, taken at constant values of 
some privileged time coordinate, should play the role as a privileged notion of
``space''. This means that said privileged time coordinate then acts as a 
(unique) global notion of ``simultaneity'' and it should be a basic element of 
the alternative space-time framework.

It should be clear from the above, that any attempt to construct a local 
version of the Hubble law and implementing this into a general geometrical 
structure, in fact necessitates the construction of a new framework of space 
and time. We show in this thesis that this new framework disposes of the 
space-time metric as a global field. So as suggested above, the mathematical 
structure of the new framework is not based on pseudo-Riemannian geometry. Thus
not all aspects of metric theory will hold in the new framework. On the other 
hand, important physical principles, previously thought to hold only for metric
theory, also hold within the new framework. For this reason, we call the new 
framework ``quasi-metric". Since the quasi-metric framework must be 
relativistic, one may always construct a {\em local} space-time metric on the 
tangent space of each event by identifying the local inertial frames with local
Lorentz frames. But rather than demanding that the set of local metrics 
constitutes a global space-time metric field, we require the less stringent 
condition that the set of local metrics constitutes a semi-global metric field.
That is, the domain of validity for the semi-global metric field is not the 
entire space-time manifold, but rather a 3-dimensional submanifold defined by 
some constant value of the privileged time coordinate. The quasi-metric 
space-time manifold may then be thought of as consisting of a family of such 
submanifolds, each of being equipped with a semi-global space-time metric 
field. The crucial fact is, that the set of such semi-global space-time metric 
fields does not necessarily constitute a single global space-time 
(non-degenerate) metric field. The reason for this, is that the affine 
connection compatible with the set of semi-global metric fields depends 
directly on the existence of a privileged time coordinate. That is, unlike the 
Levi-Civita connection, the affine connection compatible with the set of 
semi-global metrics cannot in general be derived from any single space-time 
(non-degenerate) metric field.

In this thesis, it is shown that the absence of a global Lorentzian metric
field and the existence of a (unique) privileged time coordinate, does not make
it necessary to give up important physical principles such as, e.g., the 
various versions of the principle of equivalence (but the strong principle of 
equivalence will not hold in its most stringent form). On the other 
hand, the generalization of non-gravitational physical laws to curved 
space-time is more complicated for the quasi-metric framework than for the 
metric framework. This should not be surprising, since non-gravitational fields
may possibly couple to fields characterizing the quasi-metric geometrical
structure in a way not possible in metric geometry, but such that terms 
representing such couplings vanish in the local inertial frames. (That is, in 
the local inertial frames, the non-gravitational physics takes its standard 
special-relativistic form just as for metric theory.) 

The fact that there is a preferred notion of space and time via a preferred 
foliation of quasi-metric space-time into space and time means that this 
foliation should be determined from the field equations, i.e., it should be 
dynamical. Also, the field equations should have no gauge freedom in this 
context, i.e., said foliation must necessarily be determined rigidly.
Moreover, the uniqueness of such a foliation is assured if the associated
spatial hypersurfaces are {\em compact} (with positive curvature at 
cosmic scales). Besides, just as for standard cosmology, at cosmic scales
it is natural to assume that said privileged class of observers should be be at
rest, on average, with respect to the cosmic rest frame. This could then be 
interpreted as indicating the existence of a ``preferred frame'' being relevant
for the local gravitational dynamics of isolated systems. It turns out that 
this is indeed the case. Now, for small isolated systems and weak gravitational
fields one may, to good approximation, ignore the global curvature of space and
rather assume that space is asymptotically flat. In this approximation, the 
privileged class of observers would not be unique. That is, with this 
approximation, there would be no topological property of space-time that might 
pick out one particular such class over another. So, rather than the class of 
observers associated with the cosmic rest frame, one may select an alternative 
class of ``privileged'' observers being at rest with respect to the barycentre 
of some isolated system. One may then transform the field equations with 
respect to the alternative space-time foliation. However, the field equations 
will not be invariant under said transformation, since they do not represent 
one ``genuine'' space-time tensor field. Hence the transformed field equations 
would then depend on the system's velocity with respect to the cosmic rest 
frame. As a consequence, one expects that in quasi-metric gravity, there 
should be some ``preferred frame''-effects, but that such should be small (of 
post-Newtonian order) if the speed of said isolated system with respect to the 
cosmic rest frame is much smaller than the speed of light.
 
The uniqueness of the privileged time coordinate implies that there exists a 
{\em set} of ``preferred'' coordinate systems especially well adapted to the 
geometrical structure of quasi-metric space-time. This is a consequence of the 
fact that the preferred time coordinate represents an ``absolute'' geometrical 
element. However, the existence of non-dynamical fields is possible in metric 
theory also. Thus there should not be any {\em a priori} objections to 
constructing theories of gravity based on the quasi-metric framework. 
Consequently, in this thesis it is shown how to construct such a theory. This 
theory corresponds with General Relativity in particular situations and may 
possibly be viable. 

When it comes to comparison of testable models based on the quasi-metric 
theory to experiment, it still remains to develop a suitable weak-field 
expansion analogous to the parametrized post-Newtonian expansion valid for 
the metric framework. Obviously this is a subject for further work. However, 
even if a suitable weak-field expansion is missing, one may still try to 
construct specific models for idealized situations. In the thesis, this is done
for certain spherically symmetric systems and it is shown that the classical
solar system tests come in just as for General Relativity. But one important 
difference is that the quasi-metric theory predicts that the gravitational 
field of the solar system is expanding according to the Hubble law whereas 
General Relativity predicts no such thing. Moreover, gravitationally bound 
bodies made of ideal gas are predicted to expand in this manner also. As 
discussed in chapter 3, chapter 4 and chapter 5 of the thesis, observational 
evidence for expanding gravitational fields so far seems to favour the 
quasi-metric theory over General Relativity. Thus the observational evidence 
seems to indicate that the recessional velocities of galaxies are 
non-kinematical in nature and moreover that space is not flat at cosmological 
scales. If the observational evidence holds up, one must have in mind that the 
quasi-metric framework in general and the predictions of expanding 
gravitational fields and certain gravitationally bound bodies in particular, 
are based on a reinterpretation of the Hubble law.

It may be surprising that a reinterpretation of the simple empirical Hubble 
law leads to nothing less than the construction of a new geometrical framework
as the basis for relativistic physics. However, from a philosophical point of
view, one should prefer a theoretical framework having the property that 
general empirical laws follow from first principles. Thus seen, the existence 
of general empirical laws which do {\em not} follow from first principles, may
be interpreted as a sign of incompleteness for any theoretical framework and 
may potentially lead to the demise of that framework.

Now that the reader has had a foretaste of what this thesis is all about,
it is useful to get a overview of the mathematical concepts used in the
thesis before starting on its main parts. 

\newpage

{\vspace*{5cm}}

\begin{center}
{\huge {\bf Preliminaries: \\\ \\Some Basic Mathematical Concepts}}
\end{center}

\newpage
{\hspace*{1cm}}

\newpage

\begin{center}
{\large {\bf Some Basic Mathematical Concepts}}
\end{center}
\begin{center}
by
\end{center}
\begin{center}
Dag {\O}stvang \\
{\em Institutt for Fysikk, Norges teknisk-naturvitenskapelige universitet, 
NTNU \\
N-7491 Trondheim, Norway}
\end{center}
\begin{abstract}
The canonical viewpoint of space-time is taken as fundamental by requiring the
existence of a global time function $t$ corresponding to a global foliation of 
space-time into a set of spatial hypersurfaces. Moreover, it is required that
space-time can be foliated into a set of timelike curves corresponding to a
family of fundamental observers and that these two foliations are orthogonal to
each other. It is shown that this leads to a ``quasi-metric'' space-time 
geometry if $t$ is taken to represent one extra degenerate time dimension.
The quasi-metric framework then consists of a differentiable manifold equipped
with a one-parameter family of Lorentzian 4-metrics ${\bf g}_t$ parameterized 
by the global time function $t$, in addition to a non-metric connection 
compatible with the family ${\bf g}_t$.
\end{abstract}
\topmargin 0pt
\oddsidemargin 5mm
\renewcommand{\thefootnote}{\fnsymbol{footnote}}

\section{Introduction}
In this thesis we construct a new relativistic theory of gravity. This theory
is compatible with a geometric framework that differs from the usual metric
framework. While there is nothing remarkable about the mathematical description
of the new ``quasi-metric'' framework (it is just ordinary differential 
geometry), the mathematics and its applications may seem unfamiliar to the 
traditional relativist who is used to working only with metric geometry. For 
this reason, we give an intuitive introduction to the basic mathematical 
concepts of the quasi-metric geometry; the physical motivation for introducing
it will have to wait until the main part of the thesis.
\section{From 3-geometry to 4-geometry}
The basic premise defining metric space-time geometry is that space-time can be
described as a 4-dimensional pseudo-Riemannian manifold. In particular, in 
metric geometry, the affine geometry follows uniquely from the space-time 
metric (barring torsion). Thus, since the affine geometry does not depend on 
any preferred time coordinate, the viewpoint is taken as fundamental, that 
there is no need to specify any particular split-up of space-time into space 
and time except for reasons of convenience. And certainly no such split-up can 
have anything to do with fundamental physics.

Yet one may wish to describe the geometry of space-time via the evolution with 
time of fields defined on a spatial hypersurface as in the ADM formalism [1] 
(this is called a canonical description of the space-time geometry). However, 
in metric theory, the choice of said spatial hypersurface is arbitrary, and its 
deformation along some arbitrary timelike vector field is totally dependent on 
the exact form of the space-time metric tensor field ${\bf g}$. In particular, 
this implies that in metric theory, one cannot impose general requirements 
that certain canonical structures should exist without simultaneously imposing 
restrictions on the form of ${\bf g}$. And from a physical point of view this 
is unacceptable since in metric theory, the affine structure follows directly 
from the space-time metric, so any restrictions on that will interfere with 
the dynamics of any potential theory of gravity.

Suppose then, that for some reason one wishes to take the canonical viewpoint 
of space-time as fundamental but such that space-time needs not be metric. 
Then there is no reason to believe that the requirement that certain canonical 
structures exist should restrict the affine geometry unduly. In particular,
one may require that a preferred time coordinate should exist. In this case,
the affine geometry does not necessarily depend exclusively on a space-time 
metric. We illustrate this in the next section. Meanwhile, we describe the 
canonical structures that are required to exist within our quasi-metric 
framework.

Mathematically, to justify the view that a canonical description of space-time 
is fundamental, such a description must involve canonical structures that are 
taken as basic. Thus such structures must always exist, and they should in 
some way be ``simpler" than canonical structures in metric theory. 
One should regard these criteria as fulfilled if one requires that the 
space-time manifold $\cal N$ always can be foliated into a particular set of 
spatial fundamental hypersurfaces (FHSs) ${\cal S}_t$ parametrized by the 
global time function $t$. To ensure that the parametrization in terms of $t$
exists independent of any space-time metric, we let $t$ represent one extra
degenerate time dimension. Besides, we require that $\cal N$ can always be 
foliated into a family of time-like curves corresponding to a set of 
fundamental observers (FOs), and such that these two foliations are orthogonal
to each other, see figure 1. 

\begin{figure}
\begin{center}
\epsfig{file=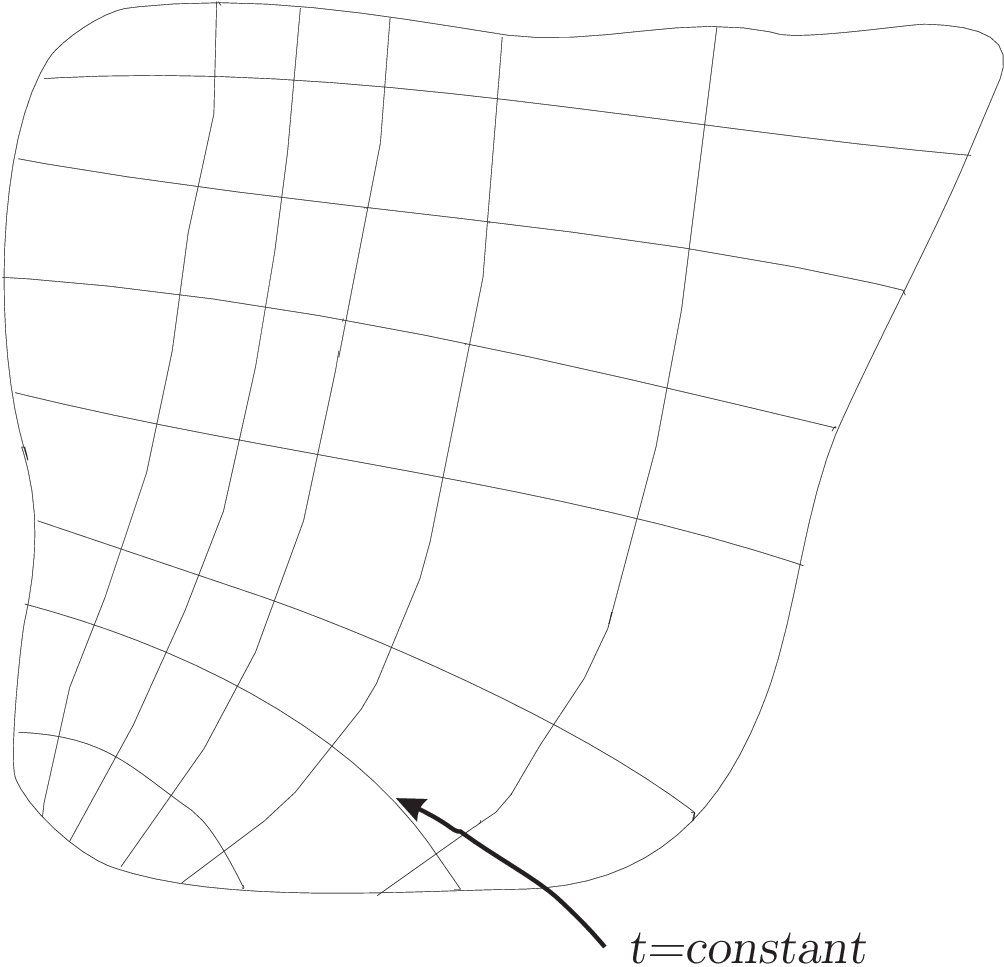,height=8cm}
\caption{{\small Basic canonical structure: space-time accepting two 
mutually orthogonal foliations.}}
\end{center}
\end{figure}

If this is going to make sense, there must be a space-time metric tensor 
available at each tangent space such that scalar products can be taken. To be 
sure that such a space-time tensor does exist, we require that on ${\cal S}_t$,
there must exist a family of scalar fields $N_t$ and a family of intrinsic 
spatial metrics ${\bf h}_t$. Moreover, there must exist a set of preferred 
coordinate systems where the time coordinate $x^0$ can be identified with the 
global time function $t$. We call such a coordinate system a {\em global time 
coordinate system} (GTCS). The coordinate motion of the FOs with respect to a 
GTCS defines a family ${\bf N}_t$ of spatial covector fields tangential to the 
FHSs. Now $N_t$, ${\bf N}_t$ and ${\bf h}_t$ uniquely determine a family of 
Lorentzian space-time metrics ${\bf g}_t$. This metric family may be thought 
of as constructed from measurements available to the FOs, if the functions 
$N_t$ are identified with the lapse functions describing the proper time 
elapsed as measured by the FOs, and if the covectors ${\bf N}_t$ are 
identified with the shift covector fields of the FOs in a GTCS. With the help 
of $N_t$ and ${\bf N}_t$, we may then construct the family of normal vector 
fields ${\bf n}_t$ being tangent vector fields to the world lines of the FOs 
and such that ${\bf h}_t({\bf n}_t,{\cdot})=0$. Thus, by construction, the 
foliation of $({\cal N},{\bf g}_t)$ into the set of world lines of the FOs is 
orthogonal to the foliation of $({\cal N},{\bf g}_t)$ into the set of FHSs, as 
asserted. To illustrate this, we may write 
\eqa
{\bf g}_t=-{\bf g}_t({\bf n}_t,{\cdot}){\otimes}
{\bf g}_t({\bf n}_t,{\cdot})+{\bf h}_t.
\ena
Note that there is no reason to believe that any affine connection compatible
with the above construction must be metric. Thus the point is to illustrate 
that it may be possible to create a geometrical structure where a space-time 
metric tensor is available at every event, but where the affine geometry does 
not necessarily follow exclusively from a space-time metric. The reason for 
this is that the existence of the global time function $t$ may be crucial when 
determining the affine connection; this possibility would not exist if the 
space-time geometry were required to be metric.
\section{Connection and curvature}
The existence of a Lorentzian space-time metric ${\bf g}{\in}{\bf g}_t$ for 
each FHS ${\cal S}{\in}{\cal S}_t$ implies that each FHS can be viewed as a 
spatial submanifold of a {\em metric} space-time manifold 
$({\cal M},{\bf g}_t)$ for constant $t$ (at least locally). We now perform a 
mathematical trick inasmuch as whenever we take this viewpoint, we may define 
a global time coordinate $x^0$ on $({\cal M},{\bf g}_t)$ such that $x^0=ct$ on 
each FHS; any coordinate system where this relationship holds is by definition 
a GTCS. We may then regard $x^0$ and $t$ as representing separate time 
dimensions such that $t=$constant on $({\cal M},{\bf g}_t)$. It is thus 
possible to view $\cal N$ as a 4-dimensional submanifold of 
${\cal M}{\times}{\bf R}$. The point of this, is that on the FHSs, we may 
take scalar products of fields defined on space-time in $({\cal M},{\bf g}_t)$,
whereas time derivatives of such fields depend on $t$ as well as on $x^0$ in 
$({\cal N},{\bf g}_t)$. Thus the existence of extra time derivatives with 
respect to $t$ of ${\bf g}_t$, means that we can find an affine connection that
is not fully determined by any single ${\bf g}{\in}{\bf g}_t$. See figure 2 
for visualization of quasi-metric space-time geometry.

\begin{figure}
\begin{center}
\epsfig{file=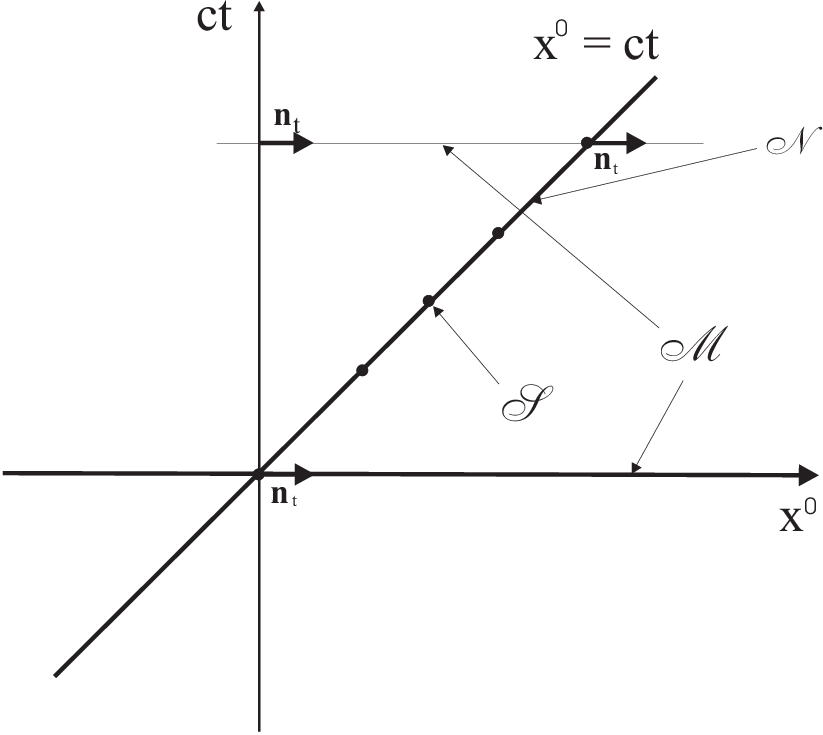,height=8cm}
\caption{The geometrical structure of quasi-metric space-time, visualized.}
\end{center}
\end{figure}

Since each ${\bf g}{\in}{\bf g}_t$ is defined on a spatial submanifold only, 
there is in general no way to extend it to all of $\cal N$. However, the 
Levi-Civita connection corresponding to each ${\bf g}{\in}{\bf g}_t$
can be calculated on each FHS by holding $t$ constant. But this family of
metric connections does not fully determine the connection compatible with
the family ${\bf g}_t$. To find the full connection compatible with the 
family ${\bf g}_t$, it is convenient to view ${\bf g}_t$ as a single 
5-dimensional degenerate metric on ${\cal M}{\times}{\bf R}$. One may then 
construct a linear, symmetric and torsion-free connection 
${\,}{\topstar{\nabla}}{\,}$ on ${\cal M}{\times}{\bf R}$ by requiring that 
this connection is compatible with the non-degenerate part of the metric 
family ${\bf g}_t$, in addition to the condition that
${\,}{\topstar{\nabla}}_{\frac{\partial}{{\partial}t}}{\bf n}_t$
vanishes. For obvious reasons, this connection is called ``degenerate", and it 
should come as no surprise that it is non-metric. One may then restrict the 
degenerate connection to $\cal N$ by just considering the submanifold 
$x^0=ct$ in a GTCS. For the exact form of the connection coefficients 
determining ${\,}{\topstar{\nabla}}{\,}$, see, e.g., [2]. 

At this point, a natural question would be if there are physical phenomena 
which are naturally modelled by the non-metric part of quasi-metric geometry. 
And as we shall see in the main part of this thesis this is indeed the case.
\section{Two simple examples}
We finish this mathematical introduction by giving two simple examples. The
first example illustrates how a family of space-time metrics can be represented
by a family of line elements, whereas the second example illustrates how 
affinely parametrized curves are represented.
\\

{\em Example 1} \\
A family of Minkowski metrics can be represented by the family of line elements
\eqa
ds_t^2=-(dx^0)^2+{\Xi}^2(t){\Big(}dx^2+dy^2+dz^2{\Big)}, 
\ena
where ${\Xi}(t)$ is a scale factor, $x^0=ct$ and $x,y,z$ are ordinary 
Cartesian coordinates. Note that the $t$-direction is treated as degenerate. 
It is easy to show that the free-fall curves obtained from the equations of 
motion (derived in chapter 1, section 5 of this thesis) and applied to equation
(2), are different than their counterparts obtained by setting $dx^0=cdt$ in 
equation (2) and using the geodesic equation compatible with the resulting 
single metric.
\\

{\em Example 2} \\
Let ${\{}t(\lambda),x^{\alpha}(\lambda){\}}$ represent an affinely 
parameterized curve, where $x^{\alpha}$ are space-time coordinates in a
GTCS. Then the tangent vector field ${\frac{\partial}{{\partial}{\lambda}}}$
along the curve may be represented by the expression
\eqa
{\frac{\partial}{{\partial}{\lambda}}}={\frac{dt}{d{\lambda}}}
{\frac{\partial}{{\partial}t}}+{\frac{dx^{\mu}}{d{\lambda}}}
{\frac{\partial}{{\partial}x^{\mu}}}.
\ena
Now the length of ${\frac{\partial}{{\partial}{\lambda}}}$ may be calculated
along the curve, e.g., for a timelike curve, 
${\bf g}_t({\frac{\partial}{{\partial}{\lambda}}},
{\frac{\partial}{{\partial}{\lambda}}})<0$ along the curve. On the other hand,
the degenerate connection taken along ${\frac{\partial}{{\partial}{\lambda}}}$ 
is
\eqa
{\topstar{\nabla}}_{{\frac{\partial}{{\partial}{\lambda}}}}=
{\frac{dt}{d{\lambda}}}{\topstar{\nabla}}_{{\frac{\partial}{{\partial}t}}}+
{\frac{dx^{\mu}}{d{\lambda}}}
{\topstar{\nabla}}_{{\frac{\partial}{{\partial}x^{\mu}}}}.
\ena
Thus we have illustrated that the degenerate part of 
${\frac{\partial}{{\partial}{\lambda}}}$ does not contribute to its length
(since ${\bf g}_t({\frac{\partial}{{\partial}t},{\cdot}}){\equiv}0$).
However, as can be seen from equation (4), the degenerate part of
${\frac{\partial}{{\partial}{\lambda}}}$ does influence parallel-transport of 
space-time objects.
\\

Hopefully, anyone who has understood the basic mathematical concepts 
underlying the quasi-metric framework, will not be so easily distracted by 
mathematical hurdles and will be in a better position to follow the derivation
of the main results presented in this thesis. Accordingly, having read this 
mathematical introduction, the reader should now be ready for the main parts 
of the thesis.
\\ [4mm]
{\bf References} \\ [1mm]
{\bf [1]} C.W. Misner, K.S. Thorne, J.A. Wheeler,
{\em Gravitation}, W.H. Freeman ${\&}$ Co. (1973). \\
{\bf [2]} D. {\O}stvang, chapter 1, section 2 of this thesis (2001). 

\newpage

{\vspace*{5cm}}

\begin{center}
{\huge {\bf Chapter One: \\\ \\ A Non-Metric Approach to
\vspace{0.5cm}\\Space, Time and Gravitation}}
\end{center}
\newpage
{\hspace*{1cm}}

\newpage

\setcounter{equation}{0}

\begin{center}
{\large {\bf A Non-Metric Approach to Space, Time and Gravitation}}
\end{center}
\begin{center}
by
\end{center}
\begin{center}
Dag {\O}stvang \\
{\em Institutt for fysikk, Norges teknisk-naturvitenskapelige universitet, 
NTNU \\ N-7491 Trondheim, Norway}
\end{center}

\begin{abstract}
This document surveys the basic concepts underlying a new geometrical framework
of space and time developed elsewhere [2]. Moreover, it is shown how this
framework is used to construct a new relativistic theory of gravity.
\end{abstract}

\section*{1{\hspace*{9mm}} Introduction}
A proper description of nature's fundamental forces involves their formulation 
as dynamical laws subject to appropriate initial conditions. This defines
the initial-value problem for whatever dynamical system one wishes to study.

The initial-value problem for gravitation generally involves what one may 
loosely call ``evolution of space with time". More precisely, given a 
one-parameter family of 3-dimensional manifolds ${\cal S}_t$, each of which is 
identified as ``space" at ``time" $t$ (such that $t=$constant on each 
${\cal S}{\in}{\cal S}_t$), the family defines an evolution of the 3-manifolds
${\cal S}_t$ with $t$. However, every conceivable theory of gravity
includes an overlying 4-dimensional geometrical structure called ``space-time"
with which the given family must be compatible. Thus it is not possible 
to analyse a given family ${\cal S}_t$ in isolation; the evolution with $t$ of 
fields that determine the relationship between the family and the overlying 
space-time must be analysed simultaneously. When this is done one recovers the 
space-time geometry.

Naturally, the structure of any space-time framework should be more general 
than any particular theory of gravity formulated within that framework. This 
implies that in general, there should exist geometrical relations between the 
space-time framework and any given family ${\cal S}_t$ that do not depend on 
the detailed form of the dynamical laws one uses. A particular set of such 
relations is called {\em kinematical} relations (see, e.g., [1]). The 
kinematical relations of different space-time frameworks are recognized by 
their sharing of a common feature; that they are not sufficient by themselves 
to determine solutions of the initial-value problem. That is, in addition to 
the initial conditions, one must supply extra fields coming from the dynamics 
at each step of the evolution of ${\cal S}_t$ if one wants to recover the 
space-time geometry. Thus the kinematical relations explicitly exhibit the 
dependence on the existence of some coupling between geometry and matter 
fields, even if the detailed nature of the coupling is not specified. This is 
what one should expect, since any relations defined as kinematical always 
should give room for appropriate dynamics.

As a first example, we take the case of the stratified non-relativistic 
space-time framework of Newton-Cartan [4]. Here, there exists a unique family
${\cal S}_t$ consisting of Euclidean spaces (and left intact by Galilean
transformations), where $t$ is Newton's absolute time. The overlying 
geometrical structure belonging to this space-time is the space-time covariant 
derivative ${\nabla}$ represented by the only non-zero connection coefficients 
${\Gamma}^{i}_{tt}{\equiv}{\Phi},_i$ (using a Cartesian coordinate system 
${\{}x^i{\}}$). Here, ${\Phi}(t,x^i)$ is a potential function and the comma 
denotes a partial derivative. The equation of motion for inertial test 
particles is the geodesic equation
\eqa
{\frac{d^2x^i}{dt^2}}+{\Gamma}^{i}_{tt}=0,
\ena
so to calculate the world lines of such test particles one must know
${\Gamma}^{i}_{tt}$ for every member of the family ${\cal S}_t$. Now we notice 
that there is a general relationship between the only non-zero components of 
the Riemann tensor and the potential function given by [4]
\eqa
R_{itjt}={\frac{{\partial}^2}{{\partial}x^i{\partial}x^j}}{\Phi}
={\frac{\partial}{{\partial}x^i}}{\Gamma}^{j}_{tt}.
\ena
That this relation is kinematical we see from the fact that, to determine the 
${\Gamma}^{i}_{tt}$ for all times and thus all information relevant for the 
motion of inertial test particles, one must explicitly supply the tensor 
$R_{itjt}$ for each time step. Furthermore, equation (2) is valid
regardless of the detailed coupling between $R_{itjt}$ and matter fields.

As a second example, we take the more realistic case of a metric relativistic 
space-time. In this case, the choice of family ${\cal S}_t$ is not unique, 
so the kinematical relations should be valid for any such choice. Furthermore,
the overlying geometrical structure belonging to space-time is the space-time
metric field, and this field induces relations between different choices of
${\cal S}_t$. Analogous to the first example, the kinematical relations are
given by (using a general spatial coordinate system and applying Einstein's 
summation convention)
\eqa
R_{i{\perp}j{\perp}}=N^{-1}{\cal L}_{N{\bf n}}K_{ij} + {K_i}^kK_{kj}
+c^{-2}a_{i{\mid}j}+c^{-4}a_ia_j,
\ena
see, e.g., [1] for a derivation. Here, {\bf a} is the 4-acceleration field of 
observers moving orthogonally to each ${\cal S}{\in}{\cal S}_t$, $N$ is their
associated lapse function field and `$\perp$' denotes projection onto the 
normal direction of ${\cal S}_t$. Furthermore, ${\cal L}_{\bf n}$ denotes the 
projected Lie derivative in the normal direction and `${\mid}$' denotes a
spatial covariant derivative intrinsic to each ${\cal S}{\in}{\cal S}_t$. The 
extrinsic curvature tensor {\bf K} describes the curvature of each 
${\cal S}{\in}{\cal S}_t$ relative to the overlying space-time curvature. 
There exists a well-known relationship between {\bf K} and the evolution in 
the normal direction of the intrinsic geometry of each 
${\cal S}{\in}{\cal S}_t$, namely (see, e.g., [1, 4])
\eqa
K_{ij}=-{\frac{1}{2}}{\pounds}_{\bf n}h_{ij},
\ena
where {\pounds}$_{\bf n}$ denotes a Lie derivative in the normal direction and
${\bf h}$ denotes the intrinsic metric of each ${\cal S}{\in}{\cal S}_t$.
Another, geometrically equivalent expression of the kinematical relations (3), 
is found (see, e.g., [1]) by projecting the Einstein space-time tensor field 
${\bf G}$ with respect to the spatial hypersurfaces ${\cal S}_t$. The 
projection of each space-time index may be a normal projection or a tangential 
projection, so for ${\bf G}$ we have 3 different projections since ${\bf G}$
is a symmetric tensor. The result is ($K{\equiv}{K^k}_k$)
\eqa
G_{{\perp}{\perp}}&=&{\frac{1}{2}}(P+K^2-K_{ks}K^{ks}), {\nonumber} \\
G_{{\perp}j}&=&(K^k_j-K{\delta}^k_j)_{{\mid}k}, {\nonumber} \\
G_{ij}&=&-N^{-1}{\cal L}_{N{\bf n}}(K_{ij}-Kh_{ij})-2{K_i}^kK_{kj}+3KK_{ij}
-{\frac{1}{2}}(K_{ks}K^{ks}+K^2)h_{ij} {\nonumber} \\
&-&c^{-2}(a_{i{\mid}j}-{a^k}_{{\mid}k}h_{ij})-c^{-4}(a_ia_j-a^ka_kh_{ij})
+H_{ij},
\ena
where ${\bf H}$ and $P$ are, respectively, the Einstein tensor and the Ricci 
scalar intrinsic to the hypersurfaces. Equation (4) in combination with 
equation (3) (or equivalently, the totally spatial projection $G_{ij}$ shown in 
equation (5), the other projections representing constraints on initial data), 
may be recognized as kinematical equations. This follows from the fact that, in
addition to initial values of $h_{ij}$ and $K_{ij}$ at an initial hypersurface,
one needs to specify the tensor field $G_{ij}$ or equivalently, the tensor 
field $R_{i{\perp}j{\perp}}$, at each step of the evolution of ${\cal S}_t$ if 
one wishes to recover the space-time metric field and thus all information 
about the geometry of space-time as predicted at subsequent hypersurfaces.
Note that this does not depend on Einstein's equations being valid (see, 
e.g., [1] for a further discussion).

In light of the two above examples, we may call the evolution represented by a 
family ${\cal S}_t$ a {\em kinematical evolution} (KE) if it depends on the
supplement of extra fields coming from an overlying geometrical structure at 
each step of the evolution. The question now is if this type of evolution
exhausts all possibilities. The reason for this concern is, that one may 
possibly imagine a type of evolution of spatial hypersurfaces that does 
{\em not} depend on the supplement of extra fields at each step of the 
evolution. Rather, an evolution of such type should be given explicitly as an 
``absolute'' property of space-time itself. It would be natural to call such a 
kind of evolution {\em non-kinematical} (NKE). Since, by construction, the NKE 
does not give room for dynamics, any geometrical framework having the 
capacity to accommodate both KE and NKE of a family ${\cal S}_t$ must be 
different from the standard metric framework. 

As illustrated by the above examples, we see from equations (2) and (3)
that the extra fields one must supply at each time step to get the KE working,
take the form of curvature tensors. On the other hand, the NKE should be
given explicitly and be obtainable as an intrinsic property of each member 
${\cal S}$ of a {\em unique} family ${\cal S}_t$. Since we want our alternative
space-time framework to be relativistic, $t$ must represent a ``preferred'' 
time coordinate and any member of the unique family ${\cal S}_t$ must 
represent a ``preferred'' notion of space. Observers moving normally to any 
${\cal S}{\in}{\cal S}_t$ get the status of preferred observers. However, the 
existence of the unique family ${\cal S}_t$ does {\em not} imply that there 
exists a preferred coordinate frame. Rather, there will exist a {\em class} of
``preferred'' coordinate systems especially well adapted to the geometry of 
quasi-metric space-time. One may expect that equations take special forms in 
any such a coordinate system.

Now to the questions; are there physical phenomena which may be suspected to 
have something to do with the NKE and if so, do we get any hint of which form 
it should take? The fact that our Universe seems to be compatible with a 
preferred time coordinate at large scales, makes us turn to cosmology in an 
attempt to answer these questions. With the possibility of a rather dramatic 
reinterpretation of the Hubble law in mind, we guess that the NKE should take 
the form of a local increase in scale with time as measured by the above
defined preferred observers. Moreover, this increase in scale should be
described via a {\em scale factor} given explicitly as a function of $t$ and
being part of the intrinsic geometry of each member ${\cal S}$ of the 
preferred family ${\cal S}_t$. That is, we guess that the Hubble law has 
nothing to do with kinematics; rather the Hubble law should be the basic 
empirical consequence of the (global) NKE.

The above guess represents the key to finding a new framework of space and time
which incorporates both KE and NKE in a natural way. We call this new
framework ``quasi-metric'' since it is not based on pseudo-Riemannian geometry
and yet the new framework is compatible with important physical 
principles previously thought to hold for metric theory only. In this thesis, 
we explore some of the possibilities of the quasi-metric framework; among other 
things we derive a new theory of gravity. The thesis is organized as follows: 
in chapter 1 we define the quasi-metric framework and list the important 
results without going through all the gory details, while chapters 2, 3, 4 and 
5 are self-contained articles containing the detailed calculations underlying 
the results listed in chapter 1.
\section*{2{\hspace*{9mm}} An alternative framework of 
space and time}
Our first problem is to arrive at a general framework describing space and time
in a way compatible with the existence of both KE and NKE of ${\cal S}_t$, as
discussed in the introduction. Since at least parts of the KE of any member 
${\cal S}$ of the preferred family ${\cal S}_t$ should be described by a 
Lorentzian space-time metric ${\bf g}$, we may assume that the KE represented 
by ${\bf g}$ is discernible at each time step. That is, at least in a 
(infinitesimally) small time interval centred at the hypersurface 
$t=$constant, it should be possible to evolve this hypersurface purely 
kinematically in the hypersurface-orthogonal direction with the help of the 
4-metric ${\bf g}$ without explicitly taking the NKE into account. But since 
this should be valid for every ${\cal S}{\in}{\cal S}_t$, to be compatible with
the existence of any NKE, different ${\cal S}{\in}{\cal S}_t$ should be 
associated with different metrics ${\bf g}{\in}{\bf g}_t$. Thus the preferred 
family ${\cal S}_t$ of 3-dimensional spatial manifolds should define a 
one-parameter family of 4-dimensional Lorentzian metrics ${\bf g}_t$, the 
domain of validity of each being exactly the hypersurface $t=$constant, but 
such that one may extrapolate each metric at least in a (infinitesimally) 
small interval centred at this hypersurface.

We may think of the parameter $t$ as representing an extra time dimension,
since fixing $t$ is equivalent to fixing a space-time metric 
${\bf g}{\in}{\bf g}_t$ on (at least part of) a space-time 4-manifold. This 
indicates that the mathematically precise definition of the quasi-metric 
framework should involve a 5-dimensional product manifold 
${\cal M}{\times}{\bf R}$ where ${\cal M}$ is a Lorentzian space-time manifold 
and ${\bf R}$ is the real line. We may then interpret $t$ as a global 
coordinate on ${\bf R}$ and the preferred family ${\cal S}_t$ as representing a
foliation into spatial 3-manifolds of a 4-dimensional submanifold ${\cal N}$ of
${\cal M}{\times}{\bf R}$. Thus ${\cal N}$ is, by definition, a 4-dimensional
space-time manifold equipped with a one-parameter family of Lorentzian metrics
${\bf g}_t$ in terms of the global time function $t$. This is the mathematical
definition of the quasi-metric space-time framework we want to use as the 
basis for constructing a new relativistic theory of gravity.

Henceforth, we refer to the members of ${\cal S}_t$ as the {\em fundamental
hypersurfaces} (FHSs). Observers always moving orthogonally to the FHSs are 
called {\em fundamental observers} (FOs). Note that the FOs are in general
non-inertial. When doing calculations, it is often necessary to define useful 
coordinate systems. A particular class of coordinate systems especially well 
adapted to the existence of the FHSs, is the set of {\em global time coordinate 
systems} (GTCSs). A coordinate system ${\{}x^{\mu}{\}}$ is a GTCS if and only
if the relationship in $\cal N$ between the time coordinate $x^0$ and the 
global time function $t$ is explicitly given by the equation $x^0=ct$. Since 
the local hypersurface-orthogonal direction in $\cal M$ is, by definition, 
physically equivalent to the direction along the world line of the local FO in
$\cal N$, the one-parameter family of metrics is most conveniently expressed 
in a GTCS. That is, in a GTCS the spatial line elements of the FHSs evolve 
along the world lines of the FOs in $\cal N$; there is no need to worry about 
the explicit evolution of the spatial geometry along alternative world lines. 
Note that there exist infinitely many GTCSs due to the particular structure 
of quasi-metric space-time, but that in general, it is not possible to find one 
which with respect to the FOs are at rest. However, in particular cases, 
further simplification is possible if we can find a {\em comoving} coordinate 
system. A comoving coordinate system is by definition a coordinate system where
the spatial coordinates are constants along the world lines of the FOs. A 
comoving coordinate system which is also a GTCS we conventionally call a 
{\em hypersurface-orthogonal coordinate system} (HOCS). Expressed in a HOCS, 
the 3-vector piece $g_{(t)0i}$ of ${\bf g}_t$ vanishes. But in general, it is 
not possible to find a HOCS; that is possible for particular cases only.

The condition that the global time function should exist everywhere, implies 
that in ${\cal M}$, each FHS must be a Cauchy hypersurface for every allowable
metric ${\bf g}{\in}{\bf g}_t$. Furthermore, each FHS must be a member of a 
family of hypersurfaces foliating ${\cal M}$ equipped with any allowable 
metric ${\bf g}{\in}{\bf g}_t$ when this metric is extrapolated off the FHS. 
Note that the hypothetical observers moving orthogonal to this family of 
hypersurfaces may be thought of as FOs with the explicit NKE ``turned off" and
that the normal curves of these hypersurfaces are in general not geodesics of 
${\bf g}{\in}{\bf g}_t$. 

Actually, it may not be possible to eliminate all the effects of the NKE just 
by holding $t$ constant. The reason for this, is the possibility that some of 
the NKE, which we will call the {\em local} NKE, is not ``realized'' 
explicitly. That is, the FOs may move in a way such that the ``expansion" 
representing the local NKE is exactly cancelled out. In other words, the FOs 
may be thought of as ``falling'' with a 3-velocity $-{\bf v}_t$ and 
simultaneously expanding with a 3-velocity ${\bf v}_t$ due to the local NKE, 
the net result being ``stationariness''. But this type of stationariness is not
equivalent to being stationary with no NKE; for example one typically gets time
dilation factors associated with the extra ``motion''. Now the effects of the
local NKE should be compensated for in the ``physical'' metric  family
${\bf g}_t$. On the other hand, said effects should in general be present 
implicitly and uncompensated for in solutions of field equations. This means 
that ${\bf g}_t$ cannot represent a {\em solution} of field equations. To 
overcome this problem, we introduce a second metric family ${\bf {\bar g}}_t$ 
on a corresponding product manifold ${\overline{\cal M}}{\times}{\bf R}$ and 
consider a 4-dimensional submanifold ${\overline{\cal N}}$ just as for the 
family ${\bf g}_t$. The crucial difference is that the family 
${\bf {\bar g}}_t$ by definition represents a solution of appropriate field 
equations. Thus when we construct such field equations, we will study the 
general properties of ${\bf {\bar g}}_t$ rather than those of ${\bf g}_t$. Any 
local non-kinematical features, disguised as kinematical features belonging to 
the metric structure of each member of the family ${\bf {\bar g}}_t$, must then
be compensated for via a transformation 
${\bf {\bar g}}_t{\rightarrow}{\bf g}_t$ involving the vector field family
${\bf v}_t$.

The family of hypersurfaces foliating ${\cal M}$ (or ${\overline{\cal M}}$) 
equipped with any allowable metric ${\bf g}{\in}{\bf g}_t$ (or ${\bf {\bar g}}
{\in}{\bf {\bar g}}_t$) defines a one-parameter family of 3-manifolds in terms 
of a suitably defined time coordinate function $x^0$. A sufficient condition 
to ensure the existence and uniqueness of the global time coordinate function 
$x^0$ on ${\cal M}$ (or ${\overline{\cal M}}$) such that the gradient field of
this coordinate is everywhere timelike, is to demand that each FHS is 
{\em compact} (without boundaries) [3]. Assuming simple topology, this should 
imply the existence of some prior 3-geometry of the FHSs; interpreted as a 
cosmological ``background'' 3-geometry, and chosen to be the geometry of the
3-sphere ${\bf S}^3$. It is expected that this 3-geometry will show up via 
specific terms in the field equations.

We now write down a general expression for the metric family 
${\bf {\bar g}}_t$, where the explicit dependence on $t$ represents the 
{\em global} NKE of the FHSs (and the other pieces of ${\bf {\bar g}}_t$). 
That is, expressed in a suitable GTCS (using Einstein's summation convention), 
the most general form allowed for the family ${\bf {\bar g}}_t$ is represented 
by the family of line elements valid on the FHSs (this may be taken as a 
definition)
\eqa
{\overline {ds}}_t^2={\bar N}_t^2{\Big \{ }
[{\bar N}_{(t)}^k{\bar N}_{(t)}^s{\tilde h}_{(t)ks}-1](dx^0)^2+
2{\frac{t}{t_0}}{\bar N}_{(t)}^k{\tilde h}_{(t)ks}dx^sdx^0+
{\frac{t^2}{t_0^2}}{\tilde h}_{(t)ks}dx^kdx^s{\Big \} }.
\ena
Here, $t_0$ is some arbitrary reference epoch (usually chosen to be the present
epoch) setting the scale of the spatial coordinates, ${\bar N}_t$ is the family
of lapse functions of the FOs and ${\frac{t_0}{t}}{\bar N}^k_{(t)}$ are the 
components of the shift vector family of the FOs in ${\overline{\cal N}}$. 
Also, ${\bf {\bar h}}_t$, with components ${\bar h}_{(t)ks}{\equiv}
{\frac{t^2}{t_0^2}}{\bar N}_t^2{\tilde h}_{(t)ks}$, is the spatial metric family 
intrinsic to the FHSs. Moreover, as a counterpart to equation (6), a general 
form for the family ${\bf g}_t$ is given by the family of line elements (using 
a GTCS)
\eqa
{ds}_t^2=[N_{(t)}^kN_{(t)}^s{\hat h}_{(t)ks}-N^2](dx^0)^2+
2{\frac{t}{t_0}}N_{(t)}^k{\hat h}_{(t)ks}dx^sdx^0+
{\frac{t^2}{t_0^2}}{\hat h}_{(t)ks}dx^kdx^s,
\ena
where the symbols have similar meanings to their (barred) counterparts in 
equation (6) (the counterpart to ${\bar h}_{(t)ks}$ is $h_{(t)ks}{\equiv}
{\frac{t^2}{t_0^2}}{\hat h}_{(t)ks}$). Note that the propagation
of sources (and test particles) is calculated by using the equations of motion 
in ${\cal N}$ (see equation (15) below). Besides, since the proper 
time as measured along a world line of a FO should not directly depend on the 
cosmic expansion, the lapse function $N$ should not depend explicitly on $t$. 
Therefore, any potential $t$-dependence of $N$ must be eliminated by 
substituting $t$ with $x^0/c$ (using a GTCS) whenever it occurs before using 
the equations of motion. In the same way, any extra $t$-dependence of 
${\bf g}_t$ coming from the transformation 
${\bf {\bar g}}_t{\rightarrow}{\bf g}_t$ must be eliminated. Consequently, any 
$t$-dependence of ${\hat h}_{(t)ks}$ will stem from that of ${\tilde h}_{(t)ks}$. 

Next, an obvious question is which sort of affine structure is induced on
${\cal N}$ by the family of metrics ${\bf g}_t$ (and analogously which affine
structure is induced on ${\overline{\cal N}}$ by ${\bf {\bar g}}_t$). To 
answer this question, it is convenient to perceive ${\bf g}_t$ as a single 
degenerate 5-dimensional metric on ${\cal M}{\times}{\bf R}$ and construct a 
torsion-free connection ${\ }{\topstar{\nabla}}{\ }$ almost compatible with 
${\bf g}_t$ (${\ }{\topstar{\nabla}}{\ }$ yields non-metricity for the 
degenerate part of ${\bf g}_t$). But in addition to almost 
metric-compatibility, i.e., 
${\,}{\topstar{\nabla}}_{\frac{\partial}{{\partial}t}}{\bf g}_t=0$, we should also 
require that the unit normal vector field family of the FHSs does not change 
when it is parallel-transported in the $t$-direction. That is, we should 
require that 
${\,}{\topstar{\nabla}}_{\frac{\partial}{{\partial}t}}{\bf n}_t$ vanishes. When a 
connection satisfying these requirements is found, we can restrict it to 
${\cal N}$ and we have the wanted affine structure. 

In order to construct ${\,}{\,}{\topstar{\nabla}}{\,}{\,}$ on 
${\cal M}{\times}{\bf R}$, the metric-preserving condition
\eqa
{\frac{\partial}{{\partial}t}}{\bf g}_t({\bf y}_t,{\bf z}_t)=
{\bf g}_t({\,}{\,}{\topstar{\nabla}}_{\frac{\partial}{{\partial}t}}
{\bf y}_t,{\bf z}_t)+
{\bf g}_t({\bf y}_t,{\topstar{\nabla}}_{\frac{\partial}{{\partial}t}}
{\bf z}_t),
\ena
involving arbitrary families of vector fields ${\bf y}_t$ and ${\bf z}_t$
in ${\cal M}$, may be used to find a candidate connection where the connection
coefficients are determined from ${\bf g}_t$ alone. It turns out that such a
candidate connection is unique and that its difference from the usual
Levi-Civita connection is determined from the only non-zero connection
coefficients containing $t$. These are given by 
${\topstar{\Gamma}}^{{\,}{\alpha}}_{{\mu}t}$ and are equal to
${\frac{1}{2}}g_{(t)}^{{\alpha}{\sigma}}{\frac{\partial}{{\partial}t}}
g_{(t){\mu}{\sigma}}$. (Here we have introduced the coordinate notation 
$g_{(t){\mu}{\sigma}}$ for the family ${\bf g}_t$.) But this candidate connection 
does in general not satisfy the criterion 
${\,}{\topstar{\nabla}}_{\frac{\partial}{{\partial}t}}{\bf n}_t=0$. However, the part
of it involving ${\bf h}_t$ does, provided that (where a comma denotes taking 
a partial derivative)
\eqa 
{\frac{\partial}{{\partial}t}}{\Big [}N_{(t)}^kN_{(t)}^s
{\hat h}_{(t)ks}{\Big ]}=0, \qquad {\Rightarrow} \qquad
N_{(t),t}^s=-{\frac{1}{2}}N_{(t)}^k{\hat h}_{(t)}^{is}{\hat h}_{(t)ik,t}.
\ena
We may thus choose the connection coefficients equal to the 
hypersurface-intrinsic part 
${\frac{1}{2}}h_{(t)}^{ik}{\frac{\partial}{{\partial}t}}h_{(t)jk}$
of the above-mentioned candidate connection coefficients. In addition, the 
shift vector field family is required to fulfil equation (9). We then have the 
unique non-zero connection coefficients
\eqa
{\topstar{\Gamma}}_{tj}^{{\,}i}={\frac{1}{2}}h_{(t)}^{is}h_{(t)sj,t}
={\frac{1}{t}}{\delta}^i_j+
{\frac{1}{2}}{\hat h}_{(t)}^{is}{\hat h}_{(t)sj,t},
{\qquad} {\topstar{\Gamma}}^{{\,}{\alpha}}_{t{\mu}}{\equiv}{\ }
{\topstar{\Gamma}}^{{\,}{\alpha}}_{{\mu}t},
\ena
valid in a GTCS. The other connection coefficients not containing $t$-indices 
are found by requiring that they should be identical to those found from the 
family of Levi-Civita connections explicitly defined by the ${\bf g}_t$.
It may be readily shown that equations (9) and (10), together with the explicit
$t$-dependence of ${\bf g}_t$ shown in equation (7), yield
\eqa
{\topstar{\nabla}}_{\frac{\partial}{{\partial}t}}{\bf g}_t=0, \qquad
{\topstar{\nabla}}_{\frac{\partial}{{\partial}t}}{\bf n}_t=0, \qquad
{\topstar{\nabla}}_{\frac{\partial}{{\partial}t}}{\bf h}_t=0,
\ena
thus the connection ${\,}{\topstar{\nabla}}{\,}$ has the desired properties.
The restriction of ${\,}{\,}{\topstar{\nabla}}{\,}{\,}$ to $\cal N$ is trivial 
inasmuch as the same formulae are valid in $\cal N$ as in 
${\cal M}{\times}{\bf R}$, the only difference being the restriction 
$x^0=ct$ in a GTCS.

Now we want to use the above defined affine structure on ${\cal N}$ to find 
equations of motion for test particles. Let ${\lambda}$ be an affine parameter 
along the world line of an arbitrary test particle. (In addition 
to the affine parameter $\lambda$, $t$ is also a (non-affine) parameter along 
any non-spacelike curve in $\cal N$.) If we define coordinate 
vector fields ${\frac{\partial}{{\partial}x^{\alpha}}}$, the coordinate 
representation of the tangent vector field 
${\frac{\partial}{{\partial}{\lambda}}}$ along the curve is given by
${\frac{dt}{d{\lambda}}}{\frac{\partial}{{\partial}t}}+
{\frac{dx^{\alpha}}{d{\lambda}}}{\frac{\partial}{{\partial}x^{\alpha}}}$. We then 
define the covariant derivative along the curve as
\eqa
{\topstar{\nabla}}_{\frac{\partial}{{\partial}{\lambda}}}{\equiv}
{\frac{dt}{d{\lambda}}}{\topstar{\nabla}}_{\frac{\partial}{{\partial}t}}
+{\frac{dx^{\alpha}}{d{\lambda}}}
{\topstar{\nabla}}_{\frac{\partial}{{\partial}x^{\alpha}}}.
\ena
A particularly important family of vector fields is the 4-velocity tangent
vector field family ${\bf u}_t$ of a curve. That is, by definition we have
\eqa
{\bf u}_t{\equiv}u_{(t)}^{\alpha}{\frac{\partial}{{\partial}x^{\alpha}}}
{\equiv}{\frac{dx^{\alpha}}{d{\tau}_t}}
{\frac{\partial}{{\partial}x^{\alpha}}},
\ena
where ${\tau}_t$ is the proper time as measured along the curve. 

The equations of motion are found by calculating the covariant derivative of 
4-velocity tangent vectors along themselves using the connection in $\cal N$. 
According to the above, this is equivalent to calculating 
${\,}{\,}{\topstar{\nabla}}{\,}{\,}{\bf u}_t$ along 
${\frac{\partial}{{\partial}{\tau}_t}}$. Using the coordinate representation 
of ${\frac{\partial}{{\partial}{\tau}_t}}$ in an arbitrary coordinate system, 
we may thus define the vector field ${\stackrel{\star}{\bf a}}$ by
\eqa
{\stackrel{\star}{\bf a}}{\equiv}{\stackrel{\star}{\nabla}}_
{\frac{\partial}{{\partial}{\tau}_t}}{\bf u}_t={\Big (}
{\frac{dt}{d{\tau}_t}}{\stackrel{\star}{\nabla}}_{\frac{\partial}{{\partial}t}}
+{\frac{dx^{\alpha}}{d{\tau}_t}}
{\nabla}_{\frac{\partial}{{\partial}x^{\alpha}}}{\Big )}{\bf u}_t{\equiv}
{\frac{dt}{d{\tau}_t}}{\stackrel{\star}{\nabla}}_{\frac{\partial}{{\partial}t}}
{\bf u}_t+{\bf a}_t,
\ena
where ${\bf a}_t$ is the ordinary 4-acceleration family. We call the vector 
field ${\stackrel{\star}{\bf a}}$ the ``degenerate" 4-acceleration. It may be 
shown that ${\stackrel{\star}{\bf a}}$ is orthogonal to ${\bf u}_t$. 

The coordinate expression for ${\stackrel{\star}{\bf a}}$ shows that this 
yields equations of motion, namely
\eqa
{\frac{d^2x^{\alpha}}{d{\lambda}^2}}+{\Big(}
{\topstar{\Gamma}}^{{\,}{\alpha}}_{t{\sigma}}{\frac{dt}{d{\lambda}}}+
{\Gamma}^{\alpha}_{(t){\beta}{\sigma}}{\frac{dx^{\beta}}{d{\lambda}}}{\Big)}
{\frac{dx^{\sigma}}{d{\lambda}}}=
{\Big(}{\frac{d{\tau}_t}{d{\lambda}}}{\Big)}^2{\stackrel{\star}{a}}^{\alpha}.
\ena
To get these equations to work, the degenerate 4-acceleration must be specified 
independently for each time step; thus they are kinematical equations.
But ${\stackrel{\star}{\bf a}}$ cannot be chosen freely.
That is, given an initial tangent vector of the world line of an arbitrary 
test particle, as we can see from equation (14), ${\stackrel{\star}{\bf a}}$ 
is uniquely determined by ${\bf a}_t$ and ${\frac{dt}{d{\tau}_t}}
{\stackrel{\star}{\nabla}}_{\frac{\partial}{{\partial}t}}{\bf u}_t$.
However, in a subsequent section we will see that 
${\stackrel{\star}{\nabla}}_{\frac{\partial}{{\partial}t}}{\bf u}_t$ vanishes
so we in fact have ${\stackrel{\star}{\bf a}}$=${\bf a}_t$.

Now ${\bf a}_t$ has the usual interpretation as an expression for the 
inertial forces experienced by the test particle. Moreover, the 
coordinate expression for ${\bf a}_t$ in ${\cal M}$ is given by
\eqa
{\frac{d^2x^{\alpha}}{d{\lambda}^2}}+{\Gamma}^{\alpha}_{(t){\sigma}{\rho}}
{\frac{dx^{\rho}}{d{\lambda}}}{\frac{dx^{\sigma}}{d{\lambda}}}{\equiv}
{\Big(}{\frac{d{\tau}_t}{d{\lambda}}}{\Big)}^2a^{\alpha}_{(t)}.
\ena
However, to find the curves of test particles, equation (15) will be 
solved rather than equation (16); the latter is insufficient for this task
since it does not take into account the dependence of ${\bf u}_t$ on $t$, 
i.e., it holds only in $\cal M$.
\section*{3{\hspace*{9mm}} The role of gravity}
We are now able to set up the postulates which must be satisfied for every
theory of gravity compatible with the quasi-metric framework. These are
(postulates for the metric framework [5] are stated in parenthesis for 
comparison)
\begin{itemize}
{\item Space-time is equipped with a metric family ${\bf g}_t$ as described
in the previous section, (metric theory: space-time is equipped with one single
Lorentzian metric),}
{\item Inertial test particles follow curves for which ${\bf a}_t=0$ 
calculated from (15), (metric theory: the curves are geodesics calculated from
(16)),}
{\item In the local inertial frames of ${\bf g}_t$, the non-gravitational 
physics is as in special relativity. (Metric theory: in the local Lorentz 
frames of the metric, the non-gravitational physics is as in special 
relativity.)}
\end{itemize}
The third postulate means that we may always find a local coordinate system
such that the connection coefficients in equation (15) vanish. Later we will 
see that in fact ${\stackrel{\star}{\bf a}}$=${\bf a}_t$, meaning that 
inertial test particles follow geodesics of ${\,}{\topstar{\nabla}}{\,}$, see
equations (50), (51) and the subsequent discussion for verification. Thus 
the local inertial frames of ${\bf g}_t$ can be identified with local Lorentz 
frames, just as in metric theory. This means that there can be no local 
consequences of curved space-time for the non-gravitational physics. But it is 
important to notice that this applies strictly locally. That is, at any finite 
scale the geodesics of ${\,}{\topstar{\nabla}}{\,}$ do not coincide with the 
geodesics of any single space-time metric as long as there is a dependence on 
$t$. This means that the non-gravitational physics at any finite scale may vary
in a way incompatible with any metric geometry. Despite this, any allowable 
theory does possess local Lorentz invariance (LLI), and furthermore
it follows that we do have local position invariance (LPI), but that
deviations from metric theory should be detectable at any finite scale. (LPI 
is by definition that the outcome of any local non-gravitational test 
experiment is independent of where or when it is performed.)

The second postulate ensures that the WEP is valid (i.e., that the free-fall
curves traced out by test particles are independent of the particles'
internal structure). The sum of LLI, LPI and the WEP is called the Einstein 
equivalence principle (EEP) [5] and it is satisfied by any metric theory
of gravity. Note that the EEP was previously thought to hold for metric 
theories only. Nevertheless, despite the differences in mathematical 
structure, the EEP should also hold for the class of quasi-metric 
gravitational theories according to the arguments made above. 

Now we want to construct a theory of gravity satisfying the above postulates. 
This particular theory should contain no extra independent gravitational 
fields other than ${\bf {\bar g}}_t$ and also no prior 4-geometry. Since the 
quasi-metric framework fulfils LLI and LPI, we might hope to construct a 
theory which also embodies the strong equivalence principle (SEP) (that the EEP
is valid for self-gravitating bodies and local gravitational test experiments 
as well as for test particles). But as we shall see, the quasi-metric framework 
predicts scale variations in space-time between non-gravitational and
gravitational systems, and this is not compatible with the SEP in its most
stringent interpretation. So we cannot ask for more than that the 
gravitational weak equivalence principle (GWEP) is satisfied (that the WEP is 
valid for self-gravitating bodies as well as for test particles). However, if 
the GWEP is not violated, this is sufficient to eliminate the most serious 
effects resulting from the violation of the SEP.

One distinct feature of our framework is the existence of a unique set of 
FHSs. This means that we can split up the metrics
${\bf {\bar g}}_t$ into space and time in a preferred way; i.e., we have a
preferred family of unit normal vector fields ${\bf {\bar n}}_t$. In addition 
we have the 3-metrics ${\bf {\bar h}}_t$ intrinsic to the FHSs. We may 
then define a family of 4-acceleration fields ${\bf {\bar a}}_{\cal F}$ by
\eqa 
c^{-2}{\bar a}_{\cal F}^{\mu} {\equiv}  
{\bar n}_{(t);{\nu}}^{\mu}{\bar n}_{(t)}^{\nu}, 
\ena
where `$;$' denotes taking a covariant derivative obtained from the metric 
connection ${\bf {\bar {\nabla}}}$ compatible with any single metric
${\bf {\bar g}}{\in}{\bf {\bar g}}_t$.

It is useful to denote a scalar product of any space-time object with
$-{\bf {\bar n}}_t$ by the symbol `${\bar {\perp}}$'; this operation defines a 
normal projection. Analogously, any projection into a FHS may be performed by 
taking scalar products with ${\bf {\bar h}}_t$. To avoid confusion about 
whether we are dealing with a space-time object or a space object intrinsic to
the FHSs, we will label space objects with a ``hat'' if necessary. That is, a 
maximal number of projections of the space-time object ${\bf {\bar A}}_t$ with 
$-{\bf {\bar n}}_t$ or ${\bf {\bar h}}_t$ will result in various space objects 
${\bf {\hat {\bar A}}}_t$ (which object will depend on the exact projections 
made). 

Thus, by repeated projections with $-{\bf {\bar n}}_t$ or 
${\bf {\bar h}}_t$, any space-time object eventually reduces to a space 
object. For example, the space-time vector field ${\bf {\bar a}}_{\cal F}$ has 
the normal projection ${\bar a}_{{\cal F}{\bar {\perp}}}{\equiv}-
{\bar a}^{\alpha}_{\cal F}{\bar n}_{(t){\alpha}}$, and the components of the 
horizontal projection are just ${\bar a}_{{\cal F}}^j$. In particular, one may 
easily show that
\eqa
{\bar a}_{{\cal F}{\bar {\perp}}}=0, {\qquad} 
c^{-2}{\bar a}_{{\cal F}j}={\frac{\partial}{{\partial}x^j}}{\ln}{\bar N}_t.
\ena
Furthermore, we define another useful quantity
\eqa
c^{-2}{\bar x}_t+c^{-1}{\bar y}_t
{\equiv}-{\frac{{\bar N}_t,_{\bar {\perp}}}{{\bar N}_t}}+
{\frac{{\bar N}_t,_t}{c{\bar N}_t^2}},
\ena
which describes how ${\bar N}_t$ changes in the direction normal to the FHSs. 
The reason why the left hand side of equation (19) consists of two terms, 
is that the first term represents the KE of ${\bar N}_t$ and the second term 
represents the NKE of ${\bar N}_t$. We also notice that the family of extrinsic 
curvature tensors ${\bf {\bar K}}_t$ corresponding to the metric family 
${\bf {\bar g}}_t$ of the type given in equation (6) has the form (in a GTCS)
\eqa
{\bar K}_{(t)ij}={\Big (}{\frac{{\bar N}_t,_{\bar {\perp}}}{{\bar N}_t}}
-{\frac{t_0}{t}}c^{-2}{\bar a}_{{\cal F}k}{\frac{{\bar N}_{(t)}^k}{{\bar N}_t}}
{\Big )}{\bar h}_{(t)ij}+{\frac{t}{2{\bar N}_tt_0}}({\bar N}_{(t)i{\mid}j}+
{\bar N}_{(t)j{\mid}i})-{\frac{1}{2}}{\frac{t^2}{t_0^2}}{\bar N}_t
{\frac{\partial}{{\partial}x^0}}{\tilde h}_{(t)ij}.
\ena
Now we want to postulate what the nature of gravity should be in our theory. 
To arrive at that, we notice that the lapse function ${\bar N}_t$ in equation 
(6) also plays the role of a spatial scale factor. This means that we may think
of ${\bar N}_t^2$ as a {\em conformal} factor in equation (6). We may then 
interpret ${\bar N}_t$ as resulting from a position-dependent units 
transformation. This indicates that the scale factor 
${\bar F}_t{\equiv}{\bar N}_tct$ of the FHSs may be thought of as representing 
a difference in scale between atomic and gravitational systems. That is, 
${\bar F}_t$ may be thought of as representing a gravitational scale measured 
in atomic units. An equivalent way of interpreting ${\bar F}_t$ is to postulate
that fixed operationally defined ``atomic'' units are considered formally 
variable throughout space-time. Since the intercomparison of operationally 
defined units only has meaning locally for objects at rest with respect to 
each other, any non-local intercomparison is purely a matter of definition [7]. 

The variation in scale between gravitational and atomic systems means that 
gravitational quantities get an extra formal variation when measured in atomic
units (and {\em vice versa}); in particular the dimensional constants of nature
may depend on this formal variation. That is, the ``constants" of nature may in 
principle depend on whether they are expressed in atomic or gravitational 
units. We notice that, when performing a units transformation, we are free to 
{\em define} as constants physical constants that cannot be related to each 
other by dimensionless numbers [6]. We exploit this fact by defining $c$ and 
Planck's constant ${\hbar}$ not to be formally variable. This means that 
measured in atomic units, the variation of gravitational lengths and times 
are identical and inverse to that of gravitational energy (or mass).

The requirement that $c$ and ${\hbar}$ are not formally variable, yields no 
restrictions on any possible relationship between the variation of units and 
physical fields. On the other hand, we may combine the unit charge $e$ with
${\hbar}$ and $c$ to form the dimensionless number $e^2/{\hbar}c$. Thus by 
{\em defining} $e$ not to be formally variable (so that gravitational charge
units will not be formally variable), we deny any possible direct connection 
between gravitation and electromagnetism. 

Formally variable atomic units represent changes of scale between 
non-gravitational and gravitational systems, and such changes of scale should
be described by the metric family ${\bf {\bar g}}_t$. Moreover,
${\bf {\bar g}}_t$ should be found from field equations. We now define the
scalar field ${\Psi}_t$; by definition ${\Psi}_t$ tells how atomic time units
vary in space-time. Any gravitational quantity is associated with ${\Psi}_t$ 
to some power describing how the quantity changes due to the formal variation 
in scale between gravitational and atomic systems. In particular, we find from 
dimensional analysis that the ``bare'' gravitational coupling parameter 
$G_t^{\rm B}$ formally has the dimension of time (or length) units squared. 
Since $G_t^{\rm B}$ is a gravitational quantity, measured in atomic units it 
will then vary as a factor ${\Psi}_t^{-2}$. Besides, since charge is not 
formally variable, $G_t^{\rm B}$ couples to charge squared (as can be seen, 
e.g., from the well-known Reissner-Nordstr\"{o}m solution in General Relativity
(GR)), or more generally, to the electromagnetic stress-energy tensor. On the 
other hand, for material sources, masses formally vary as ${\Psi}_t$, but this 
is not measurable in non-gravitational experiments. Now the ``screened'' 
gravitational coupling parameter $G_t^{\rm S}$ couples to mass, and measured for
material sources, said formal variation of mass means that $G_t^{\rm S}$ 
effectively varies as ${\Psi}_t^{-1}$. Thus we must have {\em two} (in general 
different) gravitational coupling parameters, so we cannot have universal 
coupling between matter and gravitation since the coupling will depend on the 
nature of the gravitational source. This represents a radical departure from 
GR. Notice that $G_t^{\rm B}$ and $G_t^{\rm S}$ will have different time 
dependences so they are not likely to be approximately equal at the present 
epoch. But by varying the composition of the gravitational source in a series 
of controlled experiments, it should in principle be possible to measure both 
$G_t^{\rm B}$ and $G_t^{\rm S}$ directly at the present epoch.

Since $G_t^{\rm B}$ and $G_t^{\rm S}$ usually occur in combination with charge or
mass, it is convenient to {\em define} $G_t^{\rm B}$ and $G_t^{\rm S}$ to take the 
constant values $G^{\rm B}$ and $G^{\rm S}$, respectively, as measured in 
(hypothetical) local gravitational experiments in an empty universe at epoch
$t_0$. But then one must distinguish between {\em active mass} $m_t$ measured
dynamically as a source of gravity and {\em passive mass} $m$ (i.e., passive 
gravitational mass or inertial mass). This means that we include the formal
variations of $G_t^{\rm B}$ or $G_t^{\rm S}$ into $m_t$, turning any active mass 
$m_t$ into a scalar field. That is, unlike $c$, ${\hbar}$ and $e$, $m_t$ gets a
formal variation when measured in atomic units. We then find that the formal 
variation of active mass goes as ${\Psi}_t^{-2}$ for the electromagnetic field
and as ${\Psi}_t^{-1}$ for material particles. In practice, this means that the
effective gravitational ``constant'' $G^{\rm eff}$ measured in a local test
experiment is predicted to depend on the composition of the gravitational 
source.

The difference between active mass and passive mass represents a violation 
of the SEP. To see if this is ruled out from experiments testing the constancy
of $G^{\rm eff}$, one may consider laboratory experiments (e.g., Cavendish 
experiments) and space experiments in the solar system (e.g., lunar laser 
ranging). In the latter class of experiments, it is assumed that any variation 
in $G^{\rm eff}$ can be found simply by replacing the gravitational constant with
an independently variable $G^{\rm eff}$ in Newton's equations of motion [5]. But 
this cannot be done within the quasi-metric framework since it would be 
inconsistent (see equation (21) below). Moreover, the global NKE of 
gravitationally bound systems cannot be neglected when testing the temporal 
constancy of $G^{\rm eff}$. (For an explicit example of this, see [10].) Due 
to the global effects of the NKE, one must also separate between bodies bound 
by atomic forces and gravitationally bound bodies when calculating the time 
variation of $G^{\rm eff}$. In particular, when testing whether or not 
$G^{\rm eff}$ changes with time, according to quasi-metric theory, the result 
should depend on the specific experiment performed. For example, in Cavendish 
experiments, where the experimental system is bound by atomic forces, one 
should expect a different result than by trying to measure the variation with 
time in the acceleration of gravity on the surface of the Earth, say. (The 
former experiment is expected to indicate that $G^{\rm eff}$ increases with time 
whereas the latter experiment is expected to indicate that $G^{\rm eff}$ 
decreases with time.) Now laboratory experiments testing the constancy of 
$G^{\rm eff}$ are nowhere near the precision level attained from the 
corresponding space experiments. But reported results from space experiments
are not very useful as direct tests of the variability of $G^{\rm eff}$ as 
predicted by quasi-metric theory since the analyses and interpretations are 
theory-dependent. And laboratory experiments are not yet sufficiently precise 
to rule out the predicted variations of $G^{\rm eff}$.

We conclude that any difference between active and passive mass-energy is a 
violation of the SEP since there are locally measurable consequences of this. 
But the fact that the locally measured $G^{\rm eff}$ is variable only via source
composition and the variation of (atomic) units (in contrast to any 
``physical" variation in $G^{\rm eff}$ independent of said causes) is important 
regarding the interpretation of experiments. So far it seems that no experiment
has been capable of ruling out said variations of $G^{\rm eff}$.

The following postulate is consistent with the above considerations:
\begin{itemize}
{\item Gravity is described by the metric family ${\bf g}_t$ (constructible 
from another metric family ${\bf {\bar g}}_t$) measured in atomic units. 
Measured in atomic units, gravitational quantities get an extra formal 
variability depending on dimension. By definition $c$, ${\hbar}$ and $e$ are 
not formally variable. A consequence of this formal variability, is the 
necessity to introduce two different gravitational coupling parameters 
$G_t^{\rm B}$ and $G_t^{\rm S}$ that couple to electromagnetic mass-energy and 
material mass-energy, respectively. But to be able to define constants 
$G^{\rm B}$ and $G^{\rm S}$, it is necessary to separate between active and 
passive mass-energy. The formal variation of atomic units in space-time is by 
definition identical to the variation of the scalar field ${\Psi}_t$. 
How ${\Psi}_t$ varies in space-time depends on the fields ${\bar a}_{{\cal F}j}$, 
${\frac{{\bar N}_t,_{\bar {\perp}}}{{\bar N}_t}}$,
${\frac{{\bar N}_t,_t}{c{\bar N}_t^2}}$ and ${\frac{1}{c{\bar N}_tt}}$.}
\end{itemize}
To get any further, we must know the formal dependence of ${\Psi}_t$ on 
${\bar a}_{{\cal F}j}$, ${\frac{{\bar N}_t,_{\bar{\perp}}}{{\bar N}_t}}$,
${\frac{{\bar N}_t,_t}{{\bar N}_t}}$ and ${\frac{1}{t}}$. From equation (6),
we notice that the scale factor ${\bar N}_tct$ of the FHSs is measured in 
atomic units. This suggests that any change in this scale factor may be 
explained solely by means of the formal variation of units. We thus define 
${\Psi}_t{\equiv}{\frac{1}{c{\bar N}_tt}}$, so we have
\eqa
m_t=
\left\{
\begin{array}{ll}
{\frac{t}{t_0}}{\bar N}_tm_0 & $for a fluid of material particles,$
\\ [1.5ex]
{\frac{t^2}{t_0^2}}{\bar N}_t^2m_0 & $for the electromagnetic field,$
\end{array}
\right.
\ena
where $m_0$ denotes a reference value in an empty universe at epoch $t_0$.
Note that the Newtonian approximation of equation (6) (which is obtained by 
neglecting the shift covector, setting the spatial scale factor equal to unity
and replacing ${\tilde h}_{(t)ij}$ with an Euclidean 3-metric), is inconsistent 
with any variation in space-time of active mass. That is, in quasi-metric 
relativity, $G^{\rm eff}$ will depend on source composition, but otherwise 
{\em must} be a constant in the Newtonian limit.

Before we try to find gravitational field equations, we first find how the 
quasi-metric local conservation laws differ from their counterparts in
metric theory. In particular, the formal variability of active mass-energy must 
be included into the total active stress-energy tensor ${\bf T}_t$ considered 
as a source of gravity. That is, the covariant divergence
${\bf {\bar {\nabla}}}{\bf {\cdot}}{\bf T}_t$ (holding $t$ constant)
should not be expected to vanish in general due to the postulates
determining the role of gravity within the quasi-metric framework. 

It may seem natural to believe that local conservation of mass-energy and 
momentum is necessarily violated since
${\bf {\bar {\nabla}}}{\bf {\cdot}}{\bf T}_t$ may be
non-zero. However, given the possibilities of coupling ${\bf T}_t$ to the 
fields ${\bar a}_{{\cal F}j}$, ${\frac{{\bar N}_t,_{\bar {\perp}}}{{\bar N}_t}}$,
${\frac{{\bar N}_t,_t}{{\bar N}_t}}$ and ${\frac{1}{t}}$ such that 
LLI and LPI hold, it is in fact possible that local conservation laws within 
the quasi-metric framework do indeed take the form 
${\bf {\bar {\nabla}}}{\bf {\cdot}}{\bf T}_t{\neq}0$. Note that the formal 
variation of the units is taken care of both in derivatives of $m_t$ and in 
derivatives such as ${\bf {\bar {\nabla}}}{\bf {\cdot}}{\bf T}_t$. 

To be consistent with classical electrodynamics in quasi-metric space-time,
light rays in electrovacuum should be null geodesics both in ${\cal N}$ and in 
${\overline{\cal N}}$. A necessary condition to fulfil this, is that the 
passive electromagnetic stress-energy tensor ${\bar {\cal T}}_t^{{\rm (EM)}}
{\equiv}{\frac{t_0^2}{t^2}}{\bar N}_t^{-2}{\bf T}_t^{{\rm (EM)}}$ in 
${\overline{\cal N}}$ is covariantly conserved for electrovacuum. Requiring 
the resulting form of ${\bf {\bar {\nabla}}}{\bf {\cdot}}{\bf T}_t^{{\rm (EM)}}$ 
to hold for general ${\bf T}_t$, the local conservation laws must take the form
(for fixed $t$)
\eqa
T_{(t){\mu};{\nu}}^{\nu}=2{\frac{{\bar N}_{t,{\nu}}}{{\bar N}_t}}
T_{(t){\mu}}^{\nu}=2c^{-2}{\bar a}_{{\cal F}s}{\hat T}_{(t){\mu}}^s
-2{\frac{{\bar N}_{t,{\bar {\perp}}}}{{\bar N}_t}}T_{(t){\bar {\perp}}{\mu}},
\ena
or equivalently, by projecting these equations with respect to the FHSs
(see, e.g., [1] for general projection formulae),
\eqa
{\cal L}_{{\bf {\bar n}}_t}T_{(t){\bar {\perp}}{\bar {\perp}}}=
{\Big (}{\bar K}_t-2{\frac{{\bar N}_t,_{\bar {\perp}}}{{\bar N}_t}}
{\Big )}T_{(t){\bar {\perp}}{\bar {\perp}}}
+{\bar K}_{(t)sk}{\hat T}_{(t)}^{sk}-{\hat T}^s_{(t){\bar {\perp}}{\mid}s},
\nonumber \\
{\frac{1}{{\bar N}_t}}{\cal L}_{{\bar N}_t{\bf {\bar n}}_t}
T_{(t)j{\bar {\perp}}}={\Big (}{\bar K}_t
-2{\frac{{\bar N}_t,_{\bar {\perp}}}{{\bar N}_t}}
{\Big )}T_{(t)j{\bar {\perp}}}-c^{-2}{\bar a}_{{\cal F}j}
T_{(t){\bar {\perp}}{\bar {\perp}}}
+c^{-2}{\bar a}_{{\cal F}s}{\hat T}_{(t)j}^s-{\hat T}^s_{(t)j{\mid}s},
\ena
where ${\bar K}_t{\equiv}{\bar K}_{(t)s}^s$. Moreover, if the only 
$t$-dependence of ${\bf T}_t$ is via the above mentioned formal variability, 
${\bf T}_t$ is locally conserved when $t$ varies as well. That is, for the 
non-metric part of the connection, i.e., as a counterpart to equation (22), an 
extra local ``conservation law'' can be found from the condition
${\stackrel{{\hbox{\tiny$\star$}}}{\bf {\bar {\nabla}}}}
_{{\!}{\!}{\frac{\partial}{{\partial}t}}}({\frac{t_0^2}{t^2}}{\bar N}_t^{-2}T_{(t){\mu}}^0)
=0$. This yields
\eqa
T_{(t){\mu}{\bar *}t}^0=-{\frac{2}{{\bar N}_t}}{\Big (}{\frac{1}{t}}+
{\frac{{\bar N}_t,_t}{{\bar N}_t}}{\Big )}T_{(t){\bar {\perp}}{\mu}},
\ena
where the symbol `${\bar *}$' denotes a degenerate covariant derivative 
compatible with the metric family ${\bf {\bar g}}_t$. Notice that, by applying 
equations (22) and (24) to a source consisting of a perfect fluid with no 
pressure (i.e., dust), and projecting the resulting equations with the quantity
${\bf {\bar g}}_t+c^{-2}{\bf {\bar u}}_t{\otimes}{\bf {\bar u}}_t$, one may
show that the dust particles move on geodesics of 
${\,}{\,}{\topstar{\bar {\nabla}}}{\,}{\,}$ in ${\overline{\cal N}}$.

Equation (23) determines the evolution in the hypersurface-orthogonal
direction of the matter density and the momentum density in 
${\overline{\cal M}}$. To find these evolutions in ${\overline{\cal N}}$, we
add the dependence on $t$. Using the projected Lie derivative 
${\topstar{\cal L}}_{{\,}{\bf {\bar n}}_t}$ valid in ${\overline{\cal N}}$, we find
\eqa
{\topstar{\cal L}}_{{\,}{\bf {\bar n}}_t}
T_{(t){\bar {\perp}}{\bar {\perp}}}=
{\cal L}_{{\bf {\bar n}}_t}T_{(t){\bar {\perp}}{\bar {\perp}}}
-{\frac{2}{c{\bar N}_t}}{\Big (}{\frac{1}{t}}+
{\frac{{\bar N}_t,_t}{{\bar N}_t}}{\Big )}T_{(t){\bar {\perp}}{\bar {\perp}}},
\ena
\eqa
{\frac{1}{{\bar N}_t}}{\topstar{\cal L}}_{{\bar N}_t{\bf {\bar n}}_t}
T_{(t){\bar {\perp}}j}={\frac{1}{{\bar N}_t}}
{\cal L}_{{\bar N}_t{\bf {\bar n}}_t}T_{(t){\bar {\perp}}j}
-(c{\bar N}_t)^{-1}[{\frac{1}{t}}+{\frac{{\bar N}_t,_t}{{\bar N}_t}}]
T_{(t){\bar {\perp}}j}.
\ena
Since ${\bf T}_t$ is the total active stress-energy tensor, it is not directly
measurable locally. Consequently, one needs to know how it relates to the total 
{\em passive stress-energy tensor} ${\cal T}_t$ in ${\cal N}$. Now the 
relationship between ${\bf T}_t$ and ${\cal T}_t$ depends on the nature of the
gravitational source. To illustrate this, consider ${\bf T}_t$ and ${\cal T}_t$
for a perfect fluid:
\eqa
{\bf T}_t=({\tilde {\varrho}}_{\rm m}+c^{-2}{\tilde p}){\bf {\bar u}}_t
{\otimes}{\bf {\bar u}}_t+{\tilde p}{\bf {\bar g}}_t, \qquad
{\cal T}_t=({\hat {\varrho}}_{\rm m}+c^{-2}{\hat p}){\bf u}_t{\otimes}{\bf u}_t
+{\hat p}{\bf g}_t,
\ena
where ${\tilde {\varrho}}_{\rm m}$ and ${\tilde p}$ are the active mass-energy 
density and the pressure, respectively, in the local rest frame of the source. 
Here, we have used the definitions ${\varrho}_{\rm m}{\sqrt{{\bar h}_t}}{\equiv}
{\hat {\varrho}}_{\rm m}{\sqrt{h_t}}$ and $p{\sqrt{{\bar h}_t}}{\equiv}
{\hat p}{\sqrt{h_t}}$, where ${\varrho}_{\rm m}$ and $p$ are the passive 
counterparts to ${\tilde {\varrho}}_{\rm m}$ and ${\tilde p}$, respectively. 
Also, ${\bar h}_t$ and $h_t$ are defined to be the determinants of 
${\bf {\bar h}}_t$ and ${\bf h}_t$, respectively. The relationship between 
${\bf T}_t$ and ${\cal T}_t$ depends on source composition since the 
relationship between ${\tilde {\varrho}}_{\rm m}$ and ${\varrho}_{\rm m}$ is 
given by
\eqa
{\tilde {\varrho}}_{\rm m}=
\left\{
\begin{array}{ll}
{\frac{t}{t_0}}{\bar N}_t{\varrho}_{\rm m} & $for a fluid of material 
particles,$
\\ [1.5ex]
{\frac{t^2}{t_0^2}}{\bar N}_t^2{\varrho}_{\rm m} & $for the electromagnetic 
field,$
\end{array}
\right.
\ena
and a similar relationship exists between ${\tilde p}$ and $p$. Note that 
more ``physically correct'' local conservation laws can be found by calculating 
${\nabla}{\cdot}{\cal T}_t$ when ${\bf g}_t$ is known. But these local 
conservation laws for passive mass-energy do not take any predetermined form.

To find field equations, we must take into account that the active 
electromagnetic stress-energy tensor ${\bf T}^{\rm (EM)}_t$ and the active 
stress-energy tensor for material sources ${\bf T}^{\rm (MA)}_t$ couple 
differently to space-time geometry. Furthermore, the requirement that we must 
have metric correspondence with Newtonian theory (and GR) for weak stationary 
fields, leads to the postulate (${\kappa}^{\rm B}{\equiv}8{\pi}G^{\rm B}/c^4$,
${\kappa}^{\rm S}{\equiv}8{\pi}G^{\rm S}/c^4$)
\eqa
2{\bar R}_{(t){\bar {\perp}}{\bar {\perp}}}=
2(c^{-2}{\bar a}_{{\cal F}{\mid}s}^s+
c^{-4}{\bar a}_{{\cal F}s}{\bar a}_{\cal F}^s-
{\bar K}_{(t)ks}{\bar K}_{(t)}^{ks}+{\cal L}_{{\bf {\bar n}}_t}{\bar K}_t)
\nonumber \\
={\kappa}^{\rm B}(T^{\rm (EM)}_{(t){\bar {\perp}}{\bar {\perp}}}+
{\hat T}_{(t)s}^{{\rm (EM)}s})+
{\kappa}^{\rm S}(T^{\rm (MA)}_{(t){\bar {\perp}}{\bar {\perp}}}+
{\hat T}_{(t)s}^{{\rm (MA)}s}),
\ena
where ${\bf {\bar R}}_t$ is the Ricci tensor family corresponding to the metric
family (6) and ${\bf {\bar K}}_t$ is the extrinsic curvature tensor family of 
the FHSs in ${\overline{\cal M}}$. The need to have consistency of equation 
(23) with the contracted Bianchi identities in the weak field limit, leads to 
the postulate of a second set of field equations having some correspondence 
with GR, namely 
\eqa
{\bar R}_{(t)j{\bar {\perp}}}+{\Big (}{\frac{{\bar h}_{(t)}^{ik}}{{\bar N}_t}}
{\frac{\partial}{{\partial}x^0}}{\bar h}_{(t)ij}{\Big )}_{{\mid}k}-
{\Big (}{\frac{{\bar h}_{(t)}^{ik}}{{\bar N}_t}}
{\frac{\partial}{{\partial}x^0}}{\bar h}_{(t)ik}{\Big )},_j
={\kappa}^{\rm B}T^{{\rm (EM)}}_{(t)j{\bar {\perp}}}
+{\kappa}^{\rm S}T^{\rm (MA)}_{(t)j{\bar {\perp}}}.
\ena
The field equations are completed by requiring that the traceless quantity 
${\bar Q}_{(t)ij}$ defined from the relationship
\eqa
{\bar G}_{(t)ij}{\equiv}-{\bar Q}_{(t)ij}-2c^{-2}{\bar a}_{{\cal F}i{\mid}j}-
2c^{-4}{\bar a}_{{\cal F}i}{\bar a}_{{\cal F}j}
-2{\bar K}_{(t)i}^{s}{\bar K}_{(t)sj}+2{\bar K}_t{\bar K}_{(t)ij} 
\nonumber \\
+{\frac{1}{3}}{\Big [}2{\bar R}_{(t){\bar {\perp}}{\bar {\perp}}}
-{\bar G}_{(t){\bar {\perp}}{\bar {\perp}}}+2c^{-2}{\bar a}_{{\cal F}{\mid}s}^s
+2c^{-4}{\bar a}_{{\cal F}}^s{\bar a}_{{\cal F}s}
+2{\bar K}_{(t)ks}{\bar K}_{(t)}^{ks}-2{\bar K}_t^2{\Big ]}{\bar h}_{(t)ij},
\ena
should vanish, where ${\bf {\bar G}}_t$ is the Einstein tensor family found 
from equation (6). Equation (31), in addition to the general expression (29) 
valid for the projection $2{\bar R}_{(t){\bar {\perp}}{\bar {\perp}}}$ and general 
expressions similar to those shown in equation (5) valid for 
${\bar G}_{(t){\bar {\perp}}{\bar {\perp}}}$ and ${\bar G}_{(t)ij}$, then yield the 
remainder quasi-metric field equation
\eqa
{\bar Q}_{(t)ij}{\equiv}{\frac{1}{{\bar N}_t}}{\cal L}_{{\bar N}_t
{\bf {\bar n}}_t}{\bar K}_{(t)ij}+
{\frac{1}{3}}{\Big [}2{\bar K}_{(t)ks}{\bar K}_{(t)}^{ks}-{\bar K}_t^2
-{\cal L}_{{\bf {\bar n}}_t}{\bar K}_t
{\Big ]}{\bar h}_{(t)ij}+{\bar K}_t{\bar K}_{(t)ij} \nonumber \\
-c^{-2}{\bar a}_{{\cal F}i{\mid}j}-
c^{-4}{\bar a}_{{\cal F}i}{\bar a}_{{\cal F}j}
+{\Big [}c^{-2}{\bar a}_{{\cal F}{\mid}s}^s
-{\frac{1}{(ct{\bar N}_t)^2}}{\Big ]}{\bar h}_{(t)ij}-{\bar H}_{(t)ij}=0,
\ena
where the prior-geometric requirement on the spatial Ricci curvature scalar 
family ${\bar P}_t$,
\eqa
{\bar P}_t=-4c^{-2}{\bar a}_{{\cal F}{\mid}s}^s
+2c^{-4}{\bar a}_{{\cal F}}^s{\bar a}_{{\cal F}s}+{\frac{6}{(ct{\bar N}_t)^2}},
\ena
ensures that equation (32) is indeed manifestly traceless. In equation (32), 
${\bf {\bar H}}_t$ denotes the spatial Einstein tensor family intrinsic to the 
FHSs.

Several comments apply to the field equations (29), (30) and (32). Firstly, 
apart from the dependence on source composition, equation (29) (and 
to a lesser degree, equation (30)) superficially looks somewhat like a subset 
of the Einstein field equations in GR. But the crucial difference is 
that the equations (29)-(33) are exactly valid only for projections with 
respect to the FHSs; for projections with respect to any other hypersurfaces 
they do not hold exactly. (That is, said equations are not invariant when 
transformed with respect to alternative foliations of ${\bf {\bar g}}_t$. This 
must be taken into account when dealing with isolated systems.) Secondly, 
equations (29) and (32) are dynamical equations whereas equations (30) and (33)
represent constraints. We thus have six (independent) dynamical equations and 
four constraint equations, just as for GR. This means that the dynamical 
structure of our field equations is somewhat similar to that of GR. That is, 
both equation sets accommodate two propagating dynamical degrees of freedom. 
Thirdly, the local conservation laws do not follow from the field equations as 
they do in GR. Rather, equation (33) ensures that there will be no gauge 
freedom in choosing lapse and shift. That is, quasi-metric space-time must be
foliated into a unique set of FHSs.
\section*{4{\hspace*{9mm}} Kinematics and 
non-kinematics of the FHSs}
In this section, we take a closer look at the evolution with time of the 
geometry of the FHSs in ${\overline{\cal N}}$ and in ${\cal N}$. To begin with,
we may {\em define} the split-up into kinematical and non-kinematical terms 
${\bar x}_t$ and ${\bar H}_t$, respectively, of the time evolution of the 
spatial scale factor ${\bar F}_t{\equiv}{\bar N}_tct$ occurring in equation 
(6). We may thus define
\eqa
c^{-2}{\bar x}_t+c^{-1}{\bar H}_t{\equiv}
(c{\bar N}_t)^{-1}{\Big (}{\frac{1}{t}}+{\frac{{\bar N}_t,_t}{{\bar N}_t}}
{\Big )}-{\frac{{\bar N}_t,_{\bar {\perp}}}{{\bar N}_t}}, 
\ena
where ${\bar H}_t$ is a quantity associated with the NKE of the FHSs. (We 
define ${\bar H}_t$ in equation (35) below.) 

In the introduction, we defined the NKE as a local increase of scale with time
of the FHSs. By definition, the local change of scale with time as 
defined by nearby FOs is entirely expressible as a change in ${\bar F}_t$. 
Thus we may define the non-kinematical scale rate ${\bar H}_t$ by
\eqa
{\bar H}_t{\equiv}
c{\bar F}_t^{-1}{\topstar{\cal L}}_{{\bf {\bar n}}_t}{\bar F}_t-
c^{-1}{\bar x}_t.
\ena
Note that we have subtracted the kinematical contribution 
$c^{-1}{\bar x}_t$ to the scale change in equation (35). Since ${\bar H}_t$ 
is a purely non-kinematical quantity, it should be available directly from the
geometry intrinsic to the FHSs. We argue in [2] that the correct form of this
equation is given by
\eqa
{\bar H}_t={\frac{1}{{\bar N}_tt}}+
c^{-1}{\sqrt{{\bar a}_{\cal F}^s{\bar a}_{{\cal F}s}}}.
\ena
Besides, we see from equation (19) that ${\bar y}_t$ represents the 
non-kinematical part of the evolution of ${\bar N}_t$ with time. Combining 
equations (19), (34) and (36), we find that
\eqa
{\bar y}_t={\bar H}_t-{\frac{1}{{\bar N}_tt}}=
c^{-1}{\sqrt{{\bar a}_{\cal F}^s{\bar a}_{{\cal F}s}}},
\ena
\eqa
c^{-2}{\bar x}_t=-{\frac{{\bar N}_t,_{\bar {\perp}}}{{\bar N}_t}}+
(c{\bar N}_t)^{-1}{\Big (}{\frac{1}{t}}+{\frac{{\bar N}_t,_t}{{\bar N}_t}}
{\Big )}-c^{-1}{\bar H}_t=
-{\frac{{\bar N}_t,_{\bar {\perp}}}{{\bar N}_t}}+
c^{-1}{\frac{{\bar N}_t,_t}{{\bar N}^2_t}}-c^{-2}
{\sqrt{{\bar a}_{\cal F}^s{\bar a}_{{\cal F}s}}}.
\ena
For cases where there is no coupling to non-gravitational fields via the
equations of motion (15) in ${\cal N}$, we now have a well-defined 
initial-value system for the metric family ${\bf {\bar h}}_t$ (or, 
equivalently, for ${\bf {\tilde h}}_t$ and ${\bar N}_t$) in 
${\overline{\cal N}}$: on an initial FHS we choose initial data ${\bar N}_t$, 
${\tilde h}_{(t)ij}$, ${\bar K}_{(t)ij}$, ${\bar N}_{(t)}^j$ and the various 
projections of ${\bf T}^{\rm (EM)}_t$ and ${\bf T}^{\rm (MA)}_t$ which satisfy the 
constraint equations (30) and (33). The evolutions ${\bar N}_t,_t$ and 
${\tilde h}_{(t)ij,t}$ are inferred separately at each time step from the effect 
of the cosmic expansion on ${\bf T}_t$; e.g., if equation (24) is violated, 
this will affect said evolutions. Moreover, ${\bar N}_{(t),t}^j$ follows from 
${\tilde h}_{(t)ij,t}$, since, as a counterpart to equation (9), the criterion 
${\,}{\topstar{{\bar \nabla}}}_{\frac{\partial}{{\partial}t}}{\bf {\bar n}}_t=0$ 
yields the condition
\eqa 
{\frac{\partial}{{\partial}t}}{\Big [}{\bar N}_{(t)}^k{\bar N}_{(t)}^s
{\tilde h}_{(t)ks}{\Big ]}=0, \qquad {\Rightarrow} \qquad
{\bar N}_{(t),t}^s=-{\frac{1}{2}}{\bar N}_{(t)}^k
{\tilde h}_{(t)}^{is}{\tilde h}_{(t)ik,t}.
\ena
We now evolve ${\bf {\bar h}}_t$ through ${\overline{\cal N}}$ using equations
(20) and (29)-(33). For each step of the evolution, we must supply the quantity
${\bar R}_{(t){\bar {\perp}}{\bar {\perp}}}$ to be able to construct ${\bar K}_{(t)ij}$
for each subsequent step. Then the non-kinematical quantities follow 
automatically via equations (36) and (37). Note that the local conservation 
laws (23), (25) and (26) are coupled to the field equations since they 
represent non-trivial evolution equations of $T_{(t){\bar {\perp}}{\bar {\perp}}}$ and
$T_{(t){\bar {\perp}}j}$. But to know the time evolution of the full stress-energy 
tensor ${\bf T}_t$, we need to know how it splits up into separate 
contributions from ${\bf T}^{\rm (EM)}_t$ and ${\bf T}^{\rm (MA)}_t$. Besides, we 
need to know the equation of state characterizing the gravitational source; 
i.e., how the passive mass-energy is related to the passive pressure.

Note that, for the initial-value system defined by the equations (4) and (5), 
it is possible to choose the lapse function $N$ freely throughout space-time 
since the evolution of $N$ is not determined by the constraint preservation 
[8]. On the other hand, for our initial-value system, ${\bar N_t}$ is 
constrained by equation (33), so it cannot be chosen freely. This is a 
consequence of the fact that equations (32) and (33) depend on the 
distinguished foliation, so that ${\bar N}_t$ and ${\bar N}_{(t)}^j$ represent 
lapse and shift for a particular set of observers; the evolution of the FHSs 
is not independent of how ${\bar N}_t$ and ${\bar N}_{(t)}^j$ evolve with time 
and there is no arbitrariness of choice involved.

But in general, the evolution scheme in ${\overline{\cal N}}$ is coupled to the
evolution scheme in $\cal N$ via the dynamical evolution of non-gravitational 
fields. This means that these evolution schemes are intertwined, and that we 
we must construct ${\bf g}_t$ for each step of the evolution in order to 
calculate the dynamics of said non-gravitational fields. The construction of
${\bf g}_t$ can be done by finding the family of 3-vector fields $-{\bf v}_t$ 
which tells how much the FOs in ${\overline{\cal N}}$ ``move'' compared to the 
FOs in $\cal N$. Expressed in a GTCS, said vector field may be defined as [2]
\eqa
v^{j}_{(t)}{\equiv}{\bar y}_t{\bar x}_{\cal F}^j, \qquad
v{\equiv}{\bar y}_t
{\sqrt{{\bar h}_{(t)ks}{\bar x}_{\cal F}^k{\bar x}_{\cal F}^s}},
\ena
where ${\bf {\bar x}}_{\cal F}$ is a 3-vector field representing ``the local 
distance from the centre of gravity'' generalized from the spherically 
symmetric case. Furthermore, ${\bf {\bar x}}_{\cal F}$ can be found from the 
equation
\eqa
{\Big [}{\bar a}_{{\cal F}{\mid}k}^k+c^{-2}{\bar a}_{{\cal F}k}
{\bar a}_{\cal F}^k{\Big ]}{\bar x}_{\cal F}^j-{\Big [}
{\bar a}_{{\cal F}{\mid}k}^j+c^{-2}{\bar a}_{{\cal F}k}{\bar a}_{\cal F}^j
{\Big ]}{\bar x}_{\cal F}^k-2{\bar a}_{\cal F}^j=0,
\ena
which has the unique solution ${\bf {\bar x}}_{\cal F}=
r{\frac{\partial}{{\partial}r}}$ for the spherically symmetric case (using a
spherical GTCS with a Schwarzschild radial coordinate $r$). This means that we 
intuitively expect ${\bf {\bar x}}_{\cal F}$ to vanish when 
${\bf {\bar a}}_{\cal F}$ does.

From equations (36) and (40), we see that the construction of ${\bf g}_t$ from 
${\bf {\bar g}}_t$ involves that part of the NKE which is not ``realized'' 
explicitly in the evolution parameter ${\bar H}_t$. That is, the quantity
${\bar y}_t$ given from equation (37) represents the {\em local} NKE 
of the FHSs and it is not included in the explicit time evolution of 
${\bar F}_t$ as can be seen from equations (35), (36) and (38). Note that $v$ 
should not depend explicitly on $t$. In [2], it is shown how to construct 
${\bf g}_t$ (or, equivalently ${\bf h}_t$, $N$ and $N_{(t)}^j$) algebraically 
from ${\bf {\bar g}}_t$ (or, equivalently, from ${\bf {\bar h}}_t$, 
${\bar N}_t$ and ${\bar N}_{(t)}^j$) and the vector field family ${\bf v}_t$ for
each step of the evolution; thus we also have a well-defined initial-value 
scheme in $\cal N$. 

Actually, transformations similar to ${\bf {\bar g}}_t{\rightarrow}{\bf g}_t$
apply to any tensor field which norm is required to remain unchanged when
${\bf {\bar g}}_t{\rightarrow}{\bf g}_t$. As an example, we list the
formulae valid for the transformation ${\bf {\bar Z}}_t{\rightarrow}{\bf Z}_t$,
where ${\bf {\bar Z}}_t$ is a rank 1 tensor field family. These transformation
formulae involve the unit vector field ${\bf {\bar e}}_{\cal F}{\equiv}
{\frac{t_0}{t}}{\bar e}_{\cal F}^s{\frac{\partial}{{\partial}x^s}}$ along the 
${\bf {\bar x}}_{\cal F}$-direction in ${\overline {\cal N}}$ and the 
corresponding covector field ${\bf {\bar {\omega}}}_{\cal F}{\equiv}
{\frac{t}{t_0}}{\bar {\omega}}_{{\cal F}s}dx^s$. The transformation formulae read
\eqa
Z_{(t)0}={\Big (}1-{\frac{v^2}{c^2}}{\Big )}{\bar Z}_{(t)0}, \qquad
Z_{(t)}^0={\Big (}1-{\frac{v^2}{c^2}}{\Big )}^{-1}{\bar Z}_{(t)}^0,
\ena
\eqa
Z_{(t)j}={\bar Z}_{(t)j}+{\frac{2{\frac{v}{c}}}{1-{\frac{v}{c}}}}
({\bar e}_{\cal F}^s{\bar Z}_{(t)s}){\bar {\omega}}_{{\cal F}j}, \qquad
{\hat Z}_{(t)}^j={\hat {\bar Z}}_{(t)}^j-
{\frac{2{\frac{v}{c}}}{1+{\frac{v}{c}}}}
({\bar {\omega}}_{{\cal F}s}{\hat {\bar Z}}_{(t)}^s){\bar e}^j_{\cal F}.
\ena
Note that it is also possible to find analogous formulae (involving the unit 
vector field ${\bf e}_{\cal F}{\equiv}{\frac{t_0}{t}}e_{\cal F}^s
{\frac{\partial}{{\partial}x^s}}$ in the ${\bf {\bar x}}_{\cal F}$-direction and 
its associated covector field 
${\bf {\omega}}_{\cal F}{\equiv}{\frac{t}{t_0}}{\omega}_{{\cal F}s}dx^s$ in 
$\cal N$) for the inverse transformation 
${\bf Z}_t{\rightarrow}{\bf {\bar Z}}_t$.

One example of a (co)vector field obeying the rules (42)-(43) is a general 
4-velocity (co)vector family ${\bf {\bar u}}_t$. Another important example
is any 4-acceleration field family ${\bf {\bar a}}_t$ determined from a
non-gravitational force (e.g., if ${\bf {\bar a}}_t$ is due to the Lorentz 
force acting on charged matter). For such cases the norm of
${\bf {\bar a}}_t$ must be invariant under the transformation. In particular 
${\bf {\bar a}}_t$ may vanish, and this means that {\em geodesic motion in 
${\overline{\cal N}}$ implies geodesic motion in ${\cal N}$.} Note that, 
${\bf {\bar a}}_{\cal F}$ does not in general transform according to the rules
(42)-(43), since ${\bf {\bar a}}_{\cal F}$ is determined from the requirement 
that the FOs move normal to the FHSs rather from some non-gravitational force.

For completeness, we also list the transformation formulae valid for the
transformation ${\bf {\bar W}}_t{\rightarrow}{\bf W}_t$, where 
${\bf {\bar W}}_t$ is a rank 2 tensor field family. We have
\eqa
W_{(t)00}={\Big (}1-{\frac{v^2}{c^2}}{\Big )}^2{\bar W}_{(t)00}, \qquad
W_{(t)}^{00}={\Big (}1-{\frac{v^2}{c^2}}{\Big )}^{-2}{\bar W}_{(t)}^{00},
\ena
\eqa
W_{(t)0j}&=&{\Big (}1-{\frac{v^2}{c^2}}{\Big )}{\Big [}{\bar W}_{(t)0j}+
{\frac{2{\frac{v}{c}}}{1-{\frac{v}{c}}}}
({\bar e}_{\cal F}^s{\bar W}_{(t)0s}){\bar {\omega}}_{{\cal F}j}{\Big ]}, 
\nonumber \\
{\hat W}_{(t)}^{0j}&=&{\Big (}1-{\frac{v^2}{c^2}}{\Big )}^{-1}{\Big [}
{\hat {\bar W}}_{(t)}^{0j}-{\frac{2{\frac{v}{c}}}{1+{\frac{v}{c}}}}
({\bar {\omega}}_{{\cal F}s}{\hat {\bar W}}_{(t)}^{0s}){\bar e}^j_{\cal F}
{\Big ]},
\ena
\eqa
W_{(t)ij}&=&{\bar W}_{(t)ij}+{\frac{2{\frac{v}{c}}}{(1-{\frac{v}{c}})^2}}
{\bar e}_{\cal F}^k({\bar {\omega}}_{{\cal F}i}{\bar W}_{(t)kj}+
{\bar W}_{(t)ik}{\bar {\omega}}_{{\cal F}j}), 
\nonumber \\
{\hat W}_{(t)}^{ij}&=&{\hat {\bar W}}_{(t)}^{ij}-
{\frac{2{\frac{v}{c}}}{(1+{\frac{v}{c}})^2}}
{\bar {\omega}}_{{\cal F}k}({\bar e}_{\cal F}^i{\hat {\bar W}}_{(t)}^{kj}+
{\hat {\bar W}}_{(t)}^{ik}{\bar e}_{\cal F}^j).
\ena
The transformation formulae shown in equations (44)-(46) may in particular
be used to find ${\bf g}_t$ from ${\bf {\bar g}}_t$; moreover they may easily 
be generalized to tensor field families of higher rank.
\section*{5{\hspace*{9mm}} Test particle motion 
revisited}
Once we know ${\bf g}_t$, we may calculate paths of arbitrary test
particles using the equations of motion (15). But before we are able to do 
such calculations, we need to have an independent expression for the degenerate
acceleration field ${\stackrel{\star}{{\bf a}}}$. In this section, we will see 
that we in fact have ${\stackrel{\star}{\bf a}}$=${\bf a}_t$.

In what follows, it is convenient to introduce the 3-velocity ${\bf w}_t$ 
(and its square $w^2$) of an arbitrary test particle relative to that of the 
local FO. In particular this is a useful quantity along the world line of the 
test particle. Thus we define (using a GTCS)
\eqa
w_{(t)}^j{\equiv}{\frac{dx^j}{d{\tau}_{\cal F}}}+{\frac{t_0}{t}}
{\frac{N^j_{(t)}}{N}}c, {\qquad}
{\topstar{\gamma}}{\equiv}(1-c^{-2}w_{(t)}^iw_{(t)i})^{-1/2}=
{\frac{d{\tau}_{\cal F}}{d{\tau}_t}},
\ena
where ${\tau}_t$ is the proper time as measured along the curve. Note that
${\bf w}_t$ is an object intrinsic to the FHSs and that the $t$-dependence of 
${\bf w}_t$ is given from the condition [2]
\eqa
{\frac{\partial}{{\partial}t}}w^2{\equiv}
(w_{(t)i}w_{(t)k}h_{(t)}^{ik}),_t=0, \qquad \Rightarrow \qquad
{\frac{\partial}{{\partial}t}}w_{(t)j}={\frac{1}{t}}w_{(t)j}
+{\frac{1}{2}}w_{(t)k}{\hat h}_{(t)}^{sk}{\hat h}_{(t)sj},_t.
\ena
One may readily show that the tangent 4-velocity family ${\bf u}_t$ may be 
split up into parts respectively normal and tangential to the FHSs, i.e.,
\eqa
{\bf u}_t={\topstar{\gamma}}(c{\bf n}_t+{\bf w}_t).
\ena
Since any 4-velocity vector ${\bf {\bar u}}_t$ in ${\overline{\cal N}}$
by construction has constant norm, we may locally calculate the transformation
${\bf {\bar u}}_t{\rightarrow}{\bf u}_t$ from equations (42) and (43). We may
then show that the norm ${\bar w}$ of the 3-velocity family
${\bf {\bar w}}_t$ (and thus ${\topstar{\gamma}}$) is invariant under the 
transformation, i.e., ${\bar w}=w$.

Using equations (48) and (49), it is now possible to calculate the quantity
${\,}{\,}{\topstar{\nabla}}_{\frac{\partial}{{\partial}t}}{\bf u}_t$ and
thereby the degenerate acceleration ${\stackrel{\star}{\bf a}}$ from
equation (14). That is, using equation (11) it is straightforward to show that
\eqa
{\stackrel{\star}{\nabla}}_{\frac{\partial}{{\partial}t}}{\bf w}_t=0,
\qquad \Rightarrow \qquad
{\stackrel{\star}{\nabla}}_{\frac{\partial}{{\partial}t}}{\bf u}_t=0,
\qquad \Rightarrow \qquad {\stackrel{\star}{\bf a}}={\bf a}_t.
\ena
The coordinate expression (15) for the equations of motion then takes the form
(this form is valid in a general coordinate system, but equation (10) is valid 
only in a GTCS)
\eqa
{\frac{d^2x^{\mu}}{d{\lambda}^2}}+{\Big(}
{\topstar{\Gamma}}^{\mu}_{t{\sigma}}{\frac{dt}{d{\lambda}}}+
{\Gamma}^{\mu}_{(t){\beta}{\sigma}}{\frac{dx^{\beta}}{d{\lambda}}}{\Big)}
{\frac{dx^{\sigma}}{d{\lambda}}} 
={\Big(}{\frac{cd{\tau}_t}{d{\lambda}}}{\Big)}^2c^{-2}a_{(t)}^{\mu}.
\ena
For timelike curves, one may parametrize the curve by proper time by
choosing ${\lambda}=c{\tau}_t$. For null curves, ${\lambda}$ must be chosen as
some other affine parameter such that ${\frac{cd{\tau}_t}{d{\lambda}}}=0$. 

The equations of motions (51) are the geodesic equations obtained from 
${\,}{\topstar{\nabla}}{\,}$. Moreover, it can be readily shown that 
equation (51) does not reduce to the geodesic equation (16) whenever the 
dependence on $t$ of the metric components cannot be neglected. That is, if 
the global part ${\frac{1}{{\bar N}_tt}}$ of the NKE (see equation (36)) is 
not negligible, inertial test particles do not move as if they were following 
geodesics of any single space-time metric.
\section*{6{\hspace*{9mm}} Discussion}
As we have seen in this survey, the realization that physical space should not
necessarily be treated as a purely kinematical field may have important 
theoretical consequences; this idea leads ultimately not merely to a new 
theory of gravity but to a new geometrical framework as the basis for 
relativistic physics. One of the reasons that makes the possible replacement 
of the currently prevailing metric framework with our alternative even 
remotely probable, is the fact that this can be done without sacrificing the 
EEP. 

Even though our theory seems well-posed and self consistent, the
question is still open if it is viable according to experimental criteria.
We show in [2] that it is at least possible to find certain limits where the 
predictions of our theory match those of GR within experimental precision.
Particular cases here are static systems with spherical symmetry; we show that 
by assuming that the {\em global} effects of the NKE are negligible,
it is possible to find a one-parameter family ${\bf {\bar g}}_t$ of identical, 
conformally flat metrics as a vacuum solution of our field equations. We
then include the {\em local} effects of the NKE into the family ${\bf g}_t$ 
which we construct from the conformally flat metrics. That is, when the global 
effects of the NKE can be neglected, the metric families ${\bf {\bar g}}_t$ and
${\bf g}_t$ can be represented by single metrics ${\bf {\bar g}}$ and ${\bf g}$
since there is no explicit dependence on $t$ in this case. Moreover, for the 
static, spherically symmetric case it turns out that ${\bf g}$ agrees with the 
Schwarzschild metric to sufficient order in the quantity $r_{\rm s}/r$, where 
$r_{\rm s}$ is the (generalized) Schwarzschild radius. However, there is 
disagreement for higher orders since neither ${\bf {\bar g}}$ nor ${\bf g}$
possesses an event horizon. This illustrates the fact that the geometrical 
structure of quasi-metric space-time does not allow the formation of black 
holes. Furthermore, we easily find a toy cosmological solution of our field 
equations; a family of metrics with the geometry ${\bf S}^3{\times}{\bf R}$ 
constitutes a vacuum solution of the field equations. (In this case the 
families ${\bf {\bar g}}_t$ and ${\bf g}_t$ are identical.) This indicates 
that our theory does not allow any physical singularities at all, since even 
at the cosmological singularity, no physical quantities necessarily diverge. 
(But we need a matter creation mechanism for a Universe with an empty 
beginning, this may perhaps be constructed along the lines described in [9].) 
Moreover, no counterparts to the Friedmann-Robertson-Walker models exist in 
quasi-metric theory; besides the empty toy model, the only allowed isotropic 
cosmological solutions of the field equations represent universes filled with 
a null fluid. This means that the usual cosmological parameters are not useful 
in quasi-metric cosmology. 

Another prediction of our theory is that the use of Newtonian dynamics for 
weak field systems has its limitations; in addition to the weak field 
criterion, systems should be ``small", so that the {\em global} effects of the 
NKE (i.e., the effects due to a ``real" expansion) can be neglected. (That is, 
even if these global effects are small locally, over large distances or long 
time spans they may not be small.) We explore some of the global effects of the
NKE in [10], where we discuss the possible relevance of the results for 
explaining the observed asymptotically non-Keplerian rotation curves of spiral 
galaxies. Furthermore, even in the solar system there may be observable 
consequences of the NKE. In [11] we discuss one such effect and compare to data
the predictions of the model considered in [10]. In addition, we show in [12] 
that the global effects of the NKE are predicted to apply inside 
gravitationally bound, perfect fluid bodies as well. In particular, 
quasi-metric theory predicts that the radii of bodies made of ideal gas should 
not be constant in time but rather change similarly to the cosmological scale 
factor. This model can be compared to data on expansion of the Earth. Also 
possible is comparison of the model to data on spin-down of the Earth's 
rotation rate.

A lot of work remains to be done before one can decide whether or not our 
theory is viable. For example, it is necessary to work out if it correctly
models binary pulsars, quasi-stellar objects, cosmological observations etc.
Should our theory turn out to be viable, it is believable that its
possible acceptance hinges on if it is able to explain the workings of the 
Universe from first principles better than competing theories. Only then will
the NKE get the status as a fundamental principle realized by nature.
\\ [4mm]
{\bf References} \\ [1mm]
{\bf [1]} K. Kucha{\v{r}}, {\em J. Math. Phys.} {\bf 17}, 792 (1976). \\
{\bf [2]} D. {\O}stvang, chapter 2 of this thesis (2001). \\
{\bf [3]} C. Gerhardt, {\em Commun. Math. Phys.} {\bf 89}, 523 (1983). \\
{\bf [4]} C.W. Misner, K.S. Thorne, J.A. Wheeler,
{\em Gravitation}, W.H. Freeman ${\&}$ Co. (1973). \\
{\bf [5]} C.M. Will, {\em Theory and Experiment in Gravitational Physics}, 
Cambridge University {\hspace*{7mm}}Press (1993). \\
{\bf [6]} C.H. Brans and R.H. Dicke, {\em Phys. Rev.} {\bf 124}, 925 (1961). \\
{\bf [7]} R.H. Dicke, {\em Phys. Rev.} {\bf 125}, 2163 (1962). \\
{\bf [8]} J. Isenberg and J. Nester, {\em Canonical Gravity}, in {\em General
Relativity and Gravitation, {\hspace*{6.2mm}}Vol. I}, Editor A. Held, Plenum 
Press (1980). \\ 
{\bf [9]} L. Parker, {\em Particle Creation by the Expansion of the Universe}, 
in {\em Handbook of} \\ {\hspace*{5.9mm}}
{\em Astronomy, Astrophysics and Geophysics, Vol. II Galaxies and Cosmology}, 
\\
{\hspace*{7mm}}Editors V.M Canuto and B.G. Elmegreen, Gordon {\&} Breach 
(1988). \\
{\bf [10]} D. {\O}stvang, chapter 3 of this thesis (2001). \\
{\bf [11]} D. {\O}stvang, chapter 4 of this thesis (2001). \\
{\bf [12]} D. {\O}stvang, chapter 5 of this thesis (2001). 

\newpage

{\vspace*{5cm}}

\begin{center}
{\huge {\bf Chapter Two: \\\ \\ On the Non-Kinematical Evolution
\vspace{0.5cm}\\of the 3-Geometry}}
\end{center}
\newpage
{\hspace*{1cm}}

\newpage

\newcounter{seceqn}[section]
\renewcommand{\theequation}
{\stepcounter{seceqn}\arabic{section}.\arabic{seceqn}}
\setcounter{section}{0}
\setcounter{equation}{0}
\begin{center}
{\large {\bf On the Non-Kinematical Evolution of the 3-Geometry}}
\end{center}
\begin{center}
by
\end{center}
\begin{center}
Dag {\O}stvang \\
{\em Institutt for fysikk, Norges teknisk-naturvitenskapelige universitet, 
NTNU \\
N-7491 Trondheim, Norway}
\end{center}
\begin{abstract}
We introduce the ``quasi-metric'' framework, consisting of a 4-dimensional 
differential space-time manifold equipped with two one-parameter families of 
Lorentzian 4-metrics parametrized by a global time function. Within this 
framework, we define the concept ``non-kinematical evolution of the 
3-geometry" and show how this is compatible with a new relativistic theory of 
gravity. It is not yet clear whether or not this theory is viable. However, 
it does agree with the classic solar system tests. Besides, the non-metric
sector of the theory leads to additional testable predictions.
\end{abstract}
\section{Introduction}
The class of space-time theories defined as metric theories of gravity 
satisfies the following postulates [1]: (1) space-time is equipped with a 
Lorentzian non-degenerate metric, (2) inertial test particles follow 
non-spacelike geodesics of this metric, and (3) in the local Lorentz frames of 
the metric, the non-gravitational physics is as in special relativity. In light
of the highly successful record of the metric framework, and with the lack of 
any obvious experimental motivation, one might wonder why anyone should bother 
to construct a non-metric alternative. The answer to this question is that, 
even though it is possible to model a wide range of phenomena within the metric 
framework, things do not always follow from first principles. If one within a
non-metric framework can do at least as well as with any metric theory, and if
additionally one gets a better connection to first principles, the non-metric 
framework should be considered as a serious alternative. The key questions 
are, of course, firstly; which physical/theoretical principles should guide 
one towards constructing a non-metric framework, and secondly; which phenomena
should be considered being sufficiently fundamental to motivate the 
formulation of such principles. 
 
In the search for a suitable non-metric framework, part of our phenomenological
motivation consists of the fact that the observable large-scale Universe seems 
to take part in a global increase of scale. Even though this observation has a 
natural interpretation within the metric framework, it does not follow from 
first principles. That is, within the metric framework, one may model many 
types of universe besides the expanding ones, e.g., collapsing universes, 
static universes or universes whose evolution at large scales cannot be 
modelled by a global scale factor. Contrary to this, we want to construct a 
non-metric framework which explains the above-mentioned observation from first 
principles. 

The above considerations suggest that a central feature of our non-metric 
framework should be the postulated existence of a specific set of 
(hypothetical) observers whose interrelationship consists at least partly of 
an increase of local scale with time. For reasons explained in section 2, this 
increase of local scale with time is called the ``non-kinematical evolution'' 
(NKE) of the 3-geometry spanned by these observers. In this thesis, we 
construct a particular type of non-metric framework which accommodates this 
concept as part of the basic geometrical structure. 

The introduction of the NKE can be done without sacrificing one of the basic 
principles defining metric theory; namely the Einstein equivalence principle 
(EEP). However, the mathematical structure of the new framework is not based
on pseudo-Riemannian geometry, so postulate (1) above does not apply. But 
despite that, the new framework is so similar to the metric framework that we 
may call it ``quasi-metric''.

This chapter is organized as follows: in section 2, we define the quasi-metric
framework in general terms; in section 3, we construct a theory of gravity
consistent with this framework; and in section 4, we find special solutions
and discuss the question of testing. A list of acronyms and mathematical 
symbols can be found in an appendix.
\section{The quasi-metric framework}
We start by briefly describing the well-known concept of ``kinematical
evolution" (KE) of the 3-geometry, also called geometrokinematics, within
the metric framework.

By slicing an arbitrary 3-dimensional (Cauchy) hypersurface ${\cal S}$ out of 
the space-time manifold $({\cal M},{\bf g})$, such that the normal vector field 
of ${\cal S}$ is everywhere timelike with respect to the space-time metric 
${\bf g}$, one splits space-time into space and time. Suppose that one 
furthermore specifies an arbitrary ``deformation" timelike vector field on 
${\cal S}$. Then, from the intrinsic metric of ${\cal S}$ and the extrinsic 
curvature tensor field and its rate of change along the deformation vector 
field, one may deduce the metric on the deformed spatial hypersurface and so 
on for successive deformed spatial hypersurfaces. This is the essence of 
geometrokinematics, and its detailed relations hold irrespectively of any 
dynamical laws. The reason is that these relations describe how a single 
space-time tensor field, sufficiently represented by the Einstein tensor field 
${\bf G}$, is projected into tangential and normal directions to ${\cal S}$, 
and this severely restricts how the projections behave under deformations of 
${\cal S}$; see, e.g., [2] for a further discussion.

The kinematical relations of $\cal M$ have one characteristic defining 
property; namely that they are not sufficient by themselves to determine the
space-time geometry. That is, in addition to the initial data on $\cal S$, one
must supply extra fields coming from the overlying space-time geometry at each
step of the deformation of $\cal S$ if one wants to predict the metric on
$\cal S$ for each time step of the deformation process. This is as it should 
be, since the geometry of space-time should be dependent on the coupling to
geometry of matter fields; only when appropriate dynamics is specified 
should it become possible to calculate the space-time geometry.

Any gravitational equations recognized as kinematical must leave room for 
dynamics without specifying the exact nature of the coupling between matter 
and geometry. This is realized geometrically by the fact that the extra fields 
that must be supplied take the form of curvature tensors. As noted above, it is
sufficient to supply the projection of ${\bf G}$ into $\cal S$ for each step of
the deformation; in itself this does not depend on how ${\bf G}$ is coupled to
matter. See [2] for a detailed discussion.

However, one might imagine that there are aspects of 3-geometry evolution which
do not depend on the continuous supplement of curvature fields coming from an 
overlying space-time metric. Rather, it would seem conceivable to try to 
construct a type of evolution which is given explicitly as some sort of
``absolute'' property of the space-time geometry itself. Besides, this type of
evolution should leave no room at all for dynamics, even in principle. Such 
aspects of 3-geometry evolution one may call ``non-kinematical" since they by 
construction are self-advancing. But to model said non-kinematical evolution, 
it would not be sufficient just to postulate the existence of some 
``preferred'' foliation of Lorenzian space-time into spatial hypersurfaces with
some explicitly given evolution. Besides the {\em ad hoc} nature of such an 
approach, the Lorentzian geometry still leaves room for gravitational dynamics,
and it would be awkward to find an explanation of why the explicitly given 
evolution should be excepted from said dynamics. This means that it is clearly 
not possible to make much sense of the NKE concept within the metric framework.
However, one may try to get around the above problem by postulating more than 
one space-time metric field; thus one may define some sort of evolution of the 
{\em space-time} metric field. Then there is no reason to expect that the NKE 
should have anything to do with the kinematical relations defined by any of 
the single metric fields.

We want to create a non-metric framework which incorporates the NKE in a 
natural way. Since this framework should include more than one space-time 
metric field and some way to evolve each of those into another, there must 
exist some evolution parameter by means of which this can be defined. This 
parameter must have an existence independent of the details of each single 
metric field. The straightforward way to define such a parameter is to equip 
the space-time manifold with a global time function; this quantity should be a
natural feature of the general geometrical structure on which we base our 
non-metric framework. We proceed by defining such a basic geometrical 
structure.

Consider a 5-dimensional differential manifold with the global structure
${\cal M}{\times}{\bf R}$, where ${\cal M}$ is a Lorentzian space-time 
manifold and ${\bf R}$ is the real line. We may then introduce local 
coordinates ${\{}x^{\epsilon}{\}}$ on ${\cal M}$ (${\epsilon}$ taking the 
values 0,1,2,3 where $x^{0}$ is the time coordinate), and a global coordinate 
$t$ on ${\bf R}$. Suppose next that we slice a 4-dimensional hypersurface 
${\cal N}$ specified by the equation $t=t(x^{\epsilon})$ out of 
${\cal M}{\times}{\bf R}$. On this slice there is required to exist a unique 
invertible functional relationship between $t$ and $x^{0}$ throughout the 
coordinate patch covered by ${\{}x^{\epsilon}{\}}$. Now we let an arbitrary point
with coordinate $t$ in ${\bf R}$ represent a space-time metric ${\bf g}_{t}$ on
${\cal M}$. Then ${\cal N}$ is defined to be a 4-dimensional space-time 
manifold equipped with a one-parameter family of Lorentzian 4-metrics 
${\bf g}_{t}$ parametrized by the global time function $t$. This general 
geometric structure is the basis of the quasi-metric framework, and it has the 
capacity to accommodate both KE and NKE of the relevant geometrical quantities.
We come back to questions of uniqueness and existence of such a geometrical 
structure in the next section.

The global time function $t$ singles out of ${\cal N}$ a set of fundamental 
3-dimensional spatial submanifolds ${\cal S}_{t}$. Each of these hypersurfaces 
is ``preferred" in some sense since each of these hypersurfaces constitutes
the domain of validity for one single member of the metric family; there
corresponds exactly one metric field ${\bf g}{\in}{\bf g}_t$ to each 
${\cal S}{\in}{\cal S}_t$. We call the hypersurfaces ${\cal S}_t$ the 
{\em fundamental hypersurfaces} (FHSs). Observers always moving orthogonally 
to the FHSs are called {\em fundamental observers} (FOs). It is important to 
have in mind that each 4-metric applies exactly only on its domain FHS. This 
means that each member of the metric family is relevant to the evolution of 
the FHSs only at the corresponding FHS. Thus it is sufficient to extrapolate 
each 4-metric to an infinitesimally small neighbourhood of the FHSs whereas 
any further extrapolation may be of dubious value physically.

Now the FHSs are subject to two types of evolution with proper time 
${\tau}_{{\cal F}}$ (depending directly on $t$), as measured by the FOs when 
traversing the one-parameter family of 4-metrics. Firstly, it is the KE of the 
FHSs as defined from its dependence on the supplement of curvature information
from $\cal N$; this information should be available from the 4-geometry defined
by each member of the metric family, with $t$ fixed. Secondly, it is the NKE 
of the FHSs; this type of evolution is given via the dependence of ${\bf g}_t$ 
on $t$. The {\em global} part of the NKE is given explicitly as a function of 
$t$ playing the role of a global scale factor of the FHSs. That is, the NKE
does not depend on the supplement of curvature information calculated from the 
metric family. From this it follows that the concept of NKE is intimately 
connected with the existence of a set of ``preferred" hypersurfaces, i.e., the 
FHSs, evolving according to the global part of the NKE.

If one considers a metric space-time manifold as given {\em a priori}, there
are in principle two different ways of splitting up space-time into space and 
time. The first one is the foliation of space-time into a family of space-like
hypersurfaces. This corresponds to choosing a time coordinate representing
``surfaces of simultaneity''. Note that the gradient field of this time 
coordinate is in general not orthogonal to the hypersurfaces. The second way 
of splitting up space-time is by foliating space-time into a congruence of 
time-like curves representing a family of observers. This corresponds to
assigning a set of constant spatial coordinates to each observer, i.e., each
curve of the congruence is characterized by a set of three numbers. There is,
however, no natural choice of simultaneity connecting the local rest spaces
of the different observers representing the family of curves. 

In quasi-metric relativity, each metric of the metric family is {\em defined}
from the local measurements of the FOs with the global time function serving 
as the (unique) choice of simultaneity. The construction of quasi-metric 
space-time as consisting of two mutually orthogonal foliations means that it 
must be possible to choose a global time coordinate for each metric of the
metric family. Consequently, it should exist a set of ``preferred'' coordinate
systems which adapts to the FHSs in a natural way. A set of such coordinate 
systems is found by first noticing that it exists, by definition of the global 
time function, a unique invertible functional relationship between $x^{0}$ and
$t$ throughout the coordinate patch covered by ${\{}x^{\epsilon}{\}}$. We may 
then define a particular type of coordinate system where this relationship is 
given explicitly by $x^{0}=ct$. Such a coordinate system where $x^0$ and $ct$ 
are identified and thus physically equivalent in ${\cal N}$, we call a 
{\em global time coordinate system} (GTCS). It is usually most convenient to 
use a GTCS, since such a coordinate system is better adapted to the 
quasi-metric space-time geometry than any alternative coordinate system. In 
particular cases, further simplification is possible if one can find a 
{\em comoving} coordinate system, i.e., a coordinate system where the spatial 
coordinates ${\{}x^i{\}}$ assigned to any FO do not vary along its world line.
That is, the spatial coordinates along such a world line do not depend on $t$ 
in ${\cal N}$. A GTCS which is also a comoving coordinate system we call a 
{\em hypersurface-orthogonal coordinate system} (HOCS). Note that in 
principle, infinitely many GTCSs exist, but that a HOCS can be found only in
particular cases.

If we define a GTCS ${\{}x^{\epsilon}{\}}$, each single metric of the 
one-parameter family can be expressed by its components
$g_{{\mu}{\nu}}(x^{\epsilon})$. The fact that we have a one-parameter
family of metrics rather than a single metric on ${\cal N}$, we express by an 
implicit dependence on the coordinates via the global time function 
$t$. In component notation, we write this formally as
$g_{(t){\mu}{\nu}}(x^{\epsilon})$, where the dependence on $t$ is put in
a parenthesis not to confuse it with the coordinate indices. By treating
$t$ as a constant, we get the components of each single metric. Note that
the off-diagonal components $g_{(t)0j}{\equiv}g_{(t)j0}$ (where the spatial
indices $j$ may take the values 1,2 or 3) vanish in a HOCS.

By postulating the existence of the family ${\bf g}_t$, we automatically define 
a family of Levi-Civita connections ${\nabla}_t$, where each connection is 
found from the corresponding metric by holding $t$ constant. But we are more
interested in the affine structure induced on $\cal N$ by the metric family 
than in the Levi-Civita connection of any single metric. Thus we want to 
define a unique covariant derivative in ${\cal N}$, given the functions 
$g_{(t){\mu}{\nu}}$. The natural way to do this is by thinking of the metric 
family as one single degenerate 5-dimensional metric on 
${\cal M}{\times}{\bf R}$, where the degeneracy manifests itself via the 
condition ${\bf g}_t({\frac{\partial}{{\partial}t}},{\cdot}){\equiv}0$. We may 
then introduce a torsion-free, metric-compatible 5-dimensional connection
${\,}{\topstar{\nabla}}{\,}$ on ${\cal M}{\times}{\bf R}$ and consider the 
restriction of ${\,}{\topstar{\nabla}}{\,}$ to $\cal N$; in this way we obtain
the wanted affine structure on $\cal N$. We name the connection 
${\,}{\topstar{\nabla}}{\,}$ the ``degenerate" connection since it is (almost)
compatible with which may be perceived as a degenerate 5-dimensional metric. 
The requirement that ${\,}{\topstar{\nabla}}{\,}$ must be metric-preserving 
reads (a comma denotes a partial derivative, and we use Einstein's summation 
convention throughout)
\eqa
{\frac{\partial}{{\partial}t}}g_{(t){\mu}{\nu}}&{\equiv}&
{\frac{\partial}{{\partial}t}}{\bf g}_t
({\frac{\partial}{{\partial}x^{\mu}}},{\frac{\partial}{{\partial}x^{\nu}}})
={\bf g}_t(\ {\topstar{\nabla}}_{{\frac{\partial}{{\partial}t}}}
{\frac{\partial}{{\partial}x^{\mu}}},{\frac{\partial}{{\partial}x^{\nu}}})+
{\bf g}_t({\frac{\partial}{{\partial}x^{\mu}}},
{\topstar{\nabla}}_{{\frac{\partial}{{\partial}t}}}
{\frac{\partial}{{\partial}x^{\nu}}}).
\ena
Moreover, we define the degenerate connection coefficients 
${\topstar{\Gamma}}^{{\,}{\alpha}}_{{\mu}t}$, 
${\topstar{\Gamma}}^{{\,}{\alpha}}_{t{\mu}}$
and ${\topstar{\Gamma}}^{{\,}{\alpha}}_{{\mu}{\nu}}$ by
\eqa
{\topstar{\nabla}}_{{\frac{\partial}{{\partial}t}}}
{\frac{\partial}{{\partial}x^{\mu}}}{\equiv}{\ }
{\topstar{\Gamma}}^{{\,}{\alpha}}_{{\mu}t}
{\frac{\partial}{{\partial}x^{\alpha}}},
{\qquad} {\topstar{\Gamma}}^{{\,}{\alpha}}_{t{\mu}}{\equiv}{\ }
{\topstar{\Gamma}}^{{\,}{\alpha}}_{{\mu}t}, {\qquad}
{\topstar{\nabla}}_{{\frac{\partial}{{\partial}x^{\nu}}}}
{\frac{\partial}{{\partial}x^{\mu}}}{\equiv}{\ }{\topstar{\Gamma}}^
{{\,}{\alpha}}_{{\mu}{\nu}}{\frac{\partial}{{\partial}x^{\alpha}}}, \qquad
{\topstar{\Gamma}}^{{\,}{\alpha}}_{{\mu}{\nu}}
{\equiv}{\Gamma}^{\alpha}_{(t){\mu}{\nu}},
\ena
where ${\Gamma}^{\alpha}_{(t){\mu}{\nu}}$ are the connection coefficients
obtained from the Levi-Civita connections ${\nabla}_t$. Note that equation 
(2.1) implies that ${\,}{\topstar{\nabla}}{\,}$ does not have torsion since it
is symmetric. This means that any counterpart to equation (2.1) obtained by,
e.g., interchanging the coordinate vector fields 
${\frac{\partial}{{\partial}t}}$ and ${{\frac{\partial}{{\partial}x^{\mu}}}}$,
does not hold in general. So ${\,}{\topstar{\nabla}}{\,}$ does not preserve 
the degenerate part of the degenerate 5-dimensional metric on 
${\cal M}{\times}{\bf R}$. Furthermore, note that other degenerate connection
coefficients than those given by equation (2.1) vanish by definition. This 
implies that the gradient of the global time function is covariantly constant, 
i.e., that ${\,}{\topstar{\nabla}}_{{\frac{\partial}{{\partial}t}}}dt=
{\topstar{\nabla}}_{{\frac{\partial}{{\partial}x^{\mu}}}}dt=0$, see [4] for an 
analogy with Newton-Cartan-theory. 

We must now find expressions for the degenerate connection coefficients
${\topstar{\Gamma}}^{{\,}{\alpha}}_{{\mu}t}$ defined in equation (2.2). If the
metric-preserving condition (2.1) were the only restriction one wanted to
lay on the degenerate connection, a natural choice of degenerate connection
coefficients would be ones totally determined from partial derivatives with
respect to $t$ of the functions $g_{(t){\mu}{\nu}}$, i.e., from terms
containing expressions like ${\frac{\partial}{{\partial}t}}g_{(t){\mu}{\nu}}$.
It can be shown that the requirement that the degenerate connection 
coefficients take this form yields that ${\topstar{\Gamma}}^{{\,}{\alpha}}_{{\mu}t}$
must be equal to ${\frac{1}{2}}g_{(t)}^{{\alpha}{\sigma}}
g_{(t){\sigma}{\mu}},_t$. But in addition to the condition (2.1), we also have
the condition that the geometrical structure of quasi-metric space-time
should not be affected by parallel-transport in the $t$-direction. That is,
we should require that the degenerate  connection ensures that the unit normal
vector field family ${\bf n}_t$ (i.e., a vector field family having the 
property ${\bf g}_t({\bf n}_t,{\bf n}_t)=-1$) of the FHSs will be 
parallel-transported in the $t$-direction. We thus require that
\eqa
{\topstar{\nabla}}_{\frac{\partial}{{\partial}t}}{\bf n}_t=0.
\ena
Now it turns out that the degenerate connection defined from the candidate
degenerate connection coefficients shown above does not satisfy equation
(2.3) in general. However, in particular cases it does, and this should be
significant.

Furthermore, it turns out that a degenerate connection which fulfills both 
equations (2.1) and (2.3) must necessarily be defined from the form the 
degenerate connection coefficients take in a GTCS. We choose such a degenerate 
connection on the grounds that it should reproduce the results of the 
above-mentioned candidate degenerate connection for those particular cases 
where it satisfies equation (2.3). It may be shown that the unique choice 
satisfying this criterion only involves the spatial metric family ${\bf h}_t$ 
intrinsic to the FHSs. We find the connection coefficients (valid in a GTCS)
\eqa
{\topstar{\Gamma}}^{{\,}{\alpha}}_{{\mu}t}=
{\frac{1}{2}}{\delta}_i^{\alpha}{\delta}^j_{\mu}h_{(t)}^{is}h_{(t)sj,t}.
\ena
Note that the spatial indices occurring in equation (2.4) make the
degenerate connection explicitly dependent on the structure of
quasi-metric space-time.

On ${\cal M}{\times}{\bf R}$, the Levi-Civita connections ${\nabla}_t$ define 
covariant derivatives along vectors tangent to $\cal M$. We extend the 
${\nabla}_t$ by applying these covariant derivatives to the vector
${\frac{\partial}{{\partial}t}}$ also. This must be done in a way that meshes 
with the degenerate connection ${\,}{\topstar{\nabla}}{\,}$ such that the 
torsion vanishes. By definition we then have (to avoid confusion we drop the 
$t$-label of ${\nabla}_t$ in expressions such as ${\nabla}_{{\bf z}_t}$)
\eqa
{\Gamma}^{\alpha}_{t{\mu}}{\equiv}{\Gamma}^{\alpha}_{{\mu}t}{\equiv}
{\ }{\topstar{\Gamma}}^{{\,}{\alpha}}_{{\mu}t}=
{\frac{1}{2}}{\delta}_i^{\alpha}{\delta}^j_{\mu}h_{(t)}^{is}h_{(t)sj,t}, \qquad
{\nabla}_{{\frac{\partial}{{\partial}x^{\mu}}}}
{\frac{\partial}{{\partial}t}}=
{\topstar{\nabla}}_{{\frac{\partial}{{\partial}x^{\mu}}}}
{\frac{\partial}{{\partial}t}}=
{\topstar{\nabla}}_{{\frac{\partial}{{\partial}t}}}
{\frac{\partial}{{\partial}x^{\mu}}},
\ena
and this makes the torsion vanish by definition, since coordinate vector fields
commute.

Any 5-dimensional vector field ${\bf z}$ on ${\cal M}{\times}{\bf R}$ can be 
split up into pieces, where one piece $z^{\mu}{\frac{\partial}{{\partial}
x^{\mu}}}$ is tangential to $\cal M$ and the other piece $z^t
{\frac{\partial}{{\partial}t}}$ is normal to $\cal M$. To take the degenerate 
connection ${\,}{\topstar{\nabla}}{\,}$ along a given direction in 
${\cal M}{\times}{\bf R}$, it is sufficient to know 
${\nabla}_{{\frac{\partial}{{\partial}x^{\nu}}}}$ and
${\,}{\topstar{\nabla}}_{{\,}{\frac{\partial}{{\partial}t}}}$ since the degenerate 
covariant derivative ${\,}{\topstar{\nabla}}_{{\,}{\bf z}}$ is defined by 
using the appropriate Levi-Civita connection along the piece tangent to 
$\cal M$ and using $z^t{\,}{\topstar{\nabla}}_{{\frac{\partial}{{\partial}t}}}$ 
along the remaining piece. In particular, we find by direct calculation that
\eqa
{\topstar{\nabla}}_{{\frac{\partial}{{\partial}t}}}{\bf g}_t&=& 
{\Big(}{\frac{\partial}{{\partial}t}}g_{(t){\mu}{\nu}}-
{\topstar{\Gamma}}^{{\,}{\alpha}}_{{\mu}t}g_{(t){\alpha}{\nu}}-
{\topstar{\Gamma}}^{{\,}{\alpha}}_{{\nu}t}g_{(t){\mu}{\alpha}}{\Big)}
dx^{\mu}{\otimes}dx^{\nu}, \nonumber \\
{\topstar{\nabla}}_{{\frac{\partial}{{\partial}t}}}{\bf n}_t&=& 
{\Big(}{\frac{\partial}{{\partial}t}}n_{(t)}^{\mu}+
{\topstar{\Gamma}}^{{\,}{\mu}}_{{\alpha}t}n_{(t)}^{\alpha}{\Big)}
{\frac{\partial}{{\partial}x^{\mu}}}.
\ena
To make the expressions in equation (2.6) vanish, we must define appropriate
dependences on $t$ of ${\bf g}_t$ and ${\bf n}_t$. Necessary conditions are 
that $g_{(t)00}$ and $n_{(t)}^0$ cannot depend on $t$, plus the requirements
$g_{(t)0j}={\frac{t}{t_0}}N_{(t)}^k{\hat h}_{(t)kj}$,
$n_{(t)}^j=-{\frac{t_0}{t}}{\frac{N_{(t)}^j}{N}}$
and $g_{(t)ij}=h_{(t)ij}{\equiv}{\frac{t^2}{t_0^2}}{\hat h}_{(t)ij}$ (expressed in a 
GTCS), where $t_0$ is some arbitrary reference epoch. In addition to these
dependences on $t$, the expressions in equation (2.6) will vanish in general if
\eqa 
{\frac{\partial}{{\partial}t}}{\Big [}N_{(t)}^kN_{(t)}^s
{\hat h}_{(t)ks}{\Big ]}=0, \qquad {\Rightarrow} \qquad
N_{(t),t}^s=-{\frac{1}{2}}N_{(t)}^k{\hat h}_{(t)}^{is}{\hat h}_{(t)ik,t}.
\ena
The above dependences on $t$ in addition to the criterion (2.7) ensure that 
members of the metric family and the unit normal vector family are 
parallel-transported along ${\frac{\partial}{{\partial}t}}$. Thus the 
degenerate connection is compatible with the metric family. This guarantees 
that different metrics at different events can be compared unambiguously.

Let $\lambda$ be some affine parameter describing the world line of an 
arbitrary test particle. The tangent vector field 
${\frac{\partial}{{\partial}{\lambda}}}$ along the curve will then have the
coordinate representation
${\frac{dt}{d{\lambda}}}{\frac{\partial}{{\partial}t}}+
{\frac{dx^{\alpha}}{d{\lambda}}}{\frac{\partial}{{\partial}x^{\alpha}}}$, 
where the ${\frac{\partial}{{\partial}x^{\alpha}}}$ are coordinate vector 
fields. Using the coordinate representation of 
${\frac{\partial}{{\partial}{\lambda}}}$, we may define 
${\,}{\topstar{\nabla}}_{\frac{\partial}{{\partial}{\lambda}}}{\equiv}
{\frac{dt}{d{\lambda}}}{\,}{\topstar{\nabla}}_{\frac{\partial}{{\partial}t}}+
{\frac{dx^{\alpha}}{d{\lambda}}}{\nabla}_{\frac{\partial}{{\partial}x^{\alpha}}}$; this 
is the definition of taking the covariant derivative along a curve in 
${\cal M}{\times}{\bf R}$. The restriction to ${\cal N}$ is easily obtained by
only considering curves in ${\cal N}$, i.e., curves where $x^0=ct$ in a GTCS.

As an example, we find the degenerate covariant derivative of a family of 
vector fields ${\bf z}_t$ (where $z_{(t)}^t{\equiv}0$) along an affinely 
parameterized curve $x^{\beta}({\lambda})$ in $\cal N$. Expressed in 
(arbitrary) coordinates, this is
\eqa
{\topstar{\nabla}}_{{\frac{\partial}{{\partial}{\lambda}}}}{\bf z}_t  =
{\Big [}{\frac{dz_{(t)}^{\mu}}{d{\lambda}}}+{\Big (}
{\topstar{\Gamma}}^{{\,}{\mu}}_{{\alpha}t}{\frac{dt}{d{\lambda}}}+
{\Gamma}^{\mu}_{(t){\alpha}{\beta}}{\frac{dx^{\beta}}{d{\lambda}}}{\Big )}
z^{\alpha}_{(t)}{\Big ]}{\frac{\partial}{{\partial}x^{\mu}}}.
\ena
Whether or not the fields ${\bf z}_t$ are parallel-transported with respect to 
the degenerate connection ${\,}{\topstar{\nabla}}{\,}$ along the above curve 
depends on whether equation (2.8) vanishes or not. 

An important special case is when a family of vector fields ${\bf u}_t$ is
identified with the 4-velocity tangent vector field of the curve, that is
\eqa
{\bf u}_t{\equiv}u^{\alpha}_{(t)}{\frac{\partial}{{\partial}x^{\alpha}}}
{\equiv}{\frac{dx^{\alpha}}{d{\tau}_t}}{\frac{\partial}{{\partial}x^{\alpha}}},
\ena
where ${\tau}_t$ is the proper time measured along the curve. Then we may 
calculate the degenerate covariant derivative of ${\bf u}_t$ along the tangent
vector field ${\frac{\partial}{{\partial}{\tau}_t}}$; this yields the 
degenerate acceleration vector field ${\stackrel{\star}{\bf a}}$ defined by
\eqa
{\stackrel{\star}{\bf a}}{\equiv}
{\topstar{\nabla}}_{\frac{\partial}{{\partial}{\tau}_t}}{\bf u}_t
={\frac{dt}{d{\tau}_t}}
{\topstar{\nabla}}_{\frac{\partial}{{\partial}t}}{\bf u}_t+
{\frac{dx^{\alpha}}
{d{\tau}_t}}{\nabla}_{\frac{\partial}{{\partial}x^{\alpha}}}{\bf u}_t{\equiv}
{\frac{dt}{d{\tau}_t}}
{\topstar{\nabla}}_{\frac{\partial}{{\partial}t}}{\bf u}_t+{\bf a}_t.
\ena
Using coordinate notation, we see that equation (2.10) yields equations of 
motion
\eqa
{\frac{d^2x^{\mu}}{d{\lambda}^2}}+{\Big (}
{\topstar{\Gamma}}^{{\,}{\mu}}_{{\alpha}t}{\frac{dt}{d{\lambda}}}+
{\Gamma}^{\mu}_{(t){\alpha}{\beta}}{\frac{dx^{\beta}}{d{\lambda}}}
{\Big )}{\frac{dx^{\alpha}}{d{\lambda}}}
{\equiv}{\Big (}{\frac{d{\tau}_t}{d{\lambda}}}{\Big )}^2{\topstar{a}}^{\mu},
\ena
where ${\frac{d{\tau}_t}{d{\lambda}}}$ is constant along the curve since
${\lambda}$ and ${\tau}_t$ are both affine parameters. We may compare equation
(2.11) to the standard coordinate expression for the family of 4-acceleration 
fields ${\bf a}_t{\equiv}{\nabla}_{{\bf u}_t}{\bf u}_t$ compatible with the
Levi-Civita connections ${\nabla}_t$, namely
\eqa
{\frac{d^{2}x^{\beta}}{d{\lambda}^{2}}}+
{\Gamma}_{(t){\sigma}{\rho}}^{\beta}{\frac{dx^{\sigma}}{d{\lambda}}}
{\frac{dx^{\rho}}{d{\lambda}}}={\Big (}{\frac{d{\tau}_t}{d{\lambda}}}{\Big )}^2
a^{\beta}_{(t)}.
\ena
However, the equations (2.12) cannot be used as equations of motion in $\cal N$
since they do not include the dependence of ${\bf u}_t$ on $t$. That is, the
equations (2.12) only hold in $\cal M$.

A geodesic for the degenerate connection ${\,}{\topstar{\nabla}}{\,}$ (or 
equivalently for the metric family) in $\cal N$ is defined as a curve along 
which ${\stackrel{\star}{\bf a}}$ vanishes. Clearly we need to specify 
${\stackrel{\star}{\bf a}}$ independently to determine which curves of 
${\cal N}$ are geodesics for ${\,}{\topstar{\nabla}}{\,}$ and which are not.
We will show later (section 3.4) that in fact
${\,}{\topstar{\nabla}}_{{\frac{\partial}{{\partial}t}}}{\bf u}_t=0$,
so we have ${\stackrel{\star}{\bf a}}$=${\bf a}_t$ from equation (2.10). Thus,
similar to the Levi-Civita connection, the geodesics for the degenerate
connection are identified with world lines of inertial test particles.
However, as we can see from equations (2.11) and (2.12), in quasi-metric
theory the paths of inertial test particles are different from their
counterparts in metric theory.

Note that, since ${\bf u}_{\cal F}=c{\bf n}_t$ is the 4-velocity field family
of the FOs, the FOs always move orthogonally to the FHSs. We may also show 
that we in general have (since ${\stackrel{\star}{\bf a}}$=${\bf a}_t$)
\eqa
{\topstar{a}}^{\mu}{\frac{dx^{\nu}}{d{\lambda}}}g_{(t){{\mu}{\nu}}}=
a_{(t)}^{\mu}{\frac{dx^{\nu}}{d{\lambda}}}g_{(t){{\mu}{\nu}}}=0, 
\ena
for any time-like curve in ${\cal N}$, thus the acceleration field family 
${\bf a}_t$ is normal to the tangent vector field family ${\bf u}_t$. This is 
necessary to ensure that the length of ${\bf u}_t$ is constant along the curve.

A simple illustrative example of a one-parameter family of 4-metrics is 
defined by the line element family (given in spherical GTCS
$(x^{0},r,{\theta},{\phi})$)
\eqa
ds^{2}_t = -(dx^{0})^{2} + 
({\frac{t}{t_{0}}})^{2}{\bigg(}{\frac{dr^{2}}
{1-{\frac{r^{2}}{{\Xi}_{0}^{2}}}}}+r^{2}d{\Omega}^{2}{\bigg)},
\ena
where $d{\Omega}^{2}{\equiv}d{\theta}^{2}+{\sin}^{2}{\theta}d{\phi}^{2}$,
and $t_{0}$ and ${\Xi}_{0}$ are constants. We see that this is a one-parameter 
family of static metrics, each one having the geometry of 
${\bf S}^{3}{\times}{\bf R}$, but with different scale factors. When doing
kinematical calculations using equation (2.14), one must of course apply the 
equations of motion (2.11). Note the difference between this situation and 
that one gets by applying the equations (2.12) to the single non-static metric 
one obtains by substituting $cdt=dx^{0}$ into equation (2.14); the geodesics of
this single metric are not equivalent to the geodesics calculated from the 
metric family (2.14). And although the null paths are the same, the role of 
$t$ as a parameter along these paths is different for the two cases. (See 
section 4.2 for some calculations.)

One immediate question is how to calculate the metric family ${\bf g}_t$, the 
obvious answer being that it should be found from appropriate field equations.
However, this does not necessarily mean that ${\bf g}_t$ should represent a 
{\em solution} of said field equations. Rather, we should take the more 
general view that it must be possible to construct the family ${\bf g}_t$ from
another family ${\bf {\bar g}}_t$ which {\em does} represent a solution of said
field equations, together with fields obtainable from the family 
${\bf {\bar g}}_t$. Why we should do this is motivated below.

We expect that solutions of field equations should contain some sort of 
non-kinematical ``expansion" as part of the geometrical structure. However, it 
may not be possible to eliminate all the effects of this ``expansion" just by 
holding $t$ constant. The reason for this is the possibility that some of 
the NKE, which we will call the {\em local} NKE, is not ``realized'' 
explicitly. That is, the FOs may move in a way such that the ``expansion" 
representing the local NKE is exactly cancelled out, the net result being 
``stationariness''. But this type of stationariness is not equivalent to being 
stationary with no NKE. Now the effects of the local NKE should be compensated 
for in the ``physical'' metric  family ${\bf g}_t$. On the other hand, said 
effects should in general be present implicitly and uncompensated for in 
solutions of field equations. This means that in general, ${\bf g}_t$ cannot 
represent a {\em solution} of field equations. Rather, the family ${\bf g}_t$ 
should be constructed from a solution family ${\bf {\bar g}}_t$. In section 2.3
we show that this construction involves a family of vector fields.
\subsection{Uniqueness and existence of the global time 
function}
In this section, we investigate which restrictions should be put on each member
of the family of metrics ${\bf {\bar g}}_t$ to ensure that they are compatible
with the quasi-metric framework. The reason why we put the restrictions on 
${\bf {\bar g}}_t$ and not directly on the ``physical" family ${\bf g}_t$, is 
that the former family by hypothesis represents a solution of field equations;
to know any general properties of solutions may be useful when constructing 
such equations. 

Obviously, we want the postulated global time function to exist everywhere. 
That is, as observed earlier, there always should exist a unique invertible 
functional relationship between the coordinate time $x^0$ and $t$ throughout 
the coordinate patch covered by an arbitrary local coordinate system 
${\{}x^{\epsilon}{\}}$. The requirement that the global time function exists,
leads to restrictions on the metric family ${\bf {\bar g}}_t$. In particular, 
extrapolating ${\bf {\bar g}}_t$ off the domain FHSs, we require that each FHS 
must be a member of a unique foliation of ${\overline{\cal M}}$ (equipped with 
any allowable metric) into a family of spatial 3-hypersurfaces, if we hold $t$ 
constant and propagate the FHS out into the surrounding 4-geometry of 
${\overline{\cal M}}$. (We write ${\overline{\cal M}}$ instead of ${\cal M}$ 
and ${\overline{\cal N}}$ instead of $\cal N$ when we deal with the family 
${\bf {\bar g}}_t$.) 

To guarantee the existence of a global time function, firstly it is important 
to ensure that the gradient field $dx^0$ of the global time coordinate $x^0$ 
in a GTCS is indeed timelike everywhere. (Otherwise a unit time-like vector 
field family ${\bf {\bar n}}_t$ would not exist everywhere on the FHSs.) This 
condition concerns the geometry of each 4-metric of ${\bf {\bar g}}_t$ in the 
sense that they cannot contain event horizons. By hypothesis, the family 
${\bf {\bar g}}_t$ should represent a solution field equations. This means 
that, if candidate field equations do indeed allow solutions containing 
event horizons, such candidates must be discarded as inconsistent with the 
quasi-metric framework. Secondly, even if the FHSs are spatial Cauchy 
hypersurfaces, one must ensure that each FHS can be seen as a member of a 
unique family of spatial 3-hypersurfaces foliating ${\overline{\cal M}}$ 
equipped with any allowable 4-geometry ${\bf {\bar g}}{\in}{\bf {\bar g}}_t$. 
A sufficient condition to ensure this, is to demand that the FHSs are 
{\em compact} (without boundaries) [3]. Thus, to guarantee that the global time
function exists, we will require that the FHSs are Cauchy and compact. This is 
also sufficient to ensure that the global time function is unique in the sense 
that it splits space-time into a unique family of FHSs. 

We now write down a general expression for the family ${\bf {\bar g}}_t$,
where the explicit dependence on $t$ is included to model the presence of a
global non-kinematical expansion. The global NKE of the FHSs should be 
described via a scale factor ${\bar F}_t$ (having the dimension of length) of
the FHSs. Since no extra arbitrary scale or parameter should be introduced when
defining ${\bar F}_t$, we must have that ${\bar F}_t{\propto}ct$. The general 
expression for ${\bf {\bar g}}_t$ is most conveniently written using a GTCS. 
That is, as observed earlier, the existence of a global time function singles 
out a preferred set of coordinate systems in ${\overline{\cal N}}$, namely 
those where the identification $x^0=ct$ is made. In what follows, we specify 
such a GTCS ${\{}x^0,x^{i}{\}}$, where the $x^{i}$ are spatial coordinates for 
$i=1,2,3$. Using the notation of the ADM-formalism [4], an arbitrary metric 
family ${\bf {\bar g}}_t$ can be expressed in a GTCS via the family of line 
elements (we use the signature convention $(-+++)$ throughout):
\eqa
{\overline {ds}}_t^2={\bar N}_t^2{\Big \{ }
[{\bar N}_{(t)}^k{\bar N}_{(t)}^s{\tilde h}_{(t)ks}-1](dx^0)^2+
2{\frac{t}{t_0}}{\bar N}_{(t)}^k{\tilde h}_{(t)ks}dx^sdx^0+
{\frac{t^2}{t_0^2}}{\tilde h}_{(t)ij}dx^idx^j{\Big \} }.
\ena
Here, ${\bar N}_t$ is the lapse function family of the FOs, the 
${\frac{t_0}{t}}{\bar N}_{(t)}^k$ are the components of the shift vector family 
of the FOs and ${\bar h}_{(t)ij}{\equiv}{\frac{t^2}{t_0^2}}{\bar N_t}^2
{\tilde h}_{(t)ij}$ are the spatial components of the spatial metric family 
(these are identical to the components of the metric family intrinsic to the 
FHSs). Moreover, from equation (2.15) we se that the scale factor of the FHSs 
is conveniently defined as ${\bar F}_t{\equiv}{\bar N}_tct$. One needs to 
specify more precisely the role of the global time function $t$ in equation 
(2.15), since one must require that
${\,}{\topstar{{\bar {\nabla}}}}_{\frac{\partial}{{\partial}t}}{\bf {\bar g}}_t$ and
${\,}{\topstar{{\bar {\nabla}}}}_{\frac{\partial}{{\partial}t}}{\bf {\bar n}}_t$ 
vanish. These requirements are fulfilled if (as a counterpart to equation 
(2.7))
\eqa 
{\frac{\partial}{{\partial}t}}{\Big [}{\bar N}_{(t)}^k{\bar N}_{(t)}^s
{\tilde h}_{(t)ks}{\Big ]}=0, \qquad {\Rightarrow} \qquad
{\bar N}_{(t),t}^s=-{\frac{1}{2}}{\bar N}_{(t)}^k
{\tilde h}_{(t)}^{is}{\tilde h}_{(t)ik,t}.
\ena
As mentioned above, to ensure that the global time function exists, we have 
restricted the 4-manifold ${\overline{\cal N}}$ in a topological sense by 
requiring the existence of compact Cauchy hypersurfaces (without boundaries). 
From equation (2.15) we see that this naturally implies the existence of a 
``basic" 3-geometry in the sense that for particular cases, the geometry 
${\bf {\bar h}}_t$ of the FHSs will be conformal to this basic 3-geometry. 
That is, the basic 3-geometry should be represented by a particular form of 
${\bf {\tilde h}}_t$. A natural choice of such a basic 3-geometry is the 
geometry of the 3-sphere ${\bf S}^3$ (with radius $ct_0$) since it is compact 
and has a simple topology. We may take this as a definition, thus any metric 
family (2.15) should contain ``prior'' 3-geometry. As we shall see later, one 
consequence of this is the appearance of one particular term in one of the 
field equations.
 
The $t$-dependence in equation (2.15), given equation (2.16), is now determined 
such that
${\,}{\topstar{{\bar {\nabla}}}}_{\frac{\partial}{{\partial}t}}{\bf {\bar g}}_t$
and
${\,}{\topstar{{\bar {\nabla}}}}_{\frac{\partial}{{\partial}t}}{\bf {\bar n}}_t$
vanish. But since ${\bar N}_t$ may possibly depend on $t$ (see above), the
affine structure on ${\overline{\cal N}}$ will be slightly more general than 
that on ${\cal N}$. That is, the connection coefficients in 
${\overline{\cal N}}$ read (compare to equation (2.5))
\eqa
{\topstar{\bar {\Gamma}}}_{t0}^{{\,}0}={\frac{{\bar N}_{t,t}}{{\bar N}_t}},
\qquad 
{\topstar{\bar {\Gamma}}}_{tj}^{{\,}i}=
{\Big (}{\frac{1}{t}}+{\frac{{\bar N}_{t,t}}{{\bar N}_t}}{\Big )}{\delta}^i_j
+{\frac{1}{2}}{\tilde h}_{(t)}^{is}{\tilde h}_{(t)sj,t}, \qquad
{\topstar{\bar {\Gamma}}}^{\alpha}_{{\mu}t}{\equiv}
{\topstar{\bar {\Gamma}}}^{\alpha}_{t{\mu}}, \qquad
{\topstar{\bar {\Gamma}}}^{\alpha}_{{\nu}{\mu}}{\equiv}
{\bar {\Gamma}}^{\alpha}_{(t){\nu}{\mu}}.
\ena
Note that the components of the inverse metric family ${\bf {\bar g}}_t^{-1}$ 
obtained from equation (2.15) are given by [4]
\eqa
{\bar g}_{(t)}^{00}=-{\bar N}_t^{-2}, {\qquad} {\bar g}_{(t)}^{0j}=
{\bar g}_{(t)}^{j0}={\frac{t_0}{t}}{\frac{{\bar N}_{(t)}^j}{{\bar N}_t^2}}, 
{\qquad} {\bar g}_{(t)}^{ij}={\bar h}_{(t)}^{ij}-
{\frac{t_0^2}{t^2}}{\frac{{\bar N}_{(t)}^i{\bar N}_{(t)}^j}{{\bar N}_t^2}},
\ena
where ${\bar h}_{(t)}^{ij}{\equiv}{\frac{t_0^2}{t^2}}{\bar N}_t^{-2}
{\tilde h}_{(t)}^{ij}$ are the components of the inverse spatial metric family 
${\bf {\bar h}}_t^{-1}$.

We close this section by noticing that the proper time ${\bar {\tau}}_{\cal F}$
of the FOs in ${\overline{\cal N}}$ is given by $d{\bar {\tau}}_{\cal F}
{\equiv}{\bar N}_t(x^{\mu}(t),t)dt$, which may in principle be integrated along
the world line of any FO. We thus see that: (1) quasi-metric space-time is
asymmetric: once time-orientation of the FHSs is chosen, $t$ must increase when 
${\bar {\tau}}_{\cal F}$ increases; (2) it is convenient to fix the
scale of ${\bar N}_t$ by setting it equal to unity for an isotropic, globally
empty toy universe (section 4); (3) it is possible to scale $t$ (and thus
${\bar {\tau}}_{\cal F}$) with a constant (this corresponds to an ordinary 
units transformation); (4) it is natural to choose $t=0$ for 
${\bar {\tau}}_{\cal F}=0$ as the beginning of time. Once this is done and
the units and the scale of ${\bar N}_t$ are fixed, $t$ is unique.
\subsection{The question of general covariance}
As mentioned above, the existence of a unique global time function implies that
space-time can be foliated into a set of ``preferred'' spatial hypersurfaces,
i.e., the FHSs. This distinguished foliation would indicate the possibility of 
defining a local ``preferred'' reference system, which with respect to the FOs
are taken to be ``at rest'', on average. Indeed the global time function 
naturally defines a class of ``preferred" coordinate systems, namely the GTCSs.
The FOs by definition always move orthogonally to the FHSs, but the FOs in 
general have non-zero 3-velocities in any GTCS. That is, restricted to GTCSs, 
the FOs are in general not stationary with respect to any particular coordinate
frame, but the FOs can still be at rest on average with respect to such a 
coordinate frame. This indicates that at cosmological scales, there is a 
general preferred (global) coordinate frame naturally associated with the 
quasi-metric geometrical framework and that said frame is in fact a particular 
GTCS. That is, at cosmological scales, on average the FOs should be at rest 
with respect to the cosmic rest frame (either defined as the average rest frame
of the cosmic substratum, or alternatively, defined as the cosmic frame where 
the cosmic relic radiation background is perceived as isotropic, on average).

The fact that the FOs should be at rest on average with respect to the cosmic
rest frame is convenient when doing quasi-metric cosmology. On the other hand, 
this is rather awkward when applying quasi-metric gravity to isolated systems.
Therefore, for isolated systems it is a convenient approximation to relax the
condition that the FHSs should be compact and replace it with the condition
that the FHSs should be asymptotically flat. This means that the global time
function $t$ will no longer unique but that there always can be found another 
time function $t'$ foliating ${\bf {\bar g}}_{t'}$ into an alternative set of 
hypersurfaces such that the isolated system is at rest on average with respect 
to observers moving orthogonally to the alternative set of hypersurfaces.
Moreover, the field equations may be transformed with respect to this new set 
of hypersurfaces. However, the field equations would not be invariant under 
said transformation; they would depend on the velocity of the isolated system 
with respect to the cosmic rest frame. In practice the 
``preferred frame''-effects introduced by said procedure should be small (of 
post-Newtonian order).

Now a question arises due to the natural existence of GTCSs within 
quasi-metric geometry. The GTCSs represent a set of preferred coordinate 
systems in quasi-metric relativity, so what happens to general covariance?
It was recognized long ago that if one with ``general covariance'' understands
``diffeomorphism-invariance", it is physically vacuous. In this sense, the 
quasi-metric framework is, of course, generally covariant. Nor does 
``general covariance" in the sense of invariance of form under general
coordinate transformations contain any general physical principle. (We do not 
expect our framework to be generally covariant in this sense, since equation
(2.15) is form invariant only for transformations from one GTCS to another.) 
Rather, the modern classification of theories depends on whether or not there 
exists ``prior geometry" [1, 4]. This concept refers to any dynamically 
relevant aspect of the geometry that cannot be changed by changing the matter 
distribution.

By construction, the quasi-metric framework guarantees the existence of prior 
geometry. After all, as mentioned above, we have restricted the geometry of 
the family ${\bf {\bar h}}_t$ to ensure that the global time function is 
unique. As we shall see, said prior geometry shows up explicitly via one 
specific term in one of the field equations. But besides that, there is no 
reason to expect that said prior geometry lays any extra restrictions on the 
dynamics of ${\bf {\bar g}}_t$.

It may be shown [15] that the local inertial frames of ${\bf g}_t$ can be 
identified with local Lorentz frames. This means that in quasi-metric theory,
there can be no local consequences of curved space-time for the 
non-gravitational physics and that local Lorentz invariance (LLI) holds for 
non-gravitational test experiments. But since the form of equation (2.15) is 
expected to hold in a GTCS only, this indicates that appropriate field 
equations can be defined only via (normal and tangential) projections of 
geometrical objects with respect to the FHSs. Such geometrical objects are 
expected to depend on the distinguished foliation of space-time into the set 
of FHSs, so neither local position invariance (LPI) nor LLI is expected to hold
exactly for local gravitational experiments testing the active aspects of 
gravity (such as Cavendish experiments). This should not, however, be a 
sufficient {\em a priori} reason to prefer metric over quasi-metric geometry 
as the basis for a new theory of gravity.

In the next section, we try to find for which general cases there will exist 
some sort of correspondence with the metric framework.
\subsection{The metric approximation}
In what follows, we describe how to construct the family ${\bf g}_t$ from the
family ${\bf {\bar g}}_t$ and a family of vector fields. We do this by first 
studying a particular example which we will call a ``metric approximation". A
metric approximation is defined as a metric family ${\bf {\bar g}}_t$ (or
${\bf g}_t$) which does not depend explicitly on the global time function $t$;
thus in effect we have only one metric. This metric necessarily represents an 
approximation since one neglects the effects of a changing global scale 
factor. This means that all the {\em global} NKE has been approximated away.
Note that metric approximations make physically sense for isolated 
systems only. 

When making a metric approximation, for particular cases it may be a further
good approximation to relax the requirement of compact FHSs, replacing it with 
the requirement that they should be asymptotically flat. If the FHSs are 
considered as asymptotically flat and embedded in asymptotically Minkowski 
space-time, this often represents a simplification and may make calculations 
easier. But having made this simplification, we still want to identify the FHSs
with the hypersurfaces $x^0=$constant, where $x^0$ is the global time 
coordinate we started out with. That is, when making metric approximations, we 
keep the unique global time coordinate $x^0$ we had to begin with and define 
the FHSs from the condition $x^0=$constant. Moreover, we may define a GTCS as a
global time coordinate system where this global time coordinate is used.
But note that, after having made a metric approximation and required that the
FHSs should be asymptotically flat, it is possible to choose other global time 
coordinates than the one we started out with. 

Metric approximations can be constructed from any one-parameter family
${\bf {\bar g}}_t$ of the form (2.15) by fixing $t$ equal to some constant 
reference epoch $t_{\rm r}$ (usually chosen to be $t_0$). (To make the resulting
single metric asymptotically flat, absorb any constant fraction of the type 
${\frac{t_{\rm r}}{t_0}}$ into the coordinates and then let 
$t_{\rm r}{\rightarrow}{\infty}$.) Thus any metric approximation may be written 
formally as ${\bf {\bar g}}_t{\rightarrow}{\bf {\bar g}}$ (or 
${\bf g}_t{\rightarrow}{\bf g}$); that is, metric families are approximated by 
single metrics. Now we will study the particular case where a GTCS exists in 
which the family ${\bf {\bar g}}_t$ does not depend on $x^0$ and the shift 
field family ${\frac{t_0}{t}}{\bar N}_{(t)}^j$ vanishes; this is called the 
{\em metrically static} case. Besides, for reasons explained later, we will 
require spherical symmetry. For this case, the metric approximation of the 
family ${\bf {\bar g}}_t$ is a single {\em static, spherically symmetric} 
metric ${\bf {\bar g}}$. That the metric ${\bf {\bar g}}$ is independent of 
time means that the FHSs are static, i.e., there is no net temporal change of 
scale anywhere on the FHSs.

But despite the absence of any global NKE for metric approximations, there 
still may be some effects of the NKE associated with the local gravitational 
field; these are the effects of the {\em local} NKE. To take into account these 
effects, it is necessary to make the transformation 
${\bf {\bar g}}{\rightarrow}{\bf g}$. That is, to construct a new metric 
${\bf g}$ from the metric ${\bf {\bar g}}$, we need to put the local NKE of 
the FHSs into the metric structure of ${\bf g}$. This means that, if the FOs 
are considered as stationary for the metric ${\bf g}$, then the stationariness 
of the FOs for the metric ${\bf {\bar g}}$ should be seen as a result of both 
the (local) non-kinematical expansion and the motion of the FOs taken together.
That is, the FOs in ${\overline{\cal N}}$ may be perceived as moving with a 
velocity $-{\bf v}$ say, with respect to the FOs in ${\cal N}$. Of course this 
motion is cancelled out when one adds another vector field ${\bf v}$ 
constructed from the local NKE. Thus one may think of the FOs in 
${\overline{\cal N}}$ as ``expanding" and ``falling" simultaneously such that 
the combination results in stationariness. However, this way of recovering 
stationariness is not equivalent to being stationary with no NKE. We may find 
the difference between ${\bf {\bar g}}$ and ${\bf g}$ by quantifying the two 
different versions of stationariness in terms of the vector field ${\bf v}$. 
As we shall see, the effect of ${\bf v}$ is that one must introduce correction 
factors or correction terms to the components of ${\bf {\bar g}}$; these 
corrections yield the wanted metric ${\bf g}$.

Before proceeding, we should choose an appropriate coordinate system. Whenever 
there is spherical symmetry with respect to one distinguished point, we 
naturally define a spherical GTCS $(x^0,r,{\theta},{\phi})$ (where $r$ is a
Schwarzschild radial coordinate) with the distinguished point at the spatial 
origin. The distinguished point plays the role as a natural ``centre of 
gravity''. Now the local NKE should take the form of a (in general different) 
uniform increase in scale for each tangent space of the FHSs. Due to the 
spherical symmetry, this increase in scale can be expressed as a purely radial,
outwards-pointing 3-vector field 
${\bf v}=v^{r}(r){\frac{\partial}{{\partial}r}}$. We thus define ${\bf v}$ by 
applying the ``Hubble law'' in {\em the tangent space} of each point of the 
FHSs. Besides, the distance $\ell$ to the spatial origin in the tangent space 
of each point on the FHSs, is given by ${\ell}=r{\sqrt{{\bar g}_{rr}}}$. The 
radial 3-vector field ${\bf v}$ and its norm $v{\equiv}{\sqrt{v^2}}$ are thus
defined by (where the ``local Hubble parameter'' ${\bar H}(r)$ is the 
``non-realized'' part of the quantity ${\bar H}_t$ defined in section 3.3)
\eqa
v^{r}(r)={\bar H}(r)r, {\qquad} 
v(r)={\bar H}(r){\ell}(r)={\bar H}(r){\sqrt{{\bar g}_{rr}(r)}}r.
\ena
We now calculate the new metric ${\bf g}$ when ${\bf {\bar g}}$ is static and 
spherically symmetric, as assumed above. For this case, the spherical GTCS 
defined above is in fact a HOCS and the metric takes the standard form
\eqa
\overline{ds}^{2} = -{\bar B}(r)(dx^0)^{2} + {\bar A}(r)dr^{2} + 
r^{2}d{\Omega}^{2},
\ena 
where ${\bar A}(r)$ and ${\bar B}(r)$ are smooth functions without zeros and 
$d{\Omega}^{2}{\equiv}d{\theta}^{2}+{\sin}^{2}{\theta}d{\phi}^{2}$.

We now try to find a new metric ${\bf g}$ from the metric ${\bf {\bar g}}$ of 
the type (2.20) and the vector field $\bf v$. From the local NKE, the FOs in 
${\overline{\cal N}}$ get a net 3-velocity 
${\bf v}=v^r{\frac{\partial}{{\partial}r}}$ in the positive radial direction 
of said spherical GTCS according to equation (2.19). To nullify this velocity, 
the FOs in ${\overline{\cal N}}$ must move with a velocity $-{\bf v}$. That is,
the static condition of the FOs in ${\overline{\cal N}}$ may be seen as the 
result of adding two opposite 3-velocities. Moreover, the FOs in ${\cal N}$ 
are also static, but in a way that does not involve any ``motion'' by 
definition. We now want to find the time and length intervals as expressed by 
the static FOs in $\cal N$. We do this for each tangent space, and taken 
together the result will be a new metric ${\bf g}$ derived from the metric 
${\bf {\bar g}}$ given in equation (2.20). And just as ${\bf {\bar g}}$ is the 
metric approximation of a spherically symmetric, metrically static family 
${\bf {\bar g}}_t$, the new metric ${\bf g}$ may be thought of as the metric 
approximation of a new spherically symmetric, metrically static family 
${\bf g}_t$.

We proceed in two steps, each step giving the same correction factors to the 
metric (2.20). Step one compensates for the effects of the 3-velocity 
$-{\bf v}$ of the ``static'' FOs in ${\overline{\cal N}}$ on the line element 
(2.20). That is, these FOs are not ``at rest'', but move inwards with velocity 
$-{\bf v}$ compared to the FOs in ${\cal N}$ when the NKE is not taken into 
account. Said effects of $-{\bf v}$ are integrated into ${\bf {\bar g}}$ to
begin with. To remove these effects, we include the effects of a compensating 
outwards 3-velocity ${\bf v}$ on said line element. In each tangent space, 
${\bf v}$ yields a correction to the radial interval due to the radial Doppler 
effect, the correction factor being 
${\Big(}{\frac{1+{\frac{v}{c}}}{1-{\frac{v}{c}}}}{\Big)}^{1/2}$. There is 
also an inverse time dilation correction factor 
$(1-{\frac{{v}^2}{c^2}})^{1/2}$ to the time interval. There are no 
correction factors for the angular intervals. 

Step two transforms the effects of the local NKE in ${\overline{\cal N}}$ into 
correction factors to the metric coefficients. In ${\overline{\cal N}}$, the 
NKE is described as a 3-velocity ${\bf v}$ in the positive radial direction for
each tangent space, and the effects of $\bf v$, as seen by the FOs in 
${\cal N}$, are taken care of by using a pair of correction factors identical 
to those in step one. So the new metric ${\bf g}$ constructed from the metric 
${\bf {\bar g}}$ given in equation (2.20) and ${\bf v}$ reads
\eqa
ds^{2} = -{\Big(}1-{\frac{{v}^{2}(r)}{c^{2}}}{\Big)}^{2}{\bar B}(r)(dx^0)^{2}+ 
{\bigg(}{\frac{1+{\frac{v(r)}{c}}}{1-{\frac{v(r)}{c}}}}
{\bigg)}^{2}{\bar A}(r)dr^{2} + r^{2}d{\Omega}^{2},
\ena
and it is clear that we must have $v(r)<c$ everywhere if the above expression 
is to make sense. Note that any metric approximation works because equation
(2.11) reduces to the geodesic equation (2.12) for metric approximations.

It must be stressed that any metric approximation is adequate only if one can
sensibly neglect the (global) NKE associated with a global scale factor. This 
observation leads to the question of which vector field family one should use 
when constructing ${\bf g}_t$ from ${\bf {\bar g}}_t$ for the general case. The 
answer clearly depends on the evolution of the FHSs. That is, the FHSs are not 
stationary as is in the above case, but will in general expand. However, the 
FHSs may in principle even contract in some places due to gravitation. This 
means that the net change of scale for each tangent space at points of each 
FHS may be positive or negative. But to construct the vector field family 
${\bf v}_t$ for the general case, we are interested in that part of the 
evolution of the FHSs which is not ``realized'' explicitly, i.e., not present
directly in ${\bf {\bar g}}_t$. When a metric approximation is sufficient, this 
non-realized part of the NKE is represented by ${\bar H}$. In the 
general case, however, the non-realized part of the evolution of the FHSs is 
not represented by ${\bar H}$ but some other quantity ${\bar y}_t$, say.

This means that for the general case, we should replace ${\bar H}_t$ with 
${\bar y}_t$. But besides this, there is a more subtle problem. Namely, that 
there is no obvious choice of ``the centre of gravity'' in the absence of 
spherical symmetry (this is why we imposed spherical symmetry when deriving 
equations (2.19) and (2.21)). That is, for the general case, one needs to find 
a 3-vector field family ${\bf {\bar x}}_{\cal F}$ representing ``a local 
distance vector field'' from some fictitious centre of gravity, and such that 
${\bf {\bar x}}_{\cal F}$ reduces to the radial coordinate vector field
$r{\frac{\partial}{{\partial}r}}$ for the spherically symmetric case. We will
define ${\bf {\bar x}}_{\cal F}$ in section 3.3. Meanwhile, we show how to 
construct ${\bf g}_t$ from ${\bf {\bar g}}_t$ and ${\bf v}_t$ when it is 
assumed that ${\bf {\bar x}}_{\cal F}$ is known.

To begin with, using the quantities ${\bar y}_t$ and ${\bf {\bar x}}_{\cal F}$, 
we may define the vector field family ${\bf v}_t$ just as we defined equation 
(2.19); this 3-vector field family represents the generalization of the vector 
field ${\bf v}$ used for spherically symmetric metric approximations. Instead 
of equation (2.19), we define (using an arbitrary GTCS)
\eqa
v^{j}_{(t)}={\bar y}_t{\bar x}_{\cal F}^j, \qquad 
v={\bar y}_t{\sqrt{{\bar g}_{(t)ks}{\bar x}_{\cal F}^k{\bar x}_{\cal F}^s}}.
\ena
Note that the explicit $t$-dependence should cancel out in the second formula 
and that we can eliminate any possible implicit $t$-dependence via ${\bar N}_t$
by making the substitution $t=x^0/c$. Also notice that the total NKE in the 
tangent space of each point of the FHSs can still be found from ${\bar H}_t$. 
But the global NKE associated with a global scale factor must be subtracted 
from ${\bar H}_t$ when constructing ${\bf v}_t$; we thus get equation (2.22) 
instead of equation (2.19). The effects of the NKE associated with ${\bf v}_t$ 
are the {\em local} effects of the NKE as mentioned above.

In what follows, we will write down the formulae for the components of 
${\bf g}_t$ which must hold for the general case. The main difference from the
spherically symmetric case, is that the ``extra motion'' of the FOs in 
${\overline{\cal N}}$ is locally along the ${\bf {\bar x}}_{\cal F}$-direction
rather than in the radial direction. But besides that, the principles behind 
the construction of ${\bf g}_t$ are similar. To begin with, we notice that the 
transformation formula should be identical to the spherically symmetric case 
when no spatial components are involved. That is, as in equation (2.21) we 
define
\eqa
g_{(t)00}={\Big (}1-{\frac{v^2}{c^2}}{\Big )}^2{\bar g}_{(t)00}.
\ena
To guess reasonable transformation formulae when spatial components are 
involved, it is convenient to define the unit vector field 
${\bf {\bar e}}_{\cal F}{\equiv}{\frac{t_0}{t}}{\bar e}_{\cal F}^s
{\frac{\partial}{{\partial}x^s}}$ and the corresponding covector field 
${\bf {\bar {\omega}}}_{\cal F}{\equiv}{\frac{t}{t_0}}{\bar {\omega}}_{{\cal F}s}
dx^s$ along ${\bf {\bar x}}_{\cal F}$. But it is in general not possible or 
practical to construct a GTCS where ${\bf {\bar e}}_{\cal F}$ is parallel to one
of the coordinate vector fields. For this reason, one expects that the general 
transformation formulae for spatial components should involve components of 
${\bf {\bar e}}_{\cal F}$ and ${\bf {\bar {\omega}}}_{\cal F}$. By trial and error
(involving the spherically symmetric case using non-spherical coordinates), we 
find the following transformation formula for the purely spatial components of 
${\bf g}_t$:
\eqa
g_{(t)ij}={\bar g}_{(t)ij}+{\frac{t^2}{t_0^2}}{\frac{4{\frac{v}{c}}}
{(1-{\frac{v}{c}})^2}}{\bar {\omega}}_{{\cal F}i}{\bar {\omega}}_{{\cal F}j}.
\ena
The transformation formula (2.24) has the correct correspondence with the
result found in equation (2.21) and it may be shown that this correspondence
holds in any GTCS. Moreover, from equations (2.23), (2.24) and consistency
requirements, we may find the remaining components of ${\bf g}_t$, namely
\eqa
g_{(t)0j}={\Big (}1-{\frac{v^2}{c^2}}{\Big )}{\Big [}{\bar g}_{(t)0j}+
{\frac{t}{t_0}}{\frac{2{\frac{v}{c}}}{1-{\frac{v}{c}}}}
({\bar e}_{\cal F}^s{\bar N}_{(t)s}){\bar {\omega}}_{{\cal F}j}{\Big ]}.
\ena
Given equations (2.23), (2.24) and (2.25), it is now straightforward to find
the transformation formulae valid for ${\bf g}_t^{-1}$.
\section{Equations of evolution}
Our goal now is to construct a theory of gravity consistent with the 
quasi-metric framework laid out in section 2. The class of such theories must 
satisfy the postulates: (1) space-time is constructed as a 4-dimensional 
submanifold ${\cal N}$ of a 5-dimensional product manifold 
${\cal M}{\times}{\bf R}$ as described in section 2. This implies that 
space-time is equipped with a global time function, an one-parameter family of 
Lorentzian 4-metrics ${\bf g}_t$ (constructible from another family 
${\bf {\bar g}}_t$ where each metric is restricted by the constraints stated 
in section 2.1), and a degenerate connection ${\,}{\topstar{\nabla}}{\,}$ 
compatible with the metric family; (2) inertial test particles follow curves 
calculated from equation (2.11) with the condition ${\bf a}_t=0$ and where the 
relationship between ${\stackrel{\star}{\bf a}}$ and ${\bf a}_t$ must be 
specified; (3) in the local inertial frames of the metric family ${\bf g}_t$, 
the non-gravitational physics is as in special relativity (SR).

Postulate (3) means that it should be possible to identify the local inertial 
frames of ${\bf g}_t$ with local Lorentz frames, see [15] for verification. 
This implies that the quasi-metric framework by itself induces no local 
consequences of curved space-time on the non-gravitational physics. Thus both 
LLI and LPI should hold within the quasi-metric framework. But for any finite 
scale, the free fall curves of ${\bf g}_t$ as obtained from from equation 
(2.11) do not coincide with the geodesics of any single space-time metric, see 
[15] for a further discussion. Furthermore, this deviation depends on position 
and on the velocity with respect to the FOs. Such dependence implies that 
non-gravitational laws may differ from their counterparts in any 
metric theory for any finite-sized observer. 

On the other hand, postulate (2) ensures that the weak equivalence principle 
(WEP) is fulfilled (i.e., that test particles' trajectories do not depend on 
the particles' internal composition). The sum of the WEP, LLI and LPI is 
called the Einstein equivalence principle (EEP); every metric theory of 
gravity satisfies the EEP [1]. The proposition that every consistent theory 
which satisfies the WEP also satisfies the EEP is called Schiff's conjecture 
(SC) [1]. Thus SC continues to hold for any theory consistent with the above 
postulates.

From equation (2.15), we see that how the scale of the FHSs varies from event 
to event is given by the scale factor ${\bar F}_t{\equiv}{\bar N}_tct$. 
One may interpret this scale factor as resulting from a position-dependent 
transformation of length units. Because any dependence on scale of physical 
laws should be experimentally determined only by the variation with position 
of dimensionless numbers, we are free to {\em define} as constants physical 
constants that cannot be related to other such constants by dimensionless 
numbers (see, e.g., [5]). Thus, by definition, the speed of light $c$ in 
vacuum and Planck's constant ${\hbar}$ are constants, i.e., independent of 
${\bar F}_t$. Moreover, since the EEP must be satisfied, the elementary charge 
$e$ must also be a constant, ensuring that the fine structure constant does not 
vary in quasi-metric space-time. Now, when writing down line elements 
describing the geometry of space-time, some set of units is implicitly assumed
used. Usually one does not separate between units defined from purely 
non-gravitational systems and other types of units. However, this must be done 
if one wants to describe position-dependent scale changes between 
non-gravitational and gravitational systems.

We are thus led to the notion that any fixed ``atomic'' units, i.e., units
operationally defined by means of non-gravitational forces only, may be
considered as formally variable throughout space-time; this applies both to
${\overline{\cal M}}$ and to ${\overline{\cal N}}$. That $c$ and ${\hbar}$ are
not formally variable by definition, implies that the formal variation of 
length and time units are similar and inverse to that of energy (or mass) 
units. Besides, since $e$ is not formally variable, there is no formal 
variation of charge units. One may expect that the formally variable units 
should represent scale changes between non-gravitational and gravitational 
systems, where said scale changes should be found from a theory of gravity.

This means that all physical laws are formulated in terms of ``variable''
atomic units. That is, gravitational quantities are considered as variable 
measured in such units. But which quantities do we define as gravitational?
Not unreasonably, we define the family ${\bf {\bar g}}_t$ as gravitational
since ${\bf {\bar g}}_t$ is by definition a solution of gravitational field
equations. Moreover, Newton's gravitational ``constant'' $G_{\rm N}$ and also
the (active) stress-energy tensor ${\bf T}_t$ considered as a source of 
gravity, should be defined as gravitational quantities. 

It is convenient to introduce the scalar field ${\Psi}_t$ which defines how 
atomic time units vary in space-time. Any gravitational quantity should then 
be associated with ${\Psi}_t$ to some power; this factor should tell how the 
quantity changes due to the formally variable units. To illustrate this point,
we note that $G_{\rm N}$ has a formal variation similar to that of time units 
squared. Since $G_{\rm N}$ is a gravitational quantity, this means that the 
extra function associated with $G_{\rm N}$ is a factor ${\Psi}_t^{-2}$. However, 
since mass is associated with a factor ${\Psi}_t$ and since this is not 
detectable in non-gravitational experiments involving material particles only, 
rather than $G_{\rm N}$ we will in effect have a ``screened'' gravitational 
coupling parameter $G_t^{\rm S}$ for material particles. Moreover, $G_t^{\rm S}$ 
is associated with a factor ${\Psi}_t^{-1}$. On the other hand, charge is not 
formally variable so we must also have a second, ``bare'' gravitational 
coupling parameter $G_t^{\rm B}$ that couples to charge squared, or more 
generally, to the electromagnetic field, such that $G_t^{\rm B}$ will be 
associated with a factor ${\Psi}_t^{-2}$. That is, measured in atomic units, 
$G_t^{\rm B}$ and $G_t^{\rm S}$ cannot be constants. But since $G_t^{\rm B}$ and 
$G_t^{\rm S}$ always occur in combination with charge or mass-energy, it is more 
convenient to {\em define} $G_t^{\rm B}$ and $G_t^{\rm S}$ to take constant values
$G^{\rm B}$ and $G^{\rm S}$, respectively, in atomic units and rather 
differentiate between {\em active mass} and {\em passive mass}. (Similarly, one
must differentiate between {\em active charge} and {\em passive charge}.) That 
is, active mass (or active charge) acts as a source of gravity and 
varies measured in atomic units. On the other hand, passive mass (inertial 
mass or passive gravitational mass) does not vary and should be used in atomic 
interactions (and similarly for passive charge). In what follows, we will 
denote any active mass $m_t$ (corresponding to a passive mass $m$) with a
$t$-index. 

Mass-energy considered as a gravitational quantity, scales via the variation of
mass units, which vary as ${\Psi}_t^{-1}$. Thus seen, active mass-energy is 
associated with a factor ${\Psi}_t$. But since we have defined the 
gravitational coupling parameters to be constant, we must explicitly include 
the variation of the gravitational coupling into the active mass-energy. This 
means that {\em for material particles, active mass-energy $m_t$ varies as 
${\Psi}_t^{-1}$ measured in atomic units}, whereas {\em for the electromagnetic 
field, active mass-energy $m_t$ varies as ${\Psi}_t^{-2}$ measured in atomic 
units}.

The question now is how ${\Psi}_t$ (or $m_t$) changes from event 
to event in ${\overline{\cal N}}$. Since line elements of the type (2.15)
are measured in atomic units, we propose that any change in the scale factor 
${\bar N}_tct$ can be explained solely by means of the formally variable 
units. This will work if ${\Psi}_t{\equiv}{\bar F}_t^{-1}=
{\frac{1}{{\bar N}_tct}}$, or equivalently, ${\Psi}_t=
{\frac{t_0{\bar N}_0}{t{\bar N}_t}}{\Psi}_0$, so that
\eqa
m_t=
\left\{
\begin{array}{ll}
{\frac{t{\bar N}_t}{t_0{\bar N}_0}}{m}_0 & $for a fluid of material particles,$
\\ [1.5ex]
{\frac{t^2{\bar N}^2_t}{t^2_0{\bar N}^2_0}}{m}_0 & $for the electromagnetic 
field,$
\end{array}
\right.
\ena
where ${\Psi}_0$, $m_0$ etc., refer to values taken at some arbitrary reference
event. By convention, we may choose ${\bar N}_0=1$, ${\Psi}_0={\frac{1}{ct_0}}$
which means that the chosen (hypothetical) reference situation is an empty
universe at epoch $t_0$ (see section 4). Note that the formal variability of
gravitational quantities should be unaffected by the construction of
${\bf g}_t$ from ${\bf {\bar g}}_t$ since this construction only 
involves the formally invariant vector fields ${\bf v}_t$.

The strong equivalence principle (SEP) is essentially a generalization of the 
EEP. That is, for a theory to satisfy the SEP, the EEP must be valid for 
self-gravitating bodies and gravitational test experiments as well as for 
non-gravitating bodies and non-gravitational test experiments, see, e.g., [1].
Since theories consistent with postulate (3) satisfy the EEP, one might
hope to find a quasi-metric theory that satisfies the SEP as well. However,
since active mass is not equal to passive mass, the SEP in its most stringent
interpretation cannot be satisfied within the quasi-metric framework; all we 
can hope for is that the gravitational weak principle of equivalence (GWEP) 
(that the WEP is valid for self-gravitating bodies as well) holds.

But the result of any laboratory experiment testing the constancy of the 
``effective'' gravitational ``constant'' $G^{\rm eff}$, should reveal only the 
dependence on source composition of $G^{\rm eff}$ and its variability via the 
formally variable units and not any extra, independent kind of variability. 
Thus we should be able to investigate the dynamical properties of 
${\bf {\bar g}}_t$ without having to worry about a ``physically" variable 
$G^{\rm eff}$, independent of said variability. Besides, the formal variability 
of $G^{\rm eff}$ should depend only on quantities constructed from pieces of the 
metric family ${\bf {\bar g}}_t$ and not on any extra gravitational fields. But
this means that we may be able to construct a theory of gravity, consistent 
with the quasi-metric framework, which satisfies the GWEP. In fact the GWEP 
holds since it may be shown that active mass dust particles really move on 
geodesics of ${\bf {\bar g}}_t$ and of ${\bf g}_t$ [15].
\subsection{Projections}
If we define a GTCS ${\{}x^{\mu}{\}}={\{}x^0,x^{i}{\}}$, this can be used to
find the coordinate representations of the orthonormal vector field basis 
${\{}{\bf {\bar n}}_t,{\bf e}_{i}{\}}$ defining the moving frame field spanned 
by the FOs in ${\overline {\cal N}}$, where 
${\bf e}_{i}{\equiv}{\frac{\partial}{{\partial}x^{i}}}$. The vector field family
${\bf {\bar n}}_t$ is identified with the unit-length vector field family 
$({\bar n}_{(t){\mu}}{\bar n}_{(t)}^{\mu}=-1)$ tangent to the world lines of the 
FOs, but ${\bf {\bar n}}_t$ is in general not parallel to the coordinate vector
field ${\frac{\partial}{{\partial}x^0}}$. For the coordinate representation of 
${\bf {\bar n}}_t$ in a GTCS, see below. We can use ${\bf {\bar n}}_t$ and its 
corresponding covector field together with the metric tensor family 
${\bf {\bar h}}_t$ intrinsic to the FHSs to project space-time objects into 
space objects intrinsic to the FHSs. That is, repeated scalar products with 
${\bf {\bar n}}_t$ eventually reduce space-time objects to space scalars, 
whereas repeated scalar products with ${\bf {\bar h}}_t$ eventually reduce 
space-time objects to space objects of the same type and rank. Combinations of 
these operations are possible. To avoid confusion about whether we are dealing 
with a space-time object or one of its spatial projections, we will label space
objects with a ``hat'' if necessary. That is, a maximal number of projections 
of the space-time object ${\bf A}_t$ may result in several different space 
objects, each of will be denoted by ${\bf {\hat A}}_t$ (if necessary, component
notation will be used to avoid ambiguity).

Component-wise, we define a label ${\bar {\perp}}$ to indicate a projection by 
$-{\bf {\bar n}}_t$, whereas spatial covariant indices indicate spatial 
projections. Note that contravariant spatial indices do {\em not} indicate 
spatial projections; here one must use the ``hat'' label to differentiate 
between the components of a space-time object and its spatial projection. As 
an example, we take the space vectors ${\hat Q}^{j}_{(t){\bar {\perp}}}{\equiv}
-{\bar h}_{(t){\beta}}^jQ^{\beta}_{(t){\alpha}}{\bar n}_{(t)}^{\alpha}$ which
are mixed projections of the the space-time tensors $Q^{\beta}_{(t){\alpha}}$.
The corresponding space covectors are $Q_{(t)j{\bar {\perp}}}=
{\hat Q}_{(t)j{\bar {\perp}}}{\equiv}-{\bar h}_{(t)j}^{\beta}
Q_{(t){\beta}{\alpha}}{\bar n}_{(t)}^{\alpha}$.

The decomposition of ${\bf {\bar g}}_t$ and its inverse with respect to each 
domain FHS is given by
\eqa
{\bar g}_{(t){\mu}{\nu}}&=&-{\bar n}_{(t){\mu}}{\bar n}_{(t){\nu}} + 
{\bar h}_{(t){\mu}{\nu}}, \nonumber \\
{\bar g}_{(t)}^{{\mu}{\nu}}&=&-{\bar n}_{(t)}^{\mu}{\bar n}_{(t)}^{\nu} + 
{\bar h}_{(t)}^{{\mu}{\nu}},
\ena
where ${\bar h}_{(t){\mu}{\nu}}$ are the components of ${\bf {\bar h}}_t$. We
see from equation (3.2) that ${\bar h}_{(t){\mu}{\nu}}$ represents the restriction
to the domain FHS of each space-time metric ${\bar g}_{(t){\mu}{\nu}}$. In a
GTCS, ${\bf {\bar h}}_t$ considered as a space-time tensor field family, has 
the mixed or time components
\eqa
{\bar h}_{(t)00}={\bar N}_{(t)k}{\bar N}_{(t)}^k, \quad
{\bar h}_{(t)j0}={\bar h}_{(t)0j}={\frac{t}{t_0}}{\bar N}_{(t)j}, \quad 
{\bar h}_{(t)0}^j={\frac{t_0}{t}}{\bar N}_{(t)}^j=
{\frac{t_0}{t}}{\bar N}^{-2}_t{\tilde h}_{(t)}^{js}{\bar N}_{(t)s},
\ena
other components involving time vanishing, whereas ${\bf {\bar n}}_t$ has the 
coordinate representation
\eqa
{\bar n}_{(t)}^0={\bar N}_t^{-1}, \qquad {\bar n}_{(t)}^j=
-{\frac{t_0}{t}}{\frac{{\bar N}_{(t)}^j}{{\bar N}_t}}, \qquad
{\bar n}_{(t)0}=-{\bar N}_t, \qquad {\bar n}_{(t)j}=0.
\ena
Using equations (3.2), (3.3) and (3.4) we may verify that 
${\bar h}_{(t)}^{{\mu}{\nu}}{\bar n}_{(t){\nu}}=
{\bar h}_{(t){\mu}{\nu}}{\bar n}_{(t)}^{\nu}=0$,
${\bar g}_{(t){\bar {\perp}}{\bar {\perp}}}=-1$, 
${\bar g}_{(t){\bar {\perp}}j}={\bar g}_{(t)j{\bar {\perp}}}=0$, 
${\bar g}_{(t)ij}={\bar h}_{(t)ij}$ and
${\bar g}_{(t)}^{ij}={\bar h}_{(t)}^{ij}-
{\frac{t_0^2}{t^2}}{\frac{{\bar N}_{(t)}^i{\bar N}_{(t)}^j}{{\bar N}_t^2}}$.

Any metric ${\bf {\bar g}}{\in}{\bf {\bar g}}_t$ defines a symmetric linear 
metric connection field ${\bf {\bar {\nabla}}}$. Note that $t$ is treated as 
a constant when using ${\bf {\bar {\nabla}}}$. In component notation, we denote
the operation of ${\bf {\bar {\nabla}}}$ on space-time objects by the 
symbol `;'. To illustrate this by an example, if ${\bf {\bar Y}}_t$ is vector 
field family, the components of the tensor field family 
${\bf {\bar {\nabla}}}{\bf {\bar Y}}_t$ are ${\bar Y}^{\alpha}_{(t);{\beta}}$. By 
projecting the connections into the FHSs, we can define the
hypersurface-intrinsic connection field family ${\bar {\triangle}}_t$ 
compatible with the hypersurface-intrinsic metric family ${\bf {\bar h}}_t$
(treating $t$ as a constant). In component notation, we denote the operation 
of ${\bar {\triangle}}_t$ on space objects by the symbol `${\mid}$'. (Since we 
never mix fields constructed from the different families ${\bf g}_t$ and 
${\bf {\bar g}}_t$, we let the symbols `;' and `${\mid}$' denote a space-time 
covariant derivative respectively a spatial covariant derivative also in 
$\cal M$. Doing otherwise would imply a serious risk of running out of 
appropriate symbols.)

Also, the manifold structure of ${\overline{\cal M}}$ (and of $\cal M$) 
naturally defines a Lie derivative {\pounds}$_{{\bf {\bar z}}_t}$ which operates on
space-time objects. We may also define a spatial Lie derivative 
${\bf {\cal L}}_{{\bf {\bar z}}_t}$ which operates only on space objects, where the 
vector field family ${\bf {\bar z}}_t$ is still treated as a space-time vector 
field family. See ref. [7] for the detailed projection formulae relating 
{\pounds}$_{{\bf {\bar z}}_t}$ and ${\bf {\cal L}}_{{\bf {\bar z}}_t}$. Analogously, 
the manifold structure of ${\overline{\cal N}}$ (and of $\cal N$) defines the 
Lie derivative ${\,}{\topstar{${\pounds}$}}_{{\,}{\bf {\bar z}}}$ and its spatial 
counterpart ${\,}{\topstar{\cal L}}_{{\,}{\bf {\bar z}}}$, where ${\bf {\bar z}}$ 
now is a 5-dimensional vector field (i.e., ${\bf {\bar z}}$ may have a nonzero 
$t$-component). The only differences between the Lie derivatives in 
${\overline{\cal M}}$ and ${\overline{\cal N}}$ arise from the $t$-dependence 
of the metric family.

Furthermore, the degenerate connection ${\,}{\topstar{\bar {\nabla}}}{\,}$ is 
compatible with the non-degenerate part of the metric family 
${\bf {\bar g}}_t$. In component notation, we use the symbol `${\bar *}$' to 
denote the operation on space-time fields of 
${\,}{\topstar{\bar {\nabla}}}{\,}$. Just as for metric space-time geometry, 
one may construct higher rank tensor fields by using these operators. As an 
example of how this is is done, take a vector field family ${\bf {\bar Y}}_t$ 
and construct the object
${\,}{\topstar{\bar {\nabla}}}{\,}{\,}{\bf {\bar Y}}_t$ defined by its 
components in a GTCS; ${\bar Y}_{(t){\bar *}i}^{\alpha}{\equiv}
{\bar Y}_{(t);i}^{\alpha}$, ${\bar Y}^{\alpha}_{(t){\bar *}0}{\equiv}
{\bar Y}^{\alpha}_{(t);0}+c^{-1}{\bar Y}^{\alpha}_{(t){\bar *}t}={\bar Y}^{\alpha}_{(t);0}
+c^{-1}{\bar Y}^{\alpha}_{(t)},_t+c^{-1}{\,}{\topstar{\bar {\Gamma}}}^{{\,}{\alpha}}
_{t{\beta}}{\,}{\bar Y}_{(t)}^{\beta}$. Since the extra terms represent the 
components of a 4-vector, 
${\,}{\topstar{\bar {\nabla}}}{\,}{\,}{\bf {\bar Y}}_t$ is a family of rank 2 
tensor fields. Similarly, the covariant divergence 
${\,}{\topstar {\bar {\nabla}}}{\,}{\cdot}{\bf {\bar Y}}_t$, or in component 
notation, ${\bar Y}^{\alpha}_{(t){\bar *}{\alpha}}{\equiv}c^{-1}{\bar Y}^0_{(t){\bar *}t}
+{\bar Y}^{\alpha}_{(t);{\alpha}}=
(c{\bar N}_t)^{-1}{\frac{\partial}{{\partial}t}}{\bar Y}_{(t){\bar {\perp}}}+
{\bar Y}^{\alpha}_{(t);{\alpha}}$, is a scalar field.

Next, we define the 4-acceleration field family ${\bf {\bar a}}_{\cal F}$ 
corresponding to the normal unit vector field family ${\bf {\bar n}}_{t}$ by
$c^{-2}{\bar a}_{\cal F}^{\mu}{\equiv}{\bar n}^{\mu}_{(t);{\nu}}{\bar n}_{(t)}^{\nu}$. 
This is a purely spatial vector field family, since 
${\bar a}_{{\cal F}{\bar {\perp}}}=-{\bar a}_{{\cal F}{\mu}}{\bar n}_{(t)}^{\mu}=
-{\bar a}_{\cal F}^{\mu}{\bar n}_{(t){\mu}}=0$. Besides, from equation (3.4) and 
the definition of ${\bf {\bar a}}_{\cal F}$, we find that
\eqa
c^{-2}{\bar a}_{{\cal F}j}={\frac{\partial}{{\partial}x^{j}}}{\ln}{\bar N}_t.
\ena
Furthermore, we define the family of extrinsic curvature tensors 
${\bf {\bar K}}_t$ describing how each FHS is embedded in the relevant 
${\bf {\bar g}}{\in}{\bf {\bar g}}_t$. That is, we define
\eqa
{\bar K}_{(t){\mu}{\nu}}{\equiv}
-{\bar h}^{\alpha}_{(t){\mu}}{\bar h}^{\beta}_{(t){\nu}}
{\bar n}_{(t){\alpha};{\beta}},
\ena
which is symmetric since there is no space-time torsion in 
${\overline{\cal M}}$. Projections show that ${\bf {\bar K}}_t$ is a purely 
hypersurface-intrinsic object, i.e.,
${\bar K}_{(t){\bar {\perp}}{\bar {\perp}}}={\bar K}_{(t){\bar {\perp}}j}=0$. It can also 
be readily shown [2, 4, 7] that ${\bf {\bar K}}_t$ may be written as 
${\bar K}_{(t){\mu}{\nu}}=-{\frac{1}{2}}${\pounds}$_{{\bf {\bar n}}_t}
{\bar h}_{(t){\mu}{\nu}}$. After some calculations, we find the spatial components
\eqa
{\bar K}_{(t)ij}={\Big (}{\frac{{\bar N}_t,_{\bar {\perp}}}{{\bar N}_t}}
-{\frac{t_0}{t}}c^{-2}{\bar a}_{{\cal F}k}{\frac{{\bar N}_{(t)}^k}{{\bar N}_t}}
{\Big )}{\bar h}_{(t)ij}+{\frac{t}{2{\bar N}_tt_0}}({\bar N}_{(t)i{\mid}j}+
{\bar N}_{(t)j{\mid}i})-{\frac{1}{2}}{\frac{t^2}{t_0^2}}{\bar N}_t
{\frac{\partial}{{\partial}x^0}}{\tilde h}_{(t)ij}, \nonumber \\ 
{\bar K}_{(t)i}^i{\equiv}{\bar K}_t=
{\frac{t_0}{t}}{\frac{{\bar N}_{(t){\mid}i}^i}{{\bar N}_t}}
+3{\Big (}{\frac{{\bar N}_t,_{\bar {\perp}}}{{\bar N}_t}}-{\frac{t_0}{t}}
c^{-2}{\bar a}_{{\cal F}k}{\frac{{\bar N}_{(t)}^k}{{\bar N}_t}}{\Big )}
-{\frac{1}{2{\bar N}_t}}{\tilde h}_{(t)}^{ij}
{\frac{\partial}{{\partial}x^0}}{\tilde h}_{(t)ij}.
\ena
Note the difference between the spatial components of 
{\pounds}$_{{\bf {\bar n}}_t}{\bar h}_{(t){\mu}{\nu}}$ and
${\cal L}_{{\bf {\bar n}}_t}{\bar h}_{(t)ij}$; the latter Lie derivative acts on 
spatial objects only and this means that ${\bf {\bar h}}_t$ should be treated 
as a tensor on space when using ${\cal L}_{{\bf {\bar n}}_t}$ but as a tensor on 
space-time when using {\pounds}$_{{\bf {\bar n}}_t}$. The relationship
\eqa
{\cal L}_{{\bf {\bar n}}_t}{\bar h}_{(t)ij}=
{\pounds}_{{\bf {\bar n}}_t}{\bar h}_{(t)ij}+c^{-2}
{\frac{t}{t_0}}{\Big (}{\bar a}_{{\cal F}j}
{\frac{{\bar N}_{(t)i}}{{\bar N}_t}}+{\bar a}_{{\cal F}i}
{\frac{{\bar N}_{(t)j}}{{\bar N}_t}}{\Big )},
\ena
follows directly from the definitions of the Lie derivatives. Relationships
similar to (3.8) can be found between the different Lie derivatives of any 
tensor field.

From equation (2.15) we can find the first derivative of ${\bf {\bar h}}_t$ 
with respect to $t$. This is
\eqa
{\frac{\partial}{{\partial}t}}{\bar h}_{(t)ij}=
2{\Big (}{\frac{1}{t}}+{\frac{{\bar N}_t,_t}{{\bar N}_t}}{\Big )}
{\bar h}_{(t)ij}+{\frac{t^2}{t_0^2}}{\bar N}_t^2{\frac{\partial}{{\partial}t}}
{\tilde h}_{(t)ij}, 
\ena
\eqa
{\frac{\partial}{{\partial}t}}{\bar h}_{(t)}^{ij}=
-2{\Big (}{\frac{1}{t}}+{\frac{{\bar N}_t,_t}{{\bar N}_t}}{\Big )}
{\bar h}_{(t)}^{ij}+{\frac{t_0^2}{t^2}}{\bar N}_t^{-2}
{\frac{\partial}{{\partial}t}}{\tilde h}_{(t)}^{ij}.
\ena
We now write down how ${\bar a}_{{\cal F}j}$ varies in the direction 
orthogonal to the FHSs. We find
\eqa
{\pounds}_{{\bf {\bar n}}_t}{\bar a}_{{\cal F}j}=
{\cal L}_{{\bf {\bar n}}_t}{\bar a}_{{\cal F}j}-
{\frac{t_0}{t}}{\Big [}{\bar a}_{{\cal F}s}
{\frac{{\bar N}_{(t)}^s}{{\bar N}_t}}{\Big ]}c^{-2}{\bar a}_{{\cal F}j},
\ena
and similar formulae for other covectors, and moreover
\eqa
{\pounds}_{{\bf {\bar n}}_t}{\bar a}_{{\cal F}j}=
{\bar N}_t^{-1}{\Big (}{\bar a}_{{\cal F}j},_0-{\frac{t_0}{t}}
[{\bar a}_{{\cal F}k}{\bar N}_{(t)}^k],_j{\Big )}, {\qquad}
c^{-2}{\frac{\partial}{{\partial}x^0}}{\bar a}_{{\cal F}j}=
{\frac{{\bar N}_t,_{j0}}{{\bar N}_t}}-
c^{-2}{\bar a}_{{\cal F}j}{\frac{{\bar N}_t,_0}{{\bar N}_t}},
\ena
\eqa
{\pounds}_{{\bf {\bar n}}_t}{\bar a}_{{\cal F}}^j=
{\bar h}_{(t)}^{sj}{\pounds}_{{\bf {\bar n}}_t}{\bar a}_{{\cal F}s}
+2{\bar a}_{{\cal F}s}{\bar K}_{(t)}^{sj},
\ena
and still furthermore we find that
\eqa
{\pounds}_{{\bf {\bar n}}_t}({\bar a}_{{\cal F}s}{\bar a}_{\cal F}^s)=
2{\bar a}_{\cal F}^s{\Big [}{\pounds}_{{\bf {\bar n}}_t}{\bar a}_{{\cal F}s}
+{\bar a}_{{\cal F}k}{\bar K}_{(t)s}^k{\Big ]}.
\ena
The projection of metric space-time covariant derivatives into spatial objects
and expressed in components, may be subject to some confusion of notation. The 
important thing to remember is that, when a component contains a metric
{\em space-time} derivative, the projectors ${\bf {\bar n}}_t$ and 
${\bf {\bar h}}_t$ are always used {\em outside} the derivative, whereas if 
the component contains a {\em space} derivative, the projectors are used 
{\em inside} the derivative. For example, a mixed projection of the rank 2 
space-time tensor field family ${\bf {\bar {\nabla}}}{\bf {\bar Y}}_t$ yields 
the components 
${\bar Y}_{(t){\bar {\perp}};j}{\equiv}-{\bar Y}_{(t){\mu};j}{\bar n}_{(t)}^{\mu}$ and 
not $-({\bar Y}_{(t){\mu}}{\bar n}_{(t)}^{\mu})_{;j}$, whereas the corresponding 
components of the spatial covector fields 
${\bf {\bar {\triangle}}}{\bf {\bar y}}_t$ are ${\bar y}_{(t){\bar {\perp}}{\mid}j}
{\equiv}-({\bar y}_{(t){\mu}}{\bar n}_{(t)}^{\mu})_{{\mid}j}$.

Formulae expressed in component notation for various projections of metric
space-time covariant derivatives of general families of space-time tensors, may
be found by first finding the projection formulae for 
${\bf {\bar {\nabla}}}{\bf {\bar Z}}_t$, where ${\bf {\bar Z}}_t$ is some 
family of space-time covector fields. One may then apply these formulae to 
every covariant index of a general family of totally covariant space-time 
tensors. Such formulae are derived in, e.g., [2] (see also [7]), and are given 
by
\eqa
{\bar Z}_{(t){\bar {\perp}};j} &=& {\bar Z}_{(t){\bar {\perp}}{\mid}j}-
{\bar K}_{(t)sj}{\hat {\bar Z}}^{s}_{(t)}, {\nonumber} \\
{\bar Z}_{(t)i;j}&=&{\bar Z}_{(t)i{\mid}j}-{\bar Z}_{(t){\bar {\perp}}}
{\bar K}_{(t)ij}, 
{\nonumber} \\
{\bar Z}_{(t){\bar {\perp}};{\bar {\perp}}} &=& 
-{\cal L}_{{\bf {\bar n}}_t}{\bar Z}_{(t){\bar {\perp}}}
-c^{-2}{\bar a}_{\cal F}^{s}{\bar Z}_{(t)s}, {\nonumber} \\
{\bar Z}_{(t)j;{\bar {\perp}}}&=& 
-{\cal L}_{{\bf {\bar n}}_t}{\bar Z}_{(t)j}+
c^{-2}{\bar a}_{{\cal F}j}{\Big (}{\frac{t_0}{t}}
{\frac{{\bar N}_{(t)}^s}{{\bar N}_t}}{\bar Z}_{(t)s}-
{\bar Z}_{(t){\bar {\perp}}}{\Big )}-{\bar K}_{(t)sj}{\hat {\bar Z}}^{s}_{(t)}.
\ena
Since we will explicitly need the projection formulae for the metric covariant
derivative of tensor fields of rank 2, we list the relevant ones.
If ${\bf {\bar W}}_t$ is a family of such fields, then we have [2]
\eqa
{\bar W}_{(t){\bar {\perp}}i;j}&=&{\bar W}_{(t){\bar {\perp}}i{\mid}j}-
{\bar K}_{(t)ij}{\bar W}_{(t){\bar {\perp}}{\bar {\perp}}}-
{\bar K}_{(t)jk}{\hat {\bar W}}^{k}_{(t)i}, 
\\ {\bar W}_{(t)ij;k}&=&{\bar W}_{(t)ij{\mid}k}-{\bar K}_{(t)ik}
{\bar W}_{(t){\bar {\perp}}j}-
{\bar K}_{(t)jk}{\bar W}_{(t){\bar {\perp}}i},  
\\ {\bar W}_{(t){\bar {\perp}}{\bar {\perp}};{\bar {\perp}}}&=&
-{\cal L}_{{\bf {\bar n}}_t}{\bar W}_{(t){\bar {\perp}}{\bar {\perp}}}
-c^{-2}{\bar a}_{{\cal F}s}{\hat {\bar W}}^{s}_{(t){\bar {\perp}}}-
c^{-2}{\bar a}_{{\cal F}}^s{\bar W}_{(t){\bar {\perp}}s},
\\ {\bar W}_{(t)j{\bar {\perp}};{\bar {\perp}}}&=&
-{\cal L}_{{\bf {\bar n}}_t}{\bar W}_{(t)j{\bar {\perp}}}+
{\Big (}c^{-2}{\bar a}_{{\cal F}j}{\frac{t_0}{t}}
{\frac{{\bar N}_{(t)}^s}{{\bar N}_t}}
-{\bar K}_{(t)j}^s{\Big )}{\bar W}_{(t)s{\bar {\perp}}} \nonumber \\
&&-c^{-2}{\bar a}_{{\cal F}s}{{\hat {\bar W}}_{(t)j}}^{s}-
c^{-2}{\bar a}_{{\cal F}j}{\bar W}_{(t){\bar {\perp}}{\bar {\perp}}},
\\ {\bar W}_{(t)ij;{\bar {\perp}}}&=&-{\cal L}_{{\bf {\bar n}}_t}
{\bar W}_{(t)ij}+{\Big (}c^{-2}{\bar a}_{{\cal F}j}{\frac{t_0}{t}}
{\frac{{\bar N}_{(t)}^k}{{\bar N}_t}}-{\bar K}_{(t)j}^k{\Big )}{\bar W}_{(t)ik}
-c^{-2}{\bar a}_{{\cal F}j}{\bar W}_{(t)i{\bar {\perp}}}
{\nonumber} \\
&&+{\Big (}c^{-2}{\bar a}_{{\cal F}i}{\frac{t_0}{t}}
{\frac{{\bar N}_{(t)}^k}{{\bar N}_t}}-{\bar K}_{(t)i}^k{\Big )}{\bar W}_{(t)kj}
-c^{-2}{\bar a}_{{\cal F}i}{\bar W}_{(t){\bar {\perp}}j}.
\ena
\subsection{The field equations}
In this section, we find local conservation laws for the total active 
stress-energy tensor ${\bf T}_t$ and express these laws in 
hypersurface-formalism. But as we shall see, the local conservation laws in 
quasi-metric theory are different from their metric counterparts. We also 
postulate field equations which must hold for the metric family 
${\bf {\bar g}}_t$. Note that in quasi-metric theory, the active
stress-energy tensor ${\bf T}_t$ is really a family of rank 2 tensor fields; 
this is what its $t$-label is meant to indicate.

By construction, the quasi-metric framework satisfies the EEP. But this does 
not imply that one should automatically use the usual rule ``comma goes to 
semicolon'' when generalizing the non-gravitational physics from flat to curved
space-time. The reason for this is, that it is possible to couple 
non-gravitational fields to first derivatives of the spatial scale factor 
${\bar F}_t$ such that the EEP still holds, see below. In particular, we find 
that the covariant divergence ${\bf {\bar {\nabla}}}{\bf {\cdot}}{\bf T}_t
{\neq}0$ (expressed in terms of formally variable units) in 
the quasi-metric framework (also, ${\,}{\bf {\topstar{\bar {\nabla}}}}{\,}
{\bf {\cdot}T}_t {\neq}0$). Rather than ${\bf T}_t$, we shall see that some 
other quantity, depending on the distinguished foliation of quasi-metric 
space-time into the family of FHSs, is covariantly conserved. It must be 
stressed that this does {\em not} necessarily imply any violation of local 
conservation laws in ${\overline{\cal N}}$; rather it means that the 
conservation laws take unfamiliar forms in quasi-metric space-time. Any direct 
coupling to curvature does of course not occur in the conservation laws, but
there is coupling to other fields, namely ${\bar a}_{{\cal F}i}$, 
${\frac{{\bar N}_t,_{\bar {\perp}}}{{\bar N}_t}}$,
${\frac{{\bar N}_t,_t}{c{\bar N}_t^2}}$ and ${\frac{1}{{\bar N}_tct}}$. 

Within the metric framework, it can be shown [1] that validity of the local 
conservation laws in the form ${\nabla}{\bf {\cdot}T}=0$ follows from 
universal coupling (that gravity couples similarly to all non-gravitational 
fields) and invariance of non-gravitational actions. Thus in any metric theory 
based on an invariant action principle, the local conservation laws in their 
usual form ${\nabla}{\bf {\cdot}T}=0$ are valid independently of the detailed 
form of the gravitational field equations. (Besides, it can be shown [1] that 
the local conservation laws applied to the stress-energy tensor for a perfect 
fluid, directly yield the Euler equations of fluid dynamics in the limit of 
weak gravitational fields and slow motions, establishing a Newtonian limit.) On
the other hand, within the QMF, the existence of two different gravitational
 ``constants'' means that we cannot have universal coupling. Besides, it is
expected that gravitational field equations take special forms when projected
with respect to the FHSs. This means that said distinguished foliation should
be a {\em dynamical} quantity determined by the gravitational field equations.
Therefore, it would not be surprising if the local conservation laws involve
said distinguished foliation via the covariantly conserved quantity. For this 
reason, expressed in a GTCS, the local conservation laws may be expected to 
take the unfamiliar form ${\bf {\bar {\nabla}}}{\bf {\cdot}T}_t{\neq}0$ for any
theory compatible with the quasi-metric framework. As a consequence, it should 
not be possible to find the field equations from an invariant action principle 
based on a Lagrangian which does not depend on said distinguished foliation. 

To be consistent with classical electrodynamics in quasi-metric space-time,
light rays in electrovacuum should be null geodesics both in ${\cal N}$ and in 
${\overline{\cal N}}$. Then ${\bar {\nabla}}{\bf {\cdot}}
{\bar {\cal T}}_t^{\rm (EM)}$ must necessarily vanish for electrovacuum, where
${\bar {\cal T}}_t^{\rm (EM)}{\equiv}{\frac{t_0^2}{t^2}}{\bar N}_t^{-2}
{\bf T}_t^{\rm (EM)}$ is the {\em passive} electromagnetic stress-energy tensor 
in ${\bar {\cal N}}$. This yields that 
${\bar {\nabla}}{\bf {\cdot}}({\bar N}^{-2}_t{\bf T}_t^{\rm (EM)})=0$ for 
electrovacuum so that for this case, 
${\bar {\nabla}}{\bf {\cdot}}{\bf T}_t^{\rm (EM)}$ takes a particular 
form. Requiring this form of ${\bar {\nabla}}{\bf {\cdot}}{\bf T}_t^{\rm (EM)}$ 
to hold for any active stress-energy tensor family ${\bf T}_t$, the in general 
covarianly conserved quantity is given by ${\bar N}^{-2}_t{\bf T}_t$ and the 
local conservation laws should thus read (for fixed $t$)
\eqa
T_{(t){\mu};{\nu}}^{\nu}=2{\frac{{\bar N}_{t,{\nu}}}{{\bar N}_t}}
T_{(t){\mu}}^{\nu}=2c^{-2}{\bar a}_{{\cal F}s}{\hat T}_{(t){\mu}}^s
-2{\frac{{\bar N}_{t,{\bar {\perp}}}}{{\bar N}_t}}T_{(t){\bar {\perp}}{\mu}}.
\ena
These equations can be projected with respect to the FHSs, and the result is
\eqa
{\cal L}_{{\bf {\bar n}}_t}T_{(t){\bar {\perp}}{\bar {\perp}}}=
{\Big (}{\bar K}_t-2{\frac{{\bar N}_t,_{\bar {\perp}}}{{\bar N}_t}}
{\Big )}T_{(t){\bar {\perp}}{\bar {\perp}}}
+{\bar K}_{(t)ks}{\hat T}_{(t)}^{ks}-{\hat T}^s_{(t){\bar {\perp}}{\mid}s},
\ena
\eqa
{\frac{1}{{\bar N}_t}}{\cal L}_{{\bar N}_t{\bf {\bar n}}_t}
T_{(t)j{\bar {\perp}}}={\Big (}{\bar K}_t
-2{\frac{{\bar N}_t,_{\bar {\perp}}}{{\bar N}_t}}
{\Big )}T_{(t)j{\bar {\perp}}}-c^{-2}{\bar a}_{{\cal F}j}
T_{(t){\bar {\perp}}{\bar {\perp}}}
+c^{-2}{\bar a}_{{\cal F}s}{\hat T}_{(t)j}^s-{\hat T}^s_{(t)j{\mid}s}.
\ena
If the only $t$-dependence of ${\bf T}_t$ is via the formal variability, 
${\bf T}_t$ is locally conserved when $t$ varies as well. That is, for the 
non-metric part of the connection, an extra local ``conservation law'' can be 
found from the condition 
${\stackrel{{\hbox{\tiny$\star$}}}{\bf {\bar {\nabla}}}}
_{{\!}{\!}{\frac{\partial}{{\partial}t}}}({\frac{t_0^2}{t^2}}{\bar N}_t^{-2}T_{(t){\mu}}^0)
=0$. A simple calculation yields
\eqa
T_{(t){\mu}{\bar *}t}^0=-{\frac{2}{{\bar N}_t}}{\Big (}{\frac{1}{t}}+
{\frac{{\bar N}_t,_t}{{\bar N}_t}}{\Big )}T_{(t){\bar {\perp}}{\mu}}.
\ena
Note that equation (3.24) will not always hold (e.g., for situations where 
there is non-negligible net energy transfer from material particles to 
electromagnetic radiation). The relation (3.21) or equivalently equations 
(3.22) and (3.23), together with equation (3.24) represent the local 
conservation laws for active mass-energy in ${\overline{\cal N}}$. Note that 
these laws are expressed in atomic units and that the laws take care of the 
fact that gravitational quantities get extra variability measured in such 
units. Also note that these local conservation laws do not depend on the 
nature of the gravitational source (i.e., their form is independent of source 
composition). Besides, due to the contracted Bianchi identities, the local 
conservation laws are identities in General Relativity (GR). In quasi-metric 
theory, as we shall see, the local conservation laws in ${\overline{\cal N}}$ 
represent real constraints, however. Thus it is expected that quasi-metric 
theory shall be quite different from GR when it comes to the way matter fields 
are explicitly coupled to the space-time geometry. 

To find quasi-metric field equations, we first notice that the need to 
have two different gravitational coupling parameters $G^{\rm B}$ and $G^{\rm S}$
means that the principle of universal gravitational coupling is violated. This
must be reflected explicitly in the field equations as separate couplings
to the active electromagnetic stress-energy tensor ${\bf T}_t^{\rm (EM)}$ and to
the active stress-energy tensor ${\bf T}_t^{\rm (MA)}$ for material sources,
respectively. Second, we must have metric correspondence with Newtonian theory 
for weak stationary fields. These conditions lead to a postulated field 
equation, namely (with ${\kappa}^{\rm B}{\equiv}{\frac{8{\pi}G^{\rm B}}{c^{4}}}$ 
and ${\kappa}^{\rm S}{\equiv}{\frac{8{\pi}G^{\rm S}}{c^{4}}}$)
\eqa
2{\bar R}_{(t){\bar {\perp}}{\bar {\perp}}}=
2(c^{-2}{\bar a}_{{\cal F}{\mid}s}^s+
c^{-4}{\bar a}_{{\cal F}s}{\bar a}_{\cal F}^s
-{\bar K}_{(t)ks}{\bar K}_{(t)}^{ks}+{\cal L}_{{\bf {\bar n}}_t}{\bar K}_t)
\nonumber \\
={\kappa}^{\rm B}(T^{\rm (EM)}_{(t){\bar {\perp}}{\bar {\perp}}}+
{\hat T}^{{\rm (EM)}s}_{(t)s})+
{\kappa}^{\rm S}(T^{\rm (MA)}_{(t){\bar {\perp}}{\bar {\perp}}}+
{\hat T}^{{\rm (MA)}s}_{(t)s}),
\ena
where ${\bf {\bar R}}_t$ is the Ricci tensor field family corresponding to
the metric family (2.15). Note that, whereas equation (3.25) is postulated
rather than derived, it is by no means arbitrary; in fact equation (3.25) 
follows naturally from a geometrical correspondence with Newton-Cartan theory
(except for the violation of universal coupling). Moreover, it is possible to 
write the first line of equation (3.25) in the form
\eqa
{\bar N}_{t{\mid}{\,}{\,}{\,}s}^{{\,}{\,}{\,}s}={\bar N}_t{\Big (}
{\bar R}_{(t){\bar {\perp}}{\bar {\perp}}}
+{\bar K}_{(t)ks}{\bar K}_{(t)}^{ks}-{\cal L}_{{\bf {\bar n}}_t}{\bar K}_t
{\Big )}.
\ena
According to the maximum principle applied to closed Riemannian 3-manifolds
(see reference [16] and references therein), no solutions of equation (3.26)
can exist on on the whole of each FHS if the right hand side does not change 
sign on each FHS. The only exception to this rule is the solution 
${\bar N}_t=$constant on each FHS for the particular case when the right hand 
side of equation (3.26) vanishes. Note that if 
${\kappa}^{\rm B}={\kappa}^{\rm S}$, (the metric approximations of) equations 
(3.25), (3.26) also hold in GR for projections on an arbitrary space-like 
hypersurface since
\eqa
{\bar G}_{(t){\bar {\perp}}{\bar {\perp}}} +{\bar G}_{(t)ks}{\bar h}_{(t)}^{ks}
=2{\bar R}_{(t){\bar {\perp}}{\bar {\perp}}}.
\ena
Here ${\bf {\bar G}}_t$ is the Einstein tensor family corresponding to the
metric family (2.15). 

Except for the non-universal coupling, the field equation (3.25) is similar to 
its counterpart among the various projections of the Einstein field equations 
in canonical GR. Now it would seem natural to postulate a second set of field 
equations, yielding a natural correspondence with GR, by adopting those
projections of the Einstein equations involving the quantity
${\bar R}_{(t)j{\bar {\perp}}}$. That is, it 
would be tempting to postulate a coupling of ${\bar R}_{(t)j{\bar {\perp}}}$ 
directly to ${\bar T}^{{\rm (EM)}}_{(t)j{\bar {\perp}}}$ 
and ${\bar T}^{{\rm (MA)}}_{(t)j{\bar {\perp}}}$. Unfortunately, this does not work
since it can be shown that this implies that a subset of the local conservation
laws and the corresponding subset of the contracted Bianchi identities (i.e., 
equations (3.22) and (3.32) below), would be inconsistent in the weak-field 
limit.

To arrive at somewhat similar field equations but such that no obvious 
inconsistencies appear, we shall take an alternative approach. First we define 
the vector field family ${\bf {\bar m}}_t$ by its components expressed in a 
GTCS, i.e.,
\eqa
{\bf {\bar m}}_t{\equiv}
-{\frac{1}{{\bar N}_t}}{\frac{\partial}{{\partial}x^0}}-{\frac{t_0}{t}}
{\frac{{\bar N}^i_{(t)}}{{\bar N}_t}}{\frac{\partial}{{\partial}x^i}}=
-{\bf {\bar n}}_t-2{\frac{t_0}{t}}{\frac{{\bar N}^i_{(t)}}{{\bar N}_t}}
{\frac{\partial}{{\partial}x^i}}, \nonumber \\ 
{\bar m}_{(t)}^{\nu}{\bar m}_{(t){\nu}}=
-1+4{\bar N}^i_{(t)}{\bar N}^k_{(t)}{\tilde h}_{(t)ik}, \qquad
{\bar m}_{(t)}^{\nu}{\bar n}_{(t){\nu}}=1.
\ena
Next we use equation (3.28) to define the space tensor family 
${\bf {\bar L}}_t$ via its components in a GTCS, i.e.,
\eqa
{\bar L}_{(t)ij}{\equiv}
-{\frac{1}{2{\bar N}_t}}{\cal L}_{{\bar N}_t{\bf {\bar m}}_t}
{\bar h}_{(t)ij}={\bar K}_{(t)ij}+
{\frac{1}{{\bar N}_t}}{\frac{\partial}{{\partial}x^0}}{\bar h}_{(t)ij}, \quad
{\bar L}_t{\equiv}{\bar K}_t+{\frac{{\bar h}_{(t)}^{ik}}{{\bar N}_t}}
{\frac{\partial}{{\partial}x^0}}{\bar h}_{(t)ik}.
\ena
One may interpret ${\bf {\bar L}}_t$ as some sort of ``time-reversed''
extrinsic curvature tensor family. The wanted field equation set, having
the properties mentioned above, is then obtained by coupling matter fields to
the quantity ${\bar L}_{(t)j{\mid}s}^s-{\bar L}_t,_j$ rather than to
${\bar R}_{(t)j{\bar {\perp}}}={\bar K}_{(t)j{\mid}s}^s-{\bar K}_t,_j$:
\eqa
{\hspace*{-3mm}}
{\bar R}_{(t)j{\bar {\perp}}}+{\Big (}{\frac{{\bar h}_{(t)}^{ik}}{{\bar N}_t}}
{\frac{\partial}{{\partial}x^0}}{\bar h}_{(t)ij}{\Big )}_{{\mid}k}-
{\Big (}{\frac{{\bar h}_{(t)}^{ik}}{{\bar N}_t}}
{\frac{\partial}{{\partial}x^0}}{\bar h}_{(t)ik}{\Big )},_j
={\bar L}_{(t)j{\mid}i}^i-{\bar L}_t,_j 
={\kappa}^{\rm B}T^{{\rm (EM)}}_{(t)j{\bar {\perp}}}
+{\kappa}^{\rm S}T^{\rm (MA)}_{(t)j{\bar {\perp}}}.
\ena
On the FHSs, equations (3.25) and (3.30) are one scalar and one 3-vector 
equation, respectively, coupled to the relevant projections of 
${\bf T}_t^{\rm (EM)}$ and ${\bf T}_t^{\rm (MA)}$. We still lack a tensorial field 
equation of rank 2 type on the FHSs, describing a potential coupling to 
$T_{(t)ij}^{\rm (EM)}$ and $T_{(t)ij}^{\rm (MA)}$ of the spatial projection
${\bar Q}_{(t)ij}$ of some geometric space-time tensor family 
${\bf {\bar Q}}_t$, say. To predict weak GR-like gravitational waves in vacuum, 
it is desirable that ${\bar Q}_{(t)ij}$ should have dynamical properties 
somewhat similar to those of ${\bar G}_{(t)ij}$. Moreover, we conveniently define
${\bar Q}_{(t)j{\bar {\perp}}}{\equiv}{\bar L}_{(t)j{\mid}s}^s-{\bar L}_t,_j$, and the
projection ${\bar Q}_{(t){\bar {\perp}}{\bar {\perp}}}$ should potentially couple to
$T_{(t){\bar {\perp}}{\bar {\perp}}}^{\rm (EM)}$ and
$T_{(t){\bar {\perp}}{\bar {\perp}}}^{\rm (MA)}$. This means that we are looking for
fully coupled field equations of the type 
${\bf {\bar Q}}_t={\kappa}^{\rm B}{\bf T}_t^{\rm (EM)}+ 
{\kappa}^{\rm S}{\bf T}_t^{\rm (MA)}$, where the projections of ${\bf {\bar Q}}_t$ 
with respect to the FHSs take special forms not exactly valid for projections 
with respect to other hypersurfaces. We must also have a counterpart to 
equation (3.27) involving ${\bf {\bar Q}}_t$, i.e.,
\eqa
{\bar Q}_{(t){\bar {\perp}}{\bar {\perp}}} +{\bar Q}_{(t)ks}{\bar h}_{(t)}^{ks}
=2{\bar R}_{(t){\bar {\perp}}{\bar {\perp}}}.
\ena
To get any further, it is useful to identify the particular cases where the 
postulated prior 3-geometry of the FHSs appears explicitly in equation (2.15).
That is, for these cases we have ${\tilde h}_{(t)ks}=S_{ks}$ in equation (2.15), 
where $S_{ks}dx^kdx^s$ is the metric of the 3-sphere (with radius equal to 
$ct_0$). Besides, if ${\bar N}^k_{(t)}=0$, we have that ${\bf {\bar g}}_t$ is 
conformal to a metric family with geometry ${\bf S}^3{\times}{\bf R}$. The 
latter is {\em metrically static}, i.e., it is static except for the global 
time dependence of the spatial geometry on $t$. More generally, if ${\bar N}_t$
has no time dependence, ${\bf {\bar g}}_t$ will be metrically static as well, 
and the extrinsic curvature will vanish, i.e., ${\bf {\bar K}}_t=0$. It is 
natural to identify these metrically static cases with vacuum solutions, where 
${\bf {\bar g}}_t$ is determined solely from ${\bar N}_t$.  Thus, for 
metrically static vacua the field equations should read 
${\bar Q}_{(t){\bar {\perp}}{\bar {\perp}}}=0$ and ${\bar Q}_{(t)ij}=0$, respectively,
and the solutions should be fully determined from the scale factor 
${\bar F}_t={\bar N}_tct$.

Now, with ${\bf {\bar H}}_t$ denoting the family of spatial Einstein tensor 
fields intrinsic to the FHSs, the equation ${\bar Q}_{(t)ij}=0$ should yield the
relationship ${\bar H}_{(t)ij}+c^{-2}{\bar a}_{{\cal F}i{\mid}j}+c^{-4}
{\bar a}_{{\cal F}i}{\bar a}_{{\cal F}j}-(c^{-2}{\bar a}_{{\cal F}{\mid}s}^s-
{\frac{1}{(ct{\bar N}_t)^2}}){\bar h}_{(t)ij}=0$, which follows directly from 
equation (2.15) for metrically static cases where ${\tilde h}_{(t)ks}=S_{ks}$. 
But the extrinsic curvature also vanishes identically for metrically static 
interiors (where we expect ${\tilde h}_{(t)ks}{\neq}S_{ks}$), so this means that 
we should have ${\bar Q}_{(t)ij}=-c^{-2}{\bar a}_{{\cal F}i{\mid}j}-
c^{-4}{\bar a}_{{\cal F}i}{\bar a}_{{\cal F}j}+(c^{-2}{\bar a}_{{\cal F}{\mid}s}^s
-{\frac{1}{(ct{\bar N}_t)^2}}){\bar h}_{(t)ij}-{\bar H}_{(t)ij}$ and from equation
(3.31) thus ${\bar Q}_{(t){\bar {\perp}}{\bar {\perp}}}=-{\frac{1}{2}}{\bar P}_t
+3c^{-4}{\bar a}_{{\cal F}s}{\bar a}_{\cal F}^s+{\frac{3}{(ct{\bar N}_t)^2}}$
for the metrically static cases (here, ${\bar P}_t$ denotes the Ricci scalar
family intrinsic to the FHSs). (The other sign for ${\bar Q}_{(t)ij}$ cannot 
be chosen since we must have that in general,
${\bar Q}_{(t){\bar {\perp}}{\bar {\perp}}}>0$ and ${\hat{\bar Q}}^s_{(t)s}>0$ for 
metrically static interiors as these quantities are expected to couple to 
suitable projections of ${\bf T}_t$.)

However, at this point a crucial problem arises due to the contracted Bianchi
identities ${\bar G}_{(t){\mu};{\nu}}^{\nu}{\equiv}0$. Projected with respect to 
the FHSs, these identities read (see, e.g., [2])
\eqa
{\cal L}_{{\bf {\bar n}}_t}{\bar G}_{(t){\bar {\perp}}{\bar {\perp}}}=
{\bar K}_t{\bar G}_{(t){\bar {\perp}}{\bar {\perp}}}+{\bar K}_{(t)}^{ks}
{\bar G}_{(t)ks}-2c^{-2}{\bar a}_{\cal F}^s{\bar G}_{(t){\bar {\perp}}s}
-{\hat {\bar G}}^s_{(t){\bar {\perp}}{\mid}s},
\ena
\eqa
{\frac{1}{{\bar N}_t}}{\cal L}_{{\bar N}_t{\bf {\bar n}}_t}
{\bar G}_{(t)j{\bar {\perp}}}={\bar K}_t{\bar G}_{(t)j{\bar {\perp}}}
-c^{-2}{\bar a}_{{\cal F}j}{\bar G}_{(t){\bar {\perp}}{\bar {\perp}}}
-c^{-2}{\bar a}_{\cal F}^s{\bar G}_{(t)js}-{\hat {\bar G}}^s_{(t)j{\mid}s}.
\ena
That is, it turns out that equations (3.23) and (3.33), in combination with the 
deduced expressions for ${\bar Q}_{(t){\bar {\perp}}{\bar {\perp}}}$ and 
${\bar Q}_{(t)ij}$ for metrically static interiors, yield the wrong Newtonian 
limit, so that equation (3.33) does not correspond with its counterpart Euler 
equation. In fact, the only way to avoid said problem while still keeping the 
relationship (3.31), is to set ${\bar Q}_{(t){\bar {\perp}}{\bar {\perp}}}=
2{\bar R}_{(t){\bar {\perp}}{\bar {\perp}}}$ and ${\bar Q}_{(t)ks}{\bar h}_{(t)}^{ks}=0$ 
together with the condition ${\hat {\bar Q}}^s_{(t)j{\mid}s}-c^{-2}{\bar a}_
{\cal F}^s{\bar Q}_{(t)js}=0$ coming from equation (3.33). However, this yields 
no possible consistent coupling of $T_{(t)ij}$ to ${\bar Q}_{(t)ij}$ given 
equation (3.23), so to avoid said problem we are forced to set 
${\bar Q}_{(t)ij}=0$. Thus there can be no extra scalar field equation besides 
equation (3.25) and also no additional tensor equation with full coupling to 
$T_{(t)ij}$. In other words, we have found that {\em it is not possible to 
construct a viable, fully coupled quasi-metric gravitational theory.} 

Nevertheless, fortunately it is still possible to have a partially coupled,
manifestly traceless field equation ${\bar Q}_{(t)ij}=0$ for the general case.
The choice of terms quadratic in extrinsic curvature for such an equation would 
seem somewhat uncertain, but this question can be resolved by a restriction 
involving a particular projection of the Weyl tensor family ${\bf {\bar C}}_t$.
That is, we require that the projection ${\bar C}_{(t){\bar {\perp}}i{\bar {\perp}}j}$
should be determined from the intrinsic geometry of the FHSs alone, with no 
explicit dependence on extrinsic curvature (or on ${\bf {\bar a}}_{\cal F}$). 
Thus we define a (unique) relationship having this property, i.e.,
\eqa
{\bar C}_{(t){\bar {\perp}}i{\bar {\perp}}j}={\tilde H}_{(t)ij}+
{\frac{1}{(ct{\bar N}_t)^2}}{\bar h}_{(t)ij},
\ena
where ${\bf {\tilde H}}_t$ is the spatial Einstein tensor family calculated 
from the metric family ${\bf {\tilde h}}_t$. Equation (3.34) is equivalent to 
a definition of ${\bar Q}_{(t)ij}$ via the equation
\eqa
{\bar G}_{(t)ij}{\equiv}-{\bar Q}_{(t)ij}-2c^{-2}{\bar a}_{{\cal F}i{\mid}j}-
2c^{-4}{\bar a}_{{\cal F}i}{\bar a}_{{\cal F}j}
-2{\bar K}_{(t)i}^{s}{\bar K}_{(t)sj}+2{\bar K}_t{\bar K}_{(t)ij}
\nonumber \\
+{\frac{1}{3}}{\Big [}2{\bar R}_{(t){\bar {\perp}}{\bar {\perp}}}
-{\bar G}_{(t){\bar {\perp}}{\bar {\perp}}}+2c^{-2}{\bar a}_{{\cal F}{\mid}s}^s
+2c^{-4}{\bar a}_{{\cal F}}^s{\bar a}_{{\cal F}s}
+2{\bar K}_{(t)ks}{\bar K}_{(t)}^{ks}-2{\bar K}_t^2{\Big ]}{\bar h}_{(t)ij},
\ena
as extrapolated from the metrically static case, combined with the general 
expressions for the projections ${\bar G}_{(t){\bar {\perp}}{\bar {\perp}}}$ and 
${\bar G}_{(t)ij}$, i.e., (see, e.g., [2])
\eqa
{\bar G}_{(t){\bar {\perp}}{\bar {\perp}}}={\frac{1}{2}}({\bar P}_t
+{\bar K}_t^2-{\bar K}_{(t)ks}{\bar K}_{(t)}^{ks}),
\ena
\eqa
{\bar G}_{(t)ij}=-{\frac{1}{{\bar N}_t}}
{\cal L}_{{\bar N}_t{\bf {\bar n}}_t}({\bar K}_{(t)ij}-
{\bar K}_t{\bar h}_{(t)ij})+3{\bar K}_t{\bar K}_{(t)ij}-
{\frac{1}{2}}({\bar K}_t^2+
{\bar K}_{(t)ks}{\bar K}_{(t)}^{ks}){\bar h}_{(t)ij} \nonumber \\
-2{\bar K}_{(t)is}{\bar K}_{(t)j}^s
-c^{-2}{\bar a}_{{\cal F}i{\mid}j}-c^{-4}{\bar a}_{{\cal F}i}
{\bar a}_{{\cal F}j}+(c^{-2}{\bar a}^s_{{\cal F}{\mid}s}
+c^{-4}{\bar a}_{\cal F}^s{\bar a}_{{\cal F}s}){\bar h}_{(t)ij}+
{\bar H}_{(t)ij}.
\ena
We then get the remainder quasi-metric field equation (equivalent to the
definition (3.34));
\eqa
{\bar Q}_{(t)ij}={\frac{1}{{\bar N}_t}}{\cal L}_{{\bar N}_t
{\bf {\bar n}}_t}{\bar K}_{(t)ij}+
{\frac{1}{3}}{\Big [}2{\bar K}_{(t)ks}{\bar K}_{(t)}^{ks}-{\bar K}_t^2
-{\cal L}_{{\bf {\bar n}}_t}{\bar K}_t
{\Big ]}{\bar h}_{(t)ij}+{\bar K}_t{\bar K}_{(t)ij} \nonumber \\
-c^{-2}{\bar a}_{{\cal F}i{\mid}j}-
c^{-4}{\bar a}_{{\cal F}i}{\bar a}_{{\cal F}j}
+{\Big [}c^{-2}{\bar a}_{{\cal F}{\mid}s}^s
-{\frac{1}{(ct{\bar N}_t)^2}}{\Big ]}{\bar h}_{(t)ij}-{\bar H}_{(t)ij}=0,
\ena
where the prior-geometric requirement on the spatial Ricci curvature scalar 
${\bar P}_t$,
\eqa
{\bar P}_t=-4c^{-2}{\bar a}_{{\cal F}{\mid}s}^s
+2c^{-4}{\bar a}_{{\cal F}}^s{\bar a}_{{\cal F}s}+{\frac{6}{(ct{\bar N}_t)^2}},
\ena
ensures that equation (3.38) is indeed manifestly traceless. Equation (3.38) 
(or equation (3.34)) determines 5 of the 10 independent components of 
${\bf {\bar C}}_t$. Besides, the components of the spatial Einstein tensor 
family ${\bf {\bar H}}_t$ intrinsic to the FHSs are given by
\eqa
{\bar H}_{(t)ij}=-c^{-2}{\bar a}_{{\cal F}i{\mid}j}-
c^{-4}{\bar a}_{{\cal F}i}{\bar a}_{{\cal F}j}+
c^{-2}{\bar a}_{{\cal F}{\mid}s}^s{\bar h}_{(t)ij}+{\tilde H}_{(t)ij}.
\ena
Note that, while equations (3.39) and (3.40) imply that 
${\tilde P}_t={\frac{6}{(ct_0)^2}}$, we have that ${\tilde H}_{(t)ij}$ is 
not necessarily equal to $-{\frac{1}{(ct_0)^2}}{\tilde h}_{(t)ij}$. This shows 
that, while there is prior 3-geometry, there is still room for two propagating 
dynamical dynamical degrees of freedom associated with the metric 
family ${\bf {\tilde h}}_t$. This is further illustrated by writing equation
(3.38) in the form (using equations (3.25) and (3.40))
\eqa
{\frac{1}{{\bar N}_t}}{\cal L}_{{\bar N}_t{\bf {\bar n}}_t}{\bar K}_{(t)ij}
+{\bar K}_t{\bar K}_{(t)ij}-{\tilde H}_{(t)ij} \nonumber \\
={\frac{1}{3}}{\Big [}{\bar R}_{(t){\bar {\perp}}{\bar {\perp}}}
-{\bar K}_{(t)ks}{\bar K}_{(t)}^{ks}+{\bar K}_t^2
-c^{-2}{\bar a}_{{\cal F}{\mid}s}^s
-c^{-4}{\bar a}_{{\cal F}}^s{\bar a}_{{\cal F}s}+{\frac{3}{(ct{\bar N}_t)^2}}
{\Big ]}{\bar h}_{(t)ij}.
\ena
Note that the quasi-metric field equations have a somewhat similar split-up as 
Einstein's field equations into dynamical equations and constraints. That is, 
equations (3.30) and (3.39) represent the 4 constraints while equations (3.25) 
and (3.41) represent the 6 (independent) dynamical equations. But in addition 
equation (3.39) eliminates any gauge freedom in the choice of lapse and shift, 
so that the evolution of the FHSs becomes unambiguous. Moreover, also similar 
to Einstein's equations, the dynamical gravitational field may be taken to be 
the spatial metric family ${\bf {\bar h}}_t$, representing two independent 
propagating dynamical degrees of freedom. However, according to equation (2.15)
it is possible to split up ${\bf {\bar h}}_t$ further into one dynamical scalar
field represented by the lapse function family ${\bar N}_t$, plus 
${\bf {\tilde h}}_t$, which is a dynamical tensor field. The two propagating 
dynamical degrees of freedom should be associated with the trace-free part of 
the latter.
\subsection{The quasi-metric initial-value problem}
Now we want to use the results of the previous section to carry out a
further investigation of the dynamical theory in ${\overline{\cal N}}$, which 
is not a metric manifold. We do this by aiming at a formulation of the theory 
as an initial-value problem for the evolution with time of the geometry of 
the FHSs. But in contrast to the standard initial-value problem for metric
theories of gravity, our non-metric counterpart should involve both KE and NKE 
of the FHSs. The dynamical theory can be extended to be valid in $\cal N$ by 
adding extra effects coming from the local NKE as illustrated in section 2.3.

To begin with, we split up the time evolution of ${\bar N}_t$ into kinematical 
and non-kinematical parts. We thus define
\eqa
c^{-2}{\bar x}_t+c^{-1}{\bar y}_t{\equiv}{\frac{{\bar N}_t,_t}{c{\bar N}_t^2}}
-{\frac{{\bar N}_t,_{\bar {\perp}}}{{\bar N}_t}},
\ena
where $c^{-2}{\bar x}_t$ and $c^{-1}{\bar y}_t$ respectively, define the KE and 
the NKE of ${\bar N}_t$ with time. On the other hand, we {\em define} the time 
evolution of the spatial scale factor ${\bar F}_t$ from the formula
\eqa
{\bar F}_t^{-1}{\topstar{\cal L}}_{{\,}{\bf {\bar n}}_t}{\bar F}_t=
{\bar F}_t^{-1}{\Big (}(c{\bar N}_t)^{-1}{\bar F}_t,_t+ 
{\cal L}_{{\bf {\bar n}}_t}{\bar F}_t{\Big )}=
{\frac{1}{c{\bar N}_t}}{\Big [}{\frac{1}{t}}+
{\frac{{\bar N}_t,_t}{{\bar N}_t}}{\Big ]}
-{\frac{{\bar N}_t,_{\bar {\perp}}}{{\bar N}_t}}{\equiv}
c^{-2}{\bar x}_t+c^{-1}{\bar H}_t,
\ena
where $c^{-2}{\bar x}_t$ represents the KE and $c^{-1}{\bar H}_t$ represents the 
NKE of the spatial scale factor. From equations (3.42) and (3.43) one easily 
finds that
\eqa
c^{-2}{\bar x}_t=-{\frac{{\bar N}_t,_{\bar {\perp}}}{{\bar N}_t}}-
c^{-1}{\bar H}_t+{\frac{1}{c{\bar N}_t}}{\Big [}{\frac{1}{t}}+
{\frac{{\bar N}_t,_t}{{\bar N}_t}}{\Big ]}, \qquad
{\bar y}_t={\bar H}_t-{\frac{1}{{\bar N}_tt}}.
\ena
To have a well-defined initial-value problem in ${\cal N}$, it remains to 
define the quantity ${\bar y}_t$. This quantity should never take negative 
values since it should contribute positively to ${\bar H}_t$ as a local 
``expansion'' stemming from the local gravitational field. Also, it should be 
defined from the spatial scale factor ${\bar F}_t$. So, since we expect that 
all the information available for definining ${\bar y}_t$ should be contained 
in the 4-acceleration field ${\bf {\bar a}}_{\cal F}$, we are led to the 
definitions
\eqa
{\bar y}_t{\equiv}c^{-1}{\sqrt{{\bar a}_{\cal F}^s{\bar a}_{{\cal F}s}}}, 
\qquad
{\bar H}_t{\equiv}{\frac{1}{{\bar N}_tt}}
+c^{-1}{\sqrt{{\bar a}_{\cal F}^s{\bar a}_{{\cal F}s}}}.
\ena
From equation (3.45) we see that ${\bar y}_t$ and ${\bar H}_t$ are indeed 
non-negative everywhere.

Now we are ready to formulate the quasi-metric initial-value problem in
${\overline{\cal N}}$. This is equivalent to formulating an evolution scheme 
for ${\bf {\bar h}}_t$ (or equivalently, for ${\bar N}_t$ and 
${\bf {\tilde h}}_t$). This may be done as follows: on an initial FHS given by 
$t=$constant, we choose initial data ${\bar N}_t$, ${\tilde h}_{(t)ij}$, 
${\bar K}_{(t)ij}$, ${\bar N}^i_{(t)}$ and the various projections of 
${\bf T}^{\rm (EM)}_t$ and ${\bf T}^{\rm (MA)}_t$ such that the constraints
given by equations (3.30) and (3.39) hold.

From the initial data, we are able to find ${\bar N}_t$ and ${\tilde h}_{(t)ij}$
on the subsequent FHS given by $t+dt=$constant (i.e., after one time step 
increment) (${\bar N}^i_{(t)}$ on this FHS is determined from the constraint 
equations). Moreover, from the local conservation laws (3.22), (3.23) and 
(3.24) (however, equation (3.24) will not necessarily hold), we are able to 
find $T^{\rm (EM)}_{(t+dt){\bar {\perp}}{\bar {\perp}}}$, 
$T^{\rm (MA)}_{(t+dt){\bar {\perp}}{\bar {\perp}}}$, $T^{\rm (EM)}_{(t+dt)j{\bar {\perp}}}$ and
$T^{\rm (MA)}_{(t+dt)j{\bar {\perp}}}$ on the subsequent FHS as long as the split-up
${\bf T}_t={\bf T}^{\rm (EM)}_t+{\bf T}^{\rm (MA)}_t$ is known. (Of course, 
equations (3.22) and (3.23) must be consistent with respectively equations 
(3.32) and (3.33)). The remaining projections of ${\bf T}^{\rm (EM)}_t$ and 
${\bf T}^{\rm (MA)}_t$ on the subsequent FHS must be inferred from the equation 
of state. Also, all quantities on the subsequent FHS must be compatible with 
equation (3.30) there. But in order to predict ${\bar K}_{(t)ij}$ on the 
subsequent FHS, or equivalently, ${\bar N}_t$ and ${\tilde h}_{(t)ij}$ on the 
next subsequent FHS given by another time step increment, i.e., 
$t+2dt=$constant, we need to solve the dynamical equations (3.25) and (3.38) on
the initial FHS. To do this, it is essential to supply the quantity 
${\bar R}_{(t){\bar {\perp}}{\bar {\perp}}}$ (and thus ${\bar G}_{(t)ij}$ given from 
equation (3.35)) on the initial FHS. When we have done this, we are able to 
advance one time step of the evolution process. We can then repeat the above 
described scheme for each successive time step.

Note that the quantities ${\bar N}_t,_t$ and ${\tilde h}_{(t)ij},_t$ must also 
be chosen on an initial FHS (the quantity ${\bar N}^i_{(t)},_t$ then follows 
from equation (2.16)), but that their values at successive FHSs do not follow 
from the dynamical equations. Rather, said values must be inferred 
independently from the effects of the cosmic expansion on ${\bf T}_t$ for each 
time step. (In particular, for isotropic cosmology, the KE of ${\bar F}_t$ is 
required to vanish as long as equation (3.24) holds, so that ${\bar x}_t=0$ 
from equation (3.44) determines ${\bar N}_t,_t$.) Lastly, note that the 
chosen equation of state characterizing the matter source on the initial FHS 
must be compatible with the initial data and with the local conservation laws.

But in general we must also have an evolution scheme in $\cal N$. The reason 
for this is that, when there is coupling to non-gravitational dynamical fields,
the dynamics of these fields depend on the equations of motion (2.11) in 
$\cal N$ (and not on the equations of motion in ${\overline{\cal N}}$). That 
is, if there is coupling to non-gravitational fields (e.g., electromagnetism), 
non-gravitational force laws in $\cal N$ enter the equations of motion and 
contribute to the evolution scheme of these fields. So when there is coupling 
between gravitational and non-gravitational dynamical fields, the evolution 
schemes in ${\overline{\cal N}}$ and $\cal N$ are intertwined.

To have an evolution scheme in $\cal N$, it is necessary to find ${\bar y}_t$
and ${\bf {\bar x}}_{\cal F}$ for each step of the evolution (see equation 
(2.22)). Since ${\bf {\bar x}}_{\cal F}$ is intended to represent a 
generalization of ``the local coordinate vector from the centre of gravity'' 
which does have meaning for the spherically symmetric case, it is natural to 
seek for an equation which only solution is ${\bf {\bar x}}_{\cal F}=
r{\frac{\partial}{{\partial}r}}$ for the spherically symmetric case. 
Furthermore, the wanted equation should be linear in ${\bf {\bar x}}_{\cal F}$ 
to ensure unique solutions, and it should involve ${\bf {\bar a}}_{\cal F}$ 
since one expects ${\bf {\bar x}}_{\cal F}$ to vanish whenever 
${\bf {\bar a}}_{\cal F}$ does. By inspection of the spherically symmetric case,
it turns out that it is possible to find an equation which has all the desired 
properties, namely
\eqa
{\Big [}{\bar a}_{{\cal F}{\mid}k}^k+c^{-2}{\bar a}_{{\cal F}k}
{\bar a}_{\cal F}^k{\Big ]}{\bar x}_{\cal F}^j-{\Big [}
{\bar a}_{{\cal F}{\mid}k}^j+c^{-2}{\bar a}_{{\cal F}k}{\bar a}_{\cal F}^j
{\Big ]}{\bar x}_{\cal F}^k-2{\bar a}_{\cal F}^j=0.
\ena
Since we may calculate ${\bf {\bar x}}_{\cal F}$ from equation (3.46) and then
${\bar y}_t$ for each step of the evolution according to the above described
evolution scheme, we are also able to construct ${\bf h}_t$, $N$ and $N^i_{(t)}$ 
for each time step along the lines shown in section 2.3, using equations 
(2.23), (2.24) and (2.25). Thus with the help of the coupled evolution schemes 
in ${\overline{\cal N}}$ and $\cal N$, we are able to find both metric families 
${\bf {\bar g}}_t$ and ${\bf g}_t$.

The initial-value problem described above, obviously differs from its GR 
counterpart in important ways. However, one not so obvious difference is the 
different roles of the quantities ${\bar N}_t$ and ${\bar N}^i_{(t)}$ in the two
evolution schemes. For the GR initial-value problem, the lapse function 
${\bar N}$ can be chosen freely at each time step (as long as the constraint 
equations are satisfied on the initial hypersurface), while the shift vector 
field ${\bar N}^i{\frac{\partial}{{\partial}x^i}}$ is somewhat restricted by 
the constraint equations. Nevertheless, the initial hypersurface may be evolved 
in many different ways into some given final hypersurface. That is, a 
Lorentzian manifold may be foliated into spatial hypersurfaces in many 
different ways; there is no general distinguished foliation, and the equations 
determining the time evolution of the hypersurfaces are identical for all 
possible foliations. On the other hand, for the quasi-metric initial-value 
problem, ${\bar N}_t$ and ${\bar N}^i_{(t)}$ are associated with a particular 
set of observers moving normally to the distinguished foliation of quasi-metric
space-time into a set of disinguished hypersurfaces (the FHSs). Moreover, the 
equations determining the time evolution of the FHSs are specific to the 
distinguished foliation. Thus the time evolution of the FHSs should be 
unambiguously described by the functions ${\bar N}_t$ and ${\bar N}^i_{(t)}$ 
for each time step.

Now we must emphasize an important point: since ${\bf T}_t$ is the total active 
stress-energy tensor and thus not directly measurable locally, one must know 
how it relates to {\em the total passive stress-energy tensor} ${\cal T}_t$ in 
${\cal N}$. The relationship between ${\bf T}_t$ and ${\cal T}_t$ depends in 
principle on the general nature of the matter source. To illustrate this, 
consider ${\bf T}_t$ and ${\cal T}_t$ for a perfect fluid:
\eqa
{\bf T}_t=({\tilde {\varrho}}_{\rm m}+c^{-2}{\tilde p}){\bf {\bar u}}_t
{\otimes}{\bf {\bar u}}_t+{\tilde p}{\bf {\bar g}}_t, \qquad
{\cal T}_t=({\hat {\varrho}}_{\rm m}+c^{-2}{\hat p}){\bf u}_t{\otimes}{\bf u}_t
+{\hat p}{\bf g}_t,
\ena
where ${\tilde {\varrho}}_{\rm m}$ is the active mass-energy density in the 
local rest frame of the fluid and ${\varrho}_{\rm m}$ is the passive 
mass-energy density measured in the local rest frame of the fluid 
(${\tilde p}$ and $p$ are the associated pressures). Also, in equation (3.47) 
we have used the definitions ${\varrho}_{\rm m}{\sqrt{{\bar h}_t}}{\equiv}
{\hat {\varrho}}_{\rm m}{\sqrt{h_t}}$ and  $p{\sqrt{{\bar h}_t}}{\equiv}
{\hat p}{\sqrt{h_t}}$, where ${\bar h}_t$ and $h_t$ are the determinants of 
the spatial metric families ${\bf {\bar h}}_t$ and ${\bf h}_t$, respectively. 
The relationship between ${\tilde {\varrho}}_{\rm m}$ and ${\varrho}_{\rm m}$ 
is given by
\eqa
{\varrho}_{\rm m}=
\left\{
\begin{array}{ll}
{\frac{t_0}{t}}{\bar N}_t^{-1}{\tilde {\varrho}}_{\rm m} & 
$for a fluid of material particles,$ \\ [1.5ex]
{\frac{t_0^2}{t^2}}{\bar N}_t^{-2}{\tilde {\varrho}}_{\rm m} & 
$for the electromagnetic field,$
\end{array}
\right.
\ena
and a similar relationship exists between ${\tilde p}$ and $p$. The reason why
the relationship between ${\varrho}_{\rm m}$ and ${\tilde{\varrho}}_{\rm m}$ is 
different for the electromagnetic field than for material sources, is that 
spectral shifts of the electromagnetic field influence its passive mass-energy 
but not its active mass-energy. On the other hand, the global NKE does not 
affect the passive mass-energy of material particles (see section 4.2). 
Finally, note that more ``physically correct'' local conservation laws than 
those shown in equations (3.22) and (3.23) may be found in $\cal N$ by 
calculating ${\bf {\nabla}}{\bf {\cdot}}{\cal T}_t$ when ${\bf g}_t$ is known. 
But these local ``physical'' conservation laws for passive mass-energy do not 
take any predetermined form.
\subsection{Equations of motion}
In section 2, we derived the equations of motion (2.11) valid for arbitrary 
test particles. However, these equations are of no practical value as long as 
we do not know the relationship between the 4-acceleration field ${\bf a}_t$ 
and the degenerate 4-acceleration field ${\stackrel{\star}{\bf a}}$. In this 
section, we show that in fact ${\stackrel{\star}{\bf a}}$=${\bf a}_t$.

For starters, it is convenient to introduce the 3-velocity field family
of a test particle moving along an arbitrary curve relative to that of the 
local FOs. We call this quantity ${\bf w}_t$, and we write its square as 
$w^2$. Thus we define (using a GTCS)
\eqa
w_{(t)}^j{\equiv}{\frac{dx^j}{d{\tau}_{\cal F}}}+{\frac{t_0}{t}}
{\frac{N^j_{(t)}}{N}}c, {\qquad}
{\topstar{\gamma}}{\equiv}(1-w^2/c^2)^{-1/2}=
{\frac{d{\tau}_{\cal F}}{d{\tau}_t}},
\ena
where ${\tau}_t$ is the proper time measured along the curve and 
${\topstar{\gamma}}$ is the time dilation factor relating the proper time 
intervals along the curve to the proper time intervals of the local FOs. 
Since ${\bf w}_t$ is required to be a vector field family tangent to the 
FHSs, the space-time scalar product of ${\bf w}_t$ and ${\bf n}_t$ must vanish.
That is, if we view ${\bf w}_t$ as a family of 4-vectors, we must have that
\eqa
w_{(t)}^0=0, \qquad w_{(t)}^{\mu}n_{(t){\mu}}=
w_{(t){\mu}}n_{(t)}^{\mu}=0, \qquad {\Rightarrow} \qquad
w_{(t)0}={\frac{t_0}{t}}N^s_{(t)}w_{(t)s}.
\ena
Note that, since the explicit dependence on $t$ of any scalar gravitational 
quantity is determined from its dimensionality when measured in atomic units, 
we must have that
\eqa
{\frac{\partial}{{\partial}t}}w^2{\equiv}
(w_{(t)s}w_{(t)k}h_{(t)}^{sk}),_t=0, \qquad \Rightarrow \qquad
{\frac{\partial}{{\partial}t}}w_{(t)j}={\frac{1}{t}}w_{(t)j}
+{\frac{1}{2}}w_{(t)k}{\hat h}_{(t)}^{sk}{\hat h}_{(t)sj},_t.
\ena
We now split up the 4-velocity field family ${\bf u}_t$ along the curve into 
tangential and normal pieces. This is done as follows:
\eqa
{\bf u}_t{\equiv}{\frac{dx^{\mu}}{d{\tau}_t}}
{\frac{\partial}{{\partial}x^{\mu}}}={\topstar{\gamma}}(c{\bf n}_t+{\bf w}_t).
\ena
From equation (2.10), we see that to find an expression for 
${\stackrel{\star}{\bf a}}$, we need to calculate the quantity
${\topstar{\nabla}}_{\frac{\partial}{{\partial}t}}{\bf u}_t$. We will use
equation (3.52) for this purpose. By direct calculation, using equations (2.4),
(2.7), (3.50) and (3.51), we find that
\eqa
{\topstar{\nabla}}_{\frac{\partial}{{\partial}t}}w_{(t){\mu}}=
{\topstar{\nabla}}_{\frac{\partial}{{\partial}t}}w_{(t)}^{\mu}=0, \qquad
\Rightarrow \qquad
{\topstar{\nabla}}_{\frac{\partial}{{\partial}t}}{\bf w}_t=0,
\ena
and furthermore we calculate
\eqa
{\topstar{\nabla}}_{\frac{\partial}{{\partial}t}}n_{(t){\mu}}=
{\topstar{\nabla}}_{\frac{\partial}{{\partial}t}}n_{(t)}^{\mu}=0, \qquad
\Rightarrow \qquad
{\topstar{\nabla}}_{\frac{\partial}{{\partial}t}}{\bf n}_t=0,
\ena
confirming the requirement (2.3). But from equations (2.10),
(3.52), (3.53) and (3.54) we get
\eqa
{\topstar{\nabla}}_{\frac{\partial}{{\partial}t}}{\bf u}_t=0,
\qquad \Rightarrow \qquad
{\stackrel{\star}{\bf a}}={\bf a}_t,
\ena
as asserted previously. Using a GTCS, we then get the final form of the 
equations of motion, namely
\eqa
{\frac{d^2x^{\mu}}{d{\lambda}^2}}+{\Big (}
{\topstar{\Gamma}}^{{\,}{\mu}}_{{\nu}t}{\frac{dt}{d{\lambda}}}+
{\Gamma}^{\mu}_{(t){\nu}{\beta}}{\frac{dx^{\beta}}{d{\lambda}}}{\Big )}
{\frac{dx^{\nu}}{d{\lambda}}} 
={\Big(}{\frac{cd{\tau}_t}{d{\lambda}}}{\Big)}^2
c^{-2}a_{(t)}^{\mu}.
\ena
Equation (3.56) is valid for both time-like and null curves. For time-like 
curves, one may suitably choose a parametrization in terms of the proper time 
measured along the curve, i.e., one may choose ${\lambda}=c{\tau}_t$. For null 
curves, $\lambda$ must be chosen to be some other affine parameter such that 
${\frac{cd{\tau}_t}{d{\lambda}}}=0$.
\section{Spherically symmetric vacua}
In this section, we solve the field equations for two metrically static, 
spherically symmetric vacuum cases. (We use the term ``metrically static'' to 
refer to cases where ${\bf {\tilde h}}_t$ and ${\bar N}_t$ do not depend on 
$x^0$ or $t$ and ${\bar N}^j_{(t)}$ vanishes in some GTCS, so that 
${\bf {\bar K}}_t=0$.)

The first case we analyze, is the rather trivial one where space-time is 
isotropic and globally free of matter; for this case, ${\bf {\bar a}}_{\cal F}$ 
vanishes identically. The constraint equation (3.30) is vacuous 
for this case whereas equation (3.25) has the solution ${\bar N}_t=$constant. 
However, from equation (3.38) we get (where ${\bf {\bar P}}_t$ is the family of 
Ricci tensors corresponding to the spatial metric family ${\bf {\bar h}}_t$)
\eqa
{\bar H}_{(t)ij}{\equiv}
{\bar P}_{(t)ij}-{\frac{{\bar P}_t}{2}}{\bar h}_{(t)ij}=
-{\frac{1}{({\bar N}_tct)^2}}{\bar h}_{(t)ij}, {\qquad} 
{\frac{{\bar P}_t}{6}}={\frac{1}{({\bar N}_tct)^2}},
\ena
thus the geometry of the FHSs is completely determined by the prior-geometric 
term in equation (3.38). Equivalently, we have that
\eqa
{\bar P}_{(t)ij}={\frac{{\bar P}_t}{3}}{\bar h}_{(t)ij},
\ena
which means that the solution is an Einstein space. It is easy to show that 
${\bf S}^{3}$ is the only compact spherically symmetric solution of (4.2) with 
${\bar P}_t>0$. Note that ${\bar N}_t$ is just an arbitrary constant, and 
that we may set ${\bar N}_t=1$ by making some appropriate choice of scale for 
the spatial coordinates (and simultaneously scale $x^0$ and $t$). This means 
that each space-time metric of the family ${\bf {\bar g}}_t$ is of the type 
${\bf S}^{3}{\times}{\bf R}$. Moreover, from equation (3.45), we have that
${\bar y}_t=0$ and ${\bar H}_t={\frac{1}{t}}$, which means that 
$c^{-1}{\bar x}_t=0$ from equation (3.44). Besides, this yields ${\bf v}_t=0$ 
from equation (2.22). Thus the FOs in ${\overline{\cal N}}$ do not move with 
respect to the FOs in ${\cal N}$, i.e., the metric families ${\bf g}_t$ and 
${\bf {\bar g}}_t$ are identical. We may then write
\eqa
N={\bar N}_t=1 {\qquad} {\Rightarrow} {\qquad} {\bar H}_t={\frac{1}{t}}. 
\ena
The one-parameter family of ${\bf S}^{3}{\times}{\bf R}$ space-time metrics 
consistent with equation (4.3) takes the form (note that the chosen GTCS
only covers half of ${\bf S}^{3}$)
\eqa
ds^{2}_t = \overline{ds}^{2}_t =-(dx^{0})^{2} + 
({\frac{t}{t_{0}}})^{2}{\bigg(}{\frac{dr^{2}}
{1-{\frac{r^{2}}{(ct_{0})^{2}}}}}+r^{2}d{\Omega}^{2}{\bigg)}, \qquad
0{\leq}r<ct_0.
\ena
We notice that equation (4.4) is of the form (2.15). Moreover, equation (4.4) 
yields that 
\eqa
{\Psi}_t={\frac{t_0}{t}}{\Psi}_{t_0}.
\ena
We also notice that it is just as valid to interpret equation (4.4) to mean 
that atomic length units are shrinking, as the usual interpretation in terms of
an increase of spatial dimensions with time. This illustrates that the family 
of line elements (4.4) expresses the gravitational scale (represented by the 
spatial coordinates in a GTCS) in terms of units determined from the atomic 
scale. Furthermore, since the gravitational scale is represented by coordinates 
belonging to a particular class of coordinate systems, this means that no 
particular {\em intrinsic} length scales can be associated with gravitation. 
In contrast to this, non-gravitational interactions define intrinsic length 
(and mass) scales.

It is important to realize that the metric family (4.4) represents vacuum, 
since we have neglected matter creation (which would violate equation (3.24)).
This means that equation (4.4) is just an approximate, not realistic toy 
cosmological model. However, we may insert ${\bar H}_t={\frac{1}{t}}=H_{0}$ for
the value of the Hubble parameter for the present epoch into equation (4.3) to 
get a realistic value for the age of the Universe. This is possible, since in 
contrast to standard cosmology, one should not expect that inclusion of matter 
into the toy model must necessarily result in any ``braking'' of the expansion.
In fact, one may show that the only possible cosmological models allowed with a 
perfectly isotropic matter distribution, are models where the source is a null 
fluid. That is, according to our theory, cosmological models with an 
isotropically distributed material source is possible only in the limit of an 
infinitely relativistic material source, see appendix C. This indicates that 
matter density has nothing to do with the time evolution of the cosmic 
expansion. Thus the age ${\bar H}_t^{-1}$ should be taken as a realistic value 
and not merely as an upper bound. Besides, another issue is that one may not 
neglect time dilation effects in the quantity ${\frac{1}{{\bar N}_tt}}$ when 
interpreting redshift measurements. We return to the question of interpretation
of spectral shifts in section 4.2.
\subsection{The conformally Minkowski case}
The globally matter-free metric family (4.4) found in the previous section 
is a natural candidate for a cosmological toy model since it is isotropic and
thus spherically symmetric about every point. On the other hand, we need a 
model for the gravitational field exterior to a spherically symmetric body
in an otherwise empty universe. In principle, we must then find a solution of
the field equations which is spherically symmetric about one particular point 
only and where the boundary conditions for this solution should be 
${\bf S}^{3}{\times}{\bf R}$, i.e., the metric family should be of the form 
(4.4) at the boundary. But such a solution does not exist on the whole 
space-time manifold, according to equation (3.26) and the subsequent 
discussion (a solution on a {\em subset} of the space-time manifold may exist, 
however). Now, for local solutions, cosmological boundary conditions cannot 
readibly be tested, so such are of secondary interest only. Therefore, to 
avoid the problems at the boundary, for the case where we can neglect the NKE 
due to a global scale factor, we shall feel free to use a static, 
asymptotically flat metric approximation. That is, we assume that the each 
metric of the metric family is identical and asymptotically Minkowski. 

From equation (4.4), we see that we get a Minkowski background if we set 
$t=t_0$ and let $t_0{\rightarrow}{\infty}$. We may then set 
${\bar N}_t({\infty})=1$. This means that
\eqa
{\Psi}_t(r)={\Psi}_t({\infty}){\bar N}_t^{-1}(r),
\ena
since the acceleration field is purely radial. Equation (4.6) means that
reference units at infinity are being used for each metric ${\bf {\bar g}}{\in}
{\bf {\bar g}}_t$. However, taking these units as those obtained in the limit 
$t{\rightarrow}{\infty}$, we see that they get singular if one 
compares to the units determined by a ${\bf S}^3{\times}{\bf R}$ background. 
That is, ${\Psi}_t$ vanishes in the limit $t{\rightarrow}{\infty}$. But this is
merely a consequence of the fact that the size of the universe is infinite 
for asymptotically flat metric approximations. The field equations and 
equations (3.39), (3.40) still work fine for these cases.

Even if asymptotically Minkowski solutions will be lacking any features arising
from the finite size of space, the discrepancy between these solutions and 
those with acceptable boundary conditions should be small to a good 
approximation for many situations. So we feel free to find an asymptotically 
Minkowski, spherically symmetric static solution and apply it in the 
appropriate circumstances. Now equation (3.25), together with the metric 
approximation of equation (3.38), yield (since, according to equation (3.40), 
${\tilde H}_{(t)ij}$ will vanish for this case)
\eqa
{\bar H}_{(t)ij}+c^{-2}{\bar a}_{{\cal F}i{\mid}j}+
c^{-4}{\bar a}_{{\cal F}i}{\bar a}_{{\cal F}j}+
c^{-4}{\bar a}_{{\cal F}k}{\bar a}_{{\cal F}}^k{\bar h}_{(t)ij}=0,
{\qquad} {\bar a}^{k}_{{\cal F}{\mid}k}+
c^{-2}{\bar a}_{\cal F}^{k}{\bar a}_{{\cal F}k}=0.
\ena
When we apply these equations to the general spherically static line element 
(2.20), we get the unique solution
\eqa
\overline{ds}^{2} = {\bigg(}{\sqrt{1+({\frac{r_{\rm s}}{2r}})^{2}}}-
{\frac{r_{\rm s}}{2r}}{\bigg)}^{2}{\bigg(}-(dx^{0})^{2}+{\Big[}1+
({\frac{r_{\rm s}}{2r}})^2{\Big]}^{-1}dr^{2}{\bigg)}+r^{2}d{\Omega}^{2},
\ena
where, analogous to the Schwarzschild solution,
$r_{\rm s}{\equiv}{\frac{2M^{\rm (EM)}G^{\rm B}}{c^{2}}}+
{\frac{2M^{\rm (MA)}G^{\rm S}}{c^{2}}}$, where $M^{\rm (EM)}$ and $M^{\rm (MA)}$
are the electromagnetic mass and the material mass of the source, respectively.
Some details of the derivation of equation (4.8) are given in an appendix.

Next, we calculate the spatial Ricci scalar ${\bar P}$ obtained from
the metric (4.8). Using equation (A.4), we find that
\eqa
{\bar P}(r)={\frac{3r_{\rm s}^{2}}{2r^{4}}}{\bigg(}{\frac{r_{\rm s}}{2r}}+
{\sqrt{1+({\frac{r_{\rm s}}{2r}})^{2}}}{\bigg)}^{2}, {\qquad}
{\bar P}({\rho})={\frac{3r_{\rm s}^{2}}{2{\rho}^{4}
(1-{\frac{r_{\rm s}}{{\rho}}})^{3}}},
\ena
where ${\rho}$ is an isotropic radial coordinate defined in equation (A.14). 
Furthermore, from equation (3.44) we have that ${\bar y}_t={\bar H}_t$
for asymptotically flat metric approximations. Besides, for the spherically 
symmetric, static case we can set ${\bar H}_t={\bar H}(r)$, so that
\eqa
{\bar H}(r)={\frac{r_{\rm s}}{2r^{2}}}{\bigg(}{\frac{r_{\rm s}}{2r}}+
{\sqrt{1+({\frac{r_{\rm s}}{2r}})^{2}}}{\bigg)}c, {\qquad}
{\bar H}({\rho})={\frac{r_{\rm s}c}{2{\rho}^{2}
(1-{\frac{r_{\rm s}}{{\rho}}})^{3/2}}}.
\ena
As can be readily seen, equation (4.8) contains no coordinate singularity, thus
there is no event horizon. This is in accordance with the requirement that a 
global time coordinate must exist. Furthermore, there is a curvature 
singularity at the spatial origin. However, this singularity is essentially due 
to the fact that the local NKE in the tangent space increases towards the 
speed of light at the spatial origin. Using equations (2.19) and (4.10), we 
find
\eqa
v(r)={\bar H}(r){\sqrt{{\bar h}_{rr}(r)}}r={\frac{r_{\rm s}}{2r}}
{\frac{c}{\sqrt{1+({\frac{r_{\rm s}}{2r}})^{2}}}},
\ena
and from equation (4.11) we see that $v(r){\rightarrow}c$ in the limit 
$r{\rightarrow}0$. As can be seen from equation (4.12) below, this has the 
effect of removing the curvature singularity present in the intrinsic geometry
of the FHSs when one constructs the geometry ${\bf h}$ from the geometry 
${\bf {\bar h}}$ as described in section 2.3. And although the space-time 
metric (4.12) is still singular at the spatial origin, this is merely a 
consequence of the fact that the metric approximation breaks down there so
that the premise of a gravitational point source is unphysical. That is, when 
the {\em global} NKE is taken into account, we see from equation (3.45) that
for $t{\neq}{\infty}$, ${\bar H}_t{\rightarrow}{\infty}$ when 
${\bar N}_t{\rightarrow}0$. This means that the global cosmic expansion 
within a local gravitational well should effectively prohibit the formation of
a gravitational point source. We interpret this to mean that gravitational 
collapse cannot, even in principle, form physical singularities when the global
NKE is taken into account. 

We now use equations (2.21), (4.8) and (4.11) to construct the metric 
$\bf g$ and this is sufficient to check if the ``classical'' solar system 
tests come in correctly. We find the unique metric 
\eqa
ds^{2}&=&-{\Big[}1+({\frac{r_{\rm s}}{2r}})^2{\Big]}^{-2}
{\bigg(}{\frac{r_{\rm s}}{2r}}+
{\sqrt{1+({\frac{r_{\rm s}}{2r}})^{2}}}{\bigg)}^{-2}(dx^0)^2 {\nonumber} \\
&+&{\Big[}1+({\frac{r_{\rm s}}{2r}})^{2}{\Big]}^{-1}{\bigg(}
{\frac{r_{\rm s}}{2r}}+
{\sqrt{1+({\frac{r_{\rm s}}{2r}})^{2}}}{\bigg)}^{2}dr^2 + r^{2}d{\Omega}^{2} 
{\nonumber} \\
&=&-{\Big(}1-{\frac{r_{\rm s}}{r}}+{\frac{3r_{\rm s}^{3}}{8r^{3}}}+
{\cdots}{\Big)}(dx^0)^{2}+{\Big(}1+{\frac{r_{\rm s}}{r}}+
{\frac{r_{\rm s}^{2}}{4r^{2}}}+{\cdots}{\Big)}dr^{2}+r^{2}d{\Omega}^{2},
\ena
which agrees with the Schwarzschild metric to sufficient order [1]. Thus the 
metric (4.12) is consistent with the four ``classical" solar system tests 
(radar time delay experiments included), and our theory seems to have the 
correct metric correspondence. (This can be directly confirmed by applying the 
equations of motion (2.12) to the metric (4.12).) We notice that, since the 
metric (4.8) is conformally flat as can be seen from equation (A.15), it has 
vanishing Weyl curvature. But since the Schwarzschild metric has vanishing 
Ricci curvature, it is the Weyl curvature of that metric which makes GR agree 
with the solar system tests. This is in contrast to our theory, where any
agreement with the solar system tests at the post-Newtonian level could not 
possibly have come about, were it not for the fact that the local NKE has been 
taken into account by constructing the metric (4.12).
\subsection{Testing the NKE-paradigm}
To determine whether or not our theory is viable, we must try to confront its
predictions with data from modern relativistic gravitational experiments. But 
first we must show that it has a consistent Newtonian limit. Thus we must find
a consistent Newtonian limit of equation (2.15) and the field equations (3.25) 
and (3.30) as applied to an isolated system (equation (3.41) yields nothing
extra).

The first problem one encounters is that said equations are valid in principle 
only for a particular foliation of ${\bf {\bar g}}_t$ into spatial 
hypersurfaces. In quasi-metric cosmology, this singles out the cosmic rest 
frame since the FOs should be at rest on average with respect to this frame.
On the other hand, the barycentre of an isolated system may have a
non-negligible speed ${\bar w}$ and thus a (approximately metrically constant) 
shift vector field that cannot be neglected (of order ${\frac{\bar w}{c}}$, 
i.e., $O(1)$), with respect to the cosmic rest frame. However, as discussed in 
section 2.2, for this case it should be a good approximation to ignore the 
global curvature of space. So one may substitute the condition 
${\tilde P}_t={\frac{6}{(ct_0)^2}}$ with the approximate alternative condition 
${\tilde P}_t=0$; then the FHSs will be asymptotically flat. Next one may
transform the field equatons with respect to an alternative frame which with
respect to the barycentre of the isolated system is at rest. The field 
equations will not be invariant under this transformation, but any deviations
should be negligible to Newtonian order. Besides, after said transformation, 
the new shift vector field should be at most of $O(3)$ in small quantities and 
thus negligible to Newtonian accuracy if said system's speed with respect to 
the cosmic rest frame is much smaller than the speed of light.

The next question is if all effects of the NKE may safely be neglected when 
making approximations at the Newtonian level of precision. As mentioned 
earlier, effects related to that part of the NKE which is integrated into the 
metric structure of ${\bf g}_t$ via ${\bf v}_t$, we refer to as {\em local} 
effects of the NKE, whereas the remainder of the NKE yields the {\em global} 
effects. In essence, Newtonian theory should correspond with the {\em metric} 
part of our theory, and not with the non-metric part, which has no Newtonian 
counterpart. Thus, even if the local NKE can be neglected at the Newtonian 
level of precision, there is no reason to think that this also applies to the 
global NKE. Therefore, a more useful approximation than the traditional 
Newtonian limit can be made by taking the Newtonian limit of equation (2.15), 
but such that the global spatial scale factor squared ${\frac{t^2}{t_0^2}}$ is 
included. That is, in this ``quasi-Newtonian'' limit, the FHSs are taken as 
flat, but non-static since the global NKE is included. Besides, all the local 
NKE is ignored in the quasi-Newtonian limit so that 
${\bf {\bar g}}_t={\bf g}_t$ (and we can thus drop the bar labels). In a 
special GTCS (see below) with Cartesian coordinates, the quasi-Newtonian metric
family has the components
\eqa
g_{(t)00}={\bar g}_{(t)00}=-1+2c^{-2}U(x^{\mu}), {\quad} g_{(t)i0}=
{\bar g}_{(t)i0}=0, {\quad} g_{(t)ij}={\bar h}_{(t)ij}={\frac{t^2}{t_0^2}}
{\delta}_{ij},
\ena
where $-U(x^{\mu})$ is the Newtonian potential. Note that equation (4.13) is 
consistent with the general metric family (2.15) since to Newtonian accuracy, 
we can neglect any contribution to ${\bar h}_{(t)ij}$ of order $c^{-2}U(x^{\mu})$,
i.e., $O(2)$. The quasi-Newtonian form (4.13) of the metric family is useful 
since it takes sufficiently care of the effects of the global NKE for weak 
gravitational fields and slow motions. Moreover, the traditional Newtonian 
limit is obtained by taking the metric approximation of equation (4.13), so in 
this limit, the FHSs are taken as flat and static. In appendix B, we use the 
metric approximation of equation (4.13) to construct a well-defined Newtonian 
limit for our theory by calculating the Newtonian approximation to the 
evolution equations (3.22) and (3.23) applied to a perfect fluid. Besides, we 
show that if the composition-dependent aspects of the postulated gravitational 
couplings can be neglected, equation (3.25) reduces to Newton's gravitational 
equation in the Newtonian limit and the metric approximation of equation (4.13)
is consistent with the equations of motion (3.56) with vanishing non-metric 
terms.

The next level of precision beyond the Newtonian limit is the post-Newtonian
approximation. Now the local motion of the FOs cannot be
ignored, thus nor can the local NKE. Thus any attempt to construct a 
general weak-field approximation formalism for quasi-metric gravity (along the 
lines of the parametrized post-Newtonian (PPN) formalism valid for metric 
theories of gravity) will meet some extra complications. This implies that 
it is not a good idea to try to apply the standard PPN-formalism to our 
quasi-metric theory. There are several reasons for this; one obvious 
reason is that the PPN-formalism neglects the non-metric aspects of 
quasi-metric relativity. So any PPN-analysis of our field 
equations is limited to their metric approximations. But even these metric 
approximations are not suitable for a standard PPN-analysis since the 
resulting PPN-metric ${\bf {\bar g}}$ is not the one to which experiments are 
to be compared, and ${\bf {\bar g}}$ will {\em not} have an acceptable set of 
PPN-parameters according to metric theory. For example, for metrically static 
systems, a PPN-analysis of our field equations yields the PPN-parameters 
${\gamma}=-1$ and ${\beta}=0$; both values are totally unacceptable for any 
viable metric theory. 

Besides, finding a more complete set of PPN-parameters for ${\bf {\bar g}}$
turns out to be problematic since the relationships assumed to hold between 
said parameters in metric theories will not necessarily hold for quasi-metric 
gravity. This typically leads to inconsistencies. Furthermore, when one 
attempts to construct a ``physical'' PPN-metric ${\bf g}$ from ${\bf {\bar g}}$
in the manner discussed in section 3.3, one gets more complications. For 
example, the transformation ${\bf {\bar g}}{\rightarrow}{\bf g}$ would turn the
PPN-parameters for ${\bf {\bar g}}$ into scalar fields rather than new 
constants, reducing the usefulness of a potential PPN-metric ${\bf g}$. Thus 
the bottom line is that a standard PPN-analysis, even limited to metric 
approximations of QMR, will fail.

Even without a standard formalism within which the theory can be compared to
experiment, we can try to pinpoint some of the ``direct" predictions due to the
global NKE of space. The fact that the NKE takes the form of an ``expansion", 
suggests that one of the obvious effects to look for is a corresponding 
{\em non-kinematical redshift} (NKR). As first noted by Synge, spectral shifts 
in GR (and other metric theories) can be analysed within a unified formalism 
[11]. This means that gravitational, Doppler and standard cosmological spectral
shifts are of the same nature; we refer to such spectral shifts as 
{\em kinematical} (in the general sense of the word). Kinematical spectral 
shifts may be blueshifts or redshifts depending on the specific circumstances, 
whereas the NKR is always a redshift. It should be simple to extend the unified
formalism described in [11] to include the NKR as well by including the 
non-metric piece of the connection. This should be so because the NKR 
essentially arises from the non-metric part of the connection just as 
kinematical spectral shifts arise from its metric part.

As a straightforward example, an easy calculation of the NKR can be executed
for the toy cosmological model (4.4). With a change to a more convenient radial 
coordinate ${\chi}$, equation (4.4) reads
\eqa
ds^{2}_t = -(dx^{0})^{2}+c^{2}t^{2}{\Big(}d{\chi}^{2}+{\sin}^{2}{\chi}
d{\Omega}^{2}{\Big)}.
\ena
We want to find the equation of a null geodesic in the ${\chi}$-direction by 
solving equation (3.56) for the metric family (4.14). To begin with, we find 
that all the relevant ordinary connection coefficients vanish, whereas the only 
relevant non-vanishing degenerate connection coefficient is 
${\,}{\,}{\topstar {\Gamma}}^{{\,}{\chi}}_{t{\chi}}={\frac{1}{t}}$. Furthermore, 
the time component of equation (3.56) yields 
${\frac{dx^{0}}{d{\lambda}}}=$constant whereas the radial component yields 
(using $t$ as a time parameter)
\eqa
{\frac{d^{2}{\chi}}{dt^{2}}} + t^{-1}{\frac{d{\chi}}{dt}}=0.
\ena
The solution of equation (4.15), subject to the condition 
${\frac{d{\chi}}{dx^0}}=(ct)^{-1}$ (which follows from equations
(3.49) and (3.51)), is 
\eqa
{\chi}(t)={\chi}_0 + {\ln}{\frac{t}{t_0}}.
\ena
Now consider two FOs with radial coordinates ${\chi}_0$ and ${\chi}_1$ 
respectively, and with identical angular coordinates. At epoch $t_0$, two 
pulses of light are emitted in the ${\chi}$-direction by the first FO. The 
pulses are separated by a short time interval ${\Delta}t_0$. At epoch $t_1$, 
the light pulses are received by the second FO, and the separation time 
interval is ${\Delta}t_1$. Using equation (4.16), we find (neglecting terms of 
higher order in small quantities)
\eqa
{\frac{{\Delta}t_1}{{\Delta}t_0}}={\frac{t_1}{t_0}},
\ena
a result identical to that of metric cosmological models, i.e., representing a 
redshift of momentum for decoupled massless particle species. The calculation 
leading to equation (4.17) can be generalized, i.e., one can in general find 
the equations of null geodesics connecting pairs of FOs and then calculate the 
frequency shifts of light pulses. The result is a superposition of the NKR and 
kinematical spectral shifts. In particular, radiation emitted from a FO in a 
gravitational well should, as seen from a FO far away, show both the 
gravitational redshift and the NKR.

Next, we can find the speed
$w{\equiv}{\sqrt{w^2}}=ct{\frac{d{\chi}}{d{\tau}_{\cal F}}}=
c^{2}t{\frac{d{\chi}}{dx^0}}$ with respect to the FOs for some inertial 
material particle emitted in the radial direction. Since 
${\frac{d{\chi}}{dx^0}}(t)={\frac{t_0}{t}}{\frac{d{\chi}}{dx^0}}(t_0)$ (which 
follows from equation (3.51)), we have $w=$constant, i.e., material particles 
do not slow down with respect to the FOs. This result is different from its 
counterpart in standard cosmology, and illustrates the different nature of the 
NKE compared to the KE. The curves traced out by material inertial test 
particles moving in the radial direction are found by solving equation (4.15) 
with the condition ${\frac{d{\chi}}{dx^0}}=c^{-2}wt^{-1}$. The solution is
\eqa
{\chi}(t)={\chi}_0+{\frac{w}{c}}{\ln}{\frac{t}{t_0}}.
\ena
This solution is consistent with the solution (4.16) in the limit
$w{\rightarrow}c$, as expected.

In quasi-metric theory, the cosmic expansion is predicted to be relevant also
for local systems, such as the solar system. Thus a natural question would be 
if this will result in detectable tidal forces. To answer this question, we may 
calculate the predicted cosmic tidal effects for a static, non-comoving, thin 
rod. The end points of the rod have spatial coordinates 
$(0,{\frac{t_0}{t}}{\lambda}^i)$ with respect to some chosen GTCS 
${\{}x^{\mu}{\}}$. To calculate the predicted stress in the rod due to the 
cosmic expansion, we define the (symmetric) {\em non-metric tidal tensor} 
${\stackrel{\star}{\bf O}}$ in ${\cal N}$ (with components 
${\,}{\topstar O}_{{\,}{\,}{\,}{\,}{\mu}}^{{\,}{\,}{\nu}}{\,}$ in said GTCS), 
defined from that part of the Riemann tensor ${\stackrel{\star}{\bf R}}$
in $\cal N$ depending on the connection coefficients 
${\topstar {\Gamma}}_{t{\mu}}^{{\,}{\nu}}$. Thus we define
\eqa
{\topstar O}_{{\,}{\,}{\,}{\,}{\mu}}^{{\,}{\,}{\nu}}{\equiv}
{\topstar {\gamma}}^2{\Big \{}{\Big (}{\frac{dt}{d{\tau}_t}}{\Big )}^2
{\,}{\topstar R}^{{\,}{\,}{\nu}}_{{\ }{\,}{\,}tt{\mu}}+
{\frac{dt}{d{\tau}_t}}{\frac{dx^0}{d{\tau}_t}}{\Big [}{\,}
{\topstar R}^{{\,}{\,}{\nu}}_{{\ }{\,}{\,}0t{\mu}}+
{\topstar R}^{{\,}{\,}{\nu}}_{{\ }{\,}{\,}t0{\mu}}{\,}{\Big ]}{\Big \}}.
\ena
Here, said tidal stress (per unit mass) in the rod is 
given by
${\,}{\topstar O}_{{\,}{\,}{\,}{\,}{\mu}}^{{\,}{\,}{\nu}}{\,}
{\frac{t_0}{t}}{\lambda}^{\mu}$. Moreover, we have
\eqa
{\topstar R}^{{\,}{\,}{\nu}}_{{\ }{\,}{\,}tt{\mu}}=
{\frac{\partial}{{\partial}t}}{\topstar {\Gamma}}_{t{\mu}}^{{\,}{\nu}}+
{\topstar {\Gamma}}_{t{\epsilon}}^{{\,}{\nu}}
{\topstar {\Gamma}}_{t{\mu}}^{{\,}{\epsilon}}=
{\delta}^{\nu}_i{\delta}^j_{\mu}{\Big \{}{\hat h}^{is}_{(t)}{\Big (}
{\frac{1}{t}}{\hat h}_{(t)sj},_t+{\frac{1}{2}}{\hat h}_{(t)sj},_{tt}{\Big )}
+{\frac{1}{4}}{\hat h}^{is}_{(t)},_t{\hat h}_{(t)sj},_t{\Big \}},
\ena
\eqa
{\topstar R}^{{\,}{\,}{\nu}}_{{\ }{\,}{\,}0t{\mu}}=
{\frac{\partial}{{\partial}t}}{\topstar {\Gamma}}_{0{\mu}}^{{\,}{\nu}}
+{\delta}^{\nu}_i{\topstar {\Gamma}}_{tk}^{{\,}i}
{\topstar {\Gamma}}_{0{\mu}}^{{\,}k}, \qquad
{\topstar R}^{{\,}{\,}{\nu}}_{{\ }{\,}{\,}t0{\mu}}=
g_{(t)}^{{\nu}{\alpha}}g_{(t){\mu}{\beta}}
{\topstar R}^{{\,}{\,}{\beta}}_{{\ }{\,}{\,}0t{\alpha}}.
\ena 
We find that for cases where ${\bf {\hat h}}_t$ does not depend on $t$ (in 
particular, for metrically static and metrically stationary cases), there will
be no tidal stress directly associated with the quasi-metric expansion. More 
generally, we see from equation (4.20) there will be tidal effects if
${\bf {\hat h}}_t$ depends on $t$, but these effects are too small to be 
detectable in practice.

So, although it would seem difficult to detect the cosmic expansion and thus
the global NKE of space directly by doing local experiments, its relevance for 
explaining astrophysical observations from first principles is quite promising.
To have any hope of doing better than metric theories in this respect, one must
focus on situations where the metric approximation is insufficient. In 
particular, astrophysical systems which are large compared to the solar system,
but yet non-relativistic in the sense of small velocities and weak 
gravitational fields, should be well-suited for testing the non-metric aspects 
of our theory. The most obvious examples of such systems are galaxies. Thus one
may have some hope that it within the NKE-paradigm may be possible explain 
observed galactic dynamics from first principles, a feat that metric theories 
seem incapable of doing [12]. We return to this issue in a follow-up paper 
[14], where it is shown that the NKE of space may also have been detected in 
the solar system.

More exotic systems which have been modelled successfully by applying GR, are
the binary pulsars [1]. In particular, within GR, the decay of orbital periods 
of such objects is naturally explained by assuming that the systems emit
gravitational radiation. To have any hope of competing, this phenomenon must
have a natural explanation within our quasi-metric theory also. But since the 
field equation (3.38) (or alternatively, equation (3.41)) has the same 
radiative limit as GR for weak gravitational waves in vacuum, it seems likely 
that binary pulsars can be successfully modelled within quasi-metric theory as 
well. Compared to the GR model, we have one extra effect though; in addition to
said orbital decay, there is also an increase in orbital periods due to the 
global cosmic expansion. The sum of these two opposite effects may be very 
different from the GR result, in particular for pulsars with large orbital 
periods. Of course, to test if quasi-metric models agree with the data, nothing
short of a quantitative analysis will be sufficient.

The fields where the difference in philosophical basis between our 
quasi-metric theory and metric theories shows up most clearly are cosmology 
and large-scale astronomy. That is, the traditional big bang models (and 
alternative models also) describe the Universe as a dynamical system which 
evolution can be deduced from field equations given appropriate initial 
conditions. Using this approach, it is not surprising that one of the holy 
grails of modern cosmology is to deduce the ``correct" big bang model from 
observational data by fitting a number of cosmological parameters. In contrast 
to this, according to our theory, the NKE of space as described by a 
non-dynamical global scale factor is a separate phenomenon independent of
scale, and thus our description of the Universe is not really that of a 
dynamical system (i.e., there is no ``cosmic dynamics''). In fact, a natural 
beginning of the Universe according to our theory is described by the empty 
universe toy model (4.4) in the limit $t{\rightarrow}0$. Thus even if this 
model has a curvature singularity at $t=0$, it does not represent a 
{\em physical} singularity since no particle world line actually originates at 
the singularity. This removes the vexing question of specifying ``natural" 
initial conditions, a question which cannot be avoided in a purely dynamical 
approach to cosmology. Instead, we get the problem of constructing a 
quantitative theory of particle creation based on the global NKE of space and 
vacuum fluctuations of non-gravitational fields. This may perhaps be done 
along the lines described in reference [10].

It remains to be seen if non-kinematically expanding universe models are
capable of giving natural explanations of the plethora of astronomical
observations which exist today; examples are helium abundances, the structure 
of the microwave background, etc. In this context, one must be aware that
interpretations of observational data made within the framework of a
non-kinematical universe model, may differ considerably from those made 
within a big bang model. The main reason for this, is that the nature of the 
expansion is different for the two model types. In particular, this means that 
within a non-kinematical model, one cannot uncritically take observed redshifts
of objects as indicators of distance. That is, any observed spectral shift 
includes not only the NKR due to the expansion on large scales of the Universe,
but also the NKR due to the intrinsic expansion of gravitational fields. (See 
[14] for an idealized model of an expanding gravitational field.) If the latter
is not negligible compared to the former, interpretations within standard 
cosmology may lead to peculiar results. And for a sufficiently strong 
gravitational field, the NKR due to the intrinsic expansion will not be 
negligible since the global part of the NKE goes as ${\frac{1}{{\bar N}_tt}}$. 
That is, for an observer far away, ${\frac{1}{{\bar N}_tt}}$ far down into the 
gravitational well appears larger than the local value of 
${\frac{1}{{\bar N}_tt}}$ because of gravitational time dilation. Thus, if an 
observer far away registers radiation escaping from the gravitational well, 
this radiation gets an extra NKR mimicking the cosmological redshift. In 
addition to the intrinsic expansion, massive objects may simultaneously undergo
gravitational collapse, resulting in an extra kinematical component of the 
observed redshift. Note that the size of such objects may not necessarily
decrease over time. Thus, it is at least possible that massive expanding and 
simultaneously collapsing objects may systematically show excess redshifts 
which should not necessarily be interpreted as distance indicators. Such 
objects cannot possibly exist in metric gravity. In this context, we notice 
that the apparent excess redshift associated with compact galaxy-like objects 
is the basic premise for the well-known ``redshift controversy'' regarding the 
nature of quasi-stellar objects [13].
\section{Conclusion}
In this thesis, we have introduced the so-called ``quasi-metric'' framework;
this is a non-metric space-time framework based on the NKE concept. 
Furthermore, we have constructed a theory of gravity compatible with this 
framework. The vital question remains; is the theory viable according to 
basic theoretical and experimental criteria?

At least our theory is relativistic, and it displays the correct Newtonian
limit for a large class of situations. However, for ``large'' systems, global
non-metric effects may not be negligible, and for such systems Newtonian theory 
does not correspond with our theory. On a higher level of precision, 
our theory corresponds with GR through metric approximations; we have 
illustrated this by constructing one for the static, spherically symmetric 
case. This means that our theory should agree with GR for the classical solar 
system tests (to post-Newtonian accuracy). However, due to the fact that our 
theory is not suitable for a standard PPN-analysis, a formal weak-field
comparison to GR is not available, see the previous section.

The theoretical structure of our theory makes it plausible that its equations 
are well-posed and self-consistent. This also indicates that its predictions 
are unique, i.e., that they do not depend on the method of calculation. 
Finally, the theory should not be regarded as complete, i.e., all aspects of
it have probably not yet been fully developed.

So we do not yet know if our theory is viable or not. However, there should
be a fair chance that the NKE concept may be relevant for a better 
understanding from first principles of the behaviour of the physical world. 
This should inspire to further work along the direction we have laid out in 
this thesis. \\ [4mm]
{\bf References} \\ [1mm]
{\bf [1]} C.M. Will, {\em Theory and Experiment in Gravitational Physics},
Cambridge University \\ {\hspace*{5.5mm}} Press (1993). \\
{\bf [2]} K. Kucha{\v{r}}, {\em Journ. Math. Phys.} {\bf 17}, 792 (1976). \\
{\bf [3]} C. Gerhardt, {\em Commun. Math. Phys.} {\bf 89}, 523 (1983). \\
{\bf [4]} C.W. Misner, K.S. Thorne, J.A. Wheeler,
{\em Gravitation}, W.H. Freeman ${\&}$ Co. (1973). \\
{\bf [5]} C.H. Brans and R.H. Dicke, {\em Phys. Rev. }{\bf 124}, 925 (1961). \\
{\bf [6]} R.H. Dicke, {\em Phys. Rev. }{\bf 125}, 2163 (1962). \\
{\bf [7]} J. Isenberg and J. Nester, {\em Canonical Gravity}, in
{\em General Relativity and Gravitation,  \\  {\hspace*{5.5mm}} Vol. I}, 
Editor A. Held, Plenum Press (1980). \\
{\bf [8]} K. Kucha{\v{r}}, {\em Journ. Math. Phys.} {\bf 17}, 777 (1976). \\
{\bf [9]} S. Weinberg, {\em Gravitation and Cosmology}, John Wiley ${\&}$ Sons,
Inc. (1972). \\
{\bf [10]} L. Parker, {\em Particle Creation by the Expansion of the 
Universe}, in {\em Handbook of \\
{\hspace*{7.5mm}}Astronomy, Astrophysics and Geophysics, Vol. II Galaxies and 
Cosmology}, \\ {\hspace*{7.5mm}}editors V.M. Canuto and B.G. Elmegreen,
Gordon $\&$ Breach (1988). \\
{\bf [11]} J.V. Narlikar, {\em Am. J. Phys.} {\bf 62}, 903 (1994). \\
{\bf [12]} V.V. Zhytnikov and J.M. Nester, {\em Phys. Rev. Lett.} {\bf 73},
2950 (1994). \\
{\bf [13]} H. Arp, {\em Quasars, Redshifts and Controversies}, Interstellar
Media (1987). \\
{\bf [14]} D. {\O}stvang, chapter 3 of this thesis (2001). \\
{\bf [15]} D. {\O}stvang, chapter 1 of this thesis (2001). \\
{\bf [16]} J. Isenberg, {\em Class. Quantum Grav.} {\bf 12}, 2249 (1993).
\appendix
\renewcommand{\theequation}{\thesection.\arabic{equation}}
\section{Derivation of a static line element}
\setcounter{equation}{0}
In this appendix, we give some details of the derivation of the solution 
(4.8). To do this, we insert the general line element (2.20) into equations
(4.7) to determine the functions ${\bar A}(r)$ and ${\bar B}(r)$. Connection 
coefficients for the line element (2.20) can be found in, e.g., [9]. The 
components of the acceleration field are (using the notation 
$'{\equiv}{\frac{\partial}{{\partial}r}}$)
\eqa
{\bar a}_{{\cal F}r}=c^{2}{\frac{{\bar B}'}{2{\bar B}}}, {\qquad} 
{\bar a}_{{\cal F}{\theta}}={\bar a}_{{\cal F}{\phi}}=0,
\ena
and the spatial covariant derivative of the acceleration field has the 
non-zero components
\eqa
{\bar a}_{{\cal F}r{\mid}r}={\frac{c^{2}}{2}}{\bigg[}
{\frac{{\bar B}''}{{\bar B}}}-({\frac{{\bar B}'}{{\bar B}}})^{2}-
{\frac{{\bar A}'{\bar B}'}{2{\bar A}{\bar B}}}{\bigg]}, {\qquad} 
{\sin}^{-2}{\theta}{\bar a}_{{\cal F}{\phi}{\mid}{\phi}}=
{\bar a}_{{\cal F}{\theta}{\mid}{\theta}}={\frac{c^{2}}{2}}
{\frac{{\bar B}'r}{{\bar A}{\bar B}}}.
\ena
The spatial covariant divergence of the acceleration field is
\eqa
{\bar a}^{k}_{{\cal F}{\mid}k}=c^{2}{\bigg(}
{\frac{{\bar B}''}{2{\bar A}{\bar B}}}-
{\frac{{\bar B}'^{2}}{2{\bar A}{\bar B}^{2}}}-
{\frac{{\bar A}'{\bar B}'}{4{\bar A}^{2}{\bar B}}}+
r^{-1}{\frac{{\bar B}'}{{\bar A}{\bar B}}}{\bigg)}.
\ena
Next, a standard calculation yields the spatial Ricci tensor ${\bf {\bar P}}$ 
for the line element (2.20), which we use to find the spatial Einstein tensor 
${\bf {\bar H}}$. The result is
\eqa
{\bar P}&=&{\frac{2}{r}}{\bigg[}r^{-1}(1-{\bar A}^{-1})+
{\frac{{\bar A}'}{{\bar A}^{2}}}{\bigg]}, 
\ena
\eqa
{\bar H}_{rr}&{\equiv}&{\bar P}_{rr}-{\frac{{\bar P}}{2}}{\bar h}_{rr}=
r^{-2}(1-{\bar A}), {\nonumber} \\ {\sin}^{-2}{\theta}{\bar H}_{{\phi}{\phi}}=
{\bar H}_{{\theta}{\theta}}&{\equiv}&
{\bar P}_{{\theta}{\theta}}-{\frac{{\bar P}}{2}}{\bar h}_{{\theta}{\theta}}=
-{\frac{r{\bar A}'}{2{\bar A}^{2}}}.
\ena
Inserting the above relations into equations (4.7) yields
\eqa
{\frac{{\bar B}''}{{\bar B}}}-
{\frac{{\bar A}'{\bar B}'}{2{\bar A}{\bar B}}}+2r^{-2}(1-{\bar A})=0,
\ena
\eqa
{\frac{{\bar B}''}{{\bar B}}}-{\frac{{\bar B}'^2}{2{\bar B}^2}}-
{\frac{{\bar A}'{\bar B}'}{2{\bar A}{\bar B}}}+
{\frac{2}{r}}{\frac{{\bar B}'}{{\bar B}}}=0,
\ena
\eqa
{\frac{{\bar A}'}{{\bar A}}}-{\frac{{\bar B}'}{{\bar B}}}-
r{\frac{{\bar B}'^2}{2{\bar B}^2}}=0. 
\ena
We can separate the variables ${\bar A}(r)$ and ${\bar B}(r)$ by inserting 
equation (A.8) into equations (A.6) and (A.7). The result is
\eqa
{\frac{{\bar A}''}{{\bar A}}}-{\frac{3{\bar A}'^{2}}{2{\bar A}^{2}}}+
r^{-1}{\bar A}'(3{\bar A}^{-1}-1)=0, 
\ena
\eqa
{\frac{{\bar B}''}{{\bar B}}}-{\frac{{\bar B}'^{2}}{{\bar B}^{2}}}-
r{\frac{{\bar B}'^{3}}{4{\bar B}^{3}}}+r^{-1}{\frac{2{\bar B}'}{{\bar B}}}=0, 
\ena
\eqa
{\bar A}=1-r{\frac{{\bar B}'}{{\bar B}}}+
r^{2}{\frac{{\bar B}'^{2}}{4{\bar B}^{2}}}.
\ena
One way of solving these equations is to introduce 
${\bar C}(r){\equiv}r{\frac{{\bar B}'}{{\bar B}}}$. One may then show that 
equation (A.10) is integrable in this new variable. This yields ${\bar B}(r)$, 
and equation (A.11) gives ${\bar A}(r)$ directly when ${\bar B}(r)$ is known. 
The result is the metric (4.8), which is unique. An alternative way of writing 
equation (4.8) can be found by using the identity
\eqa
{\sqrt{1+({\frac{r_{\rm s}}{2r}})^{2}}}-{\frac{r_{\rm s}}{2r}}{\equiv}
{\bigg(}{\sqrt{1+({\frac{r_{\rm s}}{2r}})^{2}}}+
{\frac{r_{\rm s}}{2r}}{\bigg)}^{-1}.
\ena
That equation (4.8) is indeed conformally flat, we can show by introducing an 
isotropic radial coordinate ${\rho}$ replacing the Schwarzschild coordinate 
$r$:
\eqa
{\rho}{\equiv}{\exp}{\bigg(}{\int}r^{-1}
{\Big[}{\sqrt{1+({\frac{r_{\rm s}}{2r}})^{2}}}-{\frac{r_{\rm s}}{2r}}{\Big]}
{\Big[}1+({\frac{r_{\rm s}}{2r}})^{2}{\Big]}^{-1/2}dr{\bigg)}.
\ena
Evaluating the integral and choosing an appropriate constant of integration,
we find that
\eqa
{\rho}=r{\bigg(}{\sqrt{1+({\frac{r_{\rm s}}{2r}})^{2}}}+
{\frac{r_{\rm s}}{2r}}{\bigg)},
{\qquad} r={\rho}{\sqrt{1-{\frac{r_{\rm s}}{\rho}}}},
\ena
and equation (4.8) expressed in isotropic coordinates reads
\eqa
\overline{ds}^{2}={\bigg(}1-{\frac{r_{\rm s}}{\rho}}{\bigg)}
{\bigg(}-(dx^{0})^{2}+d{\rho}^{2}+{\rho}^{2}d{\Omega}^{2}{\bigg)},
\ena
which is conformally flat.
\renewcommand{\theequation}{\thesection.\arabic{equation}}
\section{The Newtonian limit}
\setcounter{equation}{0}
To find Newtonian approximations of the local conservation laws (3.21) in 
${\overline{\cal N}}$, or equivalently, of the evolution laws (3.22) and 
(3.23). In a static GTCS at rest with respect to the gravitational system's 
centre of gravity, we can use the metric approximation of the
quasi-Newtonian form (4.13) of the metric family with Cartesian coordinates, 
i.e., ${\bar g}_{00}{\approx}-1+2c^{-2}U$, ${\bar g}_{i0}{\approx}0$ and 
${\bar g}_{ij}{\approx}{\bar h}_{ij}{\approx}{\delta}_{ij}$, where $-U$ is the 
Newtonian potential. Furthermore, to Newtonian order, the extrinsic curvature 
tensor vanishes and the active and passive stress-energy tensors are equal. 
For a perfect fluid we then have, to Newtonian order, that
\eqa
T^{00}={\varrho}_{\rm m}c^{2}, {\qquad} {T^{0}}_{j}={\varrho}_{\rm m}c
{\bar w}_j, {\qquad}
{T^{i}}_{j}={\varrho}_{\rm m}{\bar w}^i{\bar w}_j+p{{\delta}^{i}}_{j}.
\ena
Also, to Newtonian accuracy, $c^{-2}U$, $c^{-2}{\bar w}^{2}$, 
$c^{-2}p/{\varrho}_{\rm m}$, $c^{-2}{\mid}{\bar a}_{\cal F}^i{\mid}/{\mid}
{\frac{\partial}{{\partial}x^i}}{\mid}$ and
${\mid}{\frac{{\bar N}_t,_{\bar {\perp}}}{{\bar N}_t}}{\mid}/{\mid}
{\frac{\partial}{{\partial}x^0}}{\mid}$ are all small, of second order in the 
small quantity ${\epsilon}$, say. To Newtonian order, one uses the 
approximations 
\eqa
{T^{0{\nu}}}_{;{\nu}}{\approx}{T^{00}},_0+{T^{0i}},_i
+{\bar {\Gamma}}^0_{00}T^{00}, \qquad
{T^{j{\nu}}}_{;{\nu}}{\approx}{T^{j0}},_0+{T^{ji}},_i
+{\bar {\Gamma}}^j_{00}T^{00}, 
\ena
this is equivalent to retaining terms not higher than one extra order in 
${\epsilon}$. The approximations made in equation (B.2) yield the same 
Newtonian limit as for the metric case.

Using equations (3.22) and (3.23) with ${\bf {\bar K}}_t=0$, keeping only 
lowest order terms plus terms one extra order higher in $\epsilon$, yields 
($t$ and the ordinary coordinate time are equivalent in the Newtonian 
approximation)
\eqa
{\frac{\partial}{{\partial}t}}{\varrho}_{\rm m}+({\varrho}_{\rm m}
{\bar w}^i),_{i}&=&O({\epsilon}^2),
{\nonumber} \\ p,_{j}+{\varrho}_{\rm m}({\bar w}_{j},_{t}+
{\bar w}_{j},_{i}{\bar w}^{i})&=&
-{\varrho}_{\rm m}{\bar a}_{{\cal F}j}+O({\epsilon}^{4})=
{\varrho}_{\rm m}U,_j + O({\epsilon}^4).
\ena
These equations are identical to the Euler equations describing 
non-relativistic fluid dynamics. Here, we have used the fact that in the 
Newtonian limit, ${\bar a}_{{\cal F}i}=-U,_i$ which follows from equation (4.13). 
Note that the velocity ${\bf {\bar w}}$ can be approximated as the velocity 
of the matter source relative to the special GTCS chosen, since using this 
coordinate system, the local motion of the FOs can be ignored at the Newtonian 
level of precision.

Next, use the metric approximation of the quasi-Newtonian form (4.13) of the 
metric family in combination with equation (B.1) to get the Newtonian limit of 
the field equation (3.25). Neglecting terms of $O({\epsilon}^2)$, equation 
(3.25) yields (using a Cartesian coordinate system)
\eqa
4{\pi}(G^{\rm B}{\varrho}^{\rm (EM)}_{\rm m}+
G^{\rm S}{\varrho}^{\rm mat}_{\rm m}){\approx}4{\pi}G_{\rm N}
{\varrho}_{\rm m}={\bar a}_{{\cal F}{\mid}k}^k=-{U_,}^k_{{\ }k},
\ena
which is Newton's gravitational equation as asserted. We have assumed that
composition-dependent aspects of equation (3.25) are small enough to justify
the approximation made in equation (B.4). Note that the field equations 
(3.30) and (3.41) do not contribute anything extra in the Newtonian limit.
Other field equations than equation (3.25) may thus be ignored in the 
Newtonian limit.

Finally, to have a well-defined Newtonian limit, we must show that our 
equations of motion (3.56) reduce to the Newtonian equations of motion in the 
Newtonian limit. But since equation (3.56) reduces to the usual geodesic 
equation for asymptotically flat metric approximations (since 
$t{\rightarrow}{\infty})$, it certainly must reduce to the Newtonian equations 
of motion in the Newtonian limit. With this result we have shown that our 
theory has a consistent Newtonian limit.
\renewcommand{\theequation}{\thesection.\arabic{equation}}
\section{Isotropic cosmological models with matter}
\setcounter{equation}{0}
Rather than the vacuum toy cosmological model considered in section 4, we may
try to construct non-vacuum toy models where the source is a perfect fluid
comoving with the FOs. That is, as a generalization of equation (4.4), 
such toy models may be described by a metric family ${\bf {\bar g}}_t$ 
of the form
\eqa
{\overline{ds}}_t^2=-{\bar N}_t^2(dx^0)^2+({\bar N}_tct)^2(d{\chi}^2
+{\sin}^2{\chi}d{\Omega}^2).
\ena
As a counterpart to the Friedmann-Robertson-Walker models in GR, it is 
reasonable to try to construct models where the matter distribution is 
isotropic. That is, we consider models where ${\bar N}_t$ does not depend on 
the spatial coordinates. We show in this appendix that the only isotropic 
models possible in quasi-metric gravity are models where the source is a null 
fluid.

It is convenient to express the stress-energy tensor for a perfect fluid 
comoving with the FOs via a ``properly scaled'' density of active mass-energy 
${\bar {\varrho}}_{\rm m}$ and the corresponding ``properly scaled'' active 
pressure ${\bar p}$, i.e.,
\eqa
T_{(t){\bar {\perp}}{\bar {\perp}}}={\tilde {\varrho}}_{\rm m}(x^0,t)c^2
{\equiv}
{\Big (}{\frac{t_0}{{\bar N}_tt}}{\Big )}^2{\bar {\varrho}}_{\rm m}(t)c^2, 
{\quad}
T_{(t){\chi}}^{\chi}=T_{(t){\theta}}^{\theta}=T_{(t){\phi}}^{\phi}=
{\tilde p}(x^0,t){\equiv}
{\Big (}{\frac{t_0}{{\bar N}_tt}}{\Big )}^2{\bar p}(t),
\ena
where the variability due to the formally variable units has been scaled out.
Note that ${\bar {\varrho}}_{\rm m}(t)$ does not depend on $t$ if there is 
negligible net energy transfer between material particles and photons and if 
there is also no net particle creation. Moreover, from equations (C.1),
(3.7) and (3.39) we find that
\eqa
{\bar K}_{(t)ij}={\frac{{\bar N}_t,_{\bar {\perp}}}{{\bar N}_t}}
{\bar h}_{(t)ij}, \qquad {\bar P}_t={\frac{6}{(ct{\bar N}_t)^2}}.
\ena
Assuming no net matter creation, the local conservation laws in 
${\overline{\cal N}}$ are sufficiently represented by equations (3.21) and 
(3.22), which yield (using equations (C.2) and (C.3))
\eqa
{\cal L}_{{\bar {\bf n}}_t}{\bar {\varrho}}_{\rm m}=
-{\frac{{\bar N}_t,_{\bar {\perp}}}{{\bar N}_t}}{\Big (}
{\bar {\varrho}}_{\rm m}-3c^{-2}{\bar p}{\Big )}.
\ena
But since any dependence of ${\tilde {\varrho}}_{\rm m}$ on $x^0$ should
be via the factor ${\frac{1}{{\bar N}_t^2}}$, we also have that
\eqa
{\cal L}_{{\bar {\bf n}}_t}{\bar {\varrho}}_{\rm m}=0.
\ena
This means that a null fluid is the only nontrivial possibility satisfying 
equation (C.4). {\em That is, in our theory no cosmological model with a 
perfectly isotropic material fluid source can exist.} Thus in our theory, 
except for the vacuum model found in section 4, the only allowable cosmological
models with perfectly isotropic FHSs are models where the source is a null 
fluid. However, a sufficiently hot material fluid source will have an equation 
of state arbitrarily close to that of a null fluid, so for this case, an 
isotropic cosmological model will be a sufficient approximation to desired 
accuracy.

To confirm that cosmological models with an isotropic null fluid source
(e.g., a pure photon gas) are indeed possible in quasi-metric gravity, we must 
solve the field equations. The field equation (3.25) applied to a photon gas
yields, after some straightforward manipulations, that (with 
${\bar {\varrho}}_{\rm m}={\bar {\varrho}}^{\rm (EM)}_{\rm m}$ for a pure photon 
gas)
\eqa
{\cal L}_{{\bf {\bar n}}_t}{\Big (}{\frac{{\bar N}_t,_{\bar {\perp}}}
{{\bar N}_t}}{\Big )}-{\Big (}{\frac{{{\bar N}_t,_{\bar {\perp}}}}{{\bar N}_t}}
{\Big )}^2={\Big (}{\frac{t_0}{{\bar N}_tt}}{\Big )}^2
{\frac{{\kappa}^{\rm B}}{6}}
({\bar {\varrho}}^{\rm (EM)}_{\rm m}c^2+3{\bar p}^{\rm (EM)}).
\ena
Equation (3.41) is identically fulfilled and thus yields nothing 
extra besides equation (C.6), which may be written as
\eqa
{\Big (}{\frac{{\bar N}_t,_0}{{\bar N}_t}}{\Big )},_0=
-{\frac{t_0^2}{t^2}}{\frac{{\kappa}^{\rm B}}{3}}
{\bar {\varrho}}^{\rm (EM)}_{\rm m}c^2,
\ena
where the right hand side of equation (C.7) does not depend on $x^0$ since
${\cal L}_{{\bar {\bf n}}_t}{\bar {\varrho}}_{\rm m}^{\rm (EM)}$ vanishes. Integrating
equation (C.7) twice and requiring continuity with the empty model found in 
section 4 in the vacuum limit ${\bar {\varrho}}^{\rm (EM)}_{\rm m}{\rightarrow}0$,
we find the solution (up to a constant factor)
\eqa
{\bar N}_t={\exp}{\Big [}-{\frac{(ct_0x^0)^2}{(ct)^2}}
{\frac{{\kappa}^{\rm B}}{6}}{\bar {\varrho}}^{\rm (EM)}_{\rm m}c^2{\Big ]}.
\ena
We see from equation (C.8) that if ${\bar {\varrho}}^{\rm (EM)}_{\rm m}$ does not
depend on $t$, ${\bar N}_t$ is constant in ${\overline{\cal N}}$. Moreover, we 
see that equation (C.7) is still fulfilled if ${\bar N}_t$ is multiplied by an 
arbitrary constant. This means that we can set ${\bar N}_t=1$ in 
${\overline{\cal N}}$ if we choose the boundary condition 
${\bar N}_{t}(t_0)=1$. Note that it is also possible to find isotropic null 
fluid models where ${\bar {\varrho}}_{\rm m}$ depends on $t$, i.e., where there 
is local creation of null particles. For such models, ${\bar N}_t$ will depend 
on $t$ in ${\overline{\cal N}}$. Also note that more general cosmological 
models with homogeneous but non-isotropic matter distributions should exist in 
quasi-metric gravity. In particular, models where the matter density on each 
FHS fluctuates around some mean density should be possible according to our 
theory since the FHSs cannot be isotropic in such cases.

The result of this appendix confirms the assertion that in quasi-metric
theory, the rate of the expansion of the Universe should not depend on 
dynamical gravitational parameters such as matter density. Rather, the cosmic
expansion must be purely non-kinematical. Compared to traditional cosmology, 
this represents a radical difference in philosophy.
\section{Symbols and acronyms}
In this appendix, we give a table of symbols and acronyms used in the text. 
Sign conventions and notations follow reference [4] if otherwise not stated.
In particular, Greek indices may take integer values $0-3$ and Latin indices 
may take integer values $1-3$. \\ [5mm]
\begin{tabular}{lp{10.8cm}}
{\bf SYMBOL} & {\bf NAME/EXPLANATION} \\ 
KE/NKE & kinematical/non-kinematical evolution (of the FHSs) \\
FOs/FHSs & fundamental observers/fundamental hypersurfaces \\
NKR & non-kinematical redshift
\end{tabular}
\newpage
\begin{tabular}{lp{10.8cm}}
GTCS & global time coordinate system \\
HOCS & hypersurface-orthogonal coordinate system \\
WEP/EEP & weak/Einstein equivalence principle \\
GWEP/SEP & gravitational weak/strong equivalence principle \\
LLI/LPI & local Lorentz/position invariance \\
SR/GR & special/general relativity \\
SC & Schiff's conjecture \\
$t=t(x^{\mu})$ & global time function, conventionally we define $t=x^{0}/c$
when using a GTCS \\
${\cal M}$/${\overline{\cal M}}$ & metric space-time manifolds obtained by 
holding $t$ constant \\
${\cal N}$ & quasi-metric space-time manifold $({\cal N},{\bf g}_t)$ \\
${\overline{\cal N}}$ & shorthand notation for $({\cal N},{\bf {\bar g}}_t)$ \\
${\bf g}_t$ or $g_{(t){\mu}{\nu}}$ & one-parameter family of
space-time metric tensors \\
${\bf {\bar g}}_t$ or ${\bar g}_{(t){\mu}{\nu}}$ & auxiliary one-parameter family 
of space-time metric tensors (${\bf {\bar g}}_t$ is found as a solution of the 
field equations) \\
${\bf n}_t$ or $n_{(t)}^{\mu}$ & unit vector field family normal to the FHSs 
in $\cal N$ \\
${\bf {\bar n}}_t$ or ${\bar n}_{(t)}^{\mu}$ & unit vector field family normal to 
the FHSs in ${\overline{\cal N}}$ \\
${\bf h}_t$ or $h_{(t)ij}$ & intrinsic metric tensor family to the FHSs in
${\cal N}$ \\
${\bf {\bar h}}_t$ or ${\bar h}_{(t)ij}$ & intrinsic metric tensor family to the 
FHSs in ${\overline{\cal N}}$ \\
$N(x^{\mu})$ & lapse function in ${\cal N}$ (any dependence of $N$ on $t$ has 
been eliminated by inserting $t=x^0/c$ using a GTCS) \\
${\bar N}_t(x^{\mu},t)$ & lapse function family in ${\overline{\cal N}}$ \\
${\bf {\hat h}}_t$ or ${\hat h}_{(t)ij}$ & scaled spatial metric tensor 
family defined by ${\bf {\hat h}}_t{\equiv}{\frac{t_0^2}{t^2}}{\bf h}_t$ \\
${\bf {\tilde h}}_t$ or ${\tilde h}_{(t)ij}$ & scaled spatial metric tensor 
family defined by ${\bf {\tilde h}}_t{\equiv}
{\frac{t_0^2}{{\bar N}_t^2t^2}}{\bf {\bar h}}_t$ \\
$N^j_{(t)}(x^{\mu},t)$ & components of the shift vector field family in 
${\cal N}$ \\
${\bar N}^j_{(t)}(x^{\mu},t)$ & components of the shift vector field family 
in ${\overline{\cal N}}$ \\
${\Gamma}^{\alpha}_{(t){\beta}{\gamma}}$ & components of the metric connections
${\nabla}_{t}$ compatible with ${\bf g}_{t}$ \\
${\bar {\Gamma}}^{\alpha}_{(t){\beta}{\gamma}}$ & components of the metric 
connections ${\bar {\nabla}}_{t}$ compatible with ${\bf {\bar g}}_{t}$ \\
${\topstar{\Gamma}}^{{\,}{\alpha}}_{t{\gamma}}$,
${\topstar{\Gamma}}^{{\,}{\alpha}}_{{\beta}{\gamma}}$ & components of the degenerate
5-dimensional connection ${\,}{\topstar{\nabla}}{\,}$ associated with the
family ${\bf g}_{t}$ \\
${\topstar{\bar {\Gamma}}}^{{\,}{\alpha}}_{t{\gamma}}$,
${\topstar{\bar {\Gamma}}}^{{\,}{\alpha}}_{{\beta}{\gamma}}$ & components of the 
degenerate 5-dimensional connection ${\stackrel{\star}{\bf {\bar {\nabla}}}}$ 
associated with the family ${\bf {\bar g}}_{t}$  \\
$()_{;{\alpha}}$ & coordinate expression for a metric covariant derivative 
compatible with a single member of ${\bf g}_t$ or ${\bf {\bar g}}_t$ as 
appropriate 
\end{tabular}
\newpage
\begin{tabular}{lp{10.6cm}}
$()_{{\bar *}{\alpha}}$ & coordinate expression for a degenerate covariant 
derivative compatible with the family ${\bf {\bar g}}_t$ \\
$()_{{\mid}j}$ & coordinate expression for a spatial covariant derivative 
compatible with ${\bf h}_t$ or ${\bf {\bar h}}_t$ as appropriate \\
${\perp}$/${\bar {\perp}}$ & projection symbols (projection 
on the normal direction to the FHSs) \\
{\pounds}$_{{\bf {\bar x}}_t}$/{\pounds}$_{{\bf x}_t}$ & Lie derivative with respect 
to ${\bf {\bar x}}_t/{\bf x}_t$ in ${\overline{\cal M}}/{\cal M}$ \\
${\topstar{\pounds}}_{{\,}{\bf {\bar x}}}/{\,}{\topstar{\pounds}}_{{\,}{\bf x}}$ & Lie 
derivative with respect to ${\bf {\bar x}}/{\bf x}$ 
in ${\overline{\cal N}}/{\cal N}$ \\
${\cal L}_{{\bf {\bar x}}_t}/{\cal L}_{{\bf x}_t}$ & projected Lie derivative with 
respect to ${\bf {\bar x}}_t/{\bf x}_t$ in ${\overline{\cal M}}/{\cal M}$ \\
${\topstar{\cal L}}_{{\,}{\bf {\bar x}}}/{\,}{\topstar{\cal L}}_{{\,}{\bf x}}$ & 
projected Lie derivative with respect to ${\bf {\bar x}}/{\bf x}$ in 
${\overline{\cal N}}/{\cal N}$ \\
${\bf a}_t$ or ${a}_{(t)}^{\mu}$ & family of metric 4-accelerations 
(any observer) in $\cal M$ \\
${\bf a}_{\cal F}$/${\bf {\bar a}}_{\cal F}$ or 
$a^{i}_{\cal F}$/${\bar a}^{i}_{\cal F}$ & four-acceleration of the FOs
in $\cal M$/${\overline{\cal M}}$ \\
${\bf {\bar x}}_{\cal F}$ or ${\bar x}^{i}_{\cal F}$ & 3-vector field family 
representing the local ``generalized distance vector to the centre of 
gravity'' \\
${\bf {\bar e}}_{\cal F}$/${\bf {\bar {\omega}}}_{\cal F}$ or 
${\frac{t_0}{t}}{\bar e}^{i}_{\cal F}$/${\frac{t}{t_0}}{\bar {\omega}}_{{\cal F}i}$ 
& unit vector/covector along ${\bf {\bar x}}_{\cal F}$ \\
${\bf w}_t/{\bf {\bar w}}_t$ or $w^i_{(t)}/{\bar w}^i_{(t)}$ & 3-velocity family 
of test particles (or fluid sources) with respect to the local FOs in 
${\cal M}/{\overline {\cal M}}$ \\
${\bf u}_t$ or $u^{\mu}_{(t)}$ & 4-velocity family of test particles in 
$\cal M$ \\
${\bf v}_t$ or $v^{i}_{(t)}$ & 3-vector field family giving the motion of FOs
in $\cal M$ compared to FOs in ${\overline{\cal M}}$ \\
${\stackrel{\star}{\bf a}}$ or ${\topstar{a}}^{\mu}$ & degenerate 
4-acceleration in $\cal N$ (any observer) \\
${\bf {\bar R}}_t$/${\bf {\bar G}}_t$/${\bf {\bar C}}_t$ & Ricci/Einstein/Weyl 
tensor family in ${\overline{\cal M}}$ \\
${\bf {\bar Q}}_t$ & foliation-defined gravitational tensor family 
in ${\overline{\cal M}}$ \\
${\bf {\bar P}}_t$/${\bf {\bar H}}_t$ & Ricci/Einstein tensor family
intrinsic to the FHSs in ${\overline{\cal M}}$ \\
${\bar P}_t$ & Ricci scalar family intrinsic to the FHSs in
${\overline{\cal M}}$ \\
${\bf {\bar K}}_t$ or ${\bar K}_{(t)ij}$ & family of extrinsic curvature 
tensors of the FHSs in ${\overline{\cal M}}$ \\
${\bar H}_t$/${\bar y}_t$& global+local/local measure of the NKE \\
${\bar x}_t$ & measure of the KE \\
$d{\tau}_t$/$\overline{d{\tau}}_t$ & proper time interval (any observer) in 
$\cal N$/${\overline{\cal N}}$ \\
$d{\tau}_{\cal F}$/$\overline{d{\tau}}_{\cal F}$ {\hspace*{1.3cm}} 
& proper time interval for a FO in $\cal N$/${\overline{\cal N}}$ \\
${\bf T}_t$ or $T^{{\mu}{\nu}}_{(t)}$  
& stress-energy tensor family as an active source of gravitation \\
$G^{\rm B}, G^{\rm S}$  
& gravitational ``constants'' (separate couplings to 
electromagnetic and material sources, respectively) \\
${\kappa}^{\rm B}, {\kappa}^{\rm S}$ & defined as ${\frac{8{\pi}G^{\rm B}}{c^4}}$,
${\frac{8{\pi}G^{\rm S}}{c^4}}$, respectively \\
${\bar F}_t{\equiv}{\bar N}_tct{\equiv}{\Psi}_t^{-1}$  
& scale factor family of the FHSs in ${\overline{\cal N}}$ \\
$-U$ {\hspace*{2.3cm}} & Newtonian potential 
\end{tabular}

\newpage

{\vspace*{5cm}}

\begin{center}
{\huge {\bf Chapter Three: \\\ \\The Gravitational Field outside a\\
\vspace{0.5cm}Spherically Symmetric,}}
\end{center}
\vspace{-0.8cm}
\begin{center}
{\huge {\bf Isolated Source in a\\
\vspace{0.5cm}Quasi-Metric Theory of Gravity}}
\end{center}
\newpage
{\hspace*{1cm}}

\newpage

\renewcommand{\theequation}{\arabic{equation}}

\renewcommand{\thesection}{\arabic{section}}

\setcounter{equation}{0}

\setcounter{section}{0}

\begin{center}
{\large {\bf The Gravitational Field outside a Spherically Symmetric, Isolated
Source in a Quasi-Metric Theory of Gravity}}
\end{center}
\begin{center}
by
\end{center}
\begin{center}
Dag {\O}stvang \\
{\em Institutt for Fysikk, Norges teknisk-naturvitenskapelige universitet, 
NTNU \\
N-7491 Trondheim, Norway}
\end{center}
\begin{abstract}
Working within the ``quasi-metric'' framework (QMF) described elsewhere [3],
we find an approximate expression for a spherically symmetric, vacuum 
gravitational field in a ${\bf S}^3{\times}{\bf R}$-background and set up 
equations of motion applying to inertial test particles moving in this field. 
It is found that such a gravitational field is not static with respect to the 
cosmic expansion; i.e., distances between circular orbits increase according 
to the Hubble law. Furthermore, it is found that the dynamically measured mass 
of the source increases with cosmic scale; this is a consequence of the fact 
that within the QMF, the cosmic expansion is not a kinematical phenomenon.
Also it is shown that, if this model of an expanding gravitational field is 
taken to represent the gravitational field of the solar system, this has no 
serious consequences for observational aspects of planetary motion.
\end{abstract}
\section{Introduction}
In metric theory, the nature of the cosmological expansion is in principle
not different from other types of motion; this follows from the most basic 
postulate in metric theory, namely that space-time can be modelled as a
pseudo-Riemannian manifold. Furthermore, it is well-known that metric theory 
predicts that local systems are hardly affected at all by the cosmological 
expansion (its effect is at best totally negligible, see, e.g., [8] and 
references therein). However, when analysing the influence of the 
cosmological expansion on local systems, there should be no reason to expect
that predictions made within the metric framework should continue to hold in 
a theory where the geometrical structure of space-time is non-metric.

Recently, a new type of non-metric space-time framework, the so-called
``quasi-metric'' framework, was presented in [3]. Also presented was an
alternative relativistic theory of gravity formulated within this framework. 
This theory correctly predicts the ''classical'' solar system tests in the case 
where an asymptotically Minkowski background is invoked as an approximation 
and the cosmological expansion is neglected. However, for reasons 
explained in [3], the theory predicts a ${\bf S}^3{\times}{\bf R}$-background 
rather than a Minkowski background for the gravitational field outside a 
spherically symmetric, isolated source. More importantly, the theory also 
suggests that the cosmological expansion should not be irrelevant for local 
systems, so it should not be neglected since it may result in the prediction of
observable phenomena. Therefore, to find how the cosmological expansion affects
local systems according to the quasi-metric theory, one must first calculate 
the spherically symmetric gravitational field with the cosmological expansion 
included. Then one must calculate how test particles move in this field by 
using the quasi-metric equations of motion describing the effects of the 
cosmological expansion on test particle motion.

Of particular interest is if the cosmological expansion has any effects on the
correspondence with Newton's theory of gravity in the non-relativistic limit.
If this is the case, one may hope to explain from first principles the
asymptotically non-Keplerian rotational curves of spiral galaxies as inferred
from observations of Doppler shifted 21-cm spectral lines of neutral hydrogen.
(See reference [1] for detailed information.)

It has been shown that it is impossible to construct a realistic {\em metric} 
theory of gravity (subject to some rather general criteria of structure) such 
that the solar system tests come in and in addition galactic observations can 
be explained without dark matter [2]. But one remaining alternative to the 
introduction of dark matter may be the possibility of constructing a 
{\em quasi-metric} relativistic theory of gravity having the desired 
properties. In this chapter, we explore if the non-metric aspects of our theory 
correctly predict galactic observations from first principles, without 
introducing extra {\em ad hoc} parameters.
\section{General equations of motion}
We start by briefly summarizing the basics of our quasi-metric theory, see 
reference [3] for details.

The geometric foundation of the quasi-metric framework consists of a 
4-dimensional space-time manifold ${\cal N}$ equipped with two one-parameter 
families of Lorentzian 4-metrics ${\bf {\bar g}}_t$ and ${\bf g}_t$
parametrized by the global time function $t$. The global time function slices 
out a family ${\cal S}_t$ of preferred spatial submanifolds; these are called 
the ``fundamental hypersurfaces'' (FHSs). To ensure the uniqueness of $t$, the 
FHSs are required to be compact (without boundaries). Observers always moving 
orthogonally to the FHSs are called ``fundamental observers'' (FOs). Each 
member of the families ${\bf {\bar g}}_t$ and ${\bf g}_t$ applies exactly on 
the fundamental hypersurface $t=$constant only, but they can be extrapolated at 
least to a (infinitesimally) small normal interval around this hypersurface. 
(Other spatial hypersurfaces than the FHSs cannot exactly be associated with 
the domains of applicability of single members of the metric families 
${\bf {\bar g}}_t$ and ${\bf g}_t$.) In component notation, we formally write 
${\bf {\bar g}}_t$ and ${\bf g}_t$ as ${\bar g}_{(t){\mu}{\nu}}$ and 
$g_{(t){\mu}{\nu}}$, respectively, where the dependence on $t$ is put in a 
parenthesis in order not to confuse it with the coordinate indices. The 
components of each single metric are obtained by treating $t$ as a constant. 
Since the theory is concerned primarily with the evolution of the geometry of 
the the FHSs along the world lines of the FOs, we consider only a particular 
set of coordinate systems; namely the {\em global time coordinate systems} 
(GTCSs). By definition, a coordinate system is a GTCS if and only if the 
relation $ct=x^0$ exists in $\cal N$ between the global time function and a 
global (ordinary) time coordinate $x^0$. 

The evolution of the scale factor of the FHSs may be thought of as consisting 
of a combination of two different contributions; the non-kinematical evolution 
(NKE) and the kinematical evolution (KE), respectively. We define these terms 
in what follows.

A peculiarity of the quasi-metric framework is that the ``physical'' metric 
family ${\bf g}_t$ in general does not represent a solution of field equations.
The reason for this, is that the explicit form of ${\bf g}_t$ should include
any {\em implicit} effects coming from the NKE. Since the gravitational field 
equations are expected to include explicit dynamics only, such implicit effects
will not be compensated for when solving the field equations. This is the 
reason why we have introduced a second family of metrics ${\bf {\bar g}}_t$, 
which by definition {\em does} represent a solution of field equations. To 
compensate for said implicit effects, the first family ${\bf g}_t$ may then be 
constructed from the second family ${\bf {\bar g}}_t$ in a way described in 
[3].

In a GTCS, the general form of the family ${\bf {\bar g}}_t$ may be represented
by the family of line elements
\eqa
{\overline{ds}}^2_t={\Big [}{\bar N}_{(t)s}{\bar N}^s_{(t)}
-{\bar N}_t^2{\Big ]}(dx^0)^2+2{\frac{t}{t_0}}{\bar N}_{(t)s}dx^sdx^0+
{\frac{t^2}{t_0^2}}{\bar N}_t^2{\tilde h}_{(t)ks}dx^kdx^s,
\ena
where ${\bar N}_t$ is the lapse function field family of the FOs and
${\frac{t_0}{t}}{\bar N}^s_{(t)}{\frac{\partial}{{\partial}x^s}}$ is the family 
of shift vector fields of the FOs. Moreover, ${\bf {\bar h}}_t$, with 
components ${\bar h}_{(t)ks}{\equiv}{\frac{t^2}{t_0^2}}{\bar N}_t^2
{\tilde h}_{(t)ks}$, is the metric family intrinsic to the FHSs (where $t_0$ is 
some arbitrary reference epoch). Note that ${\bf {\bar g}}_t$ must represent 
a solution of field equations. Their detailed form can be found in [3].

The NKE of the FHSs is defined to take the form of an increase with time of 
the local scale factor ${\bar F}_t{\equiv}{\bar N}_tct$ of the FHSs as 
determined by the FOs using the metric family ${\bf {\bar g}}_t$.
The NKE is represented by the scalar field family ${\bar H}_t$ defined by
\eqa
{\bar H}_t{\equiv}c{\bar F}_t^{-1}{\topstar{\cal L}}_{{\,}{\bf {\bar n}}_t} 
{\bar F}_t-c^{-1}{\bar x}_t
={\frac{1}{{\bar N}_tt}}+{\bar y}_t, \qquad
{\bar y}_t{\equiv}{\frac{c}{{\bar N}_t}}
{\sqrt{{\bar N}_t,_k{\bar N}_t,_s{\bar h}_{(t)}^{ks}}},
\ena
where ${\topstar{\cal L}}_{{\,}{\bf {\bar n}}_t}$ denotes a (projected) Lie 
derivative in the direction normal to the FHSs. Here, the term 
${\frac{1}{{\bar N}_tt}}$ represents the {\em global} NKE due to the explicit 
presence of the scale factor ${\bar F}_t$ in equation (1), whereas the term 
${\bar y}_t$ is coming from the local gravitational field and represents the 
{\em local} NKE; its existence is the reason why we must construct the metric 
family ${\bf g}_t$ to find the proper equations of motion. Note that the term
${\bar y}_t$ has no explicit relationship to equation (1). There is also a 
kinematical evolution (KE) of the ${\bar F}_t$; this is described by the 
quantity $c^{-1}{\bar x}_t$. The overall evolution of ${\bar F}_t$ is given by 
(note that a global term ${\frac{1}{{\bar N}_tct}}$ still occurs explicitly 
here but that there is no term $c^{-1}{\bar y}_t$)
\eqa
{\bar F}_t^{-1}{\topstar{\cal L}}_{{\,}{\bf {\bar n}}_t} 
{\bar F}_t=-{\frac{{\bar N}_t,_{\bar {\perp}}}{{\bar N}_t}}+
(c{\bar N}_t)^{-1}{\Big (}{\frac{1}{t}}+
{\frac{{\bar N}_t,_t}{{\bar N}_t}}{\Big )},
\ena 
where the projection symbol ${\bar {\perp}}$ denotes a scalar product with
$-{\bf {\bar n}}_t$, which is a family of (negative) unit normal vector fields 
to the FHSs. 

Note the characteristic property of our theory that there are systematic 
scale changes between gravitational and atomic systems; this causes 
gravitational quantities to exhibit an extra formal variation when measured in 
atomic units (and {\em vice versa}). This may be thought of formally as if 
fixed operationally defined atomic units vary in space-time. Since $c$ and 
Planck's constant ${\hbar}$ by definition are not formally variable, the formal 
variation of time units is equal to that of length units and inverse to that 
of mass units. Also, since the fine structure constant is dimensionless, it 
cannot be formally variable, so the elementary charge $e$ cannot be formally 
variable. That is, charge units are not formally variable. We represent the 
formal variation of atomic time units by the scalar field 
${\Psi}_t{\equiv}{\bar F}_t^{-1}$. This means that, measured in atomic units, 
the formal variability of gravitational quantities with the dimension of time 
or length goes as ${\Psi}_t^{-1}$. In particular, the ``bare'' gravitational 
coupling parameter $G^{\rm B}_t$ has an effective dimension of length squared. 
Since $G^{\rm B}_t$ couples to charge squared, or more generally, to 
electromagnetic mass-energy, the formal variation of $G^{\rm B}_t$ goes as 
${\Psi}_t^{-2}$. On the other hand, the ``screened'' gravitational coupling 
parameter $G^{\rm S}_t$ couples to material mass-energy (which formal 
variability is not locally measurable). That is, $G^{\rm S}_t$ couples to 
material mass-energy and gets a formal variation like ${\Psi}_t^{-1}$ [3]. By 
definition, the corresponding gravitational constants $G^{\rm B}$ and $G^{\rm S}$ 
do not vary in space-time, but we then have to differentiate between 
{\em active mass} $m_t$ as an active source of gravitation and 
{\em passive mass} $m$ (passive gravitational mass or inertial mass). Active 
electromagnetic mass-energy varies formally as ${\Psi}_t^{-2}$, whereas active 
material mass-energy varies formally as ${\Psi}_t^{-1}$, measured in atomic 
units. See [3] for more details. The formal evolution of ${\Psi}_t$ in the 
normal direction to the FHSs follows from equation (3), i.e.,
\eqa
{\topstar{\cal L}}_{{\,}{\bf {\bar n}}_t}{\Psi}_t=
-{\bar F}_t^{-2}{\topstar{\cal L}}_{{\,}{\bf {\bar n}}_t}{\bar F}_t=
{\Big (}{\frac{{\bar N}_t,_{\bar {\perp}}}{{\bar N}_t}}-
{\frac{1}{c{\bar N}_t}}[{\frac{1}{t}}+{\frac{{\bar N}_t,_t}{{\bar N}_t}}]
{\Big )}{\Psi}_t.
\ena
Next, on $({\cal N},{\bf g}_t)$ there applies a linear, symmetric ``degenerate" 
connection ${\,}{\,}{\topstar{\nabla}}{\,}{\,}$. This connection is called 
degenerate due to the fact that the family ${\bf g}_t$ may be perceived as one
single degenerate 5-dimensional metric on a product manifold 
${\cal M}{\times}{\bf R}$ whereof $({\cal N},{\bf g}_t)$ is a 4-dimensional 
submanifold. (Here, $\cal M$ is a Lorentzian space-time manifold and 
$\bf R$ is the real line.) We may then introduce a torsion-free, 
metric-compatible 5-dimensional connection 
${\,}{\,}{\topstar{\nabla}}{\,}{\,}$ with the property that
\eqa
{\topstar{\nabla}}_{\frac{\partial}{{\partial}t}}{\bf g}_t=0, \qquad
{\topstar{\nabla}}_{\frac{\partial}{{\partial}t}}{\bf n}_t=0, \qquad
{\topstar{\nabla}}_{\frac{\partial}{{\partial}t}}{\bf h}_t=0,
\ena
on ${\cal M}{\times}{\bf R}$ and consider the restriction of 
${\,}{\,}{\topstar{\nabla}}{\,}{\,}$ to $\cal N$. (Here, the quantities
${\bf h}_t$ and ${\bf n}_t$ are defined similarly to their respective 
counterparts ${\bf {\bar h}}_t$ and ${\bf {\bar n}}_t$ in 
$({\cal N},{\bf {\bar g}}_t)$.) It can be shown [3] that in a GTCS, the 
components which do not vanish identically of the degenerate connection field 
are given by (a comma denotes a partial derivative, and we use Einstein's 
summation convention throughout)
\eqa
{\topstar{\Gamma}}^{{\,}{\alpha}}_{t{\mu}}{\equiv}
{\frac{1}{2}}{\delta}^{\alpha}_i{\delta}^j_{\mu}h_{(t)}^{is}h_{(t)sj,t}
{\equiv}{\topstar{\Gamma}}^{{\,}{\alpha}}_{{\mu}t}, {\qquad}
{\topstar{\Gamma}}^{{\,}{\alpha}}_{{\nu}{\mu}}{\equiv}{\frac{1}{2}}
g_{(t)}^{{\alpha}{\sigma}}{\Big(}g_{(t){\sigma}{\mu}},_{\nu}
+g_{(t){\nu}{\sigma}},_{\mu}-g_{(t){\nu}{\mu}},_{\sigma}{\Big)}{\equiv}
{\Gamma}^{\alpha}_{(t){\nu}{\mu}}.
\ena
The general equations of motion for test particles are identical to the
geodesic equation obtained from ${\,}{\,}{\topstar{\nabla}}{\,}{\,}$. In a 
GTCS, they take the form (see [3] for a derivation)
\eqa
{\frac{d^2x^{\mu}}{d{\lambda}^2}}+{\Big(}
{\topstar{\Gamma}}^{{\,}{\mu}}_{t{\nu}}{\frac{dt}{d{\lambda}}}+
{\Gamma}^{\mu}_{(t){\beta}{\nu}}{\frac{dx^{\beta}}{d{\lambda}}}{\Big)}
{\frac{dx^{\nu}}{d{\lambda}}} 
={\Big(}{\frac{cd{\tau}_t}{d{\lambda}}}{\Big)}^2c^{-2}a_{(t)}^{\mu},
\ena
where $d{\tau}_t$ is the proper time as measured along the curve, ${\lambda}$ 
is some general affine parameter and ${\bf a}_t$ is the 4-acceleration as 
measured along the curve.
\subsection{Special equations of motion}
In this chapter, we analyse the equations of motion (7) in the case of a 
uniformly expanding gravitational field in vacuum exterior to an isolated, 
spherically symmetric source in an isotropic, compact spatial background. 
Moreover, we assume that the source and the FOs are at rest with respect to 
some GTCS, and consequently that the shift vector field vanishes (this 
assumption excludes any potential ``preferred frame''-effects, see [3]). 
Furthermore, we require that ${\bar N}_t$ is independent of $x^0$ and $t$; 
i.e., that the only explicit time dependence is via the explicit dependence on 
$t$ of ${\bar F}_t$ (using the chosen GTCS). We call this a 
``metrically static'' case. For metrically static cases, the field equations 
yield that ${\bf {\tilde h}}_t$ must be equal to the metric of the 3-sphere 
${\bf S}^3$ (with radius equal to $ct_0$) [3]. This scenario may be taken as a 
generalization of the analogous case with a Minkowski background (that case 
was analysed in [3]) since the Minkowski background is not a part of our 
theory but rather invoked as an approximation being useful in particular cases.

We start by making a specific {\em ansatz} for the form of ${\bf {\bar g}}_t$.
Introducing a spherically symmetric GTCS ${\{}x^0,r,{\theta},{\phi}{\}}$, 
where $r$ is a Schwarzschild radial coordinate, we assume that the 
metric families ${\bf {\bar g}}_t$ and ${\bf g}_t$ can be represented by
line element families in the form (compatible with equation (1))
\eqa
c^2{\overline{d{\tau}}}^2_t={\bar{B}}(r)(dx^0)^2-({\frac{t}{t_0}})^2{\Big(}
{\bar{A}}(r)dr^2+r^2d{\Omega}^2{\Big)}, {\nonumber} \\
c^2d{\tau}^2_t=B(r)(dx^0)^2-({\frac{t}{t_0}})^2{\Big(}
A(r)dr^2+r^2d{\Omega}^2{\Big)},
\ena
where $d{\Omega}^2{\equiv}d{\theta}^2+{\sin}^2{\theta}d{\phi}^2$ and $t_0$ is
some arbitrary reference epoch. Note that the spatial coordinate system 
covers only half of ${\bf S}^3$, thus the range of the radial coordinate is 
$r<ct_0$ only. The functions ${\bar{A}}(r)$ and ${\bar{B}}(r)$ may be 
calculated from the field equations; we tackle this problem in the next 
section. The functions $A(r)$ and $B(r)$ may be found from ${\bf {\bar g}}_t$ 
and ${\bar y}_t$ as shown in [3].

We now calculate the metric and the degenerate connection coefficients from
the metric family ${\bf g}_t$ given in equation (8). A straightforward 
calculation yields (using the notation $'{\hspace*{2mm}}{\equiv}
{\frac{\partial}{{\partial}r}}$) 
\eqa
{\Gamma}^r_{(t)rr}&=&{\frac{A'(r)}{2A(r)}}, {\qquad}
{\Gamma}^r_{(t){\theta}{\theta}}=-{\frac{r}{A(r)}}, {\qquad}
{\Gamma}^r_{(t){\phi}{\phi}}={\Gamma}^r_{(t){\theta}{\theta}}{\sin}^2{\theta},
{\nonumber} \\
{\Gamma}^r_{(t)00}&=&({\frac{t_0}{t}})^2{\frac{B'(r)}{2A(r)}}, 
{\qquad}{\Gamma}^{\theta}_{(t)r{\theta}}={\Gamma}^{\theta}_{(t){\theta}r}=
{\frac{1}{r}}, {\qquad} {\Gamma}^{\theta}_{(t){\phi}{\phi}}=
-{\sin}{\theta}{\cos}{\theta}, {\nonumber} \\
{\Gamma}^{\phi}_{(t)r{\phi}}&=&{\Gamma}^{\phi}_{(t){\phi}r}={\frac{1}{r}},
{\qquad}{\Gamma}^{\phi}_{(t){\phi}{\theta}}=
{\Gamma}^{\phi}_{(t){\theta}{\phi}}={\cot}{\theta}, {\qquad}
{\Gamma}^0_{(t)0r}={\Gamma}^0_{(t)r0}={\frac{B'(r)}{2B(r)}},
\ena
\eqa
{\topstar{\Gamma}}^{{\,}r}_{tr}=
{\topstar{\Gamma}}^{{\,}{\theta}}_{t{\theta}}=
{\topstar{\Gamma}}^{{\,}{\phi}}_{t{\phi}}={\frac{1}{t}}.
\ena
In what follows, we use the equations of motion (7) to find the paths of 
inertial test particles moving in $({\cal N},{\bf g}_t)$. Since ${\bf a}_t$ 
vanishes for inertial test particles, we get the relevant equations 
by inserting the expressions (9) and (10) into equation (7). This yields
(making explicit use of the fact that $cdt=dx^0$ in a GTCS)
\eqa
{\frac{d^{2}r}{d{\lambda}^{2}}} + 
{\frac{A'(r)}{2A(r)}}{\Big(}{\frac{dr}{d{\lambda}}}{\Big)}^2 -
{\frac{r}{A(r)}}{\bigg[}{\Big(}{\frac{d{\theta}}{d{\lambda}}}{\Big)}^{2} + 
{\sin}^{2}{\theta}{\Big(}{\frac{d{\phi}}{d{\lambda}}}{\Big)}^{2}{\bigg]}
{\nonumber} \\
+({\frac{t_0}{t}})^2
{\frac{B'(r)}{2A(r)}}{\Big(}{\frac{dx^0}{d{\lambda}}}{\Big)}^{2} + 
{\frac{1}{ct}}{\frac{dr}{d{\lambda}}}{\frac{dx^0}{d{\lambda}}}=0,
\ena
\eqa
{\frac{d^{2}{\theta}}{d{\lambda}^{2}}} +
{\frac{2}{r}}{\frac{d{\theta}}{d{\lambda}}}{\frac{dr}{d{\lambda}}} -
{\sin}{\theta}{\cos}{\theta}{\Big(}{\frac{d{\phi}}{d{\lambda}}}{\Big)}^{2} + 
{\frac{1}{ct}}{\frac{d{\theta}}{d{\lambda}}}{\frac{dx^0}{d{\lambda}}} = 0,
\ena
\eqa
{\frac{d^{2}{\phi}}{d{\lambda}^{2}}} +
{\frac{2}{r}}{\frac{d{\phi}}{d{\lambda}}}{\frac{dr}{d{\lambda}}} +
2{\cot}{\theta}{\frac{d{\phi}}{d{\lambda}}}{\frac{d{\theta}}{d{\lambda}}} + 
{\frac{1}{ct}}{\frac{d{\phi}}{d{\lambda}}}{\frac{dx^0}{d{\lambda}}} = 0,
\ena
\eqa
{\frac{d^{2}x^0}{d{\lambda}^{2}}} +
{\frac{B'(r)}{B(r)}}{\frac{dx^0}{d{\lambda}}}{\frac{dr}{d{\lambda}}} =0.
\ena
If we restrict the motion to the equatorial plane, equation (12) becomes 
vacuous, and equation (13) reduces to
\eqa
{\frac{d^{2}{\phi}}{d{\lambda}^{2}}} +
{\frac{2}{r}}{\frac{d{\phi}}{d{\lambda}}}{\frac{dr}{d{\lambda}}} +
{\frac{1}{ct}}{\frac{d{\phi}}{d{\lambda}}}{\frac{dx^0}{d{\lambda}}} = 0.
\ena
Dividing equation (15) by ${\frac{d{\phi}}{d{\lambda}}}$, we find (assuming
that ${\frac{d{\phi}}{d{\lambda}}}{\neq}0$)
\eqa
{\frac{d}{d{\lambda}}}{\bigg[}{\ln}{\Big(}{\frac{d{\phi}}{d{\lambda}}}{\Big)} 
+ {\ln}{\Big(}r^{2}{\frac{t}{t_0}}{\Big)}{\bigg]} = 0.
\ena
We thus have a constant of the motion, namely
\eqa
J{\equiv}{\frac{t}{t_0}}r^2{\frac{d{\phi}}{d{\lambda}}}.
\ena
Dividing equation (14) by ${\frac{dx^0}{d{\lambda}}}$, yields
\eqa
{\frac{d}{d{\lambda}}}{\bigg[}{\ln}{\Big(}{\frac{dx^0}{d{\lambda}}}{\Big)} + 
{\ln}B(r){\bigg]}=0.
\ena
Equation (18) yields another constant of the motion, which we can absorb into 
the definition of ${\lambda}$ such that a solution of equation (18) is [4]
\eqa
{\frac{dx^0}{d{\lambda}}}={\frac{1}{B(r)}}.
\ena
Multiplying equation (11) by 
${\frac{2t^2A(r)}{t_0^2}}{\frac{dr}{d{\lambda}}}$ and 
using the expressions (17) and (19), we find that
\eqa
{\frac{d}{d{\lambda}}}{\bigg[}
{\frac{t^2A(r)}{t_0^2}}{\Big(}{\frac{dr}{d{\lambda}}}{\Big)}^2-
{\frac{1}{B(r)}}+{\frac{J^2}{r^2}}{\bigg]}=0,
\ena
thus a constant $E$ of the motion is defined by
\eqa
{\frac{t^2A(r)}{t_0^2}}{\Big(}{\frac{dr}{d{\lambda}}}{\Big)}^2-
{\frac{1}{B(r)}}+{\frac{J^2}{r^2}}{\equiv}-E.
\ena
Equation (21) may be compared to an analogous expression obtained for the 
static isotropic gravitational field in the metric framework [4]. Inserting 
the formulae (17), (19) and (21) into equation (8) and using the fact that we
can formally write $dx^0=cdt$ in a GTCS when traversing the family of metrics, 
we find that
\eqa
c^2d{\tau}_t^2=Ed{\lambda}^2.
\ena
Thus our equations of motion (7) force $d{\tau}_t/d{\lambda}$ to be
constant, quite similarly to the case when the total connection is metric, as 
in the metric framework. From equation (22), we see that we must have
$E=0$ for photons and $E>0$ for material particles.

We may eliminate the parameter ${\lambda}$ from equations (17), (19), (21) and 
(22) and alternatively use $t$ as a time parameter. This yields
\eqa
{\frac{t}{t_0}}r^2{\frac{d{\phi}}{cdt}}=B(r)J,
\ena
\eqa
({\frac{t}{t_0}})^2A(r)B^{-2}(r){\Big(}{\frac{dr}{cdt}}{\Big)}^2-
{\frac{1}{B(r)}}+{\frac{J^2}{r^2}}{\equiv}-E,
\ena
\eqa
d{\tau}_t^2=EB^2(r)dt^2.
\ena
We may integrate equations (23) and (24) to find the time history 
$(r(t),{\phi}(t))$ along the curve if the functions $A(r)$ and $B(r)$ 
are known.

For the static vacuum metric case with no NKE, one can solve the geodesic
equation for particles orbiting in circles with different radii, and from
this find the asymptotically Keplerian nature of the corresponding rotational
curve [4]. In our case, we see from equations (23) and (24) that we can find 
circle orbits as solutions; such orbits have the property that the orbital 
speed $B^{-1/2}(r){\frac{t}{t_0}}r{\frac{d{\phi}}{dt}}$ is independent of $t$.
\section{The dynamical problem}
In this section, we write down equations which must be solved to find the 
unknown functions ${\bar A}(r)$ and ${\bar B}(r)$. (These equations follow
directly from the field equations.) We then solve these equations 
approximately and use this to find approximations to the functions $A(r)$ and 
$B(r)$ as well.

The field equations determining the metric family ${\bf {\bar g}}_t$ of the
type shown in equation (8) may be set up exactly as for the corresponding 
metric approximation case with a Minkowski background treated in [3]. The main 
difference is the explicit time dependence present in equation (8). However, 
this time dependence is only via the global scale factor. From equations (1), 
(2) and (8) we get (since ${\bar N}_t={\sqrt{{\bar B}(r)}}$ is independent of 
$t$)
\eqa
{\frac{{\partial}{\bar h}_{(t)ij}}{{\partial}t}}=
{\frac{2}{t}}{\bar h}_{(t)ij}, {\qquad} {\bar{y}}_t=
{\frac{t_0}{t}}{\frac{{\bar B}'(r)c}{2{\bar B}(r){\sqrt{{\bar A}(r)}}}}.
\ena
Besides, after some tedious but straightforward calculations, the field 
equations yield the ordinary differential equation
\eqa
{\Big(}1-{\frac{r^2}{{\bar B}(r)c^2t_0^2}}{\Big)}
{\frac{{\bar B}''(r)}{{\bar B}(r)}}&-&
{\Big(}1-{\frac{2r^2}{{\bar B}(r)c^2t_0^2}}{\Big)}
{\frac{{\bar B}'^2(r)}{{\bar B}^2(r)}} {\nonumber} \\ 
&+&{\frac{2}{r}}{\Big(}1-{\frac{3r^2}{2{\bar B}(r)c^2t_0^2}}{\Big)}
{\frac{{\bar B}'(r)}{{\bar B}(r)}}-{\frac{r}{4}}
{\frac{{\bar B}'^3(r)}{{\bar B}^3(r)}}=0,
\ena
\eqa
{\bar A}(r)={\Big(}1-{\frac{r^2}{{\bar B}(r)c^2t_0^2}}{\Big)}^{-1}
{\Big[}1-r{\frac{{\bar B}'(r)}{{\bar B}(r)}}+{\frac{r^2}{4}}
{\frac{{\bar B}'^2(r)}{{\bar B}^2(r)}}{\Big]}.
\ena
Before we try to solve equation (27), it is important to notice that no
solution of it can exist on a whole FHS (except the trivial solution 
${\bar B}=$constant), according to the maximum principle applied to a closed 
Riemannian 3-manifold. The reason for this is the particular form of the field
equations, see reference [3] and references therein for justification. This 
means that {\em in quasi-metric theory, isolated systems cannot exist except 
as an approximation.} 

Even if a non-trivial solution of equation (27) does not exist on a whole FHS, 
we may try to find a solution valid in some finite region of a FHS. That is, 
we want to find a solution in the region $r_{\rm sf}<r{\ll}ct_0$, where 
$r_{\rm sf}$ is the coordinate radius of the source. We do not specify any 
particular boundary conditions at the boundary ${\Xi}_0{\equiv}ct_0$, since the
approximation made by assuming an isolated system is physically reliable only 
if ${\frac{r}{{\Xi}_0}}{\ll}1$. Furthermore, it may turn out to be difficult 
to find a useful exact solution of equation (27). Thus we rather try to 
find an approximate solution in the form of a series expansion, i.e., a 
perturbation around the analogous problem in a Minkowski background. That is, 
in contrast to the analogous case with a Minkowski background, there exists the
extra scale ${\Xi}_0$ in addition to the generalized Schwarzschild radius 
$r_{\rm s0}{\equiv}{\frac{2(M_{t_0}^{\rm (EM)}G^{\rm B}+M_{t_0}^{\rm (MA)}G^{\rm S})}
{c^2}}$ determined by the active electromagnetic mass $M_{t_0}^{\rm (EM)}$ and the
active material mass $M_{t_0}^{\rm (MA)}$ of the central object. (The extra index 
used in $r_{\rm s0}$ refers to the time $t_0$ and is necessary since 
gravitational quantities get an extra formal variability measured in atomic 
units.) Since we try to model the gravitational field exterior to 
galactic-sized objects, we may assume that the typical scales involved are 
determined by ${\frac{r}{{\Xi}_0}}\raisebox{-2pt}{\,$\stackrel{>}{\sim}$\,}
{\frac{r_{\rm s0}}{r}}$; this criterion tells how to compare the importance of 
the different terms of the series expansion.

One may straightforwardly show that an approximate solution of equation (27) 
is given by
\eqa
{\bar B}(r)&=&1-{\frac{r_{\rm s0}}{r}}+{\frac{r_{\rm s0}^2}{2r^2}}+
{\frac{r_{\rm s0}r}{2{\Xi}_0^2}}-{\frac{r_{\rm s0}^3}{8r^3}}+{\cdots} 
{\qquad} {\Rightarrow} {\nonumber} \\
{\bar A}(r)&=&1-{\frac{r_{\rm s0}}{r}}+
{\frac{r_{\rm s0}^2}{4r^2}}+{\frac{r^2}{{\Xi}_0^2}}+{\cdots},
\ena
and furthermore this yields, using equations (2) and (26), that
\eqa
{\bar y}_t={\frac{t_0}{t}}{\frac{r_{\rm s0}c}{2r^2}}
[1+{\frac{r_{\rm s0}}{2r}}+O({\frac{r^2_{\rm s0}}{r^2}})],
\qquad {\bar H}_t={\bar y}_t+{\frac{1}{{\sqrt{{\bar B}(r)}}t}}.
\ena
To construct the family ${\bf g}_t$ as described in [3], we need the quantity
(${\bf {\bar x}}_{\cal F}=r{\frac{\partial}{{\partial}r}}$)
\eqa
v(r){\equiv}{\bar y}_t{\sqrt{{\bar h}_{(t)ik}{\bar x}_{\cal F}^i
{\bar x}_{\cal F}^k}}={\bar y}_tr{\sqrt{{\bar h}_{(t)rr}}}=
{\frac{r_{\rm s0}c}{2r}}[1+O({\frac{r_{\rm s0}^2}{r^2}})].
\ena
We notice that $v(r)$ does not depend on $t$. The functions $A(r)$ and $B(r)$ 
are found from the following relations, valid for the spherically symmetric 
case [3] 
\eqa
A(r)={\Big(}{\frac{1+{\frac{v(r)}{c}}}{1-{\frac{v(r)}{c}}}}{\Big)}^2
{\bar A}(r), {\qquad}
B(r)={\Big(}1-{\frac{v^2(r)}{c^2}}{\Big)}^2{\bar B}(r).
\ena
From equations (29) and (32), we may then write
\eqa
ds^2_t&=&-{\bigg(}1-{\frac{r_{\rm s0}}{r}}
+{\frac{r_{\rm s0}r}{2{\Xi}_0^2}}+{\frac{3r_{\rm s0}^3}{8r^3}}+{\cdots}
{\bigg)}(dx^0)^2 \nonumber \\
&&+({\frac{t}{t_0}})^2{\bigg(}{\Big\{}1+{\frac{r_{\rm s0}}{r}}+
{\frac{r^2}{{\Xi}_0^2}} + {\frac{r^2_{\rm s0}}{4r^2}}+{\cdots}
{\Big\}}dr^2+r^2d{\Omega}^2{\bigg)}.
\ena
This expression represents the wanted metric family as a series expansion. 
Note in particular the fact that all spatial dimensions expand, whereas the 
corresponding Newtonian potential $-U=-{\frac{c^2r_{\rm s0}}{2r}}$ (to Newtonian
order) remains constant for a fixed FO. This means that the proper radius of 
any circle orbit (i.e., with $r$ constant) increases but such that the orbit 
speed remains constant. That is, {\em the mass of the central object as 
measured by distant orbiters increases to exactly balance the effect on circle
orbit speeds of expanding circle radii.} This is not as outrageous as it 
may seem due to the extra formal variation of atomic units built into our 
theory, rather it is a consequence of the fact that the coupling between 
matter and geometry depends directly on this formal variation [3]. 

What is measured by means of distant orbiters is not the ``bare'' mass 
$M^{\rm (EM)}_t+M^{\rm (MA)}_t$ itself, but rather the combination 
$M^{\rm (EM)}_tG^{\rm B}+M^{\rm (MA)}_tG^{\rm S}$. We have, however, {\em defined} 
$G^{\rm B}$ and $G^{\rm S}$ to be constants. It turns out that the variation 
of $M^{\rm (EM)}_tG^{\rm B}+M^{\rm (MA)}_tG^{\rm S}$ with time as inferred from 
equation (33) is a linear increase with $t$, i.e., identical to the formal 
variation of the active material mass $M^{\rm (MA)}_t$ with $t$. But since the 
formal variation of the active electromagnetic mass $M^{\rm (EM)}_t$ goes as 
$t^2$, to be consistent with the metrically static condition, there must be a 
net loss proportional to $t^{-1}$ of photon energy. Such a loss is an effect of
the cosmic expansion on the electromagnetic field (i.e., a ``cosmic 
redshift''), provided that the size of the gravitational source expands 
according to the Hubble law. All this means that the dynamically measured mass 
increase should not be taken as an indication of actual particle creation, but 
that the general dynamically measured mass scale for material sources should 
change via a linear increase of $M^{\rm (MA)}_t$ with $t$. Also, the combination 
of cosmic redshift and formal increase of photon energy yields a linear 
increase of $M^{\rm (EM)}_t$ with $t$, although the source experiences a net loss
of photon energy (net energy transfer between material particles and photons is 
assumed to be negligible).

The dynamical measurement of the mass of the central object, by means of 
distant orbiters, does not represent a local test experiment. Nevertheless, the 
dynamically measured mass increase thus found, is just as ``real" as the 
expansion in the sense that neither should be neglected for extended scales.
This must be so, since in quasi-metric relativity, the global scale increase 
and the dynamically measured mass increase are two different aspects of the 
same basic phenomenon.

We finish this chapter by exploring which kinds of free-fall orbits we get 
from equation (33) and the equations of motion. We also discuss observable
consequences thereof.
\section{Discussion}
To begin with, we find the shape of the rotational curve as defined by the
orbital speeds $w(r)$ of the circle orbits. Since equation (24) has no time 
dependence for such orbits, we can do a standard calculation [4] and the 
result is 
\eqa
w(r){\equiv}B^{-1/2}(r)
{\frac{t}{t_0}}r{\frac{d{\phi}}{dt}}={\sqrt{\frac{B'(r)r}{2B(r)}}}c.
\ena
However, when we apply equation (34) to the metric family (33), we get a result
essentially identical to the standard Keplerian rotational curve; the only 
effect of the dynamically measured mass increase and the non-kinematical 
expansion is to increase the scale but such that the shape of the rotational 
curve is unaffected. It is true that $B(r)$ as found from equation (33) 
contains a term linear in $r$ in addition to terms falling off with increasing 
$r$; in reference [5] it is shown that a suitable linear term may be 
successfully used to model the asymptotically non-Keplerian rotational curves 
of spiral galaxies. Unfortunately, the numerical value of the linear term found
from equation (33) is too small by a factor of order $10^{-10}$ to be able to 
match the data. So at least the simple model considered in this chapter, is 
unable to obtain the asymptotically non-Keplerian rotational curves of spiral 
galaxies from first principles. 

Another matter is how the time dependence of the equations of motion will
affect the time histories and shapes of more general orbits than the circle 
orbits. Clearly time histories will be affected, as can be seen directly from
equation (24). However, to see if this is valid for shapes as well, we may 
insert equation (23) into equation (24) to obtain $r$ as a function of 
${\phi}$. This yields
\eqa
{\frac{A(r)}{r^4}}{\Big(}{\frac{dr}{d{\phi}}}{\Big)}^2+{\frac{1}{r^2}}-
{\frac{1}{J^2B(r)}}=-{\frac{E}{J^2}},
\ena
and this equation is identical to that valid for the metric case [4]. Thus the 
{\em shapes} of free-fall orbits are unaffected by the non-metric aspects of 
equation (33). 
\subsection{Expanding space and the solar system}
If one neglects the gravitational effects of the galaxy, one may try to apply 
the metric family (33) to the solar system (by using it to describe the 
gravitational field of the Sun). That is, if we treat the solar system as 
approximately isolated, we can use the metric family (33) to describe the 
gravitational field of the Sun. Moreover, the solar system is so small that we 
can neglect any dependence on ${\Xi}_0$. The errors made by neglecting terms 
depending on ${\Xi}_0$ in equation (33) are insignificant, since the typical 
scales involved for the solar system are determined by 
${\frac{r}{{\Xi}_0}}\raisebox{-2pt}{\,$\stackrel{<}{\sim}$\,}
{\frac{r^3_{\rm s0}}{r^3}}$. Equation (33) then takes the form
\eqa
ds^2_t=-{\Big(}1-{\frac{r_{\rm s0}}{r}}+O({\frac{r^3_{\rm s0}}{r^3}}){\Big)}
(dx^0)^2+({\frac{t}{t_0}})^2{\Big(}{\{}1+{\frac{r_{\rm s0}}{r}}+
O({\frac{r^2_{\rm s0}}{r^2}}){\}}dr^2+r^2d{\Omega}^2{\Big)}.
\ena
Equation (35) shows that the shapes of orbits are unaffected by the cosmic
expansion; this means that all the classic solar system tests come in just 
as for the analogous metric case [3]. However, we get at least one 
extra prediction (irrespective of whether the galactic gravitational field 
can be neglected or not); from equation (36) we see that the effective 
distance between the Sun and any planet has been smaller in the past. That is,
the spatial coordinates are comoving rather than static, thus the gravitational
field of the Sun is not static (measured in atomic units). For example, the 
distance between the Sun and the Earth at the time of its formation may have 
been almost 50{\%} smaller than today. But since gravitationally bound bodies 
made of ideal gas are predicted to expand according to quasi-metric theory 
[10], a small Earth-Sun distance should not be incompatible with 
palaeo-climatic data, since the Sun is expected to have been smaller and thus 
dimmer in the past. Actually, since neither the temperature at the centre of 
the Sun (as estimated from the virial theorem) nor the radiation energy 
gradient times the mean free path of a photon depend on $t$, the cosmic 
luminosity evolution of the Sun should be determined from the cosmic expansion 
of its surface area as long as the ideal gas approximation is sufficient. And 
this luminosity evolution exactly balances the effects of an increasing 
Earth-Sun distance on the effective solar radiation received at the Earth.

However, an obvious question is if the predicted effect of the expansion on 
the time histories of non-relativistic orbits is compatible with the observed 
motions of the planets. In order to try to answer this question, it is 
illustrating to calculate how the orbit period of any planet depends on $t$. 
For simplicity, consider a circular orbit $r=R=$constant. Equation (23) then 
yields
\eqa
{\frac{d{\phi}}{dt}}={\frac{t_0}{t}}B(R)R^{-2}Jc.
\ena
Now integrate equation (37) one orbit period $T<<t$ (i.e., from $t$ to $t+T$). 
The result is
\eqa
T(t)=t({\exp}{\Big [}{\frac{T_{\rm GR}}{t_0}}{\Big ]}-1)={\frac{t}{t_0}}
T_{\rm GR}(1+{\frac{T_{\rm GR}}{2t_0}}+{\cdots}), {\qquad} 
T_{\rm GR}{\equiv}{\frac{2{\pi}R^2}{cJB(R)}},
\ena
where $T_{\rm GR}$ is the orbit period as predicted from General Relativity. 
From equation (38) we see that (sidereal) orbit periods are predicted to 
increase linearly with cosmic scale, i.e., 
\eqa
T(t)={\frac{t}{t_0}}T(t_0), {\qquad} {\frac{dT}{dt}}={\frac{T(t_0)}{t_0}},
\ena
and such that any ratio between periods of different orbits remains constant.
In particular, equation (39) predicts that the (sidereal) year $T_{\rm E}$ 
should be increasing with about 2.5 ms per year and the martian (sidereal) 
year $T_{\rm M}$ should be increasing with about 4.7 ms per martian year at the 
present epoch. This should be consistent with observations, since the observed 
difference in the synodical periods of Mars and the Earth is accurate to about 
5 ms.

To compare predictions coming from equation (36) against timekeeping data, one 
must also take into account the cosmological contribution to the spin-down of 
the Earth. If one assumes that the gravitational source of the exterior field 
(36) is stable with respect to internal collapse (as for a source made of
ideal gas), i.e., that possible instabilities generated by the expansion can 
be neglected, one may model this source as a uniformly expanding sphere [10].
Due to the cosmic expansion of gravitational fields, the angular momenta of 
test particles moving in the exterior field (36) increase linearly with cosmic 
scale, and this should also apply to the (passive) angular momentum $L_{\rm sb}$ 
of a spherically symmetric, metrically static source body with (coordinate) 
radius $R_{\rm sb}$, that is
\eqa
L_{\rm sb}(t)={\frac{t}{t_0}}L_{\rm sb}(t_0), {\qquad} 
{\frac{dL_{\rm sb}}{dt}}={\frac{1}{t}}L_{\rm sb}=(1+O(2))HL_{\rm sb},
\ena
where the term $O(2)$ is of post-Newtonian order and where the locally 
measured Hubble parameter $H$ is defined by $H{\equiv}{\frac{1}{Nt}}$, or
equivalently (${\tau}_{\cal F}$ is the proper time of the local FO),
\eqa
H{\equiv}{\frac{t_0}{t}}{\frac{d}{d{\tau}_{\cal F}}}({\frac{t}{t_0}})=
{\frac{ct_0}{t}}{\Big (}{\sqrt{B(r)}}{\Big )}^{-1}{\frac{d}{dx^0}}
({\frac{t}{t_0}})={\Big (}{\sqrt{B(r)}}t{\Big )}^{-1}.
\ena
Since the moment of inertia $I{\propto}MR_{\rm sb}^2$, where $M$ is the passive 
mass, we must have that (neglecting terms of post-Newtonian order)
\eqa
{\frac{dR_{\rm sb}}{dt}}=HR_{\rm sb}, {\qquad}
{\frac{d{\omega}_{\rm sb}}{dt}}=-H{\omega}_{\rm sb}, {\qquad} 
{\frac{dT_{\rm sb}}{dt}}=HT_{\rm sb},
\ena
where ${\omega}_{\rm sb}$ is the spin circle frequency and $T_{\rm sb}$ is the 
spin period of the source body. (To show equation (42), use the definition 
$L_{\rm sb}{\equiv}I{\omega}_{\rm sb}$.) The spin period of a spherically 
symmetric body made of ideal gas should thus increase linearly with $t$ due to 
the cosmic expansion. Does this apply to the Earth as well? The Earth is 
not made of ideal gas, so the cosmic expansion may induce instabilities, 
affecting its spin period $T_{\rm sE}$. However, here we assume that the Earth's
mantle is made of a material which may be approximately modelled as a perfect 
fluid obeying an equation of state close to linear. Then, if this assumption 
holds, the effects of instabilities, averaged over long time spans, should be 
negligible to a good approximation. We may also assume that there is no 
significant tidal friction, since given the cosmic contribution, this would be 
inconsistent with the observed mean acceleration of the Moon ${\dot n}_{\rm m}$ 
(see equation (46) below). We then get
\eqa
{\frac{dT_{\rm sE}(t)}{dt}}=HT_{\rm sE}(t), \qquad \Rightarrow \qquad
T_{\rm sE}(t)={\frac{t}{t_0}}T_{\rm sE}(t_0).
\ena
From equation (43), we may estimate a cosmic spin-down of the Earth at the 
present epoch of about 0.68 ms per century (using $H{\sim}2.5{\times}10^{-18}
$ s$^{-1}$). To see if this is consistent with the assumption that the dominant
contribution is due to cosmic effects, we may compare to historical 
observations indicating a lengthening of the day of about 1.4 ms per century 
averaged over the last 1000 years [12] and of about 1.7 ms per century averaged
over the last 2700 years [13]. But the historical observations depend on 
an assumed value for ${\dot n}_{\rm m}$ obtained from fitting lunar laser 
ranging (LLR) data to a Newtonian model of the Moon's motion, making these 
values theory-dependent. This theory-dependence also affects the analysis of
palaeo-tidal records obtained from sedimentary tidal rhythmities [11]. However,
if the quasi-metric prediction of ${\dot n}_{\rm m}$ (see equation (46) below) 
rather than the LLR value is used as an input to the historical observations, 
one gets a result close to the quasi-metric prediction of the spin-down of the 
Earth obtained from equation (43).

Another quantity that can be calculated from equation (43) is the number of 
days $N_{\rm y}$ in one (sidereal) year $T_{\rm E}$. This is found to be constant 
since
\eqa
T_{\rm E}=N_{\rm y}T_{\rm sE}={\frac{t}{t_0}}T_{\rm E}(t_0), {\qquad} 
\Rightarrow \qquad {\frac{dN_{\rm y}}{dt}}=0.
\ena
A similar calculation applies to the number of the days $N_{\rm m}$ in one 
(sidereal) month $T_{\rm m}$, i.e.,
\eqa
T_{\rm m}=N_{\rm m}T_{\rm sE}={\frac{t}{t_0}}T_{\rm m}(t_0), {\qquad}  
{\Rightarrow} \qquad {\frac{dN_{\rm m}}{dt}}=0.
\ena
That $N_{\rm y}$ and $N_{\rm m}$ are predicted to be constant, is not in agreement 
with standard (theory-dependent) interpretations of palaeo-geological data 
[7, 11]. (The predicted constancy of the ratio $N_{\rm y}/N_{\rm m}$ agrees well 
with a standard interpretation of the data, though.) To see why, the assumption
that $T_{\rm E}$ is constant is routinely used in the interpretation of tidal 
rhythmities and fossil coral growth data; in particular this applies to [11].
Another assumption used in the data interpretation is that active masses do 
not vary with time.

The apparent constancy of the sidereal year (as indicated by astronomical 
observations of the Sun and Mercury since about 1680 [9]), represents the 
observational basis for adopting the notion that ephemeris time (i.e., the time
scale obtained from the observed motions of the Sun and the planets) is equal 
to atomic time (plus a conventional constant), but different from so-called 
universal time (any time scale based on the rotation of the Earth). However, 
from equations (39) and (43) we see that according to quasi-metric relativity, 
ephemeris time should be scaled with a factor ${\frac{t}{t_0}}$ compared to 
atomic time, and averaged over long time spans, it should be equal to universal 
time. But within the Newtonian framework, any secular changes in the Earth-Moon
system are interpreted in terms of tidal friction (and external perturbations),
so seemingly secular inconsistencies between different time scales may be 
blamed on the variable rotation of the Earth. In practice this means 
introducing leap seconds. Given the fact that leap seconds are routinely used 
to adjust the length of the year, the predicted relationship between the 
different time scales should be consistent with observations. In particular,
the extra time corresponding to an increasing year as predicted from our model
may easily be hidden into the declining number of days in a year as predicted
from standard theory. 

As mentioned above, the value of the so-called mean acceleration of the Moon,
${\dot n}_{\rm m}{\equiv}{\frac{d}{dt}}n_{\rm m}$ (where $n_{\rm m}$ is the mean 
geocentric angular speed of the Moon as observed from its motion), is a very 
important quantity for calculating the evolution of the Earth-Moon system. 
From equation (45) we see that quasi-metric theory predicts that
\eqa
n_{\rm m}(t)={\frac{d{\phi}_{\rm m}}{dt}}={\frac{t_0}{t}}n_{\rm m}(t_0), 
\qquad \Rightarrow \qquad 
{\dot n}_{\rm m}{\equiv}{\frac{d}{dt}}n_{\rm m}=-Hn_{\rm m},
\ena
and inserting the observed value 0.549$^{{\prime}{\prime}}$/s for $n_{\rm m}$ at 
the present epoch, we get the corresponding cosmological contribution to 
${\dot n}_{\rm m}$, namely about $-13.6^{{\prime}{\prime}}$ per century$^2$. This
value may be compared to the value $-$13.68$^{{\prime}{\prime}}$ per 
century$^2$ obtained from fitting LLR data to a lunar model [14]. Note that 
this second value is the {\em total} mean acceleration, wherein other modelled
(positive) contributions, such as planetary perturbations, are included. When 
the other contributions are removed, one deduces a {\em tidal} contribution 
${\dot n}_{\rm tid}=-$25.8$^{{\prime}{\prime}}$ per century$^2$ to ${\dot n}_{\rm m}$
[14]. Similar results for the tidal contribution to ${\dot n}_{\rm m}$ inferred 
from LLR data have been found in, e.g., [15] (${\dot n}_{\rm tid}
=-$25.9$^{{\prime}{\prime}}$ per century$^2$). However, there is no need for a 
tidal term as a free parameter in the quasi-metric model. Besides, the 
non-tidal contributions to ${\dot n}_{\rm m}$ can also be ignored since they are
based on observations having an alternative explanation from the effect of the 
cosmic expansion on the Earth's orbit. And a model containing only the cosmic 
contribution fits the LLR data well.

We may also use Hubble's law directly to calculate the secular recession 
${\dot a}_{\rm qmr}$ of the Moon due to the global cosmic expansion; this yields 
about 3.0 cm per year, whereas the value ${\dot a}_{\rm tid}$ inferred from LLR 
is about 3.8 cm per year. However, the LLR value is model-dependent. From 
Kepler's third law we get the model interrelationship
\eqa
{\dot a}_{\rm tid}={\frac{2}{3}}
{\frac{{\dot n}_{\rm tid}}{{\dot n}_{\rm qmr}}}{\dot a}_{\rm qmr},
\ena
where ${\dot n}_{\rm qmr}$ is the cosmic contribution given from equation (46). 
By inserting the numerical values given above, we see that this is 
quite consistent with the LLR data.

To round off this chapter: we have seen that the predicted effects of the 
cosmic expansion on the solar system gravitational field have a number of 
observable consequences, none of which is shown to be in conflict with 
observation so far. That is, it seems that at this time, no model-independent 
evidence exists that may rule out the possibility that planetary orbits expand 
according to the Hubble law.
\\ [4mm]
{\bf References} \\ [1mm]
{\bf [1]} J. Binney and S. Tremaine, {\em Galactic Dynamics}, Princeton 
University Press (1987). \\
{\bf [2]} V.V. Zhytnikov and J.M. Nester, {\em Phys.\ Rev.\ Lett.} {\bf 73},
2950 (1994).\\
{\bf [3]} D. {\O}stvang, chapter 2 of this thesis (2001). \\
{\bf [4]} S. Weinberg, {\em Gravitation and Cosmology}, John Wiley ${\&}$ Sons,
Inc. (1972). \\
{\bf [5]} P.D. Mannheim, {\em Foundations of Physics} {\bf 24}, 487 (1994). \\
{\bf [6]} C.W. Misner, K.S. Thorne, J.A. Wheeler,
{\em Gravitation} W.H. Freeman ${\&}$ Co. (1973). \\
{\bf [7]} P. Wesson, {\em Cosmology and Geophysics}, Adam Hilger Ltd, (1978).\\
{\bf [8]} F.I. Cooperstock, V. Faraoni, D.N. Vollick, {\em ApJ} {\bf 503}, 61
(1998). \\
{\bf [9]} R.G. Hipkin, in {\em Palaeogeophysics} (editor S.K. Runcorn),
Academic Press (1970). \\
{\bf [10]} D. {\O}stvang, chapter 5 of this thesis (2001). \\
{\bf [11]} G.E. Williams, {\em Geophysical Research Letters} {\bf 24}, 421
(1997). \\
{\bf [12]} L.V. Morrison and F.R. Stephenson, {\em Observations of Secular
and Decade \\ \hspace*{6.3mm} Changes in the Earth's Rotation}, in {\em Earth 
Rotation: Solved and Unsolved \\ \hspace*{6.3mm}
Problems} (editor A. Cazenave) Reidel Publishing Company (1986). \\
{\bf [13]} F.R. Stephenson and L.V. Morrison, {\em Phil. Trans. R. Soc.
Lond.} {\bf A351}, 165 (1995). \\
{\bf [14]} J. Chapront, M. Chapront-Touz{\'{e}}, G. Francou, {\em A${\&}$A}
{\bf 343}, 624 (1999). \\
{\bf [15]} J.O. Dickey {\em et al.,} {\em Science} {\bf 265}, 482 (1994).

\newpage

{\vspace*{5cm}}

\begin{center}
{\huge {\bf Chapter Four: \\\ \\An Assumption of a Static\\
\vspace{0.5cm}Gravitational Field
}}
\end{center}
\vspace{-0.8cm}
\begin{center}
{\huge {\bf Resulting in an\\
\vspace{0.5cm}Apparently Anomalous Force}}
\end{center}
\newpage
{\hspace*{1cm}}

\newpage

\setcounter{equation}{0}

\setcounter{section}{0}

\begin{center}
{\large {\bf An Assumption of a Static Gravitational Field Resulting in an 
Apparently Anomalous Force}}
\end{center}
\begin{center}
by
\end{center}
\begin{center}
Dag {\O}stvang \\
{\em Institutt for Fysikk, Norges teknisk-naturvitenskapelige universitet, 
NTNU \\
N-7491 Trondheim, Norway}
\end{center}
\begin{abstract}
According to the ``quasi-metric'' space-time framework (QMF) developed 
elsewhere [1], the apparently anomalous force acting on the Pioneer 10/11, 
Galileo and Ulysses spacecraft as inferred from radiometric data [2, 9], may 
be naturally explained as resulting from an extra time delay (compared to 
standard theory), of the radio signals sent to and received from the 
spacecraft. The extra time delay originates from the cosmic expansion in the 
solar system as predicted from the QMF, via a piece of the quasi-metric affine 
connection having no counterpart in standard theory. That is, we show that the 
illusion of an anomalous acceleration of the right size, and acting towards 
the observer, arises as a consequence of the mismodelling of null paths in 
standard theory. The apparently anomalous acceleration is of size $cH$ [2, 9] 
(where $H$ is the Hubble parameter) as predicted by a simple non-static model 
of the solar system gravitational field [3].
\end{abstract}
\section{Introduction}
Some time ago, a detailed analysis of the observed versus the calculated orbits
of the Pioneer 10, Pioneer 11, Ulysses and Galileo spacecraft was published 
[2]. The main result was that the observed radiometric data did not agree with
calculations based on standard theory; rather the data indicated the existence 
of an ``anomalous", constant acceleration towards the Sun.

A short summary of the results presented in [2] is as follows: for the Pioneer
spacecraft, the Doppler frequency shift of the radio carrier wave was recorded 
and analysed to determine the spacecraft's orbits. Two independent analyses of
the raw data were performed. Both showed an anomalous acceleration towards the
Sun, of respectively $(8.09{\pm}0.20){\times}10^{-8}$ cm/s$^2$ and 
$(8.65{\pm}0.03){\times}10^{-8}$ cm/s$^2$ for Pioneer 10. For Pioneer 11 only 
one result is given; an anomalous acceleration of 
$(8.56{\pm}0.15){\times}10^{-8}$ cm/s$^2$ towards the Sun. The acceleration 
did not vary between $40-60$ astronomical units, within a sensitivity of
$2{\times}10^{-8}$ cm/s$^2$.

For the Galileo and Ulysses spacecraft, one also got ranging data in addition
to the Doppler data. (Ranging data are generated by cross-correlating a 
phase-modulated signal with a ground duplicate and measuring the time delay.) 
For Ulysses, one had to model the solar radiation pressure in addition to any
constant anomalous acceleration. By doing this, it was found that Ulysses was
influenced by an anomalous acceleration of size
$(12{\pm}3){\times}10^{-8}$ cm/s$^2$ towards the Sun, consistent with both
Doppler and ranging data. For Galileo, the corresponding result was an
anomalous acceleration of $(8{\pm}3){\times}10^{-8}$ cm/s$^2$ towards the Sun.

Very recently, a new comprehensive study of the anomalous acceleration was
published [9], including a total error budget for the Pioneer 10 data 
analysis. The new result reported in [9], is an ``experimental'' anomalous 
acceleration of $(7.84{\pm}0.01){\times}10^{-8}$ cm/s$^2$ towards the Sun, and
including bias and uncertainty terms the final value, this becomes
$(8.74{\pm}0.94){\times}10^{-8}$ cm/s$^2$. For Pioneer 11, an experimental
value of $(8.55{\pm}0.02){\times}10^{-8}$ cm/s$^2$ was given. Also other new 
results, such as annular and diurnal variations in the anomalous acceleration,
were reported in [9].

An interpretation of these results according to the standard general 
relativistic model, indicates the existence of an anomalous, time-independent 
force acting on the spacecraft. However, there are problems with this 
interpretation since according to the planetary ephemerides, there is no 
indication that such a force acts on the orbits of the planets; the 
hypothetical force thus cannot be of gravitational origin without violating 
the weak principle of equivalence. Thus it is speculated that the effect is 
due to anisotropic radiation of waste heat from the radioactive thermal 
generators aboard the spacecraft; the design of the spacecraft is such that 
waste heat may possibly be scattered off the back of the high gain antennae in
directions preferentially away from the Sun [4]. Moreover, besides possible 
anisotropic scattering, an estimate shows that the specific arrangement of 
waste heat radiators on the surface of the spacecraft may perhaps cause 
sufficient anisotropy in the radiative cooling to explain the data [5].
However, it seems that these explanations have been effectively refuted 
[6-9]. Other possible explanations, such as gas leaks, have been proposed 
[6, 9], but so far it seems that no satisfactory explanation based
on well-known physics exists.

However, it is an intriguing fact that the size of the anomalous acceleration
is of the order $cH$ for all the spacecraft, where $H$ is the Hubble 
parameter. Since this seems to be too much of a coincidence, one may suspect 
that the data indicate the existence of new physics rather than a prosaic 
explanation based on standard theory. This has been duly noted by others,
see [9] and references therein. But to be acceptable, any non-standard 
explanation should follow naturally from a general theoretical framework. In 
this chapter we show that such an explanation can be found, thus the data may 
indeed be taken as evidence for new physics.
\section{A quasi-metric model}
In reference [1], we defined the so-called ``quasi-metric'' space-time 
framework (QMF); this framework is non-metric since it is not based on 
pseudo-Riemannian geometry. Moreover, in reference [3] we introduced a model of 
the gravitational field outside a spherically symmetric, isolated source as 
predicted by a particular quasi-metric theory of gravity developed in [1]. 
According to this theory, it was found in [3] that at scales similar to the
size of the solar system, such a gravitational field can be expressed by the 
one-parameter family ${\bf g}_t$ of Lorentzian 4-metrics
\eqa
ds_t^2=-B(r)(dx^0)^2+({\frac{t}{t_0}})^2{\Big (}A(r)dr^2
+r^2d{\Omega}^2{\Big )},
\ena
where $r$ is a comoving Schwarzschild radial coordinate and 
$d{\Omega}^2{\equiv}d{\theta}^2+{\sin}^2{\theta}d{\phi}^2$ is the squared 
solid angle line element. Furthermore, $t$ is the global time function; its 
value is zero at the beginning of the Universe. In equation (1), $t$ is to be 
viewed as a parameter; the relationship between $t$ and the space-time 
coordinates is given by $x^0=ct$ and $t_0$ represents some arbitrary reference
epoch setting the scale of the spatial coordinates. The reason that one must 
separate between $ct$ and $x^0$ in (1), is that the {\em affine connection} 
compatible with the family ${\bf g}_t$ is non-metric. That is, although $ct$ 
and $x^0$ both can be interpreted as time coordinates, the components of the 
affine connection containing $ct$ is not equivalent to their counterparts 
containing $x^0$. See [1] for a further discussion. The locally measured
Hubble parameter $H$ as calculated from (1) reads [3]
\eqa
H(r,t)={\frac{ct_0}{t}}{\Big(}{\sqrt{B(r)}}{\Big)}^{-1}
{\frac{d}{dx^0}}({\frac{t}{t_0}})={\Big(}{\sqrt{B(r)}}t{\Big)}^{-1}.
\ena
General equations of motion are obtained from the geodesic equation using the 
non-metric connection [1]. But these equations of motion {\em cannot} be 
obtained from the geodesic equation using any metric connection. Moreover, as 
shown in [3], special equations of motion for inertial test particles moving in
the particular metric family (1) take the form (due to the spherical symmetry 
we can restrict the motion to the equatorial plane ${\theta}={\pi}/2$)
\eqa
({\frac{t}{t_0}})^2{\frac{A(r)}{B^2(r)}}({\frac{dr}{cdt}})^2-
{\frac{1}{B(r)}}+{\frac{J^2}{r^2}}=-E,
\ena
\eqa
{\frac{t}{t_0}}r^2{\frac{d{\phi}}{cdt}}=B(r)J,
\ena
\eqa
d{\tau}_t^2=-c^{-2}ds_t^2=EB^2(r)dt^2,
\ena
where $J$ and $E$ are constants of the motion. By setting the scale factor 
${\frac{t}{t_0}}=1$ in equations (1), (3), (4) and (5), we recover the 
equations of motion for inertial test particles moving in a spherically 
symmetric, static gravitational field as obtained from General Relativity 
(GR). Note that $E=0$ for photons and $E>0$ for material particles, which may 
readily be seen from equation (5). 

The functions $A(r)$ and $B(r)$ may be found as series expansions by solving
the field equations, this is done approximately in [3]. For our purposes,
we include terms to post-Newtonian order but not higher. Then we have
\eqa
A(r)=1+{\frac{r_{\rm s0}}{r}}+O({\frac{r_{\rm s0}^2}{r^2}}), {\nonumber} \\
B(r)=1-{\frac{r_{\rm s0}}{r}}+O({\frac{r_{\rm s0}^3}{r^3}}),
\ena
where $r_{\rm s0}$ is the (generalized) Schwarzschild radius at the arbitrary 
epoch $t_0$.

We now explore some of the differences between the non-static system described
by equation (1) and the corresponding static system obtained by setting 
${\frac{t}{t_0}}=1$ in equation (1) and using GR. To begin with, we notice that
the {\em shapes} of free fall orbits (expressed, e.g., as functions of the type 
$r({\phi})$) are identical for the two cases [3]. Moreover, it can be shown 
that the time dependence present in equations (3), (4) and (5) does not lead 
to easily observable perturbations in the paths of non-relativistic particles 
compared to the static case [3]. However, as we now illustrate, if one 
considers null paths, potential observable consequences do appear if one treats 
$r$ as a static coordinate rather than as a comoving one. To simplify matters, 
we consider purely radial motion, i.e., $J=0$ (one may easily generalize to
$J{\neq}0$). Since $E=0$ for photons, we get from equation (3) that radial null
curves are described by the equation
\eqa
{\frac{dr}{dt}}={\pm}{\frac{ct_0}{t}}{\sqrt{\frac{B(r)}{A(r)}}}
={\pm}{\frac{ct_0}{t}}{\sqrt{1-{\frac{2r_{\rm s0}}{r}}+
O({\frac{r^2_{\rm s0}}{r^2}}}})={\pm}{\frac{ct_0}{t}}
{\Big(}1-{\frac{r_{\rm s0}}{r}}+O({\frac{r^2_{\rm s0}}{r^2}}){\Big)},
\ena
the choice of sign depending on whether the motion is outwards or inwards. By 
integrating equation (7) to lowest order we find an extra delay, as compared to 
standard theory, in the time it takes an electromagnetic signal to travel from 
an object being observed to the observer. To lowest order, this extra time 
delay is ${\frac{HR^2}{2c^2}}$, where $R$ is the radial coordinate distance 
between the object and the observer and $H$ is the Hubble parameter as given 
from equation (2). Also, the fact that the scale factor in equation (1) 
increases with time, implies that our model predicts an extra redshift, as 
compared to standard static models, in the Doppler data obtained from any 
object emitting electromagnetic signals. To lowest order, this extra redshift 
corresponds to a ``Hubble'' redshift ${\frac{HR}{c}}$. 

But the velocity at any given time of an observed object cannot 
model-independently be split up into one ``ordinary" piece and one ``Hubble" 
piece. This means that there is no direct way to identify the predicted extra 
redshift in the Doppler data. Similarly, at any given time there is no direct 
way to sort out the predicted extra time delay when determining the distance 
to the object. Rather, to test whether the gravitational field is static or 
not, one should do observations over time and compare the observed motion to a 
model. In a model, one uses a coordinate system and to calculate coordinate 
motion, one needs coordinate accelerations. Accordingly, we construct the 
``properly scaled coordinate acceleration" quantity $a_{\rm c}$. For photons, 
this is
\eqa
a_{\rm c}{\equiv}{\frac{t}{t_0}}{\frac{\sqrt{A(r)}}{B(r)}}{\frac{d^2r}{dt^2}}
&=&{\mp}{\frac{c}{t}}+{\frac{t_0}{t}}{\frac{r_{\rm s0}c^2}{r^2}}+
O({\frac{r_{\rm s0}^2c^2}{r^3}}) {\nonumber} \\
&&={\mp}cH+{\frac{t_0}{t}}{\frac{r_{\rm s0}c^2}{r^2}}
+O({\frac{r_{\rm s0}^2c^2}{r^3}}).
\ena
The point with this is to show that, by treating the comoving coordinate system 
as a static one and using GR, an ``anomalous" term ${\mp}cH$ will be missed 
when modelling coordinate accelerations of photons. We see that the sign of 
the anomalous term is such that the anomalous acceleration is oriented in the 
opposite direction to that of the motion of the photons. This means that to 
sufficient accuracy, treating the comoving coordinate system as a static one, 
is equivalent to introducing a variable ``effective'' velocity of light 
$c_{\rm eff}$ equal to
\eqa
c_{\rm eff}=c(1-{\int}_{{\!}{\!}{\!}t}^{t+T}{\frac{dt'}{t'}})=
c(1-HT+O((HT)^2)),
\ena
where $T$ is the light time along the null path. The change of $c_{\rm eff}$
with $T$ then yields an anomalous acceleration
\eqa
a_{\rm an}={\frac{d}{dT}}c_{\rm eff}=-cH+O(cH^2T),
\ena
along the line of sight to any observed object. That is, if the comoving
coordinate system is treated as a static one, the coordinate motion of any 
object will be observed to slow down by an extra amount if the light time, or 
equivalently, the distance to the observer, increases and to speed up by an 
extra amount if the distance decreases. Hence, judging from its coordinate 
motion, it would seem as if the object were influenced by an anomalous force 
directed towards the observer.

By integrating the anomalous acceleration over the total observation time 
${\cal T}{\ll}t$, the data would indicate an ``anomalous" speed
\eqa
w_{\rm an}={\int}_{{\!}{\!}{\!}t}^{t+{\cal T}}a_{\rm an}dt'=-cH{\cal T}
+O(cH^2{\cal T}^2),
\ena
towards the observer, if one compares observations to a model where the 
coordinates are static rather than comoving. That is, the coordinate motion of 
any object observed over time indicates an anomalous blueshift compared to a 
model where the gravitational field is static.

Now, since any observer is typically located at the Earth, the quasi-metric
model in fact predicts that the illusory anomalous acceleration, as inferred 
from data analysis based on GR, should be directed towards the Earth rather 
than towards the Sun. But any directional differences will almost average out 
over time if the observed object moves approximately radially and is located 
well beyond the Earth's orbit. However, compared to a model where the anomalous
acceleration is towards the Sun, even if directional differences nearly 
average out, there remains a cumulative difference. We will estimate the
size of this difference below. If, on the other hand, the line of sight to the 
observed object (e.g., a planet) deviates significantly from the radial 
direction, observations should not be consistent with an anomalous acceleration
directed towards the Sun. Rather, the direction of the anomalous acceleration 
expressed in Sun-centred coordinates would appear to be a complicated function
of time.

To estimate the predicted differences between a model where the anomalous 
acceleration is towards the Sun and the result given by equation (10),
for the case where the observed object moves approximately radially in the
ecliptic plane and is located well beyond the Earth's orbit, it is convenient 
to define the average anomalous acceleration 
${\langle}a_{\rm c}{\rangle}_{\rm in}$ of a photon away from the Sun during the 
time of flight $T$ from the object to the observer. Thus we define
\eqa
{\langle}a_{\rm c}{\rangle}_{\rm in}{\equiv}{\frac{1}{T}}
{\int}_{{\!}{\!}{\!}t}^{t+T}a_{\rm c}dt'={\frac{c}{T}}{\Big [}
{\int}_{{\!}{\!}{\!}t}^{t+T_1}-
{\int}_{{\!}{\!}{\!}t+T_1}^{t+T}{\Big ]}{\frac{dt'}{t'}}=
cH{\Big (}2{\frac{T_1}{T}}-1+O(HT){\Big )},
\ena
where $T_1$ is the moment when the photon crosses a plane through the centre
of the Sun normal to a line connecting the object and the Sun. (If the Earth
is at the same side of this plane as the object, $T_1=T$.) Similarly, we can 
define the average anomalous acceleration
${\langle}a_{\rm c}{\rangle}_{\rm out}=-{\langle}a_{\rm c}{\rangle}_{\rm in}$ 
of a photon towards the Sun during the time of flight from the observer to 
the object. We are now able to estimate the predicted difference 
${\delta}a_{\rm an}$ between a model where the anomalous acceleration is towards
the Earth and one where it is towards the Sun. We get
\eqa
{\delta}a_{\rm an}=-cH-{\langle}a_{\rm c}{\rangle}_{\rm out}=
-2cH(1-{\frac{T_1}{T}}+O(HT)).
\ena
This function has a minimum at solar conjunction and vanishes when the observed
object and the Earth are at the same side of the Sun. One may show that
${\delta}a_{\rm an}$ can be written approximately as a truncated sine function 
where all positive values are replaced by zero. The period is equal to one 
year and the amplitude is approximately $2cH{\frac{R_{\rm e}}{R_{\rm o}}}$, where 
$R_{\rm o}$ and $R_{\rm e}$ are the radial coordinates of the object and of the 
Earth, respectively.

Anderson {\em et al.} [9] (see also [6]), found an annual perturbation on top 
of the anomalous acceleration $a_{\rm P}$ of Pioneer 10. (They claim to see 
such a perturbation for Pioneer 11 also.) Interestingly, the perturbation 
consistently showed minima near solar conjunction [9] (i.e., the absolute 
value of the anomalous acceleration reached maximum near solar conjunction). 
They fitted an annual sine wave to the velocity residuals coming from the 
annular perturbation term, using data from Pioneer 10 when the spacecraft was 
about 60 AU from the Sun. Taking the derivative with respect to time, they 
then found the amplitude of the corresponding acceleration; it was found to 
be $a_{\rm a.t.}=(0.215{\pm}0.022){\times}10^{-8}$ cm/s$^2$. We may compare this 
to the corresponding amplitude in ${\delta}a$ calculated above. We find an 
amplitude of about $0.29{\times}10^{-8}$ cm/s$^2$, close enough to be roughly 
consistent with the data. Besides the annual term, a diurnal term was also 
found in the velocity residuals [9], where the corresponding acceleration 
amplitude $a_{\rm d.t.}$ is large compared to $a_{\rm P}$. But over one year, the 
contribution from the diurnal term averages out to insignificance (less than 
$0.03{\times}10^{-8}$ cm/s$^2$ [9]).

To find the trajectories of non-relativistic particles, we may set 
$E{\equiv}1-{\frac{w^2}{c^2}}$, where ${\frac{w^2}{c^2}}$ is small. 
Then equation (3) yields
\eqa
{\frac{dr}{dt}}={\pm}{\frac{ct_0}{t}}{\sqrt{{\frac{B(r)}{A(r)}}{\Big (}1+
({\frac{w^2}{c^2}}-1)B(r){\Big )}}}=
{\pm}{\frac{ct_0}{t}}{\sqrt{{\frac{r_{\rm s0}}{r}}+{\frac{w^2}{c^2}}
+O({\frac{r_{\rm s0}^2}{r^2}})}},
\ena
and the properly scaled coordinate acceleration for non-relativistic particles
is
\eqa
a_{\rm c}={\mp}{\frac{c}{t}}{\sqrt{{\frac{r_{\rm s0}}{r}}+{\frac{w^2}{c^2}}
+O({\frac{r_{\rm s0}^2}{r^2}})}}-{\frac{t_0}{t}}{\frac{r_{\rm s0}c^2}{2r^2}}
+O({\frac{r_{\rm s0}^2c^2}{r^3}}).
\ena
We see that for non-relativistic particles, the effect on coordinate
accelerations of treating the comoving coordinates as static ones and using 
GR, is a factor ${\sqrt{{\frac{r_{\rm s0}}{r}}+{\frac{w^2}{c^2}}}}$ smaller than
the corresponding effect for photons. This means that the trajectories of 
non-relativistic particles do not depend crucially on the fact that the 
gravitational field is non-static. 

On the other hand, the paths of photons depend more significantly on whether 
the gravitational field is static or not, and this yields the illusion of an 
anomalous acceleration. That is, if one receives electromagnetic signals from 
some freely falling object located, e.g., in the outer parts of the solar 
system, the coordinate acceleration of the object as inferred from the signals
should not agree with the ``real" coordinate acceleration of the object, if one
treats the comoving coordinates as static ones. Rather, from equation (10) we 
see that it would seem as if the object were influenced by an attractive 
anomalous acceleration of size $cH$. The relevance of this is apparent when 
modelling the orbits of spacecraft and comparing to data obtained from radio 
signals received from the spacecraft; in particular this applies to the 
analyses performed in [2], [6] and [9]. An extra bonus for the model 
considered in this paper, is that it predicts small deviations during the year 
if the data are compared to a model where the anomalous acceleration is 
directed towards the Sun rather than towards the Earth. And as we have
seen, this prediction seems to be consistent with the data.
\subsection{Cosmic expansion and the PPN-formalism}
Orbit analysis of objects moving in the solar system must be based on some
assumptions of the nature of space-time postulated to hold there. The standard
framework used for this purpose is the parametrized post-Newtonian (PPN)
formalism applicable for most metric theories of gravity. But the standard 
PPN-framework does not contain any terms representing ``expanding space'' via a
global scale factor as shown in equation (1), since it is inherently assumed 
that the solar system is, for all practical purposes, decoupled from the cosmic
expansion. One may try to overcome this by inventing some other sense of 
``expanding space'', where the scale factor varies in space rather than in 
time. But such a model must necessarily be different from our quasi-metric 
model, and we show below that it cannot work. Thus, to illustrate the 
inadequacy of the PPN-framework to model expanding space, we now consider a 
specific model where suitable terms are added by hand in the metric. We may 
then compare to the change in light time obtained from our quasi-metric model.

One may try a post-Newtonian metric of the type
\eqa
ds^2=-{\Big(}1-{\frac{r_{\rm s}}{r}}+{\frac{2H_0r}{c}}+
O({\frac{r_{\rm s}^3}{r^3}}){\Big)}(dx^0)^2+
{\Big(}1+{\frac{r_{\rm s}}{r}}-{\frac{2H_0r}{c}}
+O({\frac{r_{\rm s}^2}{r^2}}){\Big)}dr^2+r^2d{\Omega}^2,
\ena
where $H_0$ is a constant, to describe expanding space within the 
PPN-framework. (To show that this metric yields a spatially variable
scale factor, transform to isotropic coordinates.) It may be readily shown
that the metric (16) yields a constant anomalous acceleration $cH_0$
towards the origin. But the problem with all metrics of this type is that 
they represent a ``real'' anomalous acceleration of gravitational origin, and 
this is observationally excluded from observations of planetary orbits [9].

Anyway, we may calculate the change in light time ${\Delta}T$ due to the 
terms containing $H_0$ in (16) by integrating a radial null path from 
$r_{\rm p}$ to $r_{\rm e}$ (let $r_{\rm p}>r_{\rm e}$, say). This yields
\eqa
{\Delta}T=-c^{-2}H_0(r_{\rm p}^2-r_{\rm e}^2)+{\cdots}{\approx}-H_0T^2,
\ena
where the light time $T$ is equal to $c^{-1}(r_{\rm p}-r_{\rm e})$ to first order 
and where the last approximation is accurate only if $r_{\rm p}$ is much larger 
than $r_{\rm e}$. Note that ${\Delta}T$ is negative; this is quite 
counterintuitive for a model representing expanding space. 

Anderson {\em et al.} [9], have considered a phenomenological model 
representing ``expanding space'' by adding a quadratic in time term to the 
light time in order to determine the coefficient of the quadratic by comparing
to data. To sufficient accuracy, this model may be represented by the 
transformation
\eqa
T{\rightarrow}(1+a_{\rm quad}t)T{\equiv}T + {\Delta}T,
\ena
where $a_{\rm quad}$ is a ``time acceleration'' term. This model fits both 
Doppler and range very well [9].

If we compare equations (17) and (18), we see that the value
$a_{\rm quad}=-H_0{\frac{T}{t}}$ corresponds to the change in light time
calculated from the metric (16). This is far too small (of order 
$10^{-30}$ s$^{-1}$ for a light time of a few hours) to be found directly from
the tracking data. Thus, the fact that $a_{\rm quad}$ was estimated to be zero 
based on the tracking data alone [9], in no way favours a constant acceleration
model over a time acceleration model. They are in fact equivalent as far as 
the data are concerned.

To compare differences in $a_{\rm quad}$, we may integrate equation (7) to lowest 
order to find the extra delay 
${\frac{H}{2c^2}}(r_{\rm p}-r_{\rm e})^2={\frac{HT^2}{2}}$ in the light time 
compared to the static case. This corresponds to a value 
$a_{\rm quad}={\frac{H^2T}{2}}$ for the time acceleration term. But a 
determination of $a_{\rm quad}$ directly from the tracking data still reflects 
the model-dependence explicitly present in the orbit determination process. 
This means that a determination of $a_{\rm quad}$ directly from the tracking 
data could in principle be consistent with the model (16) and not with our 
quasi-metric model. However, the quantity $a_{\rm quad}$ is so small that it is 
not feasible to check this. But the fact that a phenomenological model of the 
type (18) works so well, should be taken to mean that the explanation of the 
anomalous acceleration given in this chapter is sufficient.
\section{Conclusion}
We conclude that a natural explanation of the data, is that the gravitational 
field of the solar system is not static with respect to the cosmic expansion.
This also explains why any orbit analysis tool based on the PPN-formalism 
is insufficient for the task and how the largest errors arise from the 
mismodelling of null paths. (In fact, using the PPN-formalism is equivalent to
introducing a variable ``effective'' velocity of light as shown in equation 
(9).) But these explanations, while not involving any {\em ad hoc} assumptions,
are based on the premise that space-time is quasi-metric. That is, rather than 
being described by one single Lorentzian metric, the gravitational field of 
the solar system should be modelled (to a first approximation) by the metric 
family shown in equation (1). From a theoretical point of view this premise is
radical; thus it is essential that the subject is further investigated to make
certain that more mundane explanations may be eliminated. However, so far no 
such explanations based on well-known physics have been found. But the facts 
that the model presented in this paper follows from first principles and fits 
the data very well, together with independent observational evidence in favour
of the prediction that the Earth-Moon system is not static with respect to the 
cosmic expansion [3], indicate that explanations based on quasi-metric 
relativity should be taken seriously.
\\ [4mm]
{\bf References} \\ [1mm]
{\bf [1]} D. {\O}stvang, chapter 2 of this thesis (2001). \\
{\bf [2]} J.D. Anderson, P.A. Laing, E.L. Lau, A.S. Liu, M.M. Nieto and
S.G. Turyshev,  \\
{\hspace*{6.3mm}}{\em Phys. Rev. Lett.} {\bf 81}, 2858 (1998)
(gr-qc/9808081). \\
{\bf [3]} D. {\O}stvang, chapter 3 of this thesis (2001).\\
{\bf [4]} J.I. Katz, {\em Phys. Rev. Lett.} {\bf 83} 1892 (1999) 
(gr-qc/9809070). \\
{\bf [5]} E.M. Murphy, {\em Phys. Rev. Lett.} {\bf 83} 1890 (1999)
(gr-qc/9810015). \\
{\bf [6]} S.G. Turyshev, J.D. Anderson, P.A. Laing, E.L. Lau, A.S. Liu and 
M.M. Nieto, \\
{\hspace*{6.3mm}} gr-qc/9903024 (1999). \\
{\bf [7]} J.D. Anderson, P.A. Laing, E.L. Lau, A.S. Liu, M.M. Nieto and
S.G. Turyshev, \\
{\hspace*{6.3mm}}{\em Phys. Rev. Lett.} {\bf 83}, 1893 (1999)
(gr-qc/9906112). \\
{\bf [8]} J.D. Anderson, P.A. Laing, E.L. Lau, A.S. Liu, M.M. Nieto and
S.G. Turyshev, \\
{\hspace*{6.3mm}}{\em Phys. Rev. Lett.} {\bf 83}, 1891 (1999)
(gr-qc/9906113). \\
{\bf [9]} J.D. Anderson, P.A. Laing, E.L. Lau, A.S. Liu, M.M. Nieto and
S.G. Turyshev,  \\
{\hspace*{6.3mm}} 
gr-qc/0104064 (2001).

\newpage

{\vspace*{5cm}}

\begin{center}
{\huge {\bf Chapter Five: \\\ \\On the Non-Kinematical\\
\vspace{0.5cm}Expansion of Gravitationally
}}
\end{center}
\vspace{-0.5cm}
\begin{center}
{\huge {\bf Bound Bodies}}

\end{center}
\newpage
{\hspace*{1cm}}

\newpage

\setcounter{equation}{0}

\setcounter{section}{0}

\begin{center}
{\large {\bf On the Non-Kinematical Expansion of Gravitationally Bound Bodies}}
\end{center}
\begin{center}
by
\end{center}
\begin{center}
Dag {\O}stvang \\
{\em Institutt for Fysikk, Norges teknisk-naturvitenskapelige universitet, 
NTNU \\
N-7491 Trondheim, Norway}
\end{center}
\begin{abstract}
It is found that one of the predictions of a so-called ``quasi-metric'' theory 
of gravity developed elsewhere [1], is that the gravitational field inside a 
metrically static, spherically symmetric, isolated body modelled as a perfect 
fluid obeying a linear equation of state (i.e., $p{\propto}{\varrho}_{\rm m}$),
should expand according to the Hubble law. That is, for such a body, its radius
should increase like the expansion of the Universe; this is the counterpart to
the similar result found for the corresponding exterior gravitational field 
[2]. On the other hand, if the body consists of a perfect fluid obeying
some other equation of state (i.e., $p{\not}{\propto}{\varrho}_{\rm m}$, as for 
a polytrope for example), the expansion will induce instabilities and the body 
cannot be metrically static; such a body will not in general expand.
\end{abstract}
\section{Introduction}
A new type of non-metric space-time framework, the so-called ``quasi-metric''
framework, was presented in [1]. Also presented was a possibly viable 
quasi-metric theory of gravity compatible with the quasi-metric framework.
This theory predicts that the nature of the Hubble expansion is different from
how it is described by metric theory, and as a consequence, the Hubble 
expansion should influence gravitationally bound systems. In particular, the 
theory predicts that the gravitational field exterior to a spherically 
symmetric, metrically static, isolated source expands according to the Hubble 
law [2], and this seems to be consistent with observations of the orbits of 
distant spacecraft [3].

Now a natural question is what happens to the gravitational fields inside
gravitationally bound bodies; does the quasi-metric theory predict that such
gravitational fields may expand, too? We show in this chapter that this is 
indeed the case. Note that this result may support an interpretation of 
geological data indicating that the Earth is expanding, see reference [4] and 
references cited therein. Moreover, an expanding Earth should cause changes in 
its spin rate; we showed in [2] that the main part of the observed secular 
spin-down of the Earth may in fact be of cosmological origin. This also applies
to the observed recession of the Moon and its observed mean acceleration.
\section{A brief survey of the quasi-metric theory}
A basic feature of the quasi-metric framework is its non-metric nature, and
this implies that the canonical description of space-time is taken as 
fundamental. That is, quasi-metric space-time is constructed as consisting of 
two mutually orthogonal foliations: on the one hand, space-time can be sliced 
up globally into a family of 3-dimensional space-like hypersurfaces (called the 
fundamental hypersurfaces (FHSs)) by the global time function $t$; on the other
hand, space-time can be foliated into a family of time-like curves everywhere 
orthogonal to the FHSs. These curves represent the world lines of a family of 
hypothetical observers called the fundamental observers (FOs), and the FHSs 
taken at $t=$constant represent a preferred notion of space. The equations of 
quasi-metric theory depend on said split-up of quasi-metric space-time into 
space and time.

The FHSs are associated with two families of Lorentzian space-time metric 
tensors ${\bf {\bar g}}_t$ and ${\bf g}_t$ in such a way that different FHSs
correspond to domains of applicability of different members of these families
in an one-to-one relationship. The metric family ${\bf {\bar g}}_t$ represents
a solution of field equations, and from ${\bf {\bar g}}_t$ one can construct 
the ``physical'' metric family ${\bf g}_t$ which is used when comparing 
predictions to experiments. The theory is not metric since the affine 
connection compatible with any metric family is non-metric.

To be able to compare theory to experiment, we have to represent the metric 
families in terms of components with respect to some coordinate system on 
space-time. It is convenient to use a coordinate system ${\{}x^{\mu}{\}}$ 
where the relationship between the time coordinate $x^0$ and the global time
function is given by $x^0=ct$; this ensures that $x^0$ is a global time
coordinate. A coordinate system with a global time coordinate of this
type, we call a global time coordinate system (GTCS). There exist infinitely 
many GTCSs.

Expressed in a GTCS, the most general form allowed for the family 
${\bf {\bar g}}_t$ can be represented by the family of line elements (this may 
be taken as a definition) [1]
\eqa
{\overline{ds}}_t^2= {\Big [}{\bar N}_{(t)s}{\bar N}^s_{(t)}-{\bar N}_t^2
{\Big ]}(dx^0)^2+2{\frac{t}{t_0}}{\bar N}_{(t)s}dx^sdx^0+{\frac{t^2}{t_0^2}}
{\bar N}_t^2{\tilde h}_{(t)ks}dx^kdx^s,
\ena
where ${\bar N}_t$ is the lapse function field family of the FOs and where
${\frac{t}{t_0}}{\bar N}_{(t)s}dx^s$ is the family of shift covector fields of 
the FOs in ${\bf {\bar g}}_t$. Moreover, $t_0$ is an arbitrary reference epoch
and ${\bf {\bar h}}_t$ with components ${\bar h}_{(t)ks}{\equiv}
{\frac{t^2}{t_0^2}}{\bar N}_t^2{\tilde h}_{(t)ks}$ is the metric family 
intrinsic to the FHSs.

The time evolution of the spatial scale factor ${\bar F}_t{\equiv}ct{\bar N}_t$
of the FHSs in the orthogonal direction may be written as (a comma denotes a 
partial derivative and the symbol ${\bar {\perp}}$ denotes a scalar product 
with the unit normal vector field family $-{\bf {\bar n}}_t$ of the FHSs)
\eqa
{\bar F}_t^{-1}[(c{\bar N}_t)^{-1}{\bar F}_t,_t-{\bar F}_t,_{\bar {\perp}}]=
(c{\bar N}_t)^{-1}{\Big (}{\frac{1}{t}}+{\frac{{\bar N}_t,_t}{{\bar N}_t}}
{\Big )}-{\frac{{\bar N}_t,_{\bar {\perp}}}{{\bar N}_t}}
{\equiv}c^{-2}{\bar x}_t+c^{-1}{\bar H}_t.
\ena
Here, $c^{-2}{\bar x}_t$ represents the kinematical contribution to 
the evolution and $c^{-1}{\bar H}_t$ represents the so-called non-kinematical 
contribution defined by
\eqa
{\bar H}_t={\frac{1}{{\bar N}_tt}}+{\bar y}_t, \qquad
{\bar y}_t{\equiv}c^{-1}{\sqrt{{\bar a}_{{\cal F}k}{\bar a}_{\cal F}^k}}, 
\qquad c^{-2}{\bar a}_{{\cal F}j}{\equiv}{\frac{{\bar N}_t,_j}{{\bar N}_t}}.
\ena
We see that the non-kinematical evolution (NKE) of the FHSs (and thus of
${\bf {\bar h}}_t$) takes the form of an ``expansion'' since ${\bar H}_t>0$. 
Besides, we see from equation (2) that the evolution of ${\bar N}_t$ with time 
may be written in the form
\eqa
{\frac{{\bar N}_t,_t}{c{\bar N}_t^2}}
-{\frac{{\bar N}_t,_{\bar {\perp}}}{{\bar N}_t}}=c^{-2}{\bar x}_t+
c^{-1}{\bar y}_t. 
\ena
The split-ups defined in equations (3) and (4) are necessary to be able to 
construct ${\bf g}_t$ from ${\bf {\bar g}}_t$ [1].

A special property of the quasi-metric theory is that operationally defined 
atomic units by definition vary in space-time. That is, gravitational 
quantities may get an extra formal variability when measured in atomic units 
(and {\em vice versa}). To quantify the notion of formally variable units, we 
introduce the scalar field ${\Psi}_t$ telling how atomic time units vary in 
space-time. Measured in atomic units, gravitational quantities then get a 
formal variability as some power of ${\Psi}_t$. By definition $c$, $\hbar$ and
$e$ are not formally variable; this means that atomic time units vary similarly 
to atomic length units and inversely to atomic mass units (charge units are not
formally variable). This implies the existence of two distinct gravitational
coupling parameters; the ``bare'' $G^{\rm B}_t$ coupling to electromagnetic field
mass-energy, and the ``screened'' $G^{\rm S}_t$ coupling to material mass-energy,
respectively. 

Neither $G^{\rm B}_t$ nor $G^{\rm S}_t$ are constants measured in atomic units. 
However, for convenience we may {\em define} corresponding constants $G^{\rm B}$
and $G^{\rm S}$, respectively. But if we do this, we must separate between 
{\em active mass}, which is measured dynamically, and {\em passive mass} 
(passive gravitational mass or inertial mass). By dimensional analysis it is 
found [1], that active electromagnetic mass-energy varies formally as 
${\Psi}_t^{-2}$ whereas active material mass-energy varies formally as 
${\Psi}_t^{-1}$ when measured in atomic units (but passive mass-energy does of 
course not vary). By definition we have ${\Psi}_t{\equiv}{\bar F}_t^{-1}$, so
that
\eqa
{\Psi}_t,_t=-{\Big (}{\frac{1}{t}}+{\frac{{\bar N}_t,_t}{{\bar N}_t}}{\Big )}
{\Psi}_t, {\qquad}
{\Psi}_t,_{\bar {\perp}}=-{\frac{{\bar N}_t,_{\bar {\perp}}}{{\bar N}_t}}
{\Psi}_t, \qquad
{\Psi}_t,_j=-c^{-2}{\bar a}_{{\cal F}j}{\Psi}_t,
\ena 
where ${\bf {\bar a}}_{\cal F}$ is the 4-acceleration of the FOs in the family
${\bf {\bar g}}_t$. Local conservation laws, involving projections into and 
normal to the FHSs of the (active) stress-energy tensor ${\bf T}_t=
{\bf T}^{\rm (EM)}_t+{\bf T}^{\rm (MA)}_t$ (split up into electromagnetic and 
material parts), take the form (in component notation) [1]
\eqa
{\cal L}_{{\bf {\bar n}}_t}T_{(t){\bar {\perp}}{\bar {\perp}}}
={\Big (}{\bar K}_t-2{\frac{{\bar N}_t,_{\bar {\perp}}}{{\bar N}_t}}
{\Big )}T_{(t){\bar {\perp}}{\bar {\perp}}}+{\bar K}_{(t)ks}{\hat T}_{(t)}^{ks}
-{\hat T}^s_{(t){\bar {\perp}}{\mid}s},
\ena
\eqa
{\frac{1}{{\bar N}_t}}
{\cal L}_{{\bar N}_t{\bf {\bar n}}_t}T_{(t)j{\bar {\perp}}}
={\Big (}{\bar K}_t-2{\frac{{\bar N}_t,_{\bar {\perp}}}{{\bar N}_t}}
{\Big )}T_{(t)j{\bar {\perp}}}-c^{-2}{\bar a}_{{\cal F}j}
T_{(t){\bar {\perp}}{\bar {\perp}}}
+c^{-2}{\bar a}_{{\cal F}s}{\hat T}_{(t)j}^s-{\hat T}^s_{(t)j{\mid}s},
\ena
where ${\cal L}_{{\bf {\bar n}}_t}$ denotes a (projected) Lie derivative of spatial
objects in the direction normal to the FHSs and $`{\mid}$' denotes a spatial 
covariant derivative. Moreover, ${\bf {\bar K}}_t$ is the extrinsic curvature 
tensor family (with trace ${\bar K}_t$) of the FHSs. (A ``hat'' denotes an 
object intrinsic to the FHSs.) The quasi-metric field equations are postulated 
in [1]. In this chapter, we apply these field equations to metrically static 
interiors. For metrically static systems, ${\bf {\bar K}}_t$ vanishes, so that 
the relevant field equations take the simplified form [1]
\eqa
2{\bar R}_{(t){\bar {\perp}}{\bar {\perp}}}=
2(c^{-4}{\bar a}_{{\cal F}k}{\bar a}_{\cal F}^k+
c^{-2}{\bar a}^k_{{\cal F}{\mid}k})
={\kappa}^{\rm B}{\Big (}T^{\rm (EM)}_{(t){\bar {\perp}}{\bar {\perp}}}+
{\hat T}^{{\rm (EM)}k}_{(t)k}{\Big )}+
{\kappa}^{\rm S}{\Big (}T^{\rm (MA)}_{(t){\bar {\perp}}{\bar {\perp}}}+
{\hat T}^{{\rm (MA)}k}_{(t)k}{\Big )},
\ena
where ${\bf {\bar R}}_t$ is the Ricci tensor family obtained from equation (1),
in addition to the equation
\eqa
{\bar Q}_{(t)ij}=-c^{-2}{\bar a}_{{\cal F}i{\mid}j}
-c^{-4}{\bar a}_{{\cal F}i}{\bar a}_{{\cal F}j}
+c^{-2}{\Big (}{\bar a}_{{\cal F}{\mid}k}^k- 
{\frac{1}{{\bar N}_t^2t^2}}{\Big )}{\bar h}_{(t)ij}-{\bar H}_{(t)ij}=0,
\ena
where ${\bf {\bar H}}_t$ is the Einstein tensor family intrinsic to the FHSs
and where ${\bf {\bar Q}}_t$ is a foliation-dependent tensor family defined in 
[1].
\section{Metrically static, spherically symmetric interiors}
As shown in [2], the gravitational field exterior to an isolated, spherically 
symmetric body can be found from the field equations. In this case, the metric 
families ${\bf {\bar g}}_t$ and ${\bf g}_t$ are spherically symmetric in a 
${\bf S}^3{\times}{\bf R}$-background. Furthermore, they are ``metrically 
static''; by definition this means that neither ${\bar N}_t$ nor $N$ (the 
lapse function field of the FOs in ${\bf g}_t$) depend on $x^0$ (or $t$) and 
that the shift vector field family appearing in equation (1) vanishes in a 
suitably chosen GTCS. It would seem reasonable to look for interior 
gravitational fields of the same form.

To find the general form of ${\bf {\bar g}}_t$ and ${\bf g}_t$ for the 
spherically symmetric, metrically static case, we start with the general 
expression (1) and introduce a spherically symmetric GTCS
${\{}x^0,r,{\theta},{\phi}{\}}$. Then it is straightforward to show that 
the metric family ${\bf {\bar g}}_t$ can be represented by the family of line 
elements  
\eqa
c^2{\overline{d{\tau}}}_t^2={\bar B}(r)(dx^0)^2-({\frac{t}{t_0}})^2
{\Big (}{\bar A}(r)dr^2+r^2d{\Omega}^2{\Big )}, 
\ena
where $d{\Omega}^2{\equiv}d{\theta}^2+{\sin}^2{\theta}d{\phi}^2$ and $x^0=ct$. 
However, $x^0$ and $ct$ must not be interchanged in equation (10) due to the 
particular form of the connection coefficients, see [1] for a discussion. 

We now seek non-vacuum solutions of the type (10) of the field equations, and
where the source is modelled as a perfect fluid. For the vacuum case, it was 
found in [2] that putting a line element ${\overline{ds}}^2_t$ of the form
(10) into the field equations results in equations separable in ${\bar A}(r)$
and ${\bar B}(r)$. It is then possible to find a series solution. And it turns
out that for non-vacuum cases, one still gets separable equations when using 
line elements of the form (10). But we now have the passive pressure $p(r,t)$ 
as a new variable in addition to ${\bar A}(r)$ and ${\bar B}(r)$. (Note that 
the passive mass density ${\varrho}_{\rm m}(r,t)$ is also a new variable but 
that it is related to $p(r,t)$ via an equation of state.) The explicit 
dependence on $p$ in the equations makes analytical calculations rather 
impracticable so the equations should be solved numerically. In the following 
we set up the relevant equations.
\subsection{Analytical calculations and a numerical recipe}
In this section, we do some analytical calculations in order to write the
equations in a form appropriate for numerical treatment. Proceeding with this, 
the definitions yield
\eqa
c^{-2}{\bar a}_{{\cal F}r}&=&{\frac{{\bar B}'}{2{\bar B}}}, {\qquad}
c^{-2}{\bar a}_{{\cal F}r{\mid}r}={\frac{{\bar B}''}{2{\bar B}}}-
{\frac{{\bar B}'^2}{2{\bar B}^2}}-
{\frac{{\bar A}'{\bar B}'}{4{\bar A}{\bar B}}}, \nonumber \\
c^{-2}{\sin}^{-2}{\theta}{\bar a}_{{\cal F}{\phi}{\mid}{\phi}}&=&
c^{-2}{\bar a}_{{\cal F}{\theta}{\mid}{\theta}}=
{\frac{r{\bar B}'}{2{\bar A}{\bar B}}}, {\qquad}
{\bar P}_t=({\frac{t_0}{t}})^2{\frac{2}{\bar A}}{\Big (}r^{-2}({\bar A}-1)
+{\frac{{\bar A}'}{r{\bar A}}}{\Big )}, 
\ena
\eqa
c^{-2}{\bar a}^k_{{\cal F}{\mid}k}&=&({\frac{t_0}{t}})^2{\Big (}
{\frac{{\bar B}''}{2{\bar A}{\bar B}}}-
{\frac{{\bar B}'^2}{2{\bar A}{\bar B}^2}}
-{\frac{{\bar A}'{\bar B}'}{4{\bar A}^2{\bar B}}}
+{\frac{{\bar B}'}{r{\bar A}{\bar B}}}{\Big )}, {\nonumber} \\
{\bar H}_{(t)rr}&=&r^{-2}(1-{\bar A}), {\qquad}
{\sin}^{-2}{\theta}{\bar H}_{(t){\phi}{\phi}}={\bar H}_{(t){\theta}{\theta}}
=-{\frac{r{\bar A}'}{2{\bar A}^2}}.
\ena
Now ${\bf T}_t^{\rm (MA)}$ varies formally as ${\Psi}_t^2$ but 
${\bf T}_t^{\rm (EM)}$ varies formally as ${\Psi}_t$ when measured in atomic 
units. However, to have a metrically static situation, the time variability of 
the right hand side of the field equations must cancel out. This is only 
possible if ${\bf T}_t^{\rm (MA)}$ and ${\bf T}_t^{\rm (EM)}$ have identical time 
evolutions, which will hold for a source expanding according to the Hubble law 
as long as any net energy transfer between photons and material particles is 
negligible. In this case there will be an extra ``cosmic redshift'' (and also 
gravitational spectral shifts) of the source's photon energy density. This 
corresponds to an extra (non-formal) factor ${\Psi}_t$ in the variability of 
${\bf T}_t^{\rm (EM)}$, so that ${\bf T}_t^{\rm (EM)}$ and ${\bf T}_t^{\rm (MA)}$ can 
be treated in the same way.

For reasons of convenience, we choose to explicitly extract the variability of
${\bf T}_t$ associated with ${\Psi}_t^2$. What is left after separating out
this variability from the active mass density is by definition {\em the 
properly scaled density of active mass} ${\bar {\varrho}}_{\rm m}{\equiv}
{\bar {\varrho}}_{\rm m}^{\rm (EM)}+{\bar {\varrho}}_{\rm m}^{\rm (MA)}$. (The 
corresponding pressure is ${\bar p}$.) For the case when the perfect fluid is 
comoving with the FOs (i.e., $T_{(t){\bar {\perp}}j}=0$),
we may then write (using equation (5))
\eqa
T_{(t){\bar {\perp}}{\bar {\perp}}}={\tilde{\varrho}}_{\rm m}c^2{\equiv}
{\frac{t_0^2}{t^2}}{\frac{{\bar {\varrho}}_{\rm m}c^2}{{\bar B}}}, \qquad
T_{(t)r}^r=T_{(t){\theta}}^{\theta}=T_{(t){\phi}}^{\phi}=
{\tilde p}{\equiv}{\frac{t_0^2}{t^2}}{\frac{{\bar p}}{{\bar B}}},
\ena
where ${\tilde {\varrho}}_{\rm m}$ is the density of active mass-energy and 
${\tilde p}$ is the active pressure as seen in the local rest frame of the 
fluid. Furthermore, using equation (13), the local conservation laws (6) and 
(7) yield (with ${\dot {\ }}{\equiv}{\frac{\partial}{{\partial}t}}$ and
$'{\equiv}{\frac{\partial}{{\partial}{r}}}$)
\eqa
{\dot {\bar {\varrho}}}_{\rm m}={\dot {\bar p}}=0, \qquad
{\bar p},_r^{\rm (EM)}={\frac{{\bar p}^{\rm (EM)}}{\bar p}}
{\bar p}', \qquad
{\bar p},_r^{\rm (MA)}={\frac{{\bar p}^{\rm (MA)}}{\bar p}}{\bar p}',
\nonumber \\
{\bar p}'=-c^{-2}{\bar a}_{{\cal F}{r}}({\bar {\varrho}}_{\rm m}c^2
-3{\bar p})=-({\bar {\varrho}}^{\rm (MA)}_{\rm m}c^2-3{\bar p}^{\rm (MA)})
{\frac{{\bar B}'}{2{\bar B}}}.
\ena
Equation (14) is valid for any metrically static perfect fluid. But to be 
able to have experimental input, it is also necessary to define 
{\em the passive mass density} ${\varrho}_{\rm m}$ and the corresponding passive
pressure $p$. In addition, we need to specify an equation of state 
$p=p({\varrho}_{\rm m})$. We see that we need an expression relating 
${\bar {\varrho}}_{\rm m}$ (or equivalently, ${\tilde {\varrho}}_{\rm m}$) to the 
passive mass-energy density ${\varrho}_{\rm m}$. Such a relationship is given 
by [1]
\eqa
{\tilde {\varrho}}_{\rm m}=
\left\{
\begin{array}{ll}
{\frac{t}{t_0}}{\bar N}_t{\varrho}_{\rm m} {\;}{\;}{\;}
$for a fluid of material particles,$ \\ [1.5ex] 
{\frac{t^2}{t_0^2}}{\bar N}_t^2{\varrho}_{\rm m} {\;}{\;}{\;}
$for the electromagnetic field,$
\end{array}
\right.
\ena
and a similar relationship exists between ${\tilde p}$ and $p$. To find how 
the active mass $m_t$ varies in space-time, note that we are free to choose 
the background value of the active mass far from the source to be $m_0$. We 
use this to define $G^{\rm B}$ and $G^{\rm S}$ as the constants measured in local 
gravitational experiments far from the source at epoch $t_0$. Then, using the 
metrically static condition we find
\eqa
m_t(r,t)={\bar B}^{1/2}(r){\frac{t}{t_0}}m_0.
\ena
Now we can insert the equations (11), (12) and (13) into the field equation 
(8) (valid for the metrically static case). We get
\eqa
{\frac{{\bar B}''}{\bar B}}-{\frac{{\bar B}'^2}{2{\bar B}^2}}
-{\frac{{\bar A}'{\bar B}'}{2{\bar A}{\bar B}}}+
{\frac{2{\bar B}'}{r{\bar B}}}={\frac{{\bar A}}{{\bar B}}}{\Big [}
{\kappa}^{\rm B}({\bar {\varrho}}^{\rm (EM)}_{\rm m}c^2+3{\bar p}^{\rm (EM)})+
{\kappa}^{\rm S}({\bar {\varrho}}^{\rm (MA)}_{\rm m}c^2+
3{\bar p}^{\rm (MA)}){\Big ]}.
\ena
Two more equations can be found from equation (9) for the spatial curvature. 
First, the $rr$-component of equation (9) gives (with ${\Xi}_0{\equiv}ct_0$)
\eqa
{\bar A}(r)={\frac{{\Big [}1-r{\frac{{\bar B}'(r)}{2{\bar B}(r)}}
{\Big ]}^2}{1-{\frac{r^2}{{\bar B}(r){\Xi}_0^2}}}}.
\ena
Second, the angular components of equation (9) yield another equation, which 
can be combined with equation (17) to give
\eqa
{\frac{{\bar A}'}{{\bar A}}}
-{\frac{{\bar B}'}{{\bar B}}}-r{\frac{{\bar B}'^2}{2{\bar B}^2}}-
{\frac{2r{\bar A}}{{\Xi}_0^2{\bar B}}}=
-r{\frac{{\bar A}}{{\bar B}}}{\Big [}{\kappa}^{\rm B}
({\bar {\varrho}}^{\rm (EM)}_{\rm m}c^2+3{\bar p}^{\rm (EM)})+{\kappa}^{\rm S}
({\bar {\varrho}}^{\rm (MA)}_{\rm m}c^2+3{\bar p}^{\rm (MA)}){\Big ]}.
\ena
We may now eliminate ${\bar A}(r)$ and ${\bar A}'(r)$ from equation (17) by 
inserting equations (18) and (19). The result is
\eqa
{\Big (}1-{\frac{r^2}{{\bar B}(r){\Xi}_0^2}}{\Big )}
{\frac{{\bar B}''(r)}{{\bar B}(r)}}
-{\Big (}1-{\frac{2r^2}{{\bar B}(r){\Xi}_0^2}}{\Big )}
{\frac{{\bar B}'^2(r)}{{\bar B}^2(r)}} 
+{\frac{2}{r}}{\Big (}1-{\frac{3r^2}{2{\bar B}(r){\Xi}_0^2}}{\Big )}
{\frac{{\bar B}'(r)}{{\bar B}(r)}} \nonumber \\
-{\frac{r{\bar B}'^3(r)}{4{\bar B}^3(r)}}=
{\frac{1}{{\bar B}(r)}}{\Big (}1-{\frac{r{\bar B}'(r)}{2{\bar B}(r)}}{\Big )}^3
{\Big [}{\kappa}^{\rm B}
({\bar {\varrho}}^{\rm (EM)}_{\rm m}c^2+3{\bar p}^{\rm (EM)})+{\kappa}^{\rm S}
({\bar {\varrho}}^{\rm (MA)}_{\rm m}c^2+3{\bar p}^{\rm (MA)}){\Big ]}.
\ena
To solve equation (20) numerically for ${\bar B}(r)$, one may proceed as 
follows. First specify the boundary conditions at the centre of the body. 
From equation (18) we easily see that ${\bar A}(0)=1$. Also, we must 
have ${\bar B}'(0)={\bar p}'(0)=0$, so ${\bar A}'(0)=0$ from equation (19). 
Furthermore, noticing that $r^{-1}{\bar B}'$ must be stationary near the centre 
of the body, we have
\eqa
{\bar B}''(0)={\lim}_{r{\rightarrow}0}{\Big [}r^{-1}{\bar B}'(r)
{\Big ]}, \qquad \Rightarrow \qquad \nonumber \\
{\bar B}''(0)={\frac{{\kappa}^{\rm B}}{3}}
{\Big (}{\bar {\varrho}}^{\rm (EM)}_{\rm m}(0)c^2+
3{\bar p}^{\rm (EM)}(0){\Big )}+{\frac{{\kappa}^{\rm S}}{3}}
{\Big (}{\bar {\varrho}}^{\rm (MA)}_{\rm m}(0)c^2+
3{\bar p}^{\rm (MA)}(0){\Big )},
\ena
where the implication follows from equation (17).

To specify any particular model, choose the boundary condition 
${\bar p}^{\rm (EM)}(0)$, ${\bar p}^{\rm (MA)}(0)$ at some arbitrary time. Also 
choose some arbitrary value ${\bar B}(0)$ as an initial value for iteration. 
It must be possible to check how well the chosen ${\bar B}(0)$ reproduces the 
boundary condition for ${\bar B}(r_{\rm sf})$ at the surface $r=r_{\rm sf}$ of the 
body. That is, to match the exterior solution we must have [2]
\eqa
{\bar B}(r_{\rm sf})=1-{\frac{r_{\rm s0}}{r_{\rm sf}}}+
{\frac{r_{\rm s0}^2}{2r_{\rm sf}^2}}+{\frac{r_{\rm s0}r_{\rm sf}}{2{\Xi}_0^2}}-
{\frac{r_{\rm s0}^3}{8r_{\rm sf}^3}}+{\cdots},
\ena
where $r_{\rm s0}$ is a constant defined as the generalized Schwarzschild 
radius of the body at the arbitrary time $t_0$. Hence, by definition we set
$r_{\rm s0}{\equiv}{\frac{2M^{\rm (EM)}_{t_0}G^{\rm B}}{c^2}}+
{\frac{2M^{\rm (MA)}_{t_0}G^{\rm S}}{c^2}}$, where
\eqa
M^{\rm (MA)}_t{\equiv}c^{-2}{\int}{\int}{\int}{\bar N}_t{\Big [}
T^{\rm (MA)}_{(t){\bar {\perp}}{\bar {\perp}}}+
{\hat T}^{{\rm (MA)}s}_{(t)s}{\Big ]}d{\bar V}_{t}
=4{\pi}{\frac{t}{t_0}}{\int}_{{\!}{\!}0}^{r_{\rm sf}}
{\frac{{\sqrt{\bar A}}}{{\sqrt{\bar B}}}}{\Big [}
{\bar {\varrho}}^{\rm (MA)}_{\rm m}+3{\bar p}^{\rm (MA)}/c^2{\Big ]}r^2dr,
\ena
and a similar formula for $M_t^{\rm (EM)}$, where the integration is taken over 
the whole body. (The particular form of $M^{\rm (MA)}_t$ and $M^{\rm (EM)}_t$
follows directly from the ``metric approximation'' [1] of equation 
(8) applied to the interior of a spherically symmetric, static source when 
extrapolated to the exact exterior solution found in [1].) Note that 
$M_{t_0}^{\rm (MA)}G^{\rm S}+M_{t_0}^{\rm (EM)}G^{\rm B}$ is the quantity measured by 
distant orbiters at epoch $t_0$ and that this quantity depends on the pressure 
as well.

To have a metrically static system, it is necessary to specify an equation of 
state of the type $p{\propto}{\varrho}_{\rm m}$ (e.g., an ideal gas), since this 
ensures that ${\bar {\varrho}}_{\rm m}$ and ${\bar p}$ are independent of $t$. 
(In practice though, it will be required that the system stays close to 
thermal equilibrium, but then there will be a net energy transfer from the 
material particles to the photons. Thus the metrically static situation will 
hold approximately only if 
${\bar {\varrho}}_{\rm m}^{\rm (EM)}{\ll}{\bar {\varrho}}_{\rm m}^{\rm (MA)}$.)
Given a suitable equation of state, we can find
${\bar {\varrho}}^{\rm (EM)}_{\rm m}(0)$ and ${\bar {\varrho}}^{\rm (MA)}_{\rm m}(0)$. 
We are now able to integrate equation (20) outwards from $r=0$, using equation 
(14) for each integration step, until the pressure vanishes. The surface 
$r=r_{\rm sf}$ of the body is now reached. If the calculated value 
${\bar B}(r_{\rm sf})$ does not match the boundary condition (22), add a constant
to the function ${\bar B}(r)$ such that ${\bar B}(r_{\rm sf})$ plus
the constant agrees with equation (22). Repeat the calculation with the new
value for ${\bar B}(0)$. Iterate until sufficient accuracy is achieved.

Once we have done the above calculations for an arbitrary time, we know the 
time evolution of the system from equations (13), (14) and (15). That 
is, a spherical gravitationally bound body consisting of a perfect fluid 
obeying an equation of state $p{\propto}{\varrho}_{\rm m}$ will expand according
to the Hubble law. But for perfect fluid bodies obeying other equations of 
state, the expansion will induce instabilities; mass currents will be set up 
and such systems cannot be metrically static. However, for the metrically 
static case, the gravitational field interior to the body will expand along 
with the fluid; this is similar to the expansion of the exterior gravitational
field found elsewhere [2]. We are now able to calculate the family 
${\bf {\bar g}}_t$ inside the body. To find the corresponding family 
${\bf g}_t$ one uses the method described in [1].
\section{Discussion}
The possibility that cosmic expansion may be relevant for gravitationally 
bound bodies is not a new idea [4]. However, it has not been possible to fit 
this concept into the standard framework of metric gravity. In this chapter, we 
have shown that in quasi-metric theory, gravitationally bound bodies may be 
expected to expand according to the Hubble law whenever they are made of a 
perfect fluid with an equation of state of the type 
$p{\propto}{\varrho}_{\rm m}$. (This is easily seen from equation (10) and the 
fact that the coordinate radius $r_{\rm sf}$ of the body is constant in time.) 
But if the linear equation of state is not a good approximation, one expects 
that the expansion should lead to instabilities in order to restore hydrostatic
equilibrium. 

Examples of this are Newtonian stars for which the equation of state takes the 
form $p{\propto}{\varrho}_{\rm m}^{\gamma}$, ${\gamma}{\geq}{\frac{6}{5}}$; 
so-called {\em polytropes} [5]. In quasi-metric theory, it is possible to model
polytropes by taking Newtonian limits of the relevant equations but such that 
the $t$-dependence remains. However, polytropes made of degenerate matter
will not be metrically static in such a description. But if the fluid-dynamical
effects on the gravitational field coming from instabilities can be neglected,
we may solve the field equations for each epoch $t$ assuming that the 
polytrope is at hydrostatic equilibrium. To do this, we take the Newtonian 
limits of equations (13), (14) and (20), assuming that the 
composition-dependent right hand side of equation (20) can be approximated with
the source ${\frac{8{\pi}}{c^2}}G_{\rm N}{\varrho}_{\rm m}$ [1], where $G_{\rm N}$ 
is Newton's constant. Equations (14) and (20) then straightforwardly yield
\eqa
{\frac{d}{dr}}{\Big (}{\frac{r^2}{{\varrho}_{\rm m}}}p'{\Big )}
=-4{\pi}G_{\rm N}r^2{\varrho}_{\rm m0},
\ena
where ${\varrho}_{\rm m0}$ is the density field ${\varrho}_{\rm m}$ at
epoch $t_0$. For polytropes, this equation can be used only for epoch $t_0$. 
However, the equation is invariant under the scale transformation 
$r{\rightarrow}({\frac{t_0}{t}})^{{\frac{1}{3{\gamma}-4}}}r{\equiv}{\ell}$,
${\varrho}_{\rm m0}{\rightarrow}({\frac{t}{t_0}})^{\frac{3}{3{\gamma}-4}}
{\varrho}_{\rm m0}{\equiv}{\varrho}_{\rm m}$, 
$G_{\rm N}{\rightarrow}{\frac{t}{t_0}}G_{\rm N}{\equiv}G_t$. 
This means that by first scale-transforming equation (24), it may be applied to
polytropes and solved for any fixed epoch $t$. It then becomes equivalent to 
its counterpart in Newtonian theory except for a variable $G_t$. Thus the usual
Newtonian analysis of polytropes [5] applies, but with $G_t$ variable. And as a
consequence, the physical radius ${\cal R}(t)=
({\frac{t_0}{t}})^{{\frac{1}{3{\gamma}-4}}}{\cal R}(t_0)$ of a polytrope made of 
degenerate matter will actually {\em shrink} with epoch (if 
${\gamma}>{\frac{4}{3}}$).

Of particular interest are polytropes for which ${\gamma}={\frac{4}{3}}$,
since such stars are models for Chandrasekhar mass white dwarfs (WDs). From 
[5] we easily find that the (passive) mass $m_{\rm c}$ and radius 
${\cal R}_{\rm c}$ of such WDs (with identical central mass densities) depend on
epoch such that $m_{\rm c}(t)=({\frac{t_0}{t}})^{3/2}m_{\rm c}(t_0)$ and 
${\cal R}_{\rm c}(t)=({\frac{t_0}{t}})^{1/2}{\cal R}_{\rm c}(t_0)$, respectively. 
Since Chandrasekhar mass WDs are believed to be progenitors of type Ia 
supernovae, one may expect that any cosmic evolution of Chandrasekhar mass WDs 
should imply a systematic luminosity evolution of type Ia supernovae over 
cosmic time scales. However, such a luminosity evolution would be inconsistent 
with their use as standard candles when determining the cosmological parameters
in standard cosmology: luminosity evolution could have serious consequences for 
an interpretation of the supernova data in terms of an accelerating cosmic 
expansion indicating a non-zero cosmological constant [6, 7]. 

Now quasi-metric theory predicts that the cosmic expansion does neither 
accelerate nor decelerate [1]. Moreover, according to quasi-metric theory, the
Chandrasekhar mass varies with epoch and this means that type Ia supernovae 
may be generated from cosmologically induced collapse of progenitor WDs.
The consequences for type Ia supernova peak luminosities due to the predicted 
evolution of progenitor WDs are not clear, however. Since the luminosity of 
type Ia supernovae comes from ${\gamma}$-disintegration of unstable nuclear 
species (mainly $^{56}$Ni) synthesized in the explosion, this luminosity could 
depend critically on the conditions of the nuclear burn. That is, the detailed 
nuclear composition synthesized in the explosion and its energetics should
depend on the C/O-ratio as well as the mass of the progenitor WD. Also the 
presence of more massive ejecta during the explosion could have an influence on
supernova luminosities and light curves. But since there is no way to know 
exactly how these observables are affected by the predicted progenitor 
evolution without doing detailed numerical simulations, it is not possible to 
say whether or not the predictions from quasi-metric theory are consistent 
with the data.
 
However, what we {\em can} easily do is to see if it is possible to construct 
a simple luminosity evolution which, in combination with the cosmological toy 
model described in [1], yields a reasonable fit to the supernova data. That is,
we may try a luminosity evolution of the form 
$L(t){\propto}({\frac{t}{t_0}})^{\epsilon}$, or equivalently 
$L(z){\propto}(1+z)^{-{\epsilon}}$ and see if the data are well fit for some 
value(s) of ${\epsilon}$. To check this, we use the postulated luminosity 
evolution to plot apparent magnitude $m_{\rm QMT}$ versus redshift for type Ia 
supernovae. The easiest way to compare this to data, is to calculate the 
difference in apparent magnitude ${\Delta}m$ between our model and that 
predicted using an empty Friedmann model. The result is
\eqa
{\Delta}m{\equiv}m_{\rm QMT}-m_{\rm MIN}&=&2.5{\log}_{10}{\Big [}
{\sin}^2{\{}{\ln}(1+z){\}}{\Big ]} \nonumber \\
&&-5{\log}_{10}{\Big [}{\sinh}{\{}{\ln}(1+z){\}}{\Big ]}-
2.5{\log}_{10}{\frac{L_{\rm QMT}}{L_{\rm MIN}}},
\ena
where $L_{\rm QMT}/L_{\rm MIN}$ represents the luminosity evolution of the source
in our quasi-metric model relative to no luminosity evolution in an
empty Friedmann model.

Comparing the fits obtained from equation (25) to data, we find that the choice
${\epsilon}={\frac{1}{2}}$, that is, $L_{\rm QMT}=(1+z)^{-1/2}L_{\rm MIN}$, fits the 
data very well. We conclude that quasi-metric cosmology combined with a simple 
luminosity evolution seems to be consistent with the data, but that a much more 
detailed model should be constructed to see if the found luminosity 
evolution has some basis in the physics of type Ia supernovae.
\\ [4mm]
{\bf References} \\ [1mm]
{\bf [1]} D. {\O}stvang, chapter 2 of this thesis (2001). \\
{\bf [2]} D. {\O}stvang, chapter 3 of this thesis (2001).\\
{\bf [3]} D. {\O}stvang, chapter 4 of this thesis (2001). \\
{\bf [4]} P. Wesson, {\em Cosmology and Geophysics}, Adam Hilger Ltd, (1978).
\\
{\bf [5]} S. Weinberg, {\em Gravitation and Cosmology}, John Wiley ${\&}$ 
Sons, Inc. (1972). \\
{\bf [6]} S. Perlmutter et al., {\em ApJ} {\bf 517}, 565 (1999). \\
{\bf [7]} A.G. Riess et al., {\em AJ} {\bf 116}, 1009 (1998).
\end{document}